\documentclass[cernpreprint, texlive=2020, UKenglish, texmf]{atlasdoc}
 
\usepackage{atlaspackage}
\usepackage{atlasbiblatex}
 
\usepackage{atlasphysics}
\usepackage{float}
\usepackage{multirow}

\graphicspath{{logos/}{figures/}}
 
\usepackage{ANA-STDM-2020-32-PAPER-defs}

\AtlasTitle{Determination of the parton distribution functions of the proton
using diverse ATLAS data from $pp$ collisions at $\sqrt{s} = 7$, 8 and 13~TeV}

\AtlasAbstract{
This paper presents an analysis at next-to-next-to-leading order in the theory of quantum chromodynamics for the determination of
a new set of proton parton distribution functions using diverse measurements in $pp$ collisions at
$\sqrt{s} = 7$, 8 and 13~TeV, performed by the ATLAS experiment at the Large Hadron Collider, together with deep inelastic scattering
data from $ep$ collisions at the HERA collider. The ATLAS data sets considered are differential cross-section
measurements of  inclusive $W^{\pm}$ and $Z/\gamma^*$ boson production, $W^{\pm}$ and $Z$ boson production
in association
with jets, $t\bar{t}$ production, inclusive jet production and direct photon
production.
In the analysis, particular attention is paid to the correlation of
systematic uncertainties within and between the various ATLAS data sets and to the impact of model,
theoretical and parameterisation
uncertainties. The resulting set of parton distribution functions is called ATLASpdf21.}

\AtlasRefCode{STDM-2020-32}
 
\PreprintIdNumber{CERN-EP-2021-239}
 
\AtlasDate{\today}

\arXivId{2112.11266}

\AtlasJournalRef{\EPJC 82 (2022) 438}
\AtlasDOI{10.1140/epjc/s10052-022-10217-z}

\hypersetup{pdftitle={ATLAS document},pdfauthor={The ATLAS Collaboration}}
 
\DeclareUnicodeCharacter{2212}{+}
\DeclareUnicodeCharacter{0301}{\'{e}}
 
\begin{document}
 
\maketitle
 
\tableofcontents
 
\clearpage
\section{Introduction}
 
Precise knowledge of the content of protons, the parton distribution functions (PDFs), is a necessary ingredient for accurate predictions of both Standard Model (SM) and beyond-the-SM (BSM) cross sections at the Large Hadron Collider (LHC).
Searching for BSM effects in the deviations of SM parameters
from their predicted SM values~\cite{Azzi:2019yne,Cepeda:2019klc} requires better knowledge of PDFs in the kinematic regions where they are already best known~\cite{Gao:2017yyd,Ethier:2020way}. This means that many effects which were previously considered negligible must now be examined seriously. In this paper, particular attention is paid to some of these effects, including theoretical scale uncertainties and the correlations of experimental systematic uncertainties between, as well as within, data sets.
 
With present theoretical knowledge, PDFs can be determined at up to next-to-next-leading order (NNLO) in perturbative quantum chromodynamics (QCD). In order to determine the PDFs to the required
precision, a wide coverage of the scale, $Q^2$, and $x$, the fraction of the proton's longitudinal momentum carried by the parton participating in the initial interaction, is required.
This is facilitated by combining new inputs with older data from various experiments and by measurement of different processes to constrain different regions of phase space, which span the kinematic region: $10^{-5} \lesssim x \lesssim 1$ and $1 \lesssim Q^2 \lesssim 10^6~\GeV{}^2$.
 
Knowledge of the PDFs of the proton, in the kinematic region relevant for the LHC,
comes mainly from the precision deep inelastic scattering (DIS) data from $ep$ collisions at the HERA
collider, which covers a broad range of $Q^2$ and $x$.
The PDF set HERAPDF2.0 was determined from HERA data alone~\cite{herapdf20}
using information from $e^{\pm} p$ neutral-current (NC) and charged-current (CC) processes. The
lower-$Q^2$ NC data constrain the low-$x$ sea-quark distribution but are not able to distinguish between
quark flavours in the sea at low~$x$, or between the down-type quarks, $\bar{d}$ and $\bar{s}$ at
any $x$. The difference between the NC $e^+p$ and
$e^-p$ cross sections at high $Q^2$, together with the high-$Q^2$ CC data,
constrains the valence distributions. The
$Q^2$ dependence measured in the data constrains the gluon distribution. Additional
information about the gluon comes from using HERA cross sections measured at various
centre-of-mass energies ($\sqrt{s}$) through the contribution of the longitudinal structure
function~\cite{sarkarFL,H1FL,ZEUSFL}.

Diverse ATLAS data sets can be used in addition to the HERA data to constrain the PDFs better. In Ref.~\cite{1612.03016}, high precision measurements of the inclusive differential
$W^{\pm}$ and $Z/\gamma^*$ boson cross sections at 7~\TeV\ were added to the HERA data, resulting in the PDF set ATLASepWZ16, which improved
on the HERAPDF2.0 set in various respects. Firstly, the strange content of the sea was determined rather
than assumed to be a fixed fraction of the light sea. Indeed, compared to previous determinations, the strange sea was found to be enhanced at low~$x$.  Secondly, the accuracy of the valence quark
distributions for $x<0.1$ was improved. The effect of adding
differential $t\bar{t}$ distributions, in the lepton\,+\,jets and dilepton channels at 8~\TeV, to the HERA data and the inclusive $W,Z/\gamma^*$ data was studied in Ref.~\cite{ATL-PHYS-PUB-2018-017}. These $t\bar{t}$ data are
complementary to the $W,Z/\gamma^*$ data
in their power to constrain PDFs because they are sensitive to the high-$x$ gluon
distribution. The resulting PDF set was called ATLASepWZtop18.
In Ref.~\cite{Vjets}, data on the production of $W$ and $Z$ bosons in
association with jets ($V$+\,jets) were added to the HERA data and inclusive $W$, $Z/\gamma^*$ data,
resulting in the ATLASepWZVjets20 PDF set. The $V$+\,jets data are sensitive to
partons at higher~$x$ than can be accessed by inclusive $W$ and $Z/\gamma^*$ data and, in particular, they
constrain the $\bar{d}$ and $\bar{s}$ quarks at higher~$x$.
 
It would clearly be advantageous to combine ATLAS $W,Z/\gamma^*$ data, $t\bar{t}$
data and $V$+\,jets data in a single QCD fit. This is done in the present paper, and
additional ATLAS data sets are also included. This paper presents the first comprehensive and comparative NNLO perturbative QCD analysis of a number of ATLAS data sets with potential sensitivity to parton distributions. The additional ATLAS data sets are described in the following. Firstly, the data on $W$~\cite{W8} production and
$Z/\gamma^*$~\cite{z3d} production  with 20.2~fb$^{-1}$ at 8~\TeV\ are added, providing further constraints on the valence PDFs and on the composition of the light-quark sea.
Secondly, the direct photon production differential cross sections with  20.2~fb$^{-1}$ at 8~\TeV\ and 3.2~fb$^{-1}$ at 13~\TeV\ are added in the form of their ratios~\cite{terron}, whereby many systematic
uncertainties cancel out. Thirdly, the $t\bar{t}$ differential cross sections in the lepton\,+\,jets channel
from 3.2~fb$^{-1}$ at 13~\TeV~\cite{1908.07305} are added and, fourthly,
inclusive jet production cross sections with 4.5~fb$^{-1}$ at 7~\TeV~\cite{1410.8857}, 20.2~fb$^{-1}$ at
8~\TeV~\cite{1706.03192} and 3.2~fb$^{-1}$ at 13~\TeV~\cite{1711.02692} are considered. These direct photon data, $t\bar{t}$ data and inclusive jet data all have impact on the gluon PDF at medium to high~$x$.
 
All ATLAS data sets considered in this study have full information about correlated systematic uncertainties.
The analysis considers systematic uncertainty correlations between, as well as within, data sets. This is important
now that several of the input data sets have systematic uncertainties deriving from jet measurements,
since these uncertainties are larger than those from the lepton measurements.
Theoretical uncertainties, such as scale uncertainties, are also
considered, including their correlations between data sets.
The resultant PDF set is called ATLASpdf21.

The structure of the paper is as follows. Section~\ref{sec:data} presents the input data sets. Section~\ref{sec:theoryframework} describes the theoretical framework of the fit.
Section~\ref{sec:method} presents the fitting methodology, including the definition of the fit $\chi^2$ and details of the parameterisation. Section~\ref{sec:results} presents ATLASpdf21 PDFs and then compares them with a fit in which correlations between data sets are not implemented. The impact
of each data set is considered and then model uncertainties and further theoretical uncertainties, including scale uncertainties, are considered. Section~\ref{sec:globaltol} presents a study of the $\chi^2$ tolerance and a comparison with other modern PDF sets.
Section~\ref{sec:conclusion} gives the summary and conclusions. Appendix~\ref{sec:scale} gives a more detailed exposition of the impact of scale uncertainties for the inclusive $W,Z/\gamma^*$ data. The impacts of inclusive jet data at different centre-of-mass energies are compared in Appendix~\ref{sec:cofmjets}.
Appendices~\ref{sec:datafitcomp} and~\ref{sec:extradatafitcomp} compare the ATLASpdf21 fit predictions with the ATLAS input data sets and with data sets from other experiments, respectively.
 
\clearpage
\section{Input data sets}
\label{sec:data}
\subsection{Description of data sets}
 
The combined $e^{\pm}p$ NC
cross-section measurements of H1 and ZEUS~\cite{herapdf20}
cover a kinematic range of  $Q^2$, defined as the negative four-momentum transfer squared,
from $0.045$~\GeV$^2$ to \num{50000}~\GeV$^2$ and of Bjorken-$x$, $x_{\mathrm{Bj}}$, which is equal to the fraction of the proton's longitudinal momentum carried by the struck parton at leading order (LO) in QCD,
from $6\times 10^{-7}$ to $0.65$. The CC cross-section measurements
cover a kinematic region $Q^2$ $\sim 300$~\GeV$^2$ to beyond $10^4$~\GeV$^2$ and
of $x$ from ${\sim}0.65$ down to ${\sim}10^{-2}$. Low-$x$ data, below $x_{\mathrm{Bj}} = 10^{-5}$,
are excluded from this analysis by
requiring $Q^2 > 10~$\GeV$^2$, motivated by the poorer fit observed in this region compared to the rest of HERA data~\cite{herapdf20}, which may reflect the need for resummation corrections at low $x$~\cite{1802.00064,1710.05935}. Full information about correlated
systematic uncertainties is provided. There are 169 sources of correlated systematic
uncertainty. Total uncertainties are below 1.5$\%$ over the $Q^2$ range of $3 < Q^2 < 500$~\GeV$^2$ and remain below 3$\%$ up to $Q^2= 3000$~\GeV$^2$.
 
The ATLAS $W,Z/\gamma^*$ differential cross-section measurements at $\sqrt{s} = 7$~\TeV\ based on an
integrated luminosity of  $4.6$~fb$^{-1}$ are used~\cite{1612.03016}. A combination of the electron and muon decay channels is used.
They access a kinematic range whereby the scale is identified with the measured boson mass
range, and the $x$ range is determined by the scale, the proton beam
energy and the measured rapidity $(y)$ ranges, such that $Q^{2} = m^{2}$ and $x= (Q/\sqrt{s})\,\mathrm{e}^{\pm y}$, which gives an $x$ range $0.001 \lesssim  x \lesssim 0.1$.\footnote{The definitions of the scale for the DIS and inclusive vector boson processes considered so far are conventional. They have been defined here in order to give an idea of the kinematic coverage of the fit.
For the other data sets considered, the scale definitions are more conveniently
given together with the description of the theoretical predictions in
Section~\ref{sec:theoryframework}.}
The  $W^{\pm}$ differential cross sections were measured
as a function of the $W$ decay lepton
pseudorapidity, $\eta_\ell$, with an experimental precision of 0.6\%--1.0\%.
The $Z/\gamma^*$ boson rapidity distribution, $y_{\mathit{Z}}$, was measured in three mass ranges: $
46<m_{\ell\ell}<66~$\GeV; $66<m_{\ell\ell}<116~$\GeV\ and $116<m_{\ell\ell}<136~$\GeV, with the two decay leptons in central-central (c-c) and central-forward (c-f) rapidity ranges,\footnote{The central-central region requires both decay leptons to have absolute rapidity less than 2.5 and the central-foward region requires one lepton with absolute rapidity less than 2.5 and the other in the range 2.5--4.9.} and
experimental uncertainties as low as $0.4\%$ for c-c rapidity and $2.3\%$ for c-f rapidity.
There are 131 sources of correlated systematic uncertainty. The low-mass off-peak
$Z/\gamma^*$ data are not used in the present
analysis because they are subject to much larger corrections for non-fixed order, parton shower $p_\mathrm{T}$-resummation effects than the other $W,Z/\gamma^*$ data~\cite{ATL-PHYS-PUB-2018-004}. These low-mass data had little impact on the ATLASepWZ16 PDFs.
 
ATLAS has published differential cross sections and forward-backward asymmetries for $Z/\gamma^*$ production at $\sqrt{s}= 8$~\TeV, based on $20.2$~fb$^{-1}$ of data~\cite{z3d}. The measurement is
presented in terms of three variables for both the c-c and c-f regions of the
dilepton rapidity, $y_{\ell\ell}$. For the c-c region, there are
12 bins of absolute rapidity ranging from 0.0 to 2.4, in intervals of $0.2$, divided into 7
bins of the dilepton mass $m_{\ell\ell}$ ranging from $46$ to $200$~\GeV\
and 6 bins of the polar angle, $\theta$, of the decay lepton in the Collins--Soper frame~\cite{PhysRevD.16.2219}.
For the c-f region, there are 5 bins of absolute rapidity ranging from $1.2$ to $3.6$
and 5 mass bins ranging from $66$ to $150$~\GeV\
in the same 6 bins of the Collins--Soper angle. The dilepton data were extracted
in both the dielectron and dimuon channels up to $|y_{\ell\ell}|=2.4$, which is extended to
$|y_{\ell\ell}|=3.6$ in the dielectron channel. The dimuon and dielectron data are combined for the
c-c region.
Measurements are used as cross sections when the acceptance of the lepton fiducial cuts with respect to the ($m_{\ell\ell}$, $|y_{\ell\ell}|$, $\cos\theta^{\ast}$) bin is greater than $95\%$, such that the leptons are constructed
with high accuracy and the cross sections are known to ${\sim}0.5\%$ accuracy. There are 184 cross-section data points
which fulfil these requirements and all of them lie in the c-c region of dilepton rapidity.
These are used for the central fit.  The selection of the bins in $m_{\ell\ell}$, $y_{\ell\ell}$
and $\theta$
ensures that the NNLO predictions for
these high acceptance bins are not highly sensitive to fiducial cuts applied to leptons~\cite{Alekhin:2021xcu}. The uncertainties of the high-$|y_{\ell\ell}|$ data range from
$2\%$ to $33\%$  and the combined central rapidity data reaches a precision ranging from  $0.4\%$, for the
$Z$-mass-peak bins at $|y_{\ell\ell}| < 1.0$, to $1.8\%$ for the high-mass bins at
$|y_{\ell\ell}|=2.4$. There are 330 sources of correlated systematic uncertainty.
The remaining lower-acceptance data are used only in cross-checks.
They are converted to 188 data points for
forward-backward asymmetries across the Collins--Soper angle in order to minimise sensitivity to acceptance cuts.
These data lie in both the c-c and c-f dilepton rapidity regions.
 
The $W$ inclusive differential cross sections, from $20.2$~fb$^{-1}$ of data at $\sqrt{s} = 8$~\TeV~\cite{W8}, are measured
as a function of muon pseudorapidity, $\eta_{\mu}$. Data are provided for
$W^+$ and $W^-$ cross sections separately as well as for the $W$-asymmetry. The separate $W^+$ and $W^-$ cross
sections are used for the fit, rather than the $W$-asymmetry.
The precision of these cross-section measurements varies from
$0.8\%$ to $1.5\%$ as a function of $\eta_{\mu}$.
The pseudorapidity covers the same central range as the $\sqrt{s} = 7$~\TeV\ inclusive $W$ data.
There are 41 sources of correlated systematic uncertainty.
 
The ATLAS differential cross-section measurements of $W$ production in association with jets ($W$+\,jets) are based on data recorded during $pp$ collisions at $\sqrt{s} = 8$~\TeV,
and a total integrated luminosity of 20.2~fb$^{-1}$,
in the electron decay channel only. Each event contains at least one jet with transverse momentum $\pt > 30~\GeV{}$ and rapidity $\lvert y\rvert < 4.4$.
The data are split into $W^{+}$ and $W^{-}$ cross sections. There are 51 sources of correlated systematic uncertainty.
As explained in Ref.~\cite{Vjets} the differential spectra of the transverse momenta of the $W^+$ and $W^-$ bosons
$p_{\mathrm{T}}^{\mathit{W}}$ are fitted in the range $25 < p_{\mathrm{T}}^{\mathit{W}} < 800~\GeV{}$.
The data precision ranges from 10\% to 25$\%$ from low to high $p_{\mathrm{T}}^{\mathit{W}}$~\cite{wjetspaper}.
 
The ATLAS differential cross-section measurements of $Z$ production in association with jets \cite{zjetspaper}
are based on data recorded during $pp$ collisions at $\sqrt{s} = 8$~\TeV,
and a total integrated luminosity of 20.2~fb$^{-1}$, in the electron decay channel only.
The measurement was performed as a function of the absolute rapidity of inclusive jets, $|y^\mathrm{jet}|$, for several bins of the transverse momentum within
$25~\GeV < p_{\mathrm{T}}^{\mathrm{jet}} < 1050~\GeV$. There are 42 sources of correlated systematic uncertainty.
The data precision ranges from ${\sim}10\%$ to ${\sim}30\%$ from low
$|y^{\mathrm{jet}}|$ and $p_{\mathrm{T}}^{\mathrm{jet}}$ to  high $|y^{\mathrm{jet}}|$ and $p_{\mathrm{T}}^{\mathrm{jet}}$.
 
The $t\bar{t}$ differential cross sections were measured at $\sqrt{s} = 8$~\TeV\ using $20.2$~fb$^{-1}$ of data in the lepton\,+\,jets~\cite{1511.04716} and
dilepton~\cite{1607.07281} decay channels.
The $t\bar{t}$ cross sections are provided as both normalised and absolute spectra. The absolute
spectra are used in the present study since they carry extra information about the normalisation of the cross-sections, which is not input anywhere else in the
fit. As detailed in Ref.~\cite{ATL-PHYS-PUB-2018-017},
in the lepton\,+\,jets channel the differential spectra that are used are the mass of
the $t\bar{t}$ pair, $m_{t\bar{t}}$, and the average top-quark transverse momentum, $p_{\mathrm{T}}^t$.
Each of these  spectra has full information
about systematic and statistical bin-to-bin correlations, including statistical correlations between the spectra. There are 55 sources of
correlated systematic uncertainty in common between the different spectra. The precision of
these lepton\,+\,jets data is ${\sim}15\%$ for $p_{\mathrm{T}}^t$ and ranges from ${\sim}10\%$ to ${\sim}20\%$ from low to high $m_{t\bar{t}}$.
In the dilepton channel,
the spectrum for the rapidity of the  $t\bar{t}$ pair, $y_{t\bar{t}}$, is used. The precision of
these data ranges from  ${\sim}5\%$ to ${\sim}15\%$ from low to high $|y_{t\bar{t}}$|.
For these dilepton channel data the correlations are provided as a total covariance
matrix.
Further studies of these top-quark data are presented in Ref.~\cite{ATL-PHYS-PUB-2018-017}.
 
The $t\bar{t}$ differential cross sections were measured at $\sqrt{s} = 13$~\TeV\ using $36$~fb$^{-1}$ of data in the lepton\,+\,jets~\cite{1908.07305} channel.   Similarly to the $t\bar{t}$ data at 8~\TeV,
the absolute spectra are used.
For the $t\bar{t}$ data at 13~\TeV, two complementary topologies are measured;
the `resolved' and `boosted''
topologies, which differ in that the decay products of the hadronically decaying top quark
are well separated for the resolved case and collimated for the boosted case.
The `boosted' topology reaches higher absolute rapidity of
the $t\bar{t}$ pair. The differential spectra used for the PDF fit are
the mass of the $t\bar{t}$ pair, $m_{t\bar{t}}$, the average top-quark transverse momentum, $p_{\mathrm{T}}^t$, the average top-quark rapidity, $y_t$, and the boosted rapidity of the $t\bar{t}$ pair, $y^\mathrm{b}_{t\bar{t}}$. These spectra were found to be the most sensitive to the gluon PDF in Ref.~\cite{1611.08609}. Each of these spectra has full information
about systematic and statistical bin-to-bin correlations, including statistical correlations between the spectra.\footnote{The final two data points, for which $p^t_\mathrm{T}>360~$\GeV, are excluded from the fitted data set.
This is because the full statistical correlation matrix could not be inverted when these bins are included. A cross-check is made in which the fit uses only two of the $t\bar{t}$ 1-D spectra
$p_\mathrm{T}^t$ and $m_{t\bar{t}}$ at 13~\TeV, for which the statistical matrix could be inverted, both including and excluding the final two  bins of the $p^t_\mathrm{T}$ spectrum. The resulting PDFs were compared to the PDFs from a fit to the 2-D $p_\mathrm{T}^t$, $m_{t\bar{t}}$ spectra.
The PDFs of these three fits are almost identical, thus the exclusion of the high $p^t_\mathrm{T}$
bins does not affect the PDF shapes significantly.}
There are 88 sources of
correlated systematic uncertainty in common between the different spectra and
statistical correlations between the spectra are implemented.  Double differential spectra in $p_{\mathrm{T}}^t$ and $m_{t\bar{t}}$ were also studied. The precision of
these lepton\,+\,jets data ranges from $15\%$ to $30\%$ for $p_{\mathrm{T}}^t$ and from $15\%$ to $22\%$ for $m_{t\bar{t}}$, with the largest uncertainties in the lowest and
highest bins. For both of the rapidity spectra the total uncertainty is ${\sim}15\%$.
 
The ratio of the cross sections for inclusive isolated-photon production at $\sqrt{s} = 8$ and 13~\TeV\ comes from integrated luminosities of $20.2$~fb$^{-1}$ and $3.2$~fb$^{-1}$ respectively. The data are presented as a function of photon transverse energy,
$E_{\mathrm{T}}^{\gamma}$, for $E_{\mathrm{T}}^{\gamma}> 125$~\GeV\  in bins of photon pseudorapidity,
$\eta^{\gamma}$, for the central region $|\eta^{\gamma}|  < 2.37$. Details of the
isolation criteria are given in Ref.~\cite{terron}. Evaluation of experimental
uncertainties in the ratio data takes into account the correlations between
the data at the two different centre-of-mass energies. The dominant source of
correlated uncertainty is the photon energy scale.
The uncertainty due to this source is much lower when considering the ratio of cross sections.
Nevertheless, 25 correlated uncertainty components from this source remain, and there are
21 further uncertainty components which are treated as correlated or uncorrelated
as specified in Ref.~\cite{terron}. It should be noted that for the present fits the
combined luminosity uncertainty at 8 and 13~\TeV\ is not used, so the
separate luminosity uncertainties may be correlated with those of other data
at these centre-of-mass energies. The experimental precision of the ratios ranges
from ${\sim}4\%$ at low $E_{\mathrm{T}}^{\gamma}$ and $|\eta^{\gamma}|$ to ${\sim}10\%$ at the largest
$E_{\mathrm{T}}^{\gamma}$ and $|\eta^{\gamma}|$.

The ATLAS measurements of inclusive jet cross sections at $\sqrt{s}= 7$~\TeV~\cite{1410.8857}, 8~\TeV~\cite{1706.03192} and 13~\TeV~\cite{1711.02692} were considered. The data are presented as a function of jet $p_{\mathrm{T}}^{\mathrm{jet}}$, in six bins of absolute rapidity
from $|y^{\mathrm{jet}}|=0.0$ to $|y^{\mathrm{jet}}| = 3.0$. Jets were reconstructed using the anti-$k_t$ algorithm with jet radius parameter $R=0.4$ and $R=0.6$
for data at 7 and 8~\TeV, and with $R=0.4$ for data at 13~\TeV. Statistical correlation matrices between bins are provided for all
jet data sets. However, it is not possible to fit inclusive jet production data at different centre-of-mass energies simultaneously, because the
full experimental systematic uncertainty correlations between these data sets have not been fully specified.
The data at 8~\TeV\ are selected for the central fit, firstly because a better understanding
of correlated systematic uncertainties has been achieved for these data than for the data at 7~\TeV, and secondly because
they are available for $R=0.6$, unlike for the data at 13~\TeV, and a larger $R$ value is considered more reliable theoretically~\cite{1807.03692}.
Details of the data set are therefore
given only for the jet data at 8~\TeV. The data at 7 and 13~\TeV\ have similar features, the main difference being in the $p_{\mathrm{T}}^{\mathrm{jet}}$  range probed. These data sets, including the alternative choice of jet radius, are used in cross-checks.
At 8~\TeV\ there are 171 jet data points
from 20.2~fb$^{-1}$ of integrated luminosity. The fitted $p_{\mathrm{T}}^{\mathrm{jet}}$ range is from 85~\GeV\ to 2.5~\TeV.
The dominant uncertainties
come from the estimation of the jet energy scale and jet energy resolution. Extensive work~\cite{1706.03192} has been done to understand
correlated sources of uncertainty. There are $320$ sources of correlated
systematic uncertainty, in addition to the luminosity uncertainty, and the jet energy scale's $\eta$ intercalibration
uncertainty is split into 250 sources.  The total
uncertainty in the central  $p_{\mathrm{T}}^{\mathrm{jet}}$ range, $300 < p_{\mathrm{T}}^{\mathrm{jet}} <600$~\GeV, is
${\sim}5\%$ for $|y^{\mathrm{jet}}| < 0.5$, rising to ${\sim}10\%$ for $|y^{\mathrm{jet}}|$ in the range 2.5--3.0. For low $p_{\mathrm{T}}^{\mathrm{jet}}$, $85 < p_{\mathrm{T}}^{\mathrm{jet}} <300$~\GeV, the uncertainty
is ${\sim}15\%$ and  for high $p_{\mathrm{T}}^{\mathrm{jet}}$, $600 < p_{\mathrm{T}}^{\mathrm{jet}} <2000$~\GeV, it is ${\sim}50\%$.
 
All the ATLAS input data sets for the QCD fit are summarised in Table~\ref{tab:inputs}.

\begin{table}\small
\caption{Summary of all the ATLAS input data sets considered in the QCD fit. It should be noted that: i) the inclusive $W$ cross-section data at 7 and 8~\TeV\ are used as $W^+$ and $W^-$ data separately, ii) the isolated photon data are used as the ratio of the 13~\TeV\ to 8~\TeV\ cross sections and iii) the inclusive jet production data at 7 and 13~\TeV\ are used only for a cross-check. \label{tab:inputs}}
\begin{center}
\begin{tabular}{lcccc}
\hline
Data set & $\sqrt{s}$ [\TeV] & Luminosity [fb$^{-1}$] & Decay channel & Observables entering the fit \\
\hline
Inclusive $W,Z/\gamma^*$~\cite{1612.03016} & 7 & ~~4.6 & $e,\mu$ combined& $\eta_{\ell}$ ($W$), $y_{\mathit{Z}}$ ($Z$) \\
Inclusive $Z/\gamma^*$~\cite{z3d} & 8 & 20.2 & $e,\mu$ combined & $\cos\theta^{\ast}$ in bins of $y_{\ell\ell}$, $m_{\ell\ell}$\\
Inclusive $W$~\cite{W8} & 8 & 20.2 & $\mu$ & $\eta_{\mu}$ \\
$W^{\pm}$\,+\,jets~\cite{wjetspaper} & 8 & 20.2 & $e$ & $p_{\mathrm{T}}^{\mathit{W}}$ \\
$Z$\,+\,jets~\cite{zjetspaper} & 8 & 20.2 & $e$ & $p_{\mathrm{T}}^{\mathrm{jet}}$ in bins of $|y^{\mathrm{jet}}|$\\
$t\bar{t}$~\cite{1511.04716,1607.07281} & 8 & 20.2 & lepton\,+\,jets, dilepton & $m_{t\bar{t}}$, $p_{\mathrm{T}}^t$, $y_{t\bar{t}}$ \\
$t\bar{t}$~\cite{1908.07305} & 13~~ & 36~~~ & lepton\,+\,jets & $m_{t\bar{t}}$, $p_{\mathrm{T}}^t$, $y_t$, $y_{t\bar{t}}^{\mathrm{b}}$ \\
Inclusive isolated $\gamma$~\cite{terron} & 8, 13 & 20.2, 3.2 & - & $E_{\mathrm{T}}^{\gamma}$ in bins of $\eta^{\gamma}$ \\
Inclusive jets~\cite{1410.8857,1706.03192,1711.02692} & 7, 8, 13  & 4.5, 20.2, 3.2 & - & $p_{\mathrm{T}}^{\mathrm{jet}}$ in bins of $|y^{\mathrm{jet}}|$ \\
\hline
\hline
\end{tabular}
\end{center}
\end{table}
 
\subsection{Correlations of uncertainties within and between data sets}
\label{sec:corrwb}
 
The correlated systematic uncertainties are applied
within each data set with the following small number of exceptions. Firstly, for the $W$+\,jets, the
$Z$\,+\,jets and the inclusive jet spectra at 8~\TeV, the systematic uncertainty due to the unfolding procedure
is treated as being
uncorrelated, both within and between these spectra.\footnote{The two systematic uncertainties
in each of the $W$+\,jets and $Z$\,+\,jets spectra related to unfolding (one related to the MC modelling and one to the size of the data samples) are fully decorrelated between spectra and bins within a single spectrum as they contain a large statistical component in both data sets owing to MC simulation statistics.} As shown in Ref.~\cite{Vjets}, this affects the $\chi^2$
of the fits to $V$+\,jets, but has little impact on the fitted PDFs. Similarly, the parton shower systematic uncertainty
is decorrelated between the $p_{\mathrm{T}}^t$ and $m_{t\bar{t}}$ spectra in the $t\bar{t}$ lepton\,+\,jets channel,
as done in Ref.~\cite{ATL-PHYS-PUB-2018-017}.
It was established that this decorrelation has a minimal effect on the PDFs,
while reducing the fit $\chi^2$ to acceptable levels. This decorrelation and the aforementioned
decorrelation of the unfolding systematic uncertainty in $V$+\,jets and inclusive jet data, can be
justified because the systematic uncertainties concerned are evaluated from
the difference of two Monte Carlo estimates, and thus do not represent well-behaved Gaussian uncertainties.
Similar conclusions were reached in a recent study in the MMHT framework~\cite{mmhttop}. Thirdly, in the inclusive jet data at 8~\TeV\ further decorrelations of such systematic
uncertainties, derived from the difference of two Monte Carlo estimates, are considered following Ref.~\cite{1706.03192}. The experimental systematic uncertainties for the jet energy scale (JES) such as the `Flavour Response',
`Multi-Jet Balance Fragmentation', `Pile-up Rho Topology',\footnote{The `Flavour Response' is the systematic uncertainty due to the response difference between quark- and gluon-induced jets, the `Multi-Jet Balance Fragmentation' represents the jet fragmentation uncertainty in the multi-jet \pT balance and the `Pile-up Rho Topology' takes into account the uncertainty in the density $\rho$ of pile-up activity in a given event.} and the `Non-Perturbative Correction' uncertainty, are not considered completely correlated between
all rapidity bins. Instead, they are split into two or three components as a function of rapidity
and $p_{\mathrm{T}}^{\mathrm{jet}}$ as specified in the various splitting options
described in the Appendix of Ref.~\cite{1706.03192}. For the central fit, the preferred
set of splitting options for $R=0.6$ is used, in which the JES `Flavour Response' is split into three components (see Table 6 of Ref.~\cite{1706.03192}).
In the present paper, this is called `Decorrelation Scenario 2' and it is chosen because it is one of the two preferred options as determined in the analysis in Ref.~\cite{1706.03192}.\footnote{The decorrelations used at next-to-leading order (NLO) in  Ref.~\cite{1706.03192}, include decorrelations of systematic uncertainties due to scale choice. These are not applied in the present NNLO analysis because scale uncertainties are much smaller. NNLO scale uncertainties for the inclusive jets are studied in Section~\ref{sec:model}.}
Alternative decorrelation scenarios are also considered in Section~\ref{sec:model}.
 
Correlations of systematic uncertainties between data sets are explained below.
The luminosity uncertainties are considered fully correlated for all data sets at the same centre-of-mass
energy. For the data sets
considered in this analysis, systematic uncertainties involving electron and  muon measurements
are small (${<}1\%$), whereas systematic uncertainties involving the jet measurements can be much
larger ($O(10\%)$). Moreover, the high-precision inclusive $W,Z/\gamma^*$ differential cross-section measurements at 7~\TeV\ and the inclusive
$Z/\gamma^*$ triple differential cross-section measurements at 8~\TeV\ both had the electron and muon channel data combined and thus the
identities of the systematic uncertainty sources are lost in the combination procedure such that the analysis cannot
correlate them with muon and electron uncertainties in other data sets (or between the inclusive
$Z/\gamma^*$ measurements at 7 and 8~\TeV). Thus, correlations of lepton uncertainties between these inclusive $W$ and $Z$
measurements and the other data sets are neglected.  This is not a significant limitation in the study of the effect of correlations between data sets
since the lepton uncertainties are small compared to the jet uncertainties, as shown in the recent study of
the impact of $V$+\,jets data on PDF fits~\cite{Vjets}. This study reverted to the use of the separate electron and muon
channel data for the inclusive $W,Z/\gamma^*$ data at 7~\TeV\ in order to correlate the electron systematic
uncertainties
with those of the $W$+\,jets data and the $Z$\,+\,jets data at 8~\TeV\ in the electron channel.
These studies produce PDFs which are barely different from those in which the electron/muon
combined data are used and lepton systematic uncertainty correlations are not applied between the data sets.
Indeed, in Section~\ref{sec:classes} of the present analysis,
it is shown that the impact of the $V$+\,jets data within the present analysis,
which uses combined electron/muon data for both inclusive $W,Z/\gamma^*$ data at 7~\TeV\ and inclusive $Z/\gamma^*$ data at 8~\TeV,
is very similar to the impact of these data in Ref.~\cite{Vjets}, where uncombined electron and muon
data are used. Thus, the use of the more accurate combined electron and muon data is preferred.
 
In this paper, the correlations of the much larger jet-measurement uncertainties are considered.
The $W$+\,jets data and the $Z$\,+\,jets data at 8~\TeV\
have common sources of jet systematic uncertainties, which are listed in Table~\ref{tab:corrVtjets}.
These sources are considered $100\%$ correlated.
Since the $t\bar{t}$
cross sections at both 8 and 13~\TeV\ consider data in the lepton\,+\,jets channel, there are some common sources of systematic
uncertainty, due to the jet measurements, between these data and the $V$+\,jets data.
These are also listed in Table~\ref{tab:corrVtjets} and are considered $100\%$ correlated between
these data sets. This table also lists correlations of some sources of smaller systematic uncertainties, in the muon measurements and diboson, single-top and $Z$\,+\,jets backgrounds, which can be correlated between the $t\bar{t}$ measurements.
It should be noted that detailed correlations between the $t\bar{t}$ data at 8~\TeV\ in the dilepton channel and other data sets cannot be applied
because the correlations for the dilepton channel are supplied as a covariance matrix and the separate individual sources of uncertainty cannot be separated. This does not have a significant effect on the fit since the dilepton data itself has only a small impact on the fit, as seen in Ref.~\cite{ATL-PHYS-PUB-2018-017}.
 
Correlations of jet systematic uncertainties between the inclusive jet data and the $V$+\,jets data and $t\bar{t}$ data in the lepton\,+\,jets channel
have been identified and are listed in Table~\ref{tab:corrVtjets}.  There is a choice to be made in the implementation of these
inter-data-set correlations. The JES `Flavour Response' and
`Pile-up Rho Topology' are part of the Jet Decorrelation Scenario 2 chosen for the central fit, and thus they have been split into three components according to jet rapidity and $p_\mathrm{T}$~\cite{1706.03192}. Consequently, there are no longer corresponding systematic sources in the $V$+\,jets data and $t\bar{t}$
data. The choice made for the central fit is to correlate only the remaining six sources (in the right-hand column of Table~\ref{tab:corrVtjets}) between the data sets.
The alternative choice of correlating these two
systematic sources with the other data sets but  not including them in the jet decorrelation scenario was also investigated. The effect on the resulting PDFs is negligible.
 
There is a further caveat on how the correlations between the inclusive jet data set and the $V$+\,jets and $t\bar{t}$ data sets are applied, namely that the last two data sets have radius $R=0.4$ jets, and hence the systematic uncertainties may not be fully
correlated. Checks were made using $100\%$ correlation and no correlation, yielding little difference between the resultant PDFs.
For the central fit a correlation of $100\%$ is used.
 
The systematic uncertainties of the inclusive jet data at different beam energies are correlated with each other, but
understanding these correlations in detail is non-trivial. In the present study, these data sets are fitted separately
and results are compared. As already stated the data at 8~\TeV\ are used for the central fit.

\begin{table}\scriptsize
\caption{Systematic uncertainties that are correlated between the $W$+\,jets data at 8~\TeV, $Z$\,+\,jets data at 8~\TeV, $t\bar{t}$ lepton\,+\,jets data at 8~\TeV, $t\bar{t}$ lepton\,+\,jets data at 13~\TeV\ and inclusive jets data at 8 and 13~\TeV\ are listed. The names of the systematic uncertainties are those found in the HEPData entries~\cite{Maguire:2017ypu}. Entries in the same row are taken as $100\%$ correlated for the $V$+\,jets and $t\bar{t}$ lepton\,+\,jets data, which all have jet radius $R=0.4$. Different degrees of correlation are considered for the inclusive jet data at $R=0.6$, because of the differing choice of jet radius. Where entries are omitted, that systematic uncertainty does not exist for that data set (denoted by `-'). The luminosity uncertainty of data sets at the same centre-of-mass energy are also fully correlated. The JES `Flavour Response' and JES `Pile-up Rho topology' are considered fully correlated with other data sets only for cross-checks. They are not correlated for the central fit because
they are part of the Decorrelation Scenario 2 which is applied to the inclusive jet measurements, as explained in the text. For this reason they are marked with the symbol $^*$.\label{tab:corrVtjets}}
\begin{center}
\resizebox{\textwidth}{!}{
\begin{tabular}{lcccccc}
\hline
Systematic uncertainty &   8~\TeV\ $W$+\,jets & 8~\TeV\ $Z$\,+\,jets & 8~\TeV\ $t\bar{t}$ lepton\,+\,jets & 13~\TeV\ $t\bar{t}$ lepton\,+\,jets & 8~\TeV\ inclusive jets \\ \cline{1-6}
Jet flavour response & JetScaleFlav2 & Flavor Response & flavres-jes  & JET29NP JET Flavour Response & syst JES Flavour Response$^*$ \\ \cline{1-6}
Jet flavour composition & JetScaleFlav1Known & Flavor Comp & flavcomp-jes  & JET29NP JET Flavour Composition & syst JES Flavour Comp \\ \cline{1-6}
Jet punchthrough & JetScalepunchT & Punch Through & punch-jes & - & syst JES PunchThrough MC15\\ \cline{1-6}
\multirow{4}{*}{Jet scale} & JetScalePileup2 &  PU OffsetMu & pileoffmu-jes & - &syst JES Pileup MuOffset\\ \cline{2-6}
& -  & PU Rho & pileoffrho-jes &JET29NP JET Pileup RhoTopology & syst JES Pileup Rho topology$^*$\\ \cline{2-6}
& JetScalePileup1 & PU OffsetNPV & pileoffnpv-jes &JET29NP JET Pileup OffsetNPV & syst JES Pileup NPVOffset\\ \cline{2-6}
& - &  PU PtTerm & pileoffpt-jes &JET29NP JET Pileup PtTerm & syst JES Pileup Pt term\\ \cline{1-6}
Jet JVF selection & JetJVFcut & JVF & jetvxfrac & - &syst JES Zjets JVF\\ \cline{1-6}
B-tagged jet scale  &  -  & btag-jes &JET29NP JET BJES Response& - & -\\ \cline{1-6}
Jet resolution & -  & jeten-res & JET JER SINGLE NP& - & -\\ \cline{1-6}
Muon scale & -  & - & mup-scale & MUON SCALE& -\\ \cline{1-6}
Muon resolution & - &  -  & muonms-res & MUON MS& -\\ \cline{1-6}
Muon identification & - & -  & muid-res & MUON ID& -\\ \cline{1-6}
Diboson cross section &  - & -  & dibos-xsec & Diboson xsec& -\\ \cline{1-6}
$Z$\,+\,jets cross section & - & -  & zjet-xsec & Zjets xsec& -\\ \cline{1-6}
Single-$t$ cross section & - & -  & singletop-xsec & st xsec& -\\
\hline
\hline
\end{tabular}}
\end{center}
\end{table}

The measurement of the direct-photon production ratio  already considered correlations between the
data at 8~\TeV\ and 13~\TeV. The photon energy scale is the largest correlated systematic
uncertainty between the two measurements. There are no further important correlations with the other data sets.
The luminosity uncertainties of
the data at 8~\TeV\ and 13~\TeV\ are not combined for the present study. Instead, the
8~\TeV\ luminosity is correlated with that of the other 8~\TeV\ data sets and the 13~\TeV\
luminosity is correlated with that of the other 13~\TeV\ data sets.

\clearpage
\section{Theoretical framework}
\label{sec:theoryframework}
\subsection{NNLO QCD and NLO electroweak predictions for each data set}
The present analysis uses the
xFitter framework~\cite{HERAFitter,HERA:2009wt,Aaron:2009kv}. This program interfaces to theoretical calculations directly or uses fast interpolation grids to make theoretical predictions for the considered processes.
The program MINUIT~\cite{minuit} is used for the minimisation. Each step is cross-checked with an independent fit program~\cite{zeusfitter}.
This section describes how these predictions are obtained for each data set.
 
For the DIS processes the light-quark coefficient functions are calculated to NNLO in QCD theory
as implemented in QCDNUM~\cite{Botje:2010ay}.
The contributions of charm and bottom quarks are calculated in the general-mass
variable-flavour-number scheme of Refs.~\cite{Thorne:1997ga,Thorne:2006qt}, known as the optimised TRVFN scheme.
 
The renormalisation and factorisation scales for the DIS processes are taken to be the conventional choices,
$\muR=\muF=\sqrt{Q^2}$, where $Q^2$ is the negative four-momentum transfer squared as already stated.
For the DIS data
electroweak (EW) effects are already unfolded to leading order, such that LO-EW corrections need not be applied, apart from the running of $\alpha$. NLO-EW corrections are not well defined and no such corrections are applied to the DIS data.
 
In the present analysis, the photon PDF within the proton is not accounted for and thus data sets with substantial
sensitivity to the photon PDF, such as very high-mass Drell-Yan data, are excluded. It is estimated~\cite{Harland-Lang:2019pla} that the photon takes only $\sim 0.3\%$ of the momentum of the proton in our kinematic range.
 
The DIS processes are the only ones for which fast NNLO QCD predictions can be made analytically, such that they can be used in an iterative fit. For the LHC processes, NNLO calculations are too time-consuming and thus fast interpolation grids are used.
However, currently only the $t\bar{t}$ production processes in the lepton\,+\,jets channel have grids available for NNLO calculations.
For the remainder of the data sets the grids are available only at NLO and $K$-factors are used to correct the QCD predictions for the differential cross sections from NLO to NNLO. These $K$-factors are calculated from the ratio of the NNLO to NLO cross sections,
with the same cuts as the data, using a fixed input PDF. These $K$-factors have very
little dependence on the choice of input PDF for this calculation, and iteration using the output PDF of the fit is not necessary.
The calculations used to construct the interpolation grids are usually to leading order (LO) in the EW part of the calculation.
Correction from LO to NLO in EW predictions is also applied by a $K$-factor technique. For some data sets, this is applied together with the QCD predictions'
correction from NLO to NNLO while
for others it is a separate calculation. Details for each data set are given below and summarised in Table~\ref{tab:kFs}.
For data sets measuring final-state jets, there are also non-perturbative corrections to account for hadronisation. These are also applied using a $K$-factor technique as specified below.\footnote{Note that the ATLAS data are all at much higher scales and the charm and bottom quarks are treated in the 5-flavour zero-mass variable flavour number scheme, for which the top is an additional heavy quark.}
 
The $W$ and $Z$ inclusive cross sections at 7~\TeV\ are calculated at fixed order, to NNLO in QCD theory and to NLO
in EW theory, as described in Ref.~\cite{1612.03016}.
The results obtained from
DYNNLO\,1.5~\cite{Catani:2007vq,Catani:2009sm} and
FEWZ\,3.1.b2~\cite{Gavin:2012sy,Li:2012wna} are compared.
A small difference between the
NNLO predictions of FEWZ and DYNNLO of up to ${\sim}1\%$ is observed for the $W$- and $Z$-peak data. The sensitivity of NNLO programs to the fiducial cuts applied to the data was
highlighted again recently~\cite{Alekhin:2021xcu}. This is considered further in Section~\ref{sec:scaleunc}.
The scales for the Drell--Yan (DY) processes are
taken to be the decay dilepton invariant mass appropriate to the centre of
the mass bin in the NC case and the $W$-boson mass in the CC case.
The xFitter package uses the \textsc{APPLgrid} code~\cite{Carli:2010rw} interfaced
to the \MCFM program~\cite{Campbell:1999ah,Campbell:2010ff}
for fast calculation of the differential $W$ and $Z/\gamma^*$
boson cross sections at NLO in QCD theory and LO in EW theory, and a $K$-factor
technique is used to correct the NLO QCD predictions to NNLO and the LO EW predictions to NLO.
These $K$-factors are within 1--2\% of unity for the $W^{\pm}$ and $Z/\gamma^*$ in the c-c
dilepton
rapidity region and within ${\sim}4\%$ of unity in the c-f dilepton rapidity region.
The high-mass sideband of $Z/\gamma^*$
production is also subject to
background from photon-induced dilepton production, which was
estimated using the \textsc{MRST2004qed} photon PDF~\cite{Martin:2004dh} and subtracted from the data. Compared to the signal, the size of this background is ${\sim}(1.5\pm 0.5)\%$.
 
The QCD predictions for the triple differential distributions of $Z/\gamma^*$ production at 8~\TeV\
are made to NLO by \MCFM interfaced to \textsc{APPLgrid}, and $K$-factors are applied for NNLO QCD and
NLO EW effects using \textsc{NNLOjet}. These $K$-factors are within ${\sim}3\%$ of unity.
The renormalisation and factorisation scales for these data are
taken to be the decay dilepton invariant mass appropriate to the centre of
the mass bin.  For the present study, Particle Data Group (PDG) values~\cite{Zyla:2020zbs} are used for the electroweak parameters in the $G_\mu$ scheme, in particular the value of $\sin^2\!\theta_{\text{W}}$ is  $\sin^2\!\theta_{\text{W}} = 0.23127$.

The QCD predictions for the $W$ cross sections at 8~\TeV\ are
calculated to NLO and the EW predictions to LO using
\textsc{MG5\_aMC@NLO}\,2.6.4 interfaced to \textsc{APPLgrid} via \textsc{aMCfast}\,1.3.0, and the NNLO QCD + NLO EW $K$-factors are calculated using DYNNLO\,1.5. These $K$-factors are
within 1\%--2\% of unity. The renormalisation and factorisation scales for these data are taken to be the $W$-boson mass.
 
The predictions for $W$+\,jets production at 8~\TeV\ are obtained at fixed order, up to NNLO in QCD theory
and at LO in
EW  theory, using the $\mathrm{N_{jetti}}$ program~\cite{WjetNNLO}.
Outputs from the \textsc{APPLgrid} code are used for fast calculation at NLO in QCD theory, and $K$-factors from the aforementioned $\mathrm{N_{jetti}}$ prediction are used to
correct this calculation to NNLO.
The renormalisation and factorisation scales are set to $\sqrt{m_W^2+\Sigma (\pt^{\mathrm{jet}})^2}$, where the
second term in the square root is a sum over the transverse momentum of each jet.
Additionally,
all non-perturbative corrections, such as for hadronisation effects, are included in
the $K$-factors.
Further $K$-factors are applied for NLO EW corrections as computed by the authors of \SHERPA~\cite{Bothmann:2019yzt}.
As in the previous ATLAS analysis of $V$+\,jets data~\cite{Vjets}, the lowest $p_\mathrm{T}^{\mathrm{jet}}$ bin, $p_\mathrm{T}^{\mathrm{jet}} < 25$~\GeV, is not used, since the calculation is effectively only at NLO for such low $p_\mathrm{T}$.
 
The NNLO QCD predictions for $Z$\,+\,jets production at 8~\TeV\ were calculated by the authors of
Ref.~\cite{ZjetNNLO}. The renormalisation and factorisation scales are set to
$\muR = \muF = \frac{1}{2}\left(\sqrt{m_{\ell\ell}^2 + p_{\text{T},\ell\ell}^2} + \Sigma p_{\text{T},\text{partons}}\right)$
where $m_{\ell\ell}$ is the invariant mass of the electron pair, $p_{\text{T},\ell\ell}$
is the transverse momentum of the electron pair and $\Sigma p_{\text{T},\mathrm{partons}}$ is the sum of
the transverse momenta of the outgoing partons. Four sets of $K$-factors are provided: the first
corrects the NLO QCD predictions to NNLO, the second
gives corrections for non-perturbative effects, the third corrects from LO to NLO in QED and the fourth applies NLO EW corrections (excluding QED radiation) as computed by \SHERPA[2.2.10]. The $K$-factors for both $W$+\,jets and $Z$\,+\,jets production are typically within $10\%$ of unity. Uncertainties in the non-perturbative corrections are supplied for the $Z$\,+\,jets predictions and these are applied as correlated systematic uncertainties. Their impact is very small. Such uncertainties are not supplied for the $W$+\,jets predictions.
 
The NNLO QCD predictions for $t\bar{t}$ production at 8~\TeV\ were calculated by the authors of Ref.~\cite{1704.08551} and are available in
the form of fast interpolation grids, \textsc{APPLgrid}~\cite{Carli:2010rw} or \textsc{fastNLO}~\cite{Kluge:2006xs,Britzger:2012bs}, for the data in the lepton\,+\,jets  channel. The
predictions for $m_{t\bar{t}}$ are given for renormalisation and factorisation scales
equal to $H_{\mathrm{T}}/4$, where
$H_{\mathrm{T}}=\sqrt{m_t^2+(p_{\mathrm{T}}^t)^2}+\sqrt{m_t^2 +(p_{\mathrm{T}}^{\bar{t}})^2}$, and the predictions for
$p_{\mathrm{T}}^t$ are given for scales equal to $m_{\mathrm{T}}/2$, where $m_{\mathrm{T}}=\sqrt{m_t^2+(p_{\mathrm{T}}^t)^2}$ and
$m_t=173.3~$\GeV\ is the pole mass. Non-perturbative corrections and their uncertainties are supplied with the data and these uncertainties are applied as correlated systematic uncertainties.
For the dilepton channel, \textsc{APPLgrid} is interfaced to \MCFM to produce
NLO grids, and a $K$-factor
technique is used to correct NLO predictions to NNLO, using $K$-factors from
Ref.~\cite{1611.08609}. The scales used for the predictions are equal to
$H_{\mathrm{T}}/4$. The NNLO/NLO $K$-factors are ${\sim}7\%$ above unity and constant for $y_{t\bar{t}}$.
The calculations are for zero width of the top mass. NLO EW corrections for the spectra are also considered, using the additional
$K$-factors given in Ref.~\cite{1705.04105}. For the 8~\TeV\ data, these
corrections can be up to $1\%$ for $y_{t\bar{t}}$ and $y_t$, up to $2\%$
for $m_{t\bar{t}}$, and up to $4\%$ for $p_{\mathrm{T}}^t$.
 
The NNLO predictions for $t\bar{t}$ production at 13~\TeV\ have been calculated by the authors of Ref.~\cite{1704.08551} in the
form of fast interpolation grids~\cite{Carli:2010rw,Kluge:2006xs,Britzger:2012bs} for the data in the lepton\,+\,jets channel~\cite{ttbarWEB}.
The predictions for $m_{t\bar{t}}$, $y_t$ and $y^b_{t\bar{t}}$
are given for renormalisation and factorisation scales
equal to $H_{\mathrm{T}}/4$  and the predictions for
$p_{\mathrm{T}}^t$ are given for scales equal to $m_{\mathrm{T}}/2$, just as for the 8~\TeV\ $t\bar{t}$ data. The only difference is
that at 13~\TeV\ the value $m_t=172.5~$\GeV\ is used for the top-quark mass, because this is the value used for the interpolation grids.
The small difference between the masses used for the central fit for 8 and 13~\TeV\ $t\bar{t}$ data is studied as part of the model uncertainties and found to be negligible (see Section~\ref{sec:model}). NLO EW corrections for the spectra are also considered, using
$K$-factors taken from Ref.~\cite{1705.04105}. For the 13~\TeV\ data, these
corrections can be up to $1\%$ for $y_{t\bar{t}}$, up to $0.5\%$ for $y_t$, up to $2.5\%$
for $m_{t\bar{t}}$, and up to $5\%$ for $p_{\mathrm{T}}^t$. Non-perturbative corrections and their uncertainties are again supplied with the data and these uncertainties are applied as correlated systematic uncertainties.
 
The NLO QCD predictions for direct photon production were taken from \MCFM interfaced to \textsc{APPLgrid}, and
$K$-factors from Ref.~\cite{cew} were applied to obtain NNLO QCD predictions.
Resummation of electroweak effects is also applied as in Ref.~\cite{becher}. The corrections from these $K$-factors are 2\%--15\% and 0.5\%--10\% for the separate 8 and 13~\TeV\ data sets, respectively, but they almost cancel out in the ratio, leaving corrections of less than 2\%.
Finally, a
photon isolation criterion is applied to the predictions to follow that applied to the data
as closely as possible (see Ref.~\cite{1802.03021}). The renormalisation and factorisation scales are set to $E_{\mathrm{T}}^{\gamma}$.
 
The NLO QCD predictions for inclusive jet production at 7, 8 and 13~\TeV\ are made with \textsc{NLOjet++} interfaced to \textsc{APPLgrid}.
The renormalisation and
factorisation scale choices $\muR=\muF= p_{\mathrm{T}}^{\mathrm{max}}$ and $\muR=\muF= p_{\mathrm{T}}^{\mathrm{jet}}$ are both available.
For the 7 and 8~\TeV\ data, predictions are available for jet radii $R=0.4$ and 0.6, while for the 13~\TeV\ data they are available for $R=0.4$ only, matching the data. The $K$-factors to convert NLO QCD predictions to NNLO have been calculated using
\textsc{NNLOjet}~\cite{nnlojet}. These $K$-factors are rather different for the two scale choices. For the 8~\TeV\ data and
scale $p_{\mathrm{T}}^{\mathrm{max}}$ they range from ${\sim}{+}10\%$ at low $p_{\mathrm{T}}$ to a few percent at high $p_{\mathrm{T}}$
for low $|y^{\mathrm{jet}}|$, but the high-$p_{\mathrm{T}}$ corrections
can become slightly negative at high $|y^{\mathrm{jet}}|$. For the scale $p_{\mathrm{T}}^{\mathrm{jet}}$, they range from ${\sim}{-}5\%$ at low $p_{\mathrm{T}}$ to a few percent positive at high
$p_{\mathrm{T}}$ for low $|y^{\mathrm{jet}}|$,  but become increasingly negative for all $p_\mathrm{T}$ at high $|y^{\mathrm{jet}}|$. The $K$-factors for the 7~\TeV\ data are very similar, and those for the 13~\TeV\ data follow the same trends while being a bit larger.
The choice of scale for the central fit is $\muR=\muF= p_\mathrm{T}^{\mathrm{jet}}$, since
this is preferred theoretically~\cite{1807.03692}, but the alternative scale choice $\muR=\muF= p_\mathrm{T}^{\mathrm{max}}$ and other
scale variations are considered in Section~\ref{sec:model}.
 
The precision of the $K$-factors, in the predictions for inclusive jet data, depends on the statistical precision of the \textsc{NNLOjet} calculations which
evaluated the NLO and NNLO cross sections. For the 8~\TeV\ inclusive jet data, the resulting $K$-factors suffer from some statistical fluctuations that can be as large as ${\sim}1\%$.\footnote{For the other processes considered in this section the statistical precision of the $K$-factors is much higher and further consideration of this source of uncertainty is not necessary.}  For the central fit, these $K$-factors are smoothed and the estimated
uncertainty in this procedure is applied as being $60\%$ correlated and $40\%$ uncorrelated between $p_\mathrm{T}$ bins within the
same rapidity bin. Alternative treatments are considered in Section~\ref{sec:model}.
 
The QCD inclusive jet predictions are corrected for NLO electroweak effects as detailed in Ref.~\cite{1706.03192}. This correction
can reach more than $10\%$ for the highest $p_{\mathrm{T}}^{\mathrm{jet}}$ in the lowest $|y^{\mathrm{jet}}|$ bin,
but decreases rapidly as $|y^{\mathrm{jet}}|$ increases.
It is less than $3\%$ for $|y^{\mathrm{jet}}| > 1.0$.
 
In order to compare
fixed-order QCD calculations with the measured inclusive jet cross sections, corrections for non-perturbative effects must
also be applied.
These are derived using LO Monte Carlo event generators. There are
significant differences between the corrections derived with different event generators, \PYTHIA~\cite{Sjostrand:2014zea} and \HERWIG~\cite{Bahr:2008pv},
and different sets of tuned parameter values.
The corrections can be up to ${\sim}10\%$ at low $p_{\mathrm{T}}^{\mathrm{jet}}$. They are different for $R=0.4$ and $R=0.6$ due to
differing
interplay of the underlying-event and hadronisation corrections, and for this reason fits to data with both jet
radii are investigated.
Values of the non-perturbative correction are taken at the centre of the uncertainty band, which is constructed as the envelope of all such corrections considered, and the systematic uncertainty of each correction value,
correlated in $p_{\mathrm{T}}^{\mathrm{jet}}$ and $y^{\mathrm{jet}}$, is taken from the spread of the band (see Ref.~\cite{1706.03192} for details).

Fits were performed using 7, 8 and 13~\TeV\ inclusive jet data separately, using both choices for the jet radius, $R=0.4$ and $R=0.6$, when available, and both choices for the scales, $p_\mathrm{T}^{\mathrm{max}}$ and $p_\mathrm{T}^{\mathrm{jet}}$, but the final choice for the central fit is 8~\TeV\ jets with $R=0.6$ and scales $\muR=\muF= p_{\mathrm{T}}^{\mathrm{jet}}$, as discussed in Section~\ref{sec:model}. Table~\ref{tab:kFs} details the code used for NNLO/NLO QCD corrections and NLO/LO EW
corrections to ATLAS data in the QCD fit, for ease of reference.
 
\begin{table}\small
\caption{Summary of code used for NLO, NNLO QCD and LO, NLO EW predictions for ATLAS data as applied in the QCD fit. For most data sets, predictions are provided at NLO in QCD and LO in EW in the form of fast interpolation grids and are corrected to NNLO QCD and NLO EW by $K$-factors. For the $t\bar{t}$ lepton\,+\,jets channel the grids are calculated at NNLO in QCD directly, so no entry appears in the column `NLO QCD code'.\label{tab:kFs}}
\begin{center}
\resizebox{\textwidth}{!}{
\begin{tabular}{lcccc}
\hline
Data set & NLO QCD code & LO EW code& NNLO QCD code& NLO EW code \\
\hline
Inclusive $W,Z/\gamma^*$~\cite{1612.03016} & \MCFM & \MCFM & DYNNLO\,1.5, FEWZ\,3.1.b2 & DYNNLO\,1.5, FEWZ\,3.1.b2 \\
Inclusive $Z/\gamma^*$~\cite{z3d} & \MCFM & \MCFM & \textsc{NNLOjet}& \textsc{NNLOjet}\\
Inclusive $W$~\cite{W8} & \textsc{MG5\_aMC@NLO}\,2.6.4 & \textsc{MG5\_aMC@NLO}\,2.6.4 & DYNNLO\,1.5 & DYNNLO\,1.5 \\
$W^{\pm}$\,+\,jets~\cite{wjetspaper} & $\mathrm{N_{jetti}}$ & $\mathrm{N_{jetti}}$ & $\mathrm{N_{jetti}}$ & \SHERPA \\
$Z$\,+\,jets~\cite{zjetspaper} & Ref.~\cite{ZjetNNLO} & Ref.~\cite{ZjetNNLO} & Ref.~\cite{ZjetNNLO} & \SHERPA \\
$t\bar{t}$ (lepton\,+\,jets)~\cite{1511.04716} & -  &Ref.~\cite{1704.08551}  & Ref.~\cite{1704.08551}  &Ref.~\cite{1705.04105} \\
$t\bar{t}$ (dilepton)~\cite{1607.07281} & \MCFM & \MCFM  & Ref.~\cite{1611.08609}  &Ref.~\cite{1705.04105} \\
$t\bar{t}$~\cite{1908.07305} & - & Ref.~\cite{1704.08551} & Ref.~\cite{1704.08551}  &Ref.~\cite{1705.04105} \\
Inclusive isolated $\gamma$~\cite{terron} & \MCFM & \MCFM & Ref.~\cite{cew}& Ref.~\cite{becher} \\
Inclusive jets~\cite{1410.8857,1706.03192,1711.02692} & \textsc{NLOjet++} & \textsc{NLOjet++} & \textsc{NNLOjet} & Ref.~\cite{1210.0438}\\
\hline
\hline
\end{tabular}}
\end{center}
\end{table}

\subsection{Scale uncertainties and sensitivity to NNLO code}
\label{sec:scaleunc}
For some of the ATLAS data sets considered, the precision of the data set is so high that it is comparable to the size of the NNLO scale uncertainties.
This is the case for the ATLAS $W$ and $Z/\gamma^*$ inclusive data
sets  at both 7 and 8~\TeV, for which the total experimental uncertainty and the scale
uncertainties both approach ${\sim}0.5\%$. In this case, the scale uncertainties are considered as additional theoretical uncertainties which are added to the $\chi^2$ calculation in the same way as experimental systematic uncertainties (see Section~\ref{sec:method}).
These scale uncertainties are evaluated as follows. The $K$-factors are evaluated for separate changes of the
renormalisation and factorisation scales by factors of 2 and 0.5.
The magnitude of the $K$-factor difference is symmetrised as $(K [\muR(2)] - K[\muR(0.5)])/2$
and $(K [\muF(2)] - K[\muF(0.5)])/2$ and its sign is preserved as positive
if the upward variation of $\muR$ or $\muF$ makes the $K$-factor increase and negative if it makes the
$K$-factor decrease. For these renormalisation and factorisation scale changes,
the ratios of the signed $K$-factor changes to the $K$-factors for the nominal scales are calculated to
obtain the fractional scale uncertainties. These can then be used in the fit
like any other systematic uncertainty, as suggested in Ref.~\cite{NNPDF:2019ubu}.
 
Due to the similarity of the $W$ and $Z$ processes, both the renormalisation scale and factorisation scale are considered correlated within
the $W, Z$ data sets at 7~\TeV\ and between the $W$ and $Z$ data sets at 8~\TeV. They
are also considered to be correlated between the $W$ and $Z$ data sets at 7 and 8~\TeV\ for the central fit. Studies of alternative approaches for the scale uncertainties are presented in Appendix~\ref{sec:scale}.
 
Scale uncertainties are accounted for in the central fit only for the $W$ and $Z$ inclusive data
sets, because the scale uncertainties are comparable to the experimental uncertainties only
in this case. For the other ATLAS data sets the experimental uncertainties are larger and scale uncertainties are
treated by repeating the fit with varied scales as discussed in Section~\ref{sec:model}.
 
As remarked earlier, there is a small difference of up to ${\sim}1\%$ between the NNLO predictions of FEWZ and DYNNLO. A fractional uncertainty can be defined as the ratio of the signed difference between the FEWZ and DYNNLO predictions to the FEWZ prediction, and applied similarly to the scale uncertainties. The inclusion of such a theoretical uncertainty makes no further difference to the fits, provided the scale uncertainties are applied.
 
Recently, next-to-next-to-next-to-leading-order (N$^{3}$LO) calculations for the total cross section of the Drell--Yan process have become available~\cite{prl125-172001}. It is
notable that the N$^{3}$LO result does not quite lie within the scale uncertainties of the
NNLO result. The N$^{3}$LO correction to the central value is ${\sim}2\%$ compared to NNLO
scale uncertainties of ${\sim}0.5\%$. Full accounting for N$^{3}$LO corrections, as well as $p_\mathrm{T}$ resummation effects which affect the prediction of $W$ and $Z$ fiducial cross sections at the level of ${\sim}0.5-1\%$~\cite{Alekhin:2021xcu}, are beyond the scope of the current fixed-order-NNLO QCD fit.

\clearpage
\section{Fit methodology}
\label{sec:method}

The DGLAP~\cite{Gribov:1972ri,Altarelli:1977zs,Dokshitzer:1977sg} evolution equations yield the PDFs
at any value of $Q^2$ given that they are parameterised
as a function of $x$ at a starting scale $Q^2_0$.
In the present analysis, this scale is
chosen to be $Q^2_0 = 1.9$~\GeV$^2$ so that it is below the
charm mass threshold $m_c^2$.
A recent combination of HERA heavy-quark data~\cite{heraf2cbnew} has
stimulated a re-analysis of the optimal values of the heavy-quark masses and their variations as used in the optimised TRVFN scheme~\cite{Thorne:2006qt}.
The resulting values are $m_c=1.41$~\GeV\ and $ m_b=4.2$~\GeV~\cite{2112.01120}, and these are used for the central fit
as applied to the DIS data.
A minimum $Q^2$ cut of $Q^2_{\mathrm{min}} = 10.0$~\GeV$^2$ is imposed on the HERA data.
All these assumptions are varied in the consideration of model uncertainties, as shown in  Section~\ref{sec:model}.
The strong coupling constant is fixed to $\alphas(\mZ) =  0.118$.
 
The quark distributions at the starting scale are represented by the generic form
\begin{equation}\nonumber
xq_i(x) = A_i x^{B_i} (1-x)^{C_i} P_i(x),
\end{equation}
where $P_i(x)= (1 + D_i x + E_i x^2 + F_ix^3)$.
The parameterised quark distributions, $xq_i$, are chosen to be
the valence quark distributions ($xu_v,~xd_v$) and the
light anti-quark distributions
($x\bar{u},~x\bar{d},~x\bar{s}$). The gluon distribution
is parameterised with the more flexible form
\begin{equation}\nonumber
xg(x) = A_g x^{B_g} (1-x)^{C_g}P_g(x)  - A'_g x^{B'_g} (1-x)^{C'_g},
\end{equation}
where $C'_g$ is set to $25$ to suppress negative contributions at high $x$.
The parameters $A_{u_v}$ and $A_{d_v}$ are fixed using
the valence-quark number sum rule, and $A_g$ is fixed using the momentum sum rule.
The normalisation and slope parameters, $A$ and $B$,
of the light-quark sea, $x\bar{u}$, $x\bar{d}$ and $x\bar{s}$, are all independent of each other, such that there is no constraint on
$x\bar{d} - x\bar{u}$, or on $\bar{s}/(\bar{d}+\bar{u})$, either in shape or in normalisation as $x \to 0$.
By default it is assumed that $xs=x\bar{s}$.
The $D,E$ and $F$ terms in the polynomial
expansion $P_i(x)$ are used only if  required
by the data, following the procedure described
in Ref.~\cite{HERA:2009wt}, whereby parameters are added only if the $\chi^2$ of the fit decreases significantly. This leads
to a 21-parameter fit with:
$P_{u_v}(x)=1+ D_{u_v} x + E_{u_v} x^2$, $P_{d_v}(x)=1+ D_{d_v} x$ and $P_{g}(x)=1+ D_{g} x $, with $P_i(x)=1$ for the light-sea PDFs.
It has been established that adding further parameters beyond the point at which there is little further change in $\chi^2$ (called the point of `saturation' of the $\chi^2$) only serves to fit noise in the data~\cite{NadolskySoper}. Thus,
addition of further parameters to the central fit is not considered. However, parameterisation uncertainties are considered in Section~\ref{sec:param}.
 
A cross-check was made using Chebyshev polynomials rather than ordinary polynomials for the $P_i(x)$ terms. There is no improvement
in $\chi^2$ for a comparable 21-parameter fit using the Chebyshev polynomials and the extracted PDFs lie within the total
uncertainty bands of the ordinary polynomial fit.

The level of agreement between the data and the predictions from a PDF parameterisation is quantified by the $\chi^2$ per degree of freedom
($\chi^2$/NDF, for NDF degrees of freedom). The definition of the $\chi^2$ is
\begin{equation} \label{eqn:chi2}
\begin{aligned}
\chi^2 = &\sum_{i,k}\left(D_i - T_i(1-\sum_{j}\gamma_{ij}b_{j})\right)C^{-1}_{\text{stat,uncor}, ik}(D_i,D_k)\left(D_k - T_k(1-\sum_{j}\gamma_{kj}b_{j})\right)\\
&+ \sum_{i} \log \frac{\delta^{2}_{i,\text{uncor}}T_i^2 + \delta^{2}_{i,\text{stat}}D_i T_i}{\delta^{2}_{i,\text{uncor}}D_i^2 + \delta^{2}_{i,\text{stat}}D_i^2}\\
&+ \sum_{j} b_j^2
\end{aligned}
\end{equation}
where $D_i$ represent the measured data, $T_i$ the corresponding theoretical prediction,
$\delta_{i,\text{uncor}}$ and $\delta_{i,\text{stat}}$ are the uncorrelated systematic uncertainties and the
statistical uncertainties of $D_i$, and the correlated systematic uncertainties, described by $\gamma_{ij}$,
are accounted for using the nuisance parameters $b_j$. The quantity $C_{\text{stat,uncor}, ik}$ is a covariance matrix for both the
statistical and uncorrelated systematic uncertainties. Summations over $i$ and $k$ run over all
data points and summation over $j$ runs over all sources of correlated systematic uncertainty.
For each data set, the first term gives the main contribution to the partial $\chi^2$ of the data set and the second term is a small bias correction term, referred to as the {\em log penalty}, which arises because the diagonal term of the matrix, $C$, is
given by $C_{ii} =\delta^{2}_{i,\text{uncor}}T_i^2 + \delta^{2}_{i,\text{stat}}D_i T_i$, with different weighting for statistical uncertainties and uncorrelated systematic uncertainties. This form of the $\chi^2$ is used as standard in HERA and ATLAS PDF
fits~\cite{herapdf20,1612.03016,ATL-PHYS-PUB-2018-017}. In reporting the $\chi^2$ for the fits, the first and second terms together give the \emph{partial} $\chi^2$ of the data set.  The third term gives the
{\em correlated} $\chi^2$ arising from the penalty for the nuisance parameters describing correlations of systematic uncertainties within and between data sets. The experimental uncertainties of the fit are first set using the usual tolerance, $T=1$, where $T^2 = \Delta\chi^2 =1$. The use of an enhanced tolerance is considered in Section~\ref{sec:globaltol}.
 
\clearpage
\section{Results}
\label{sec:results}
In this section the ATLASpdf21 PDF set is presented. The impact of variations of the central choice of fit settings and parameterisation is discussed in Section~\ref{sec:uncertainties}.
Table~\ref{tab:chisqfinal} gives the total $\chi^2$ per degree of freedom, $\chi^2/\mathrm{NDF}$, of the fit using all data sets and the $\chi^2$ per data point,
\emph{partial} $\chi^2/\mathrm{NDP}$ for NDP data points, of each data set.
The correlated terms (term 3) are shown separated into groups with common systematic
correlations. In order to evaluate the separate contributions of the data sets to this correlated term, the fit is run with its
final parameters fixed for each data set separately. These values follow the total correlated terms in brackets, in order of their appearance in the
table. The quality of the fit to the HERA data is $\chi^2/\mathrm{NDP} =1.14$, comparable to that of HERAPDF2.0, so that there is no tension
between the ATLAS data and HERA data. The quality of the fit to the ATLAS $W,Z$ data is $\chi^2/\mathrm{NDP} =1.44$, and to the ATLAS
$V$+\,jets, $t\bar{t}$ and inclusive jet data it is $\chi^2/\mathrm{NDP} =1.40$. The quality of fit to the ATLAS direct photon data is
$\chi^2/\mathrm{NDP} = 0.7$. These $\chi^2$ values are comparable to those obtained by the global PDF fits for similar data sets, but indicate a need to consider the appropriate $\chi^2$ tolerance of the fit. Both of these points are discussed further in Section~\ref{sec:globaltol}.

\begin{table}[t]
\caption{$\chi^2$ contributions for the all data sets entering
the PDF fit. The \emph{partial} $\chi^2$ for the individual data sets are given with respect to the number of data points (NDP).
They represent the addition of terms 1 and 2 in Eq.~(\ref{eqn:chi2}).
The correlated terms (term 3) are shown separated into groups with common systematic
correlations. The total value of the correlated term for these groups is also split into the
additon of the separate contributions in the order in which they are given in the table.
\label{tab:chisqfinal}
}
\begin{center}
\begin{tabular}{lcccc}
\hline
\hline
Total $\chi^2/\mathrm{NDF}$&2010/1620\\
\hline
\hline
HERA $\chi^2/\mathrm{NDP}$&1112/1016\\
HERA correlated term&50\\
\hline
ATLAS $W,Z$ 7~\TeV $\chi^2/\mathrm{NDP}$ &68/55 \\
ATLAS $Z/\gamma^*$ 8~\TeV $\chi^2/\mathrm{NDP}$ & 208/184\\
ATLAS $W$ 8~\TeV $\chi^2/\mathrm{NDP}$& 31/22\\
ATLAS $W$ and $Z/\gamma^*$ 7 and 8~\TeV &\\
correlated term & $71=(38+33)$ \\
\hline
ATLAS direct $\gamma$ 13/8~\TeV $\chi^2/\mathrm{NDP}$&27/47 \\
ATLAS direct $\gamma$ 13/8~\TeV &\\
correlated term & 6\\
\hline
ATLAS $V$+\,jets 8~\TeV $\chi^2/\mathrm{NDP}$&105/93~~\\
ATLAS $t\bar{t}$ 8~\TeV $\chi^2/\mathrm{NDP}$&13/20 \\
ATLAS $t\bar{t}$ 13~\TeV $\chi^2/\mathrm{NDP}$ & 25/29\\
ATLAS inclusive jets 8~\TeV $\chi^2/\mathrm{NDF}$&207/171\\
ATLAS $V$+\,jets 8~\TeV\ and&\\
$t\bar{t}$ + jets 8,13~\TeV\ and&\\
$R = 0.6$ inclusive jets 8~\TeV\ correlated term& $87=(16+9+21+41)$ \\
\hline
\hline
\end{tabular}
\end{center}
\end{table}
 
\subsection{Comparison of PDFs with and without inclusion of correlations between data sets}
\begin{figure*}
\begin{centering}
\includegraphics[width=0.48\textwidth]{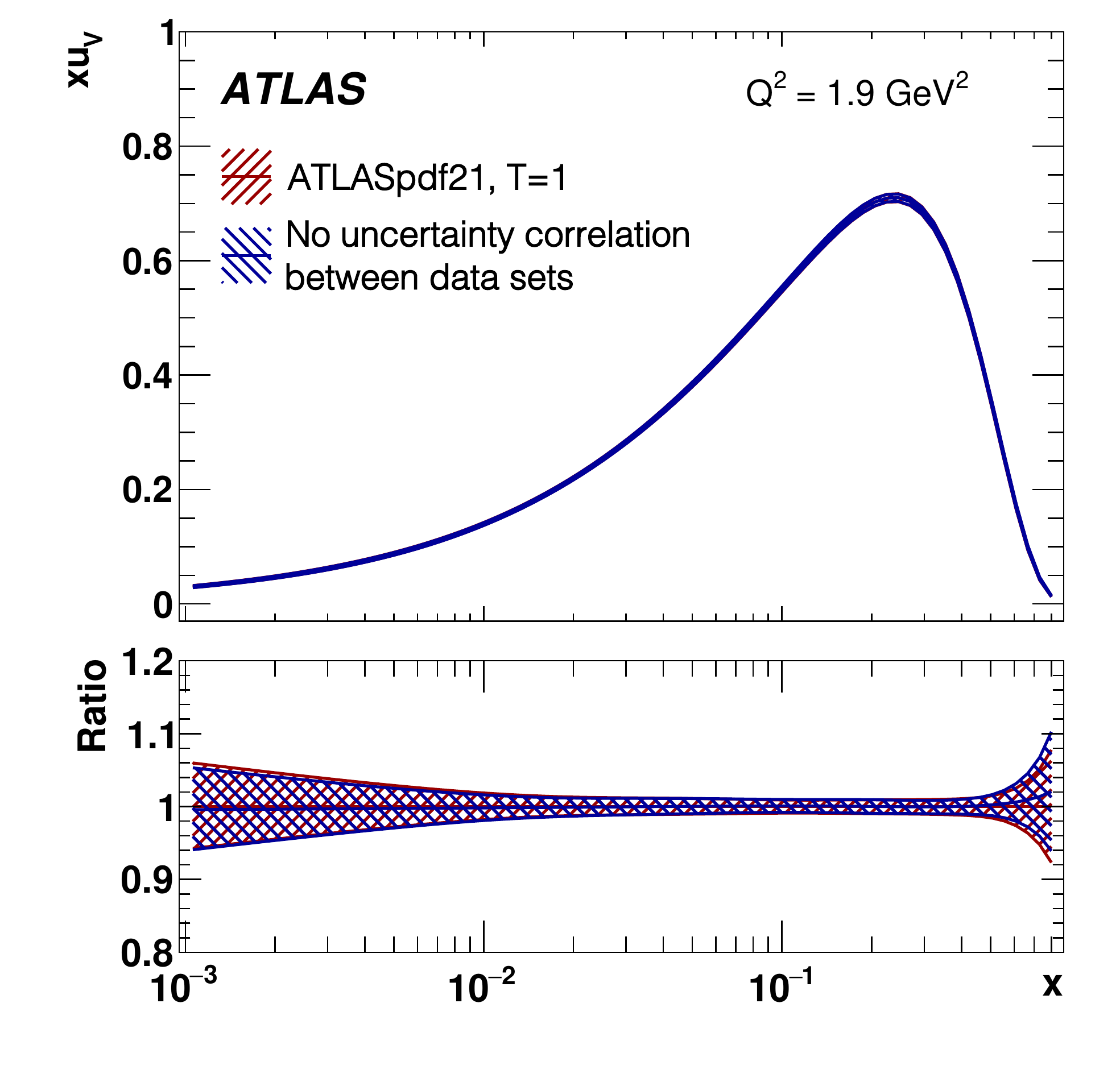}
\includegraphics[width=0.48\textwidth]{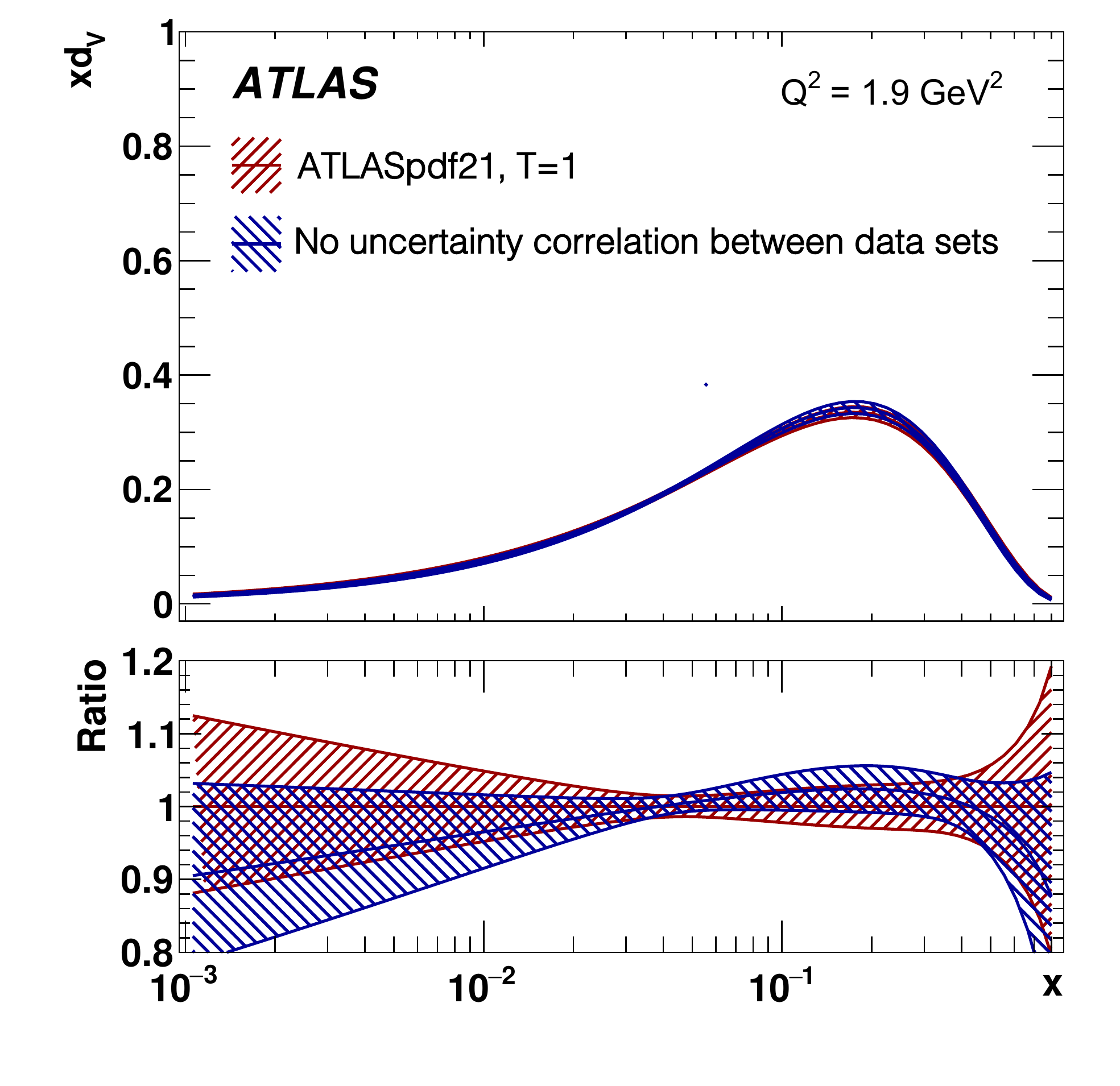}
\includegraphics[width=0.48\textwidth]{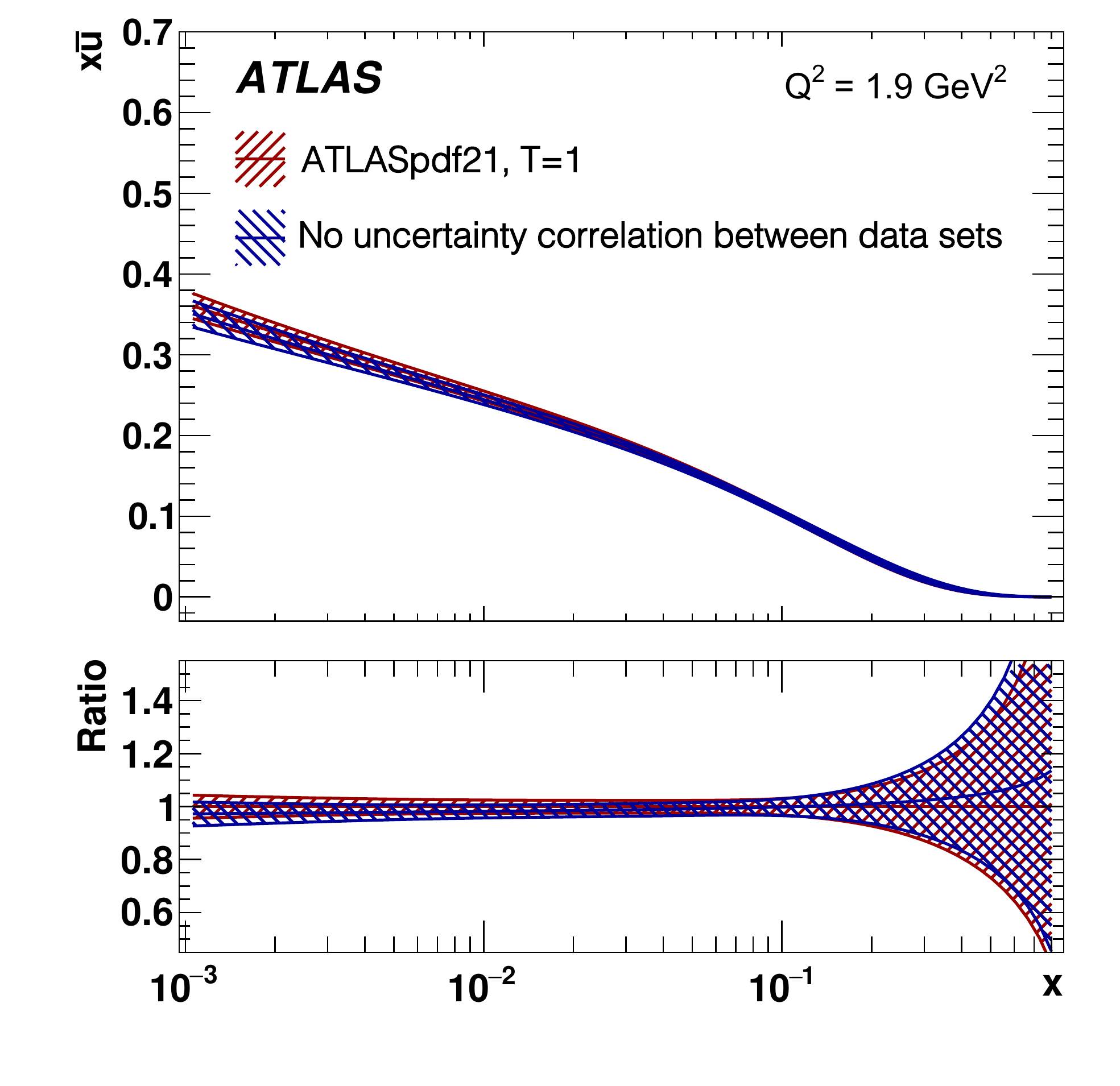}
\includegraphics[width=0.48\textwidth]{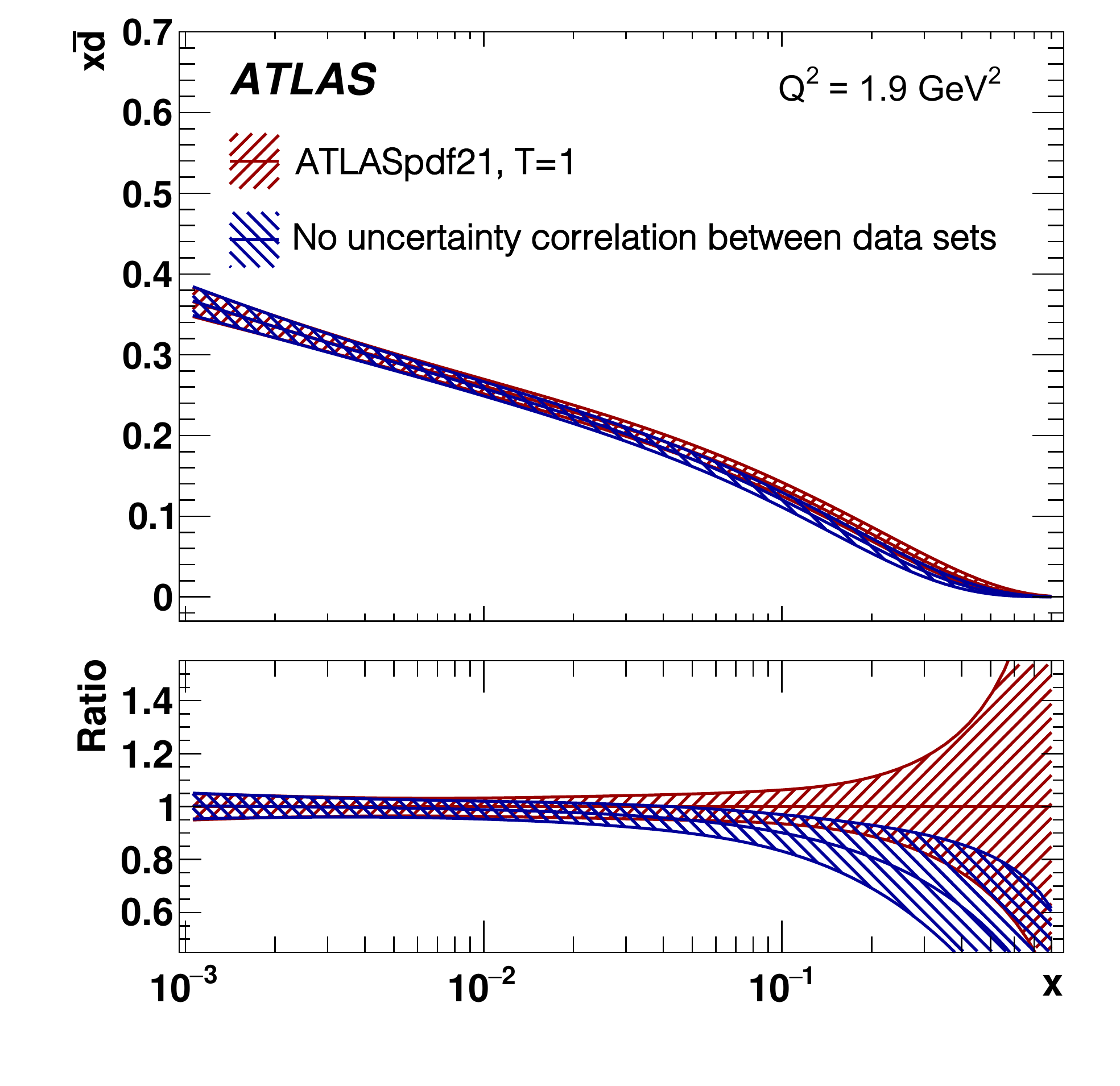}
\caption{ ATLASpdf21 PDFs comparing those extracted from a fit in which correlations of systematic uncertainties
between data sets are applied, with those extracted from a fit in which only the luminosity uncertainties for each centre-of-mass energy
are correlated between data sets. Only experimental uncertainties are shown, evaluated with tolerance $T=1$. Top left: $xu_v$. Top right: $xd_v$. Bottom left: $x\bar{u}$. Bottom right: $x\bar{d}$.
The lower panels show the comparison as a ratio to the default ATLASpdf21 PDF.
\label{fig:corr1}
}
\end{centering}
\end{figure*}

\begin{figure*}
\begin{centering}
\includegraphics[width=0.48\textwidth]{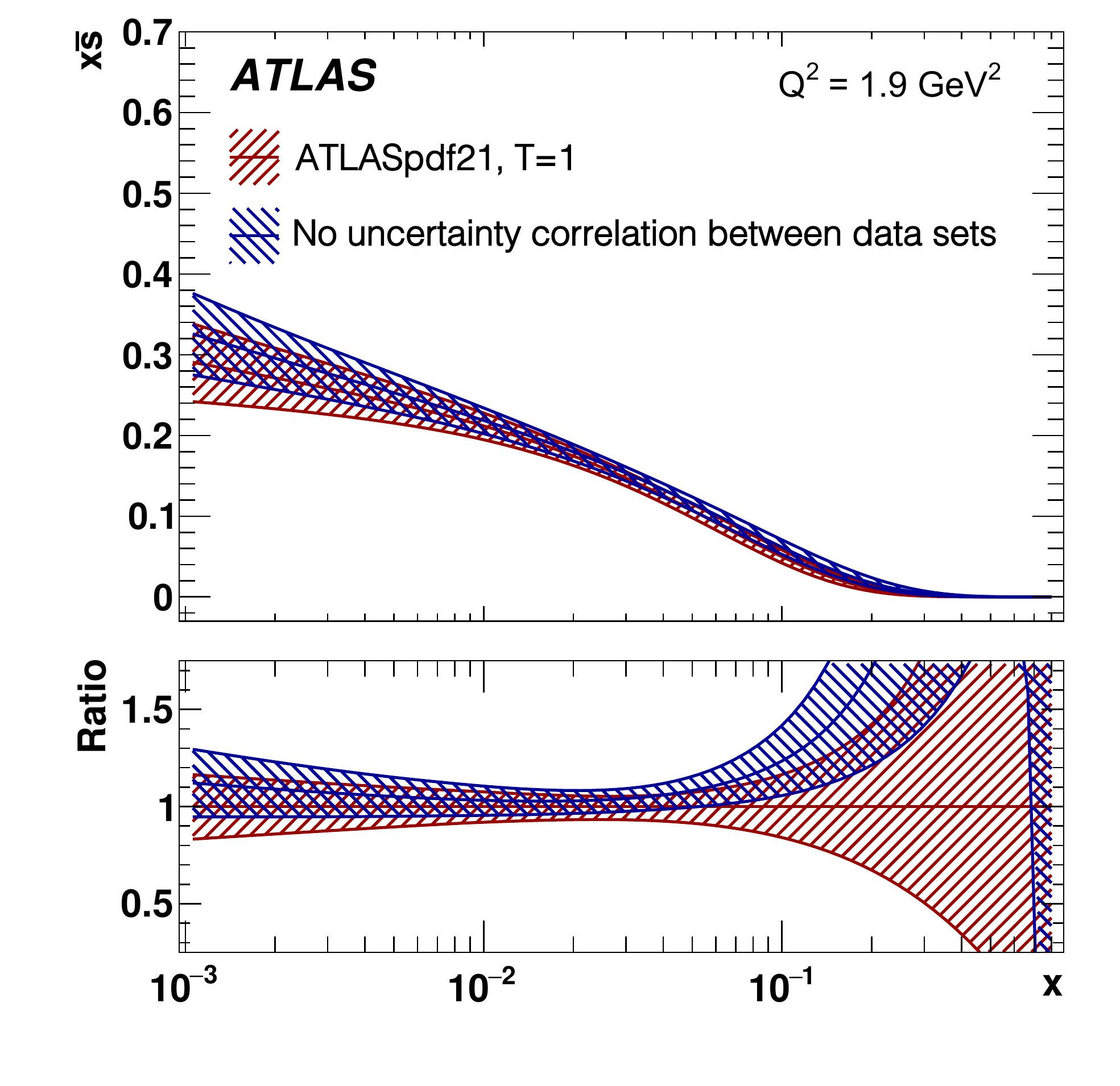}
\includegraphics[width=0.48\textwidth]{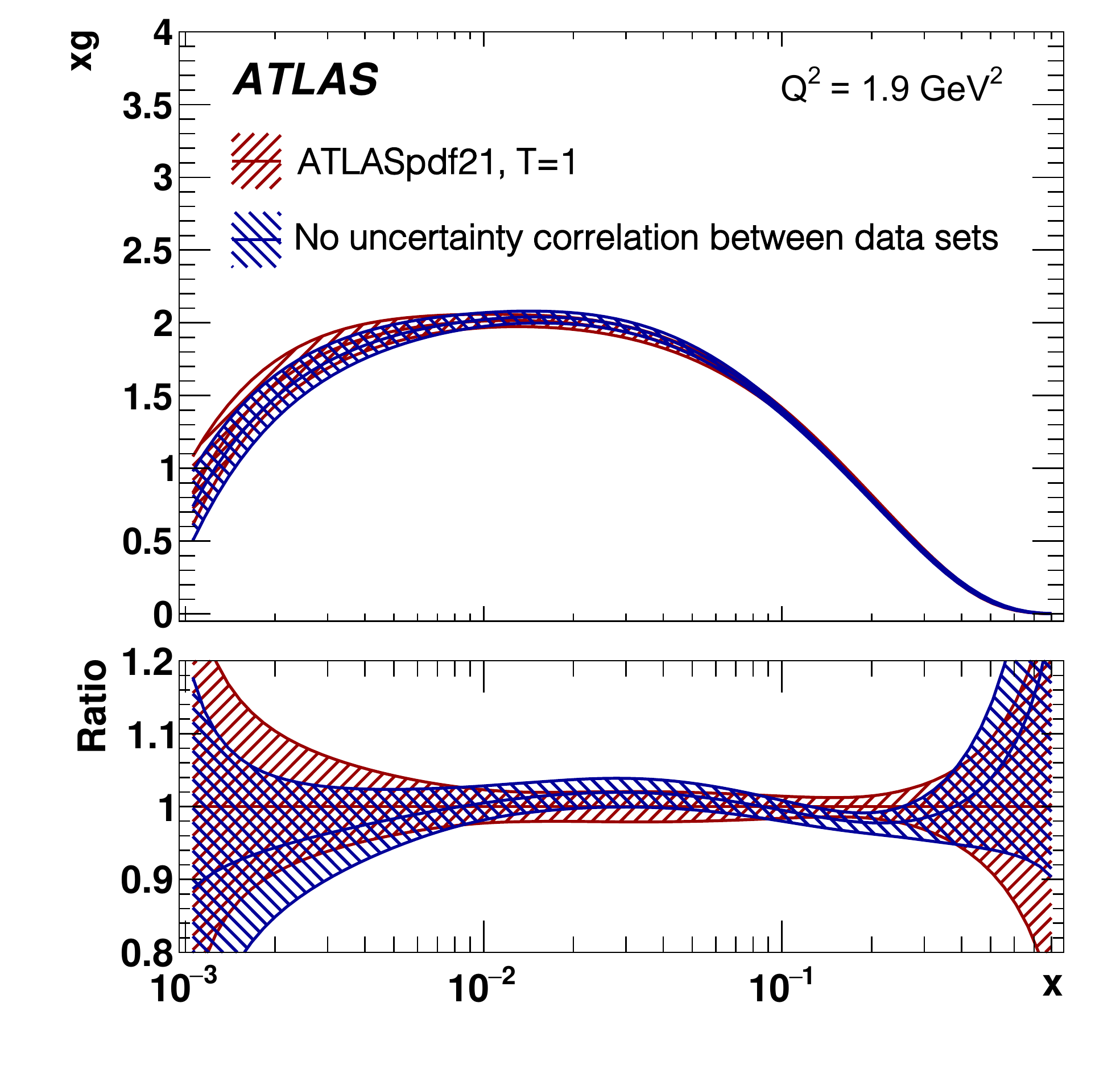}
\includegraphics[width=0.48\textwidth]{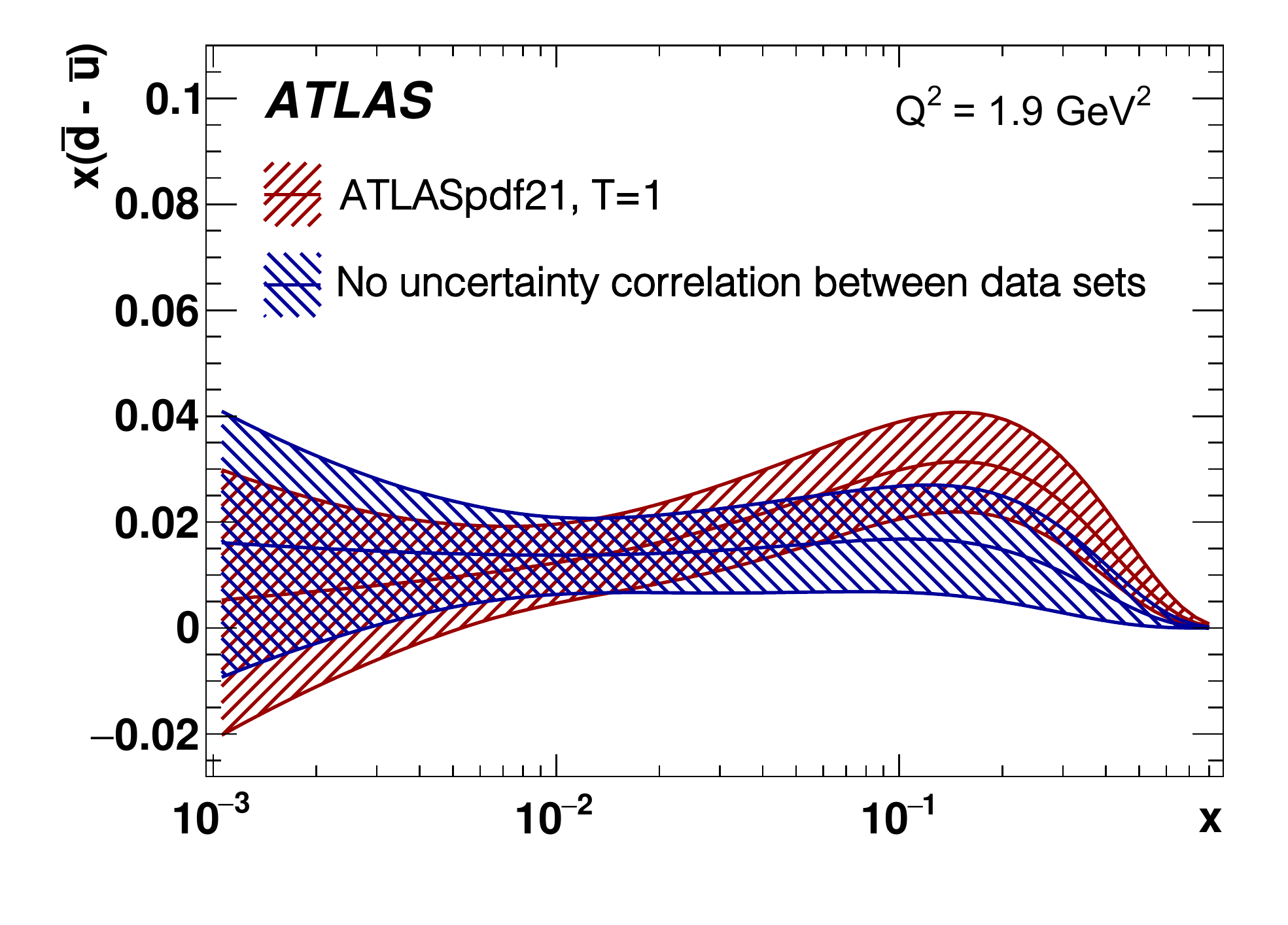}
\includegraphics[width=0.48\textwidth]{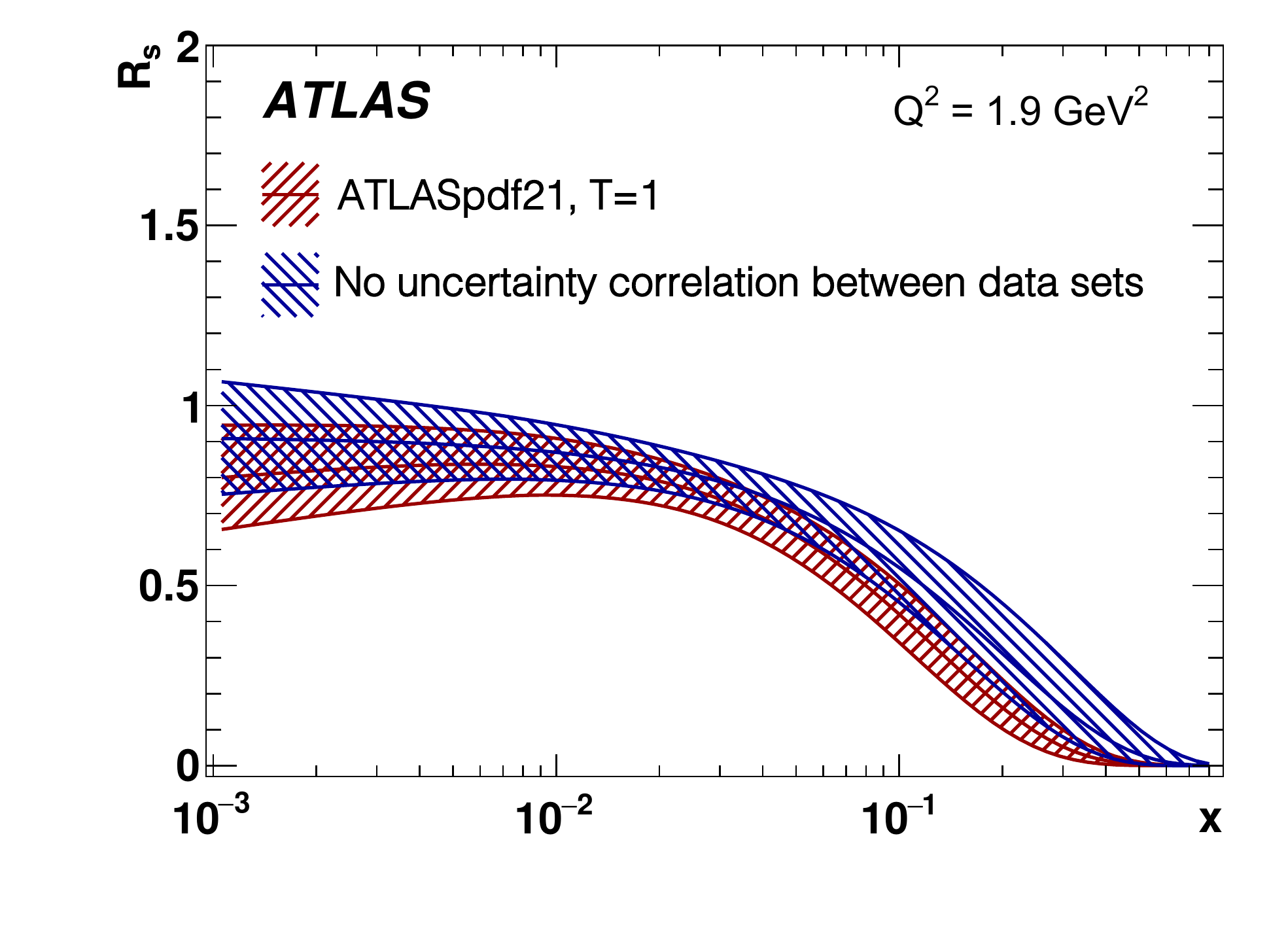}
\caption{ ATLASpdf21 PDFs comparing those extracted from a fit in which correlations of systematic uncertainties between data sets are applied, with those
extracted from a fit in which only the luminosity uncertainties for each centre-of-mass energy are correlated between data sets. Only experimental uncertainties are shown, evaluated with tolerance $T=1$.
Top left: $x\bar{s}$. Top right: $xg$. Bottom left: $x(\bar{d}-\bar{u})$. Bottom right: $R_s$.
For the $x\bar{s}$ and $xg$ plots, the lower panels show the comparison as a ratio to the default ATLASpdf21 PDF.
\label{fig:corr2}
}
\end{centering}
\end{figure*}

The ATLASpdf21 PDFs at $Q^2 = 1.9$~\GeV$^2$ (the starting scale)  are shown in Figures~\ref{fig:corr1} and \ref{fig:corr2}. Only experimental uncertainties with tolerance $T=1$ are shown for the comparisons throughout this Section. Full uncertainties including model and parameterisation variations are considered for the ATLASpdf21 fit in Section~\ref{sec:uncertainties}.
 
Figure~\ref{fig:corr2} also shows the ratio $R_s=x(s + \bar{s})/x(\bar{u}+\bar{d})$ and the difference $x(\bar{d}-\bar{u})$. In this section, all PDFs are presented with uncertainties evaluated using the standard
$\chi^2$ tolerance $T^2 = \Delta\chi^2 = 1$. Enhanced tolerance is considered in Section~\ref{sec:globaltol}.
Also shown in these figures are the PDFs from a fit in which correlations of systematic uncertainties between the $V$+\,jets, $t\bar{t}$ (lepton\,+\,jets) and inclusive jets data sets
are not applied, with the exception of the luminosities for the data at the same centre-of-mass energy.\footnote{The $W$+\,jets and $Z$\,+\,jets data sets are always considered as a single data set with correlations applied and these correlations are maintained in the present exercise. Consideration of the impact of correlations between the $W$+\,jets and $Z$\,+\,jets data sets was studied in Ref.~\cite{Vjets}.}
The systematic uncertainties which are
correlated are specified in Table~\ref{tab:corrVtjets} and their treatment for the central fit is discussed in
Section~\ref{sec:corrwb}.
 
The differences
between the extracted PDFs with and without inter-data-set correlations are mostly in the $d$-type sector and can reach ${\sim}20\%$ for $x\bar{d}$ at large $x$. The difference $x(\bar{d}-\bar{u})$ and ratio $R_s$, shown in the bottom half of
Figure~\ref{fig:corr2}, are particularly sensitive to the use of these correlations.
 
\begin{figure*}
\begin{centering}
\includegraphics[width=0.48\textwidth]{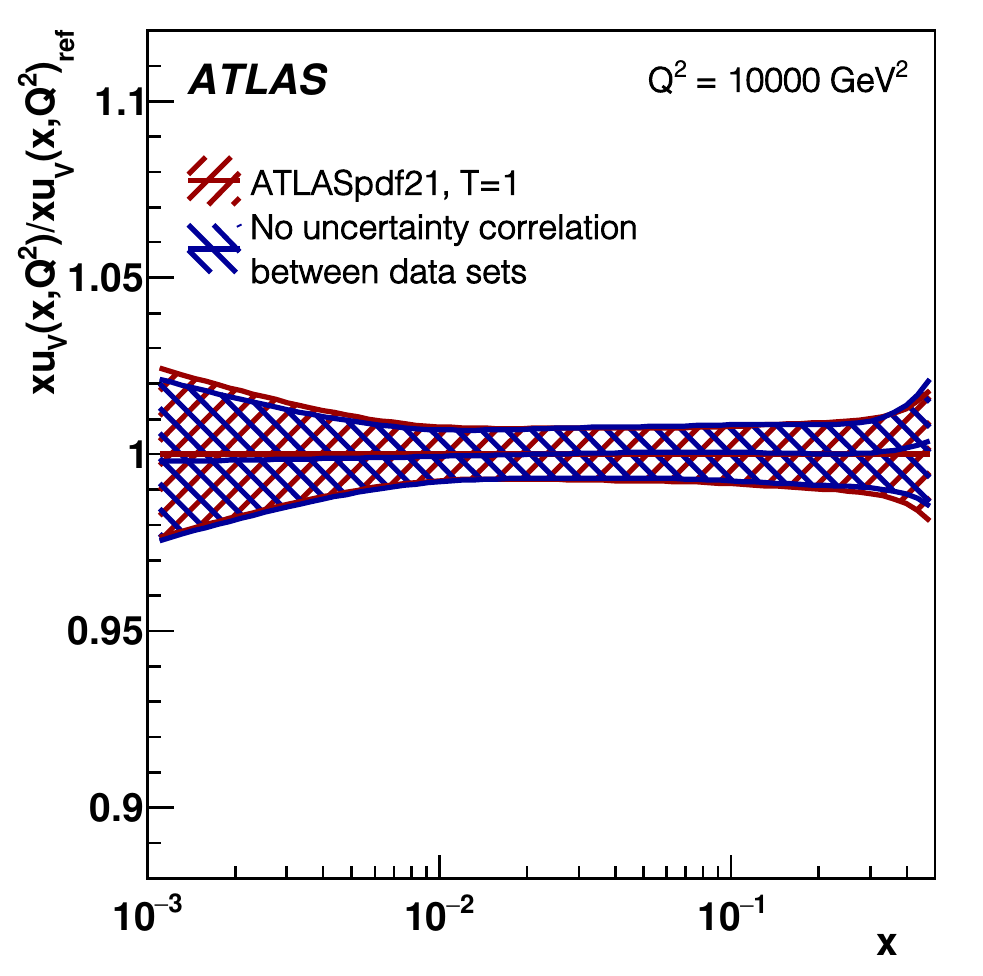}
\includegraphics[width=0.48\textwidth]{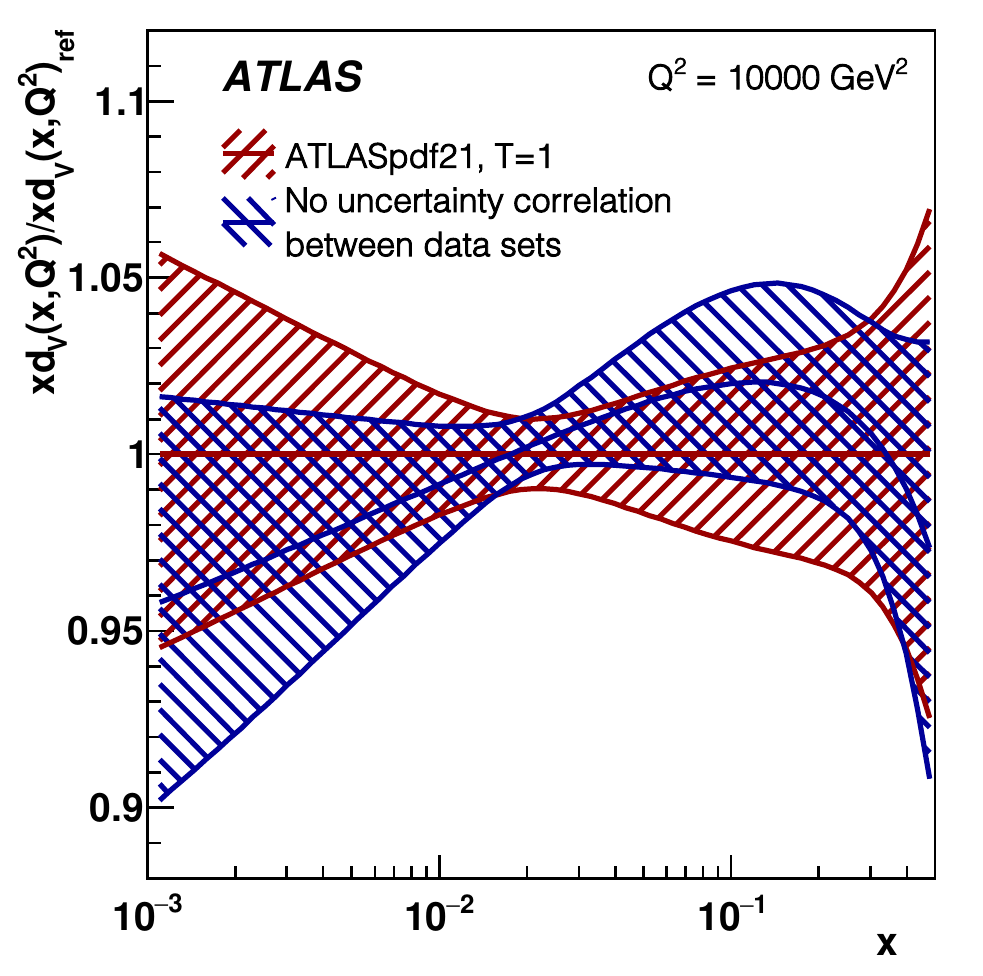}
\includegraphics[width=0.48\textwidth]{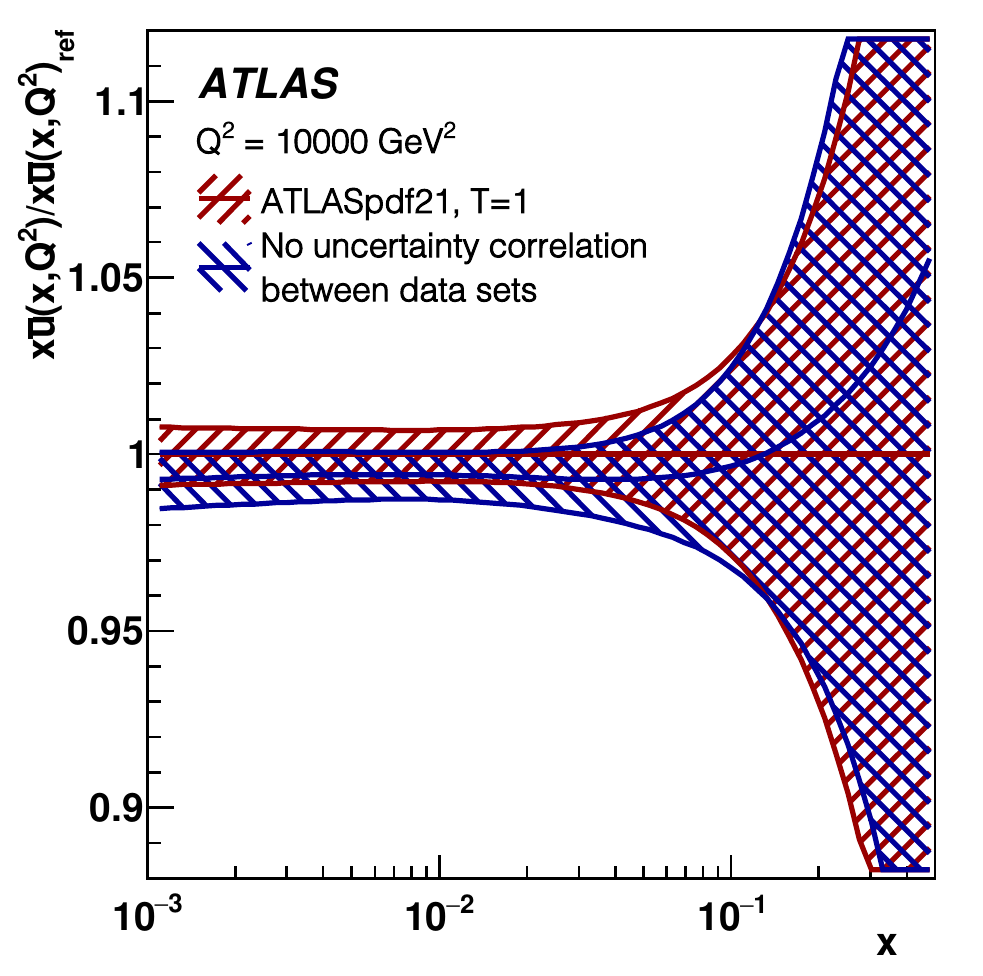}
\includegraphics[width=0.48\textwidth]{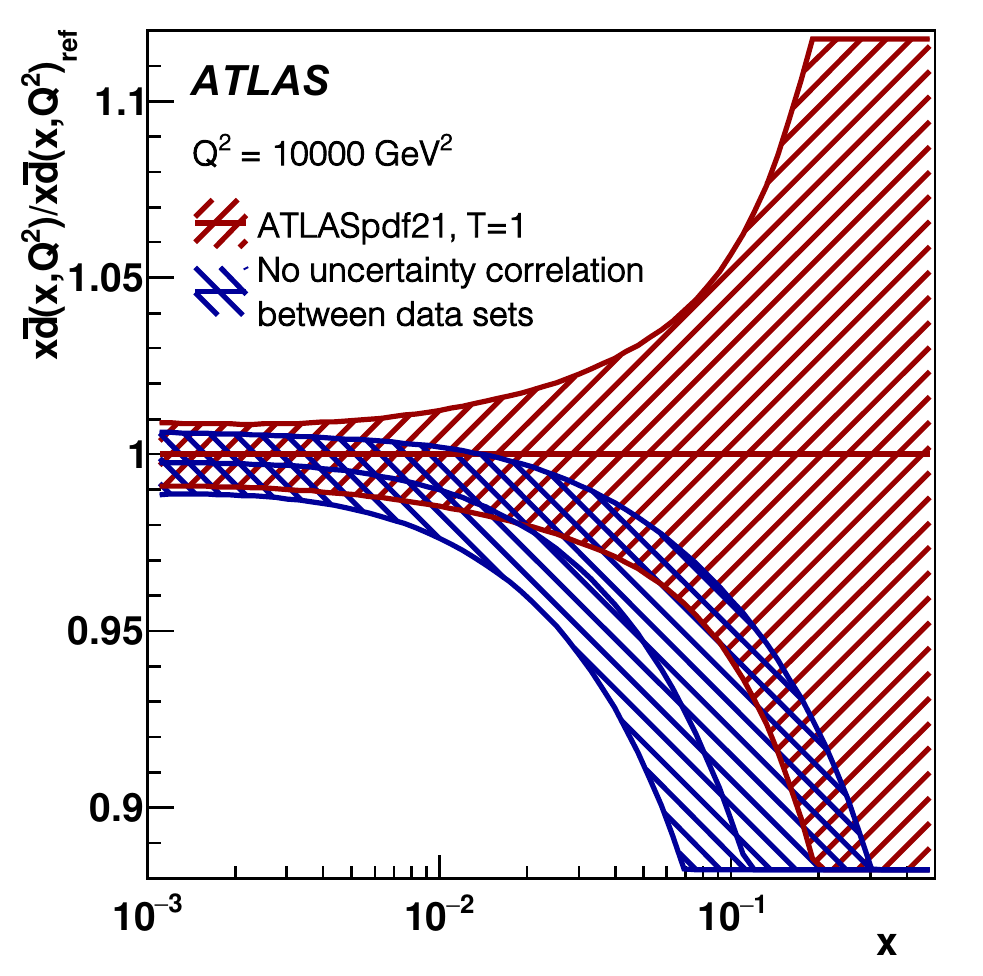}
\caption{Ratios of ATLASpdf21 PDFs extracted from a fit including correlations of systematic uncertainties between data sets to those extracted from a fit in which only the luminosity uncertainties for each centre-of-mass energy are correlated between data sets, at scale $Q^2 = \num{10000}$~\GeV$^2$.  Only experimental uncertainties are shown, evaluated with tolerance $T=1$. Top left: $xu_v$. Top right: $xd_v$. Bottom left: $x\bar{u}$. Bottom right: $x\bar{d}$.
\label{fig:corr1rat}
}
\end{centering}
\end{figure*}

\begin{figure*}
\begin{centering}
\includegraphics[width=0.48\textwidth]{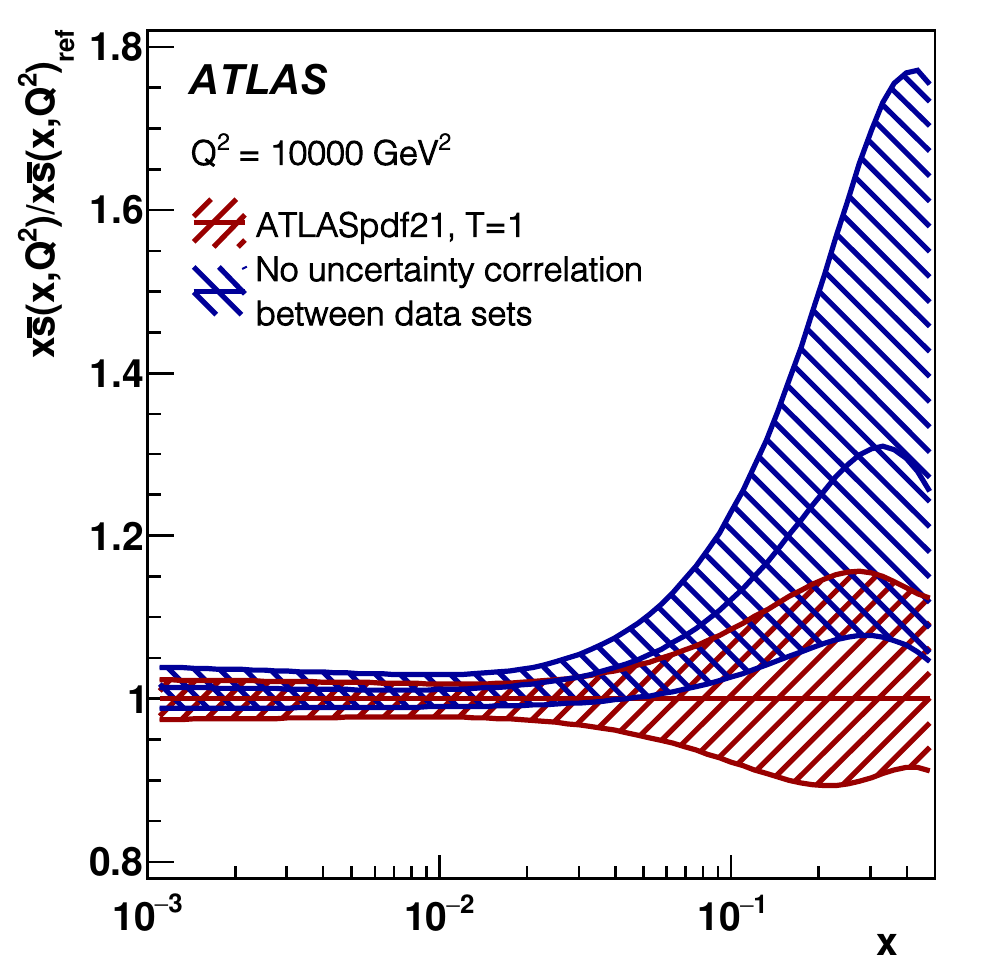}
\includegraphics[width=0.48\textwidth]{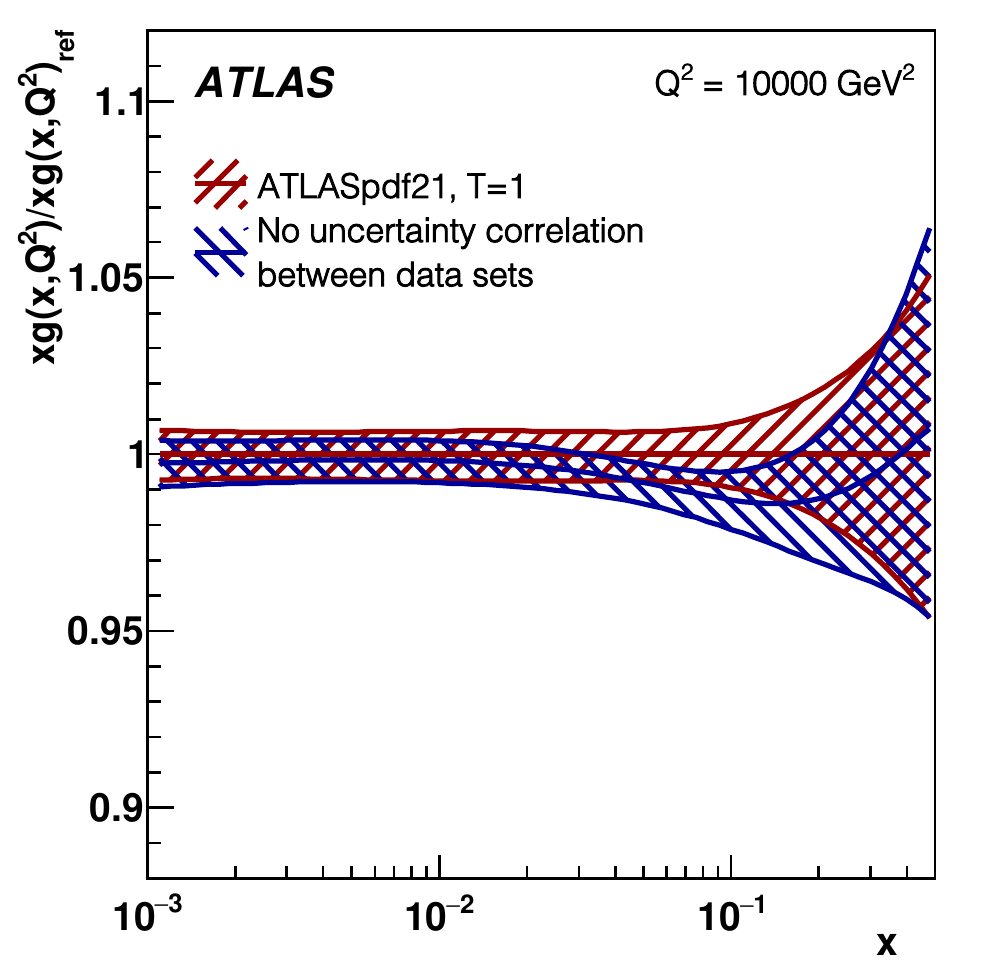}
\caption{Ratios of ATLASpdf21 PDFs extracted from a fit including correlations of systematic uncertainties between data sets to those extracted from a fit in which only the luminosity uncertainties for each centre-of-mass energy are correlated between data sets, at scale $Q^2 = \num{10000}$~\GeV$^2$. Only experimental uncertainties are shown, evaluated with tolerance $T=1$. Left: $xs$.
Right: $xg$.
\label{fig:corr2rat}
}
\end{centering}
\end{figure*}

It is important to examine these differences at a scale relevant for precision LHC physics and this is illustrated in Figures~\ref{fig:corr1rat} and~\ref{fig:corr2rat} where the ratios of the PDFs with and without the correlations are shown at $Q^2 = \num{10000}$~\GeV$^2$. Although these differences are not large compared to current experimental precision, they can nevertheless be important if $O(1\%)$ accuracy is to be a realistic goal for PDFs~\cite{2109.02653}.
 
These plots also allow a comparison of the size of the current experimental uncertainty of the PDFs with and without inclusion of inter-data-set correlations. The sizes are generally similar but the uncertainty in the high-$x$ $\bar{d}$, strange and gluon PDFs is larger if such correlations are not taken into account.

\clearpage
 
\subsection{Impact of each data set}
\label{sec:classes}
 
In this section the impact of each data set is considered. Only experimental uncertainties with tolerance $T=1$ are shown for these
comparisons. Full uncertainties including model and parameterisation variations are considered for the ATLASpdf21 fit in Section~\ref{sec:uncertainties}.
 
\subsubsection{Impact of $W,Z$ inclusive data}
 
\begin{figure*}[t]
\begin{centering}
\includegraphics[width=0.48\textwidth]{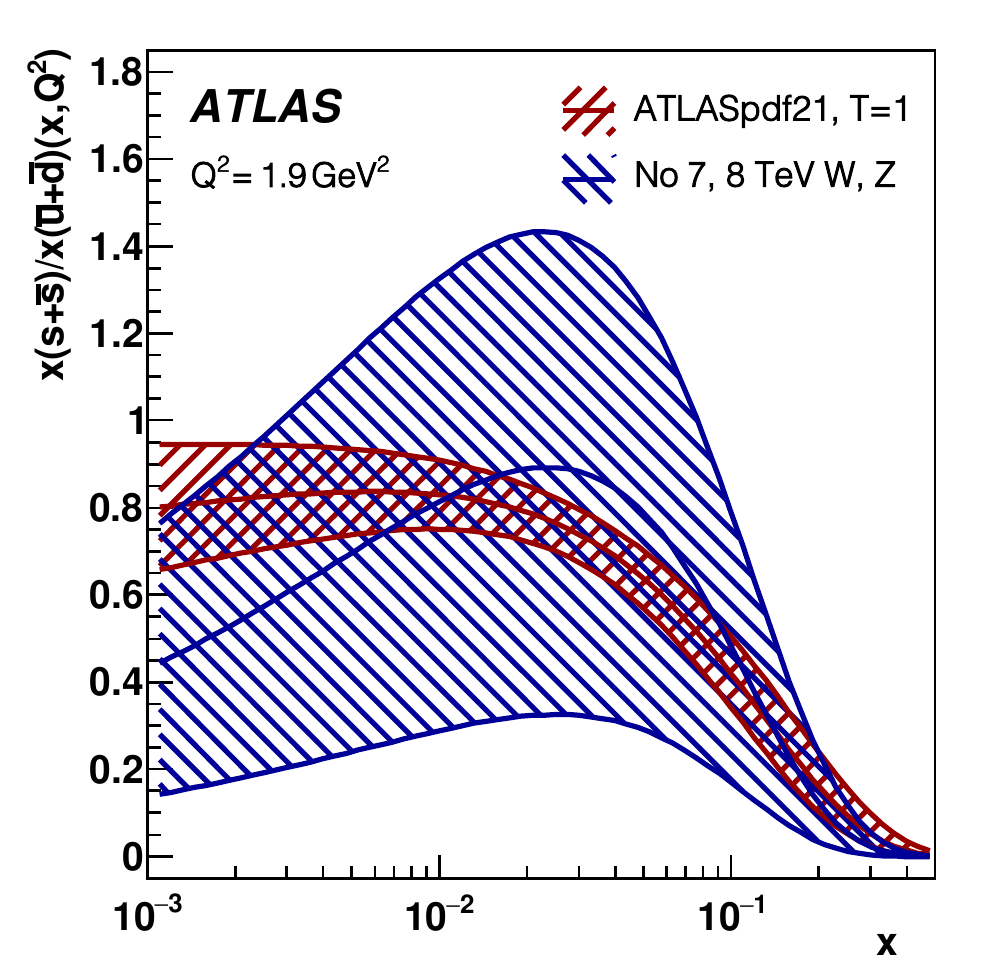}
\includegraphics[width=0.48\textwidth]{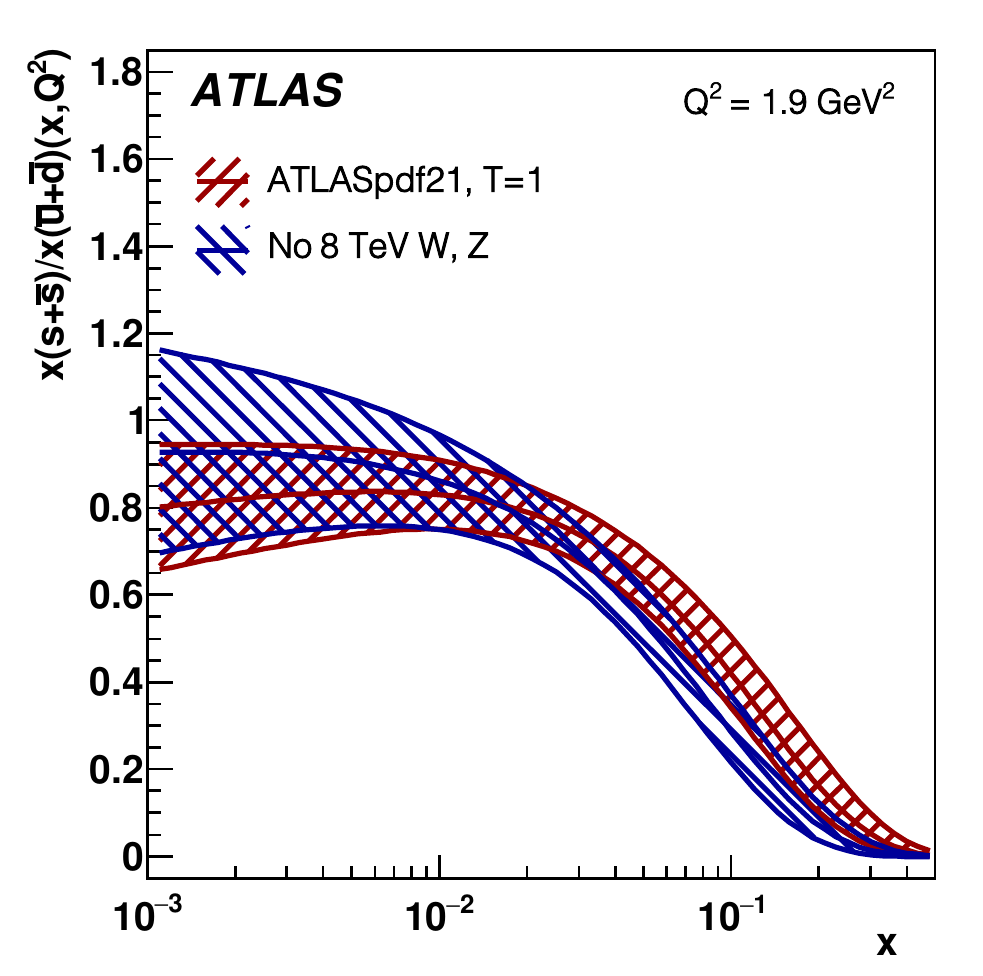}
\caption{ The PDF ratio $R_s=x(s+\bar{s})/x(\bar{u}+\bar{d})$ from ATLASpdf21 compared with $R_s$ for fits not including some of the $W,Z$ data sets. Only experimental uncertainties are shown, evaluated with tolerance $T=1$. Left: not including $W,Z$ data at both 7 and 8~\TeV. Right: not including $W,Z$ data at 8~\TeV.
\label{fig:noWZ78}
}
\end{centering}
\end{figure*}

Figure~\ref{fig:noWZ78} shows the ratio $R_s$ for the ATLASpdf21
fit and compared with a fit in which the inclusive $W,Z$ data at 7 and 8~\TeV\ are removed (left-hand plot), as well as to a fit in which only
$W,Z$ data at 8~\TeV\ are removed  (right-hand plot). It is clear that without $W,Z$ inclusive data the ratio $R_s$
cannot be determined reliably. Once $W,Z$ data at 7~\TeV\ are input the determination improves considerably, but the inclusive $W,Z$ data at 8~\TeV\ still
add information.
 
\begin{figure*}[t!]
\begin{centering}
\includegraphics[width=0.48\textwidth]{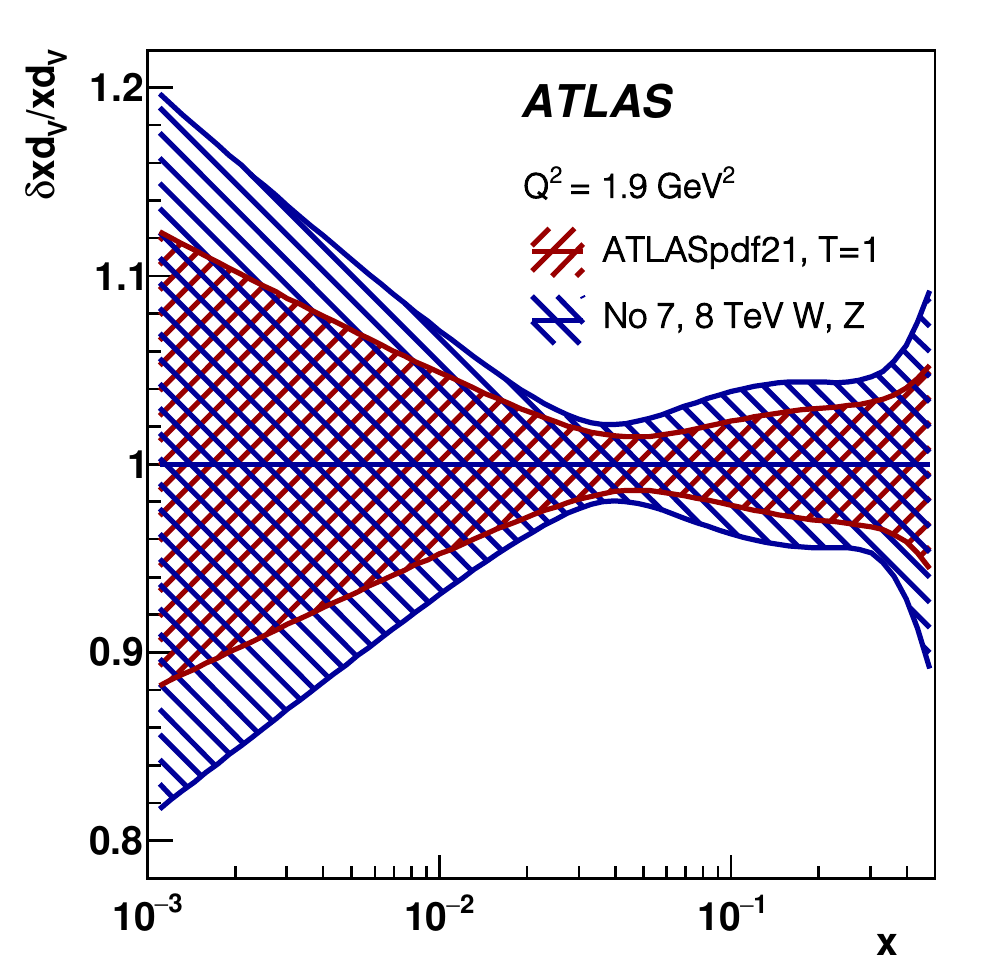}
\includegraphics[width=0.48\textwidth]{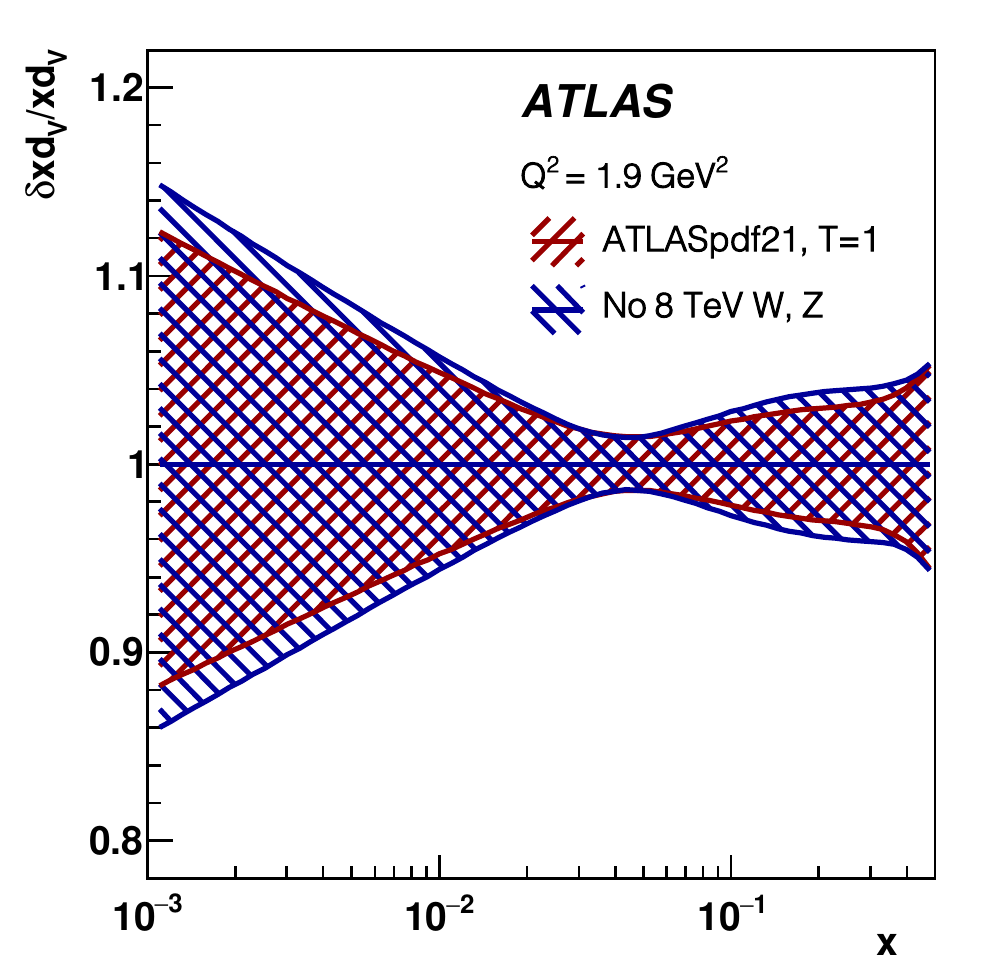}
\includegraphics[width=0.48\textwidth]{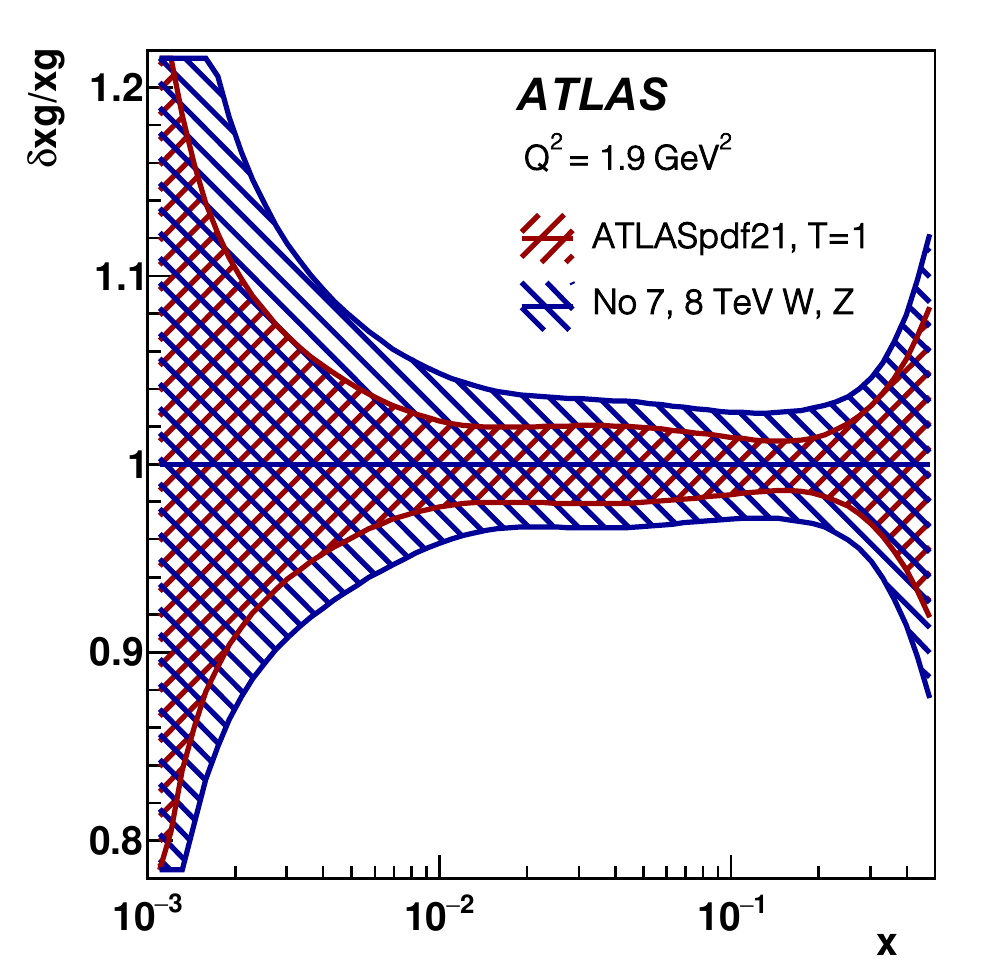}
\includegraphics[width=0.48\textwidth]{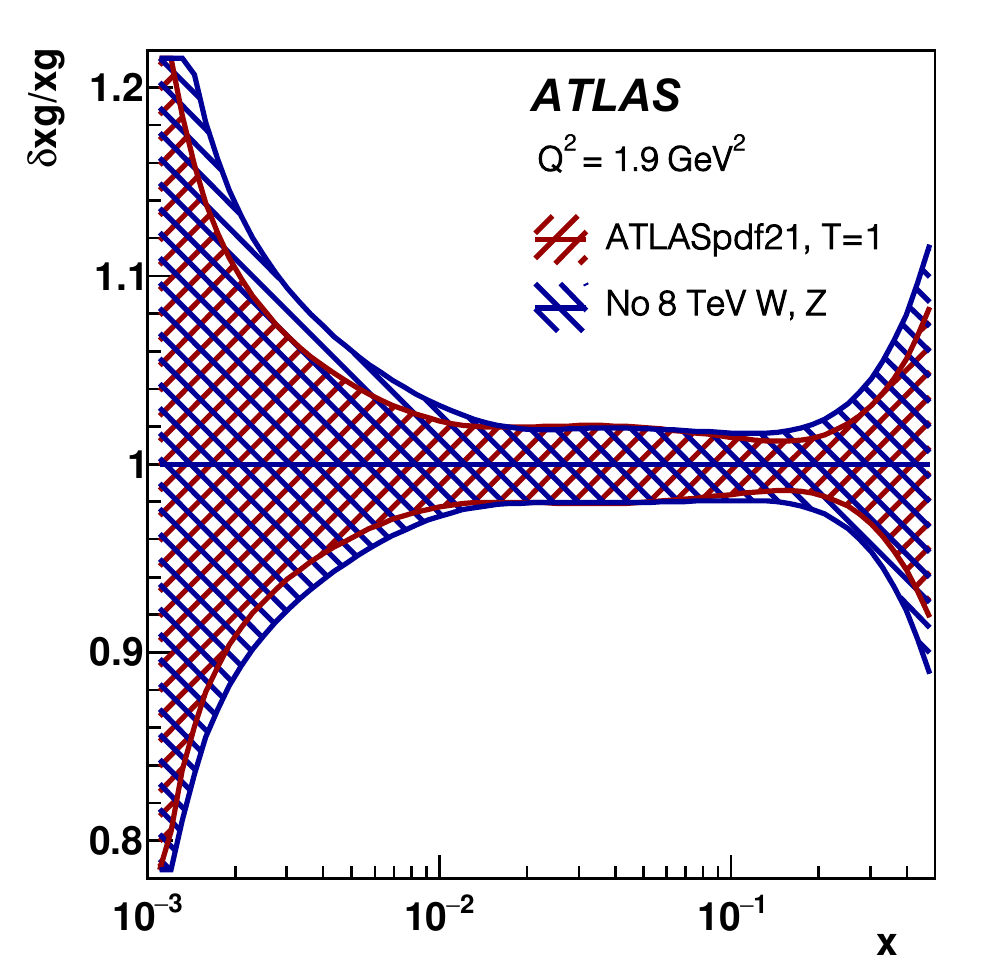}
\caption{Relative uncertainties in ATLASpdf21 $xd_v$ and $xg$ compared with fits not including some of the $W,Z$ data sets. Only experimental uncertainties are shown, evaluated with tolerance $T=1$. Top: $xd_v$ uncertainties, (left) not including inclusive $W,Z$  data at both 7 and 8~\TeV, (right) not including inclusive $W,Z$ data at 8~\TeV. Bottom: $xg$ uncertainties, (left) not including inclusive $W,Z$  data at both 7 and 8~\TeV, (right) not including inclusive $W,Z$ data at 8~\TeV.
\label{fig:noWZ78valglu}
}
\end{centering}
\end{figure*}

In contrast, the valence and gluon PDFs are still reasonably well determined without any $W,Z$ data but the input of these data decreases their uncertainties significantly, as illustrated for the $xd_v$ and $xg$  PDFs on the left-hand side of Figure~\ref{fig:noWZ78valglu}.
On the right-hand side of Figure~\ref{fig:noWZ78valglu} the decrease in the uncertainties of the $xd_v$ and $xg$ PDFs from removing only the $W,Z$ data taken at 8~\TeV\ is illustrated, showing that the major decrease comes from retaining the $W,Z$ data taken at 7~\TeV.
 
\begin{figure*}
\begin{centering}
\includegraphics[width=0.48\textwidth]{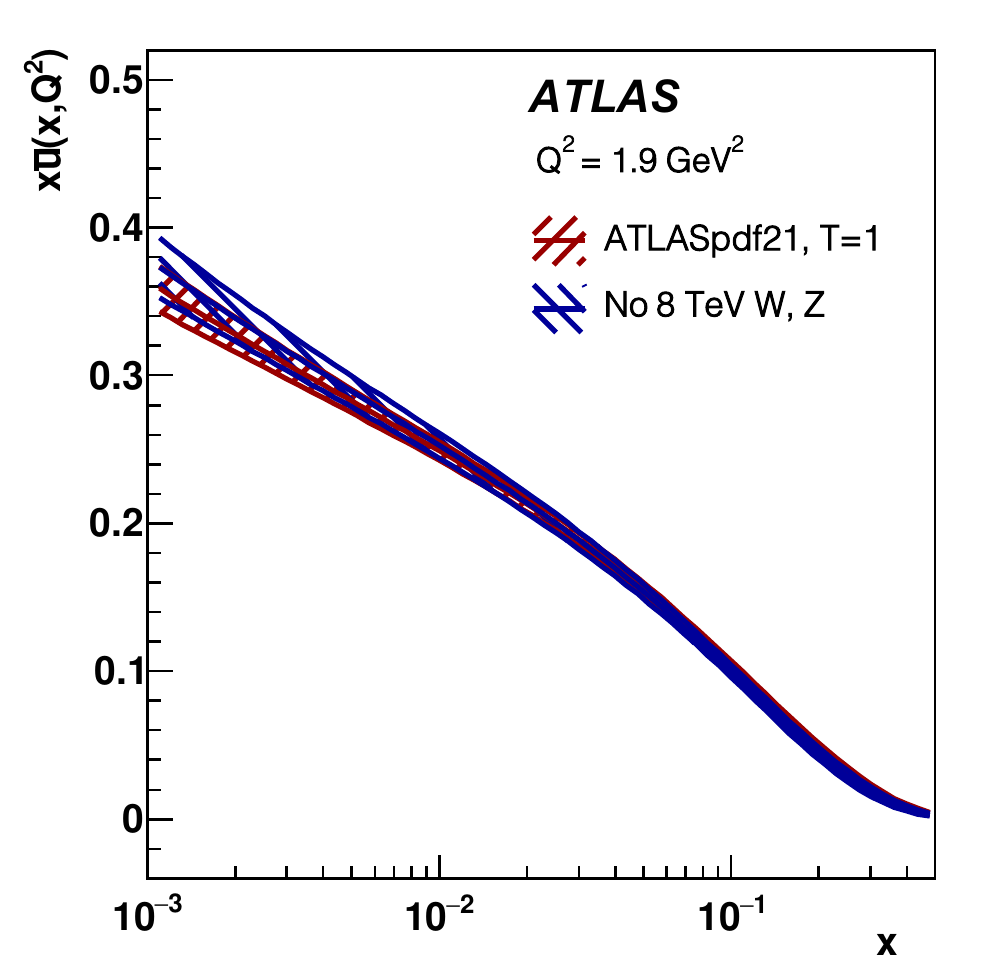}
\includegraphics[width=0.48\textwidth]{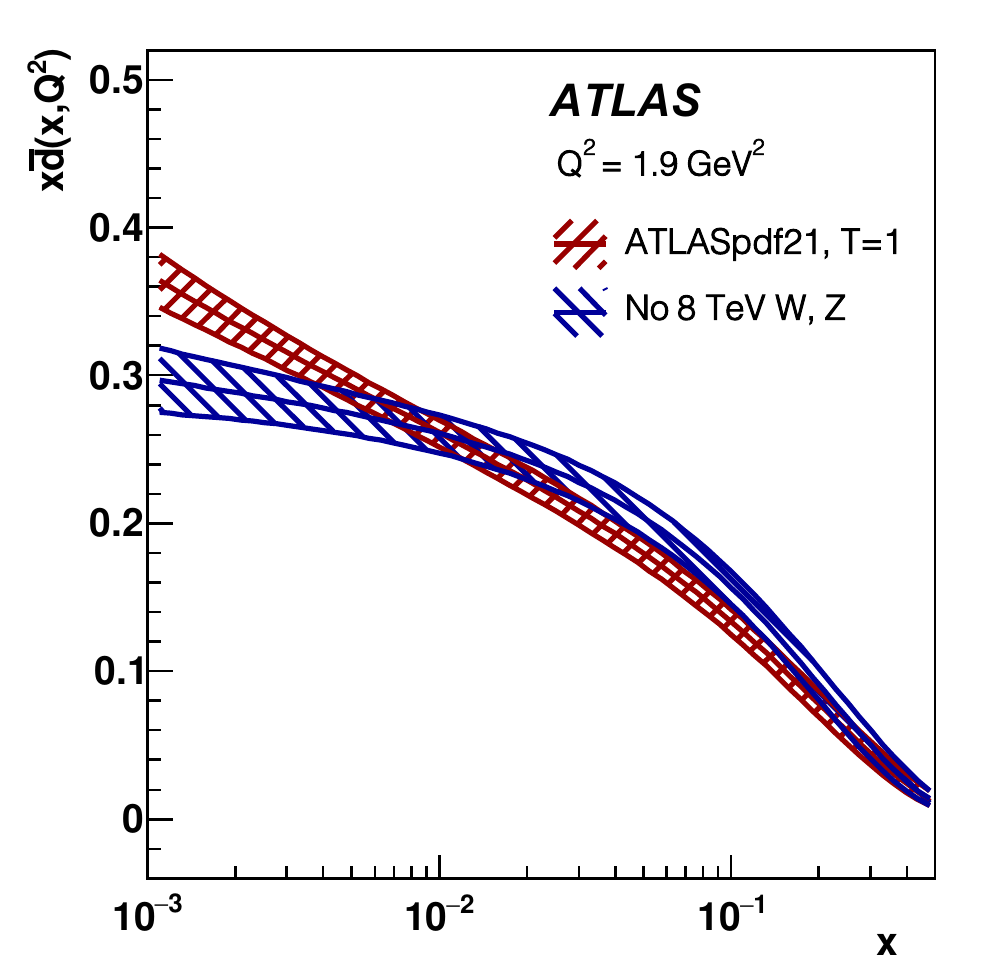}
\includegraphics[width=0.48\textwidth]{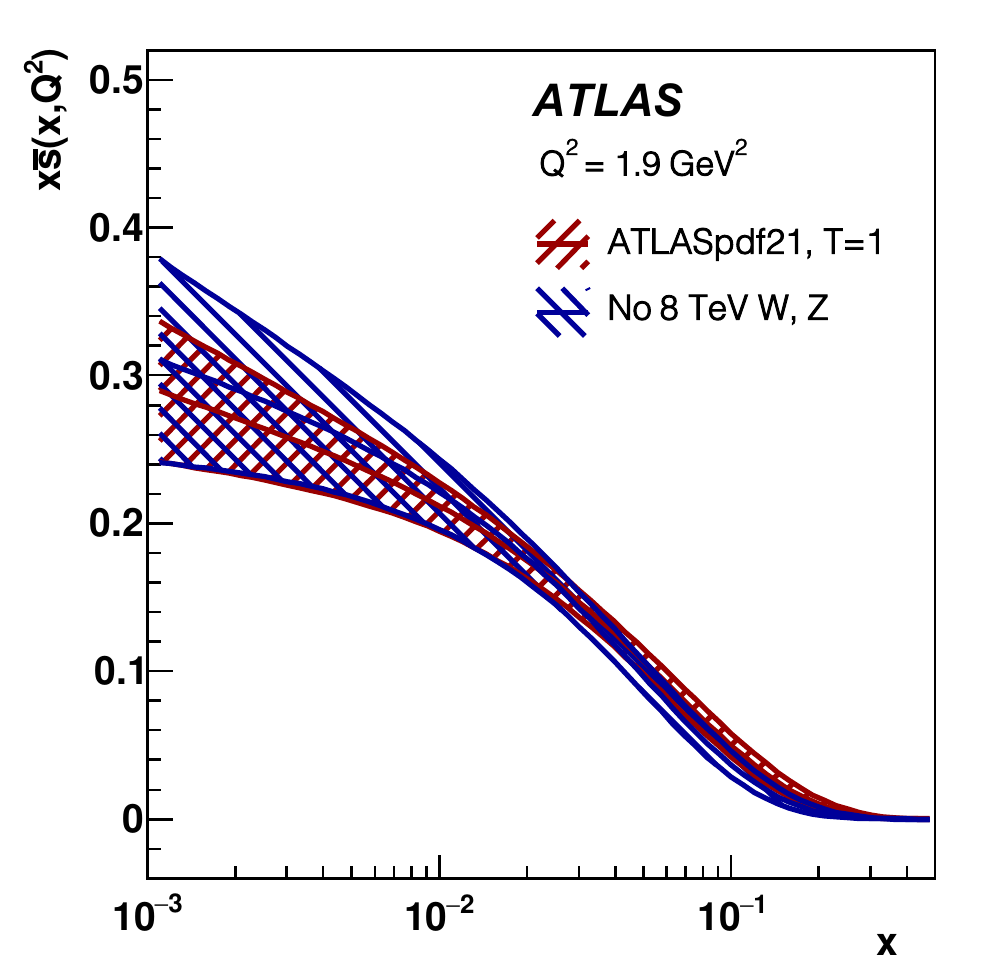}
\caption{ATLASpdf21 PDFs compared with those from a fit not including the $W,Z$ data at 8~\TeV. Only experimental uncertainties are shown, evaluated with tolerance $T=1$. Top left: $x\bar{u}$. Top right: $x\bar{d}$. Bottom: $x\bar{s}$.
\label{fig:noWZ8sea}
}
\end{centering}
\end{figure*}

However, the $W,Z$ data taken at 8~\TeV\ have a major role to play in ensuring that $x\bar{u} \sim x\bar{d}$ holds at
low~$x$, even though this constraint is not imposed. Without them, one observes $x\bar{d} < x\bar{u}$ at low~$x$, as seen in Figure~\ref{fig:noWZ8sea}. These data
also somewhat reduce the low-$x$ strange distribution and harden the high-$x$ strange distribution, while softening the high-$x$ $x\bar{d}$ distribution, as also shown in Figure~\ref{fig:noWZ8sea}.
 
There is mild tension between the $W,Z$ data at 8~\TeV\ and the $W,Z$ data at 7~\TeV.
The \emph{partial} $\chi^2/\mathrm{NDP}$ for the $W,Z$ data at 7~\TeV\ decreases from $68/55$ to $50/55$ if the $W,Z$ data at 8~\TeV\ are excluded from the fit,
and the \emph{partial} $\chi^2/\mathrm{NDP}$ for the $W,Z$ data at 8~\TeV\ decreases from $239/206$ to $222/206$
if the $W,Z$ data at 7~\TeV\ are excluded from the fit.  These increases in $\chi^2$ are most
pronounced for the 7~\TeV\ c-c data around the $Z$ mass-peak  (66--116~\GeV) and for the mass bins around the $Z$ peak in 8~\TeV\ data. As already remarked, theoretical scale uncertainties for
$W,Z$ data at both 7 and 8~\TeV\ are added to the fit uncertainties.
If these uncertainties are not added the tension between $W,Z$ data at 7 and 8~\TeV\ increases.
The \emph{partial} $\chi^2/\mathrm{NDP}$ for $W,Z$ data at 7~\TeV\ increases to $80/55$ and the \emph{partial} $\chi^2/\mathrm{NDP}$ for $W,Z$ data at 8~\TeV\ increases to $268/206$ if both data sets are included in the fit and scale uncertainties are not applied.
The differences between the PDFs are not large, whether or not scale uncertainties are applied, compared to the current experimental precision. However, consideration of such theoretical uncertainties is important if accuracy to 1\% is the ultimate goal for the PDFs.
N$^{3}$LO calculations~\cite{2107.09085} indicate that higher-order corrections are likely to be larger than our current estimate of scale uncertainties.
A further study of scale uncertainties is given in Appendix~\ref{sec:scale}.

\subsubsection{Impact of \boldmath$V$+\,jets data}
 
\begin{figure*}
\begin{centering}
\includegraphics[width=0.48\textwidth]{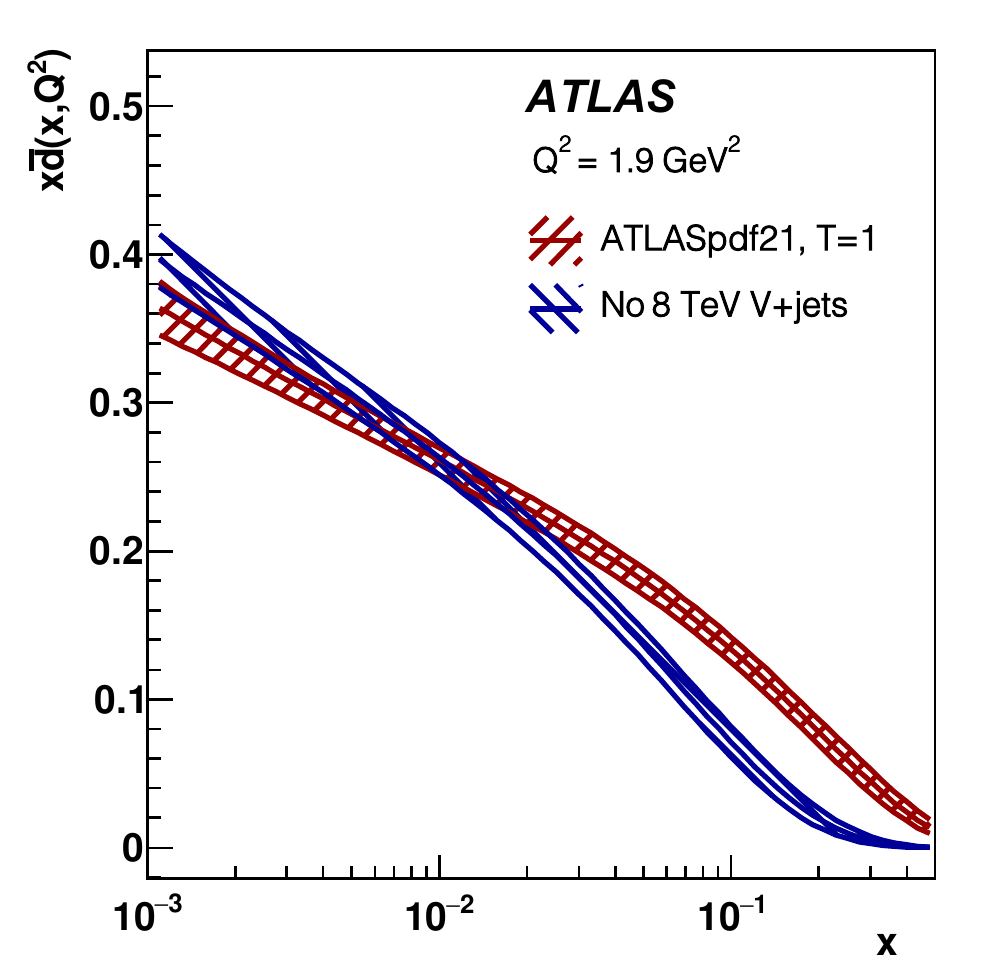}
\includegraphics[width=0.48\textwidth]{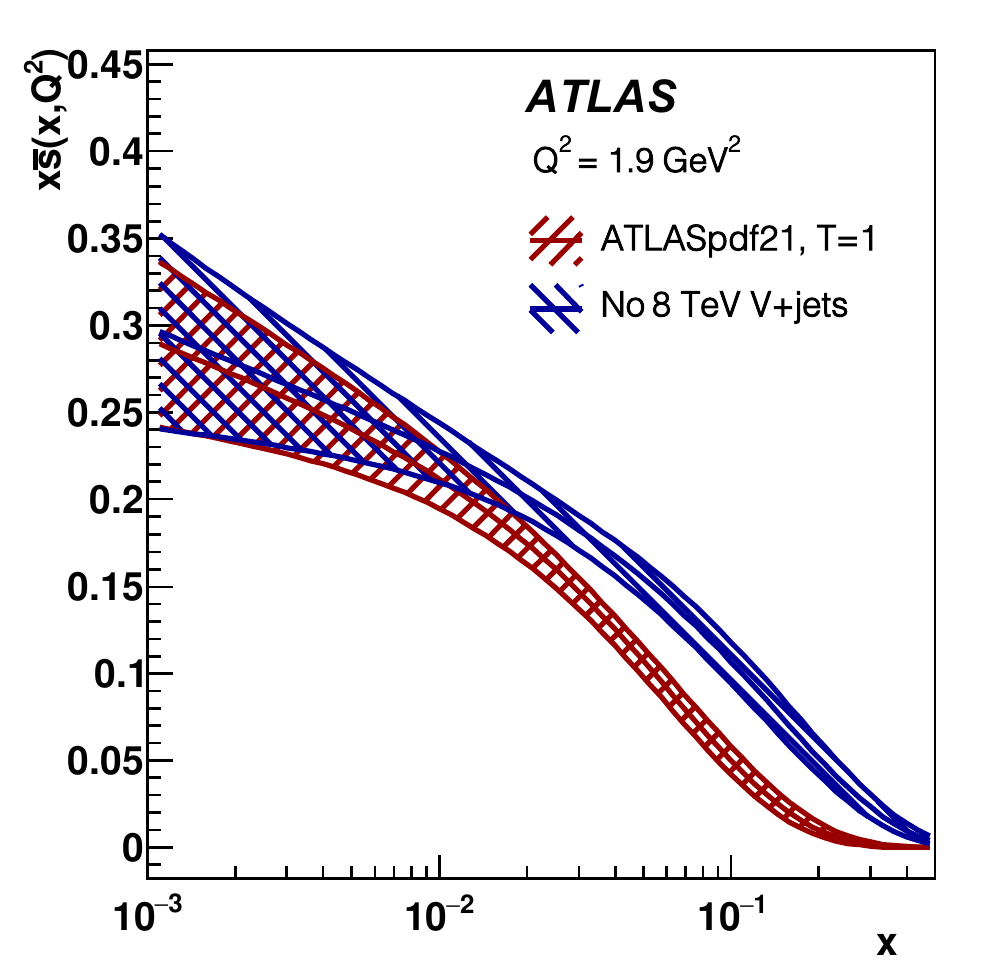}
\includegraphics[width=0.48\textwidth]{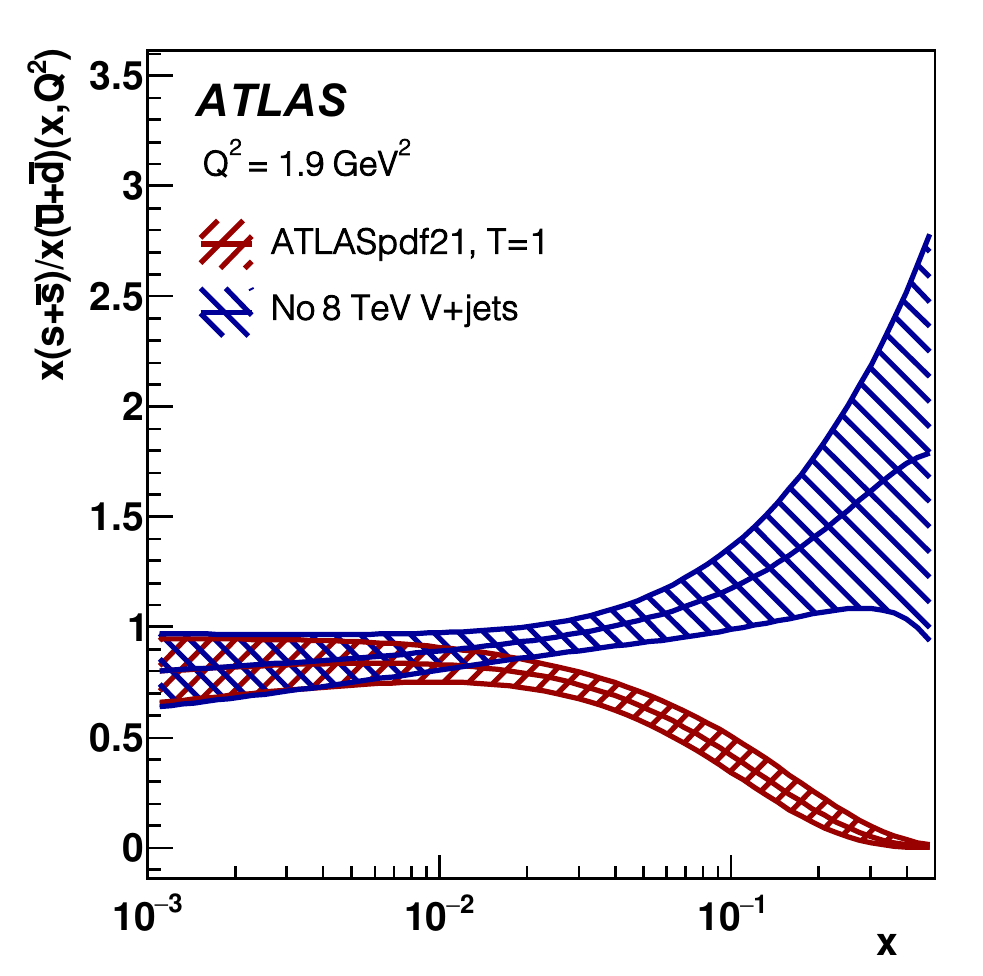}
\includegraphics[width=0.48\textwidth]{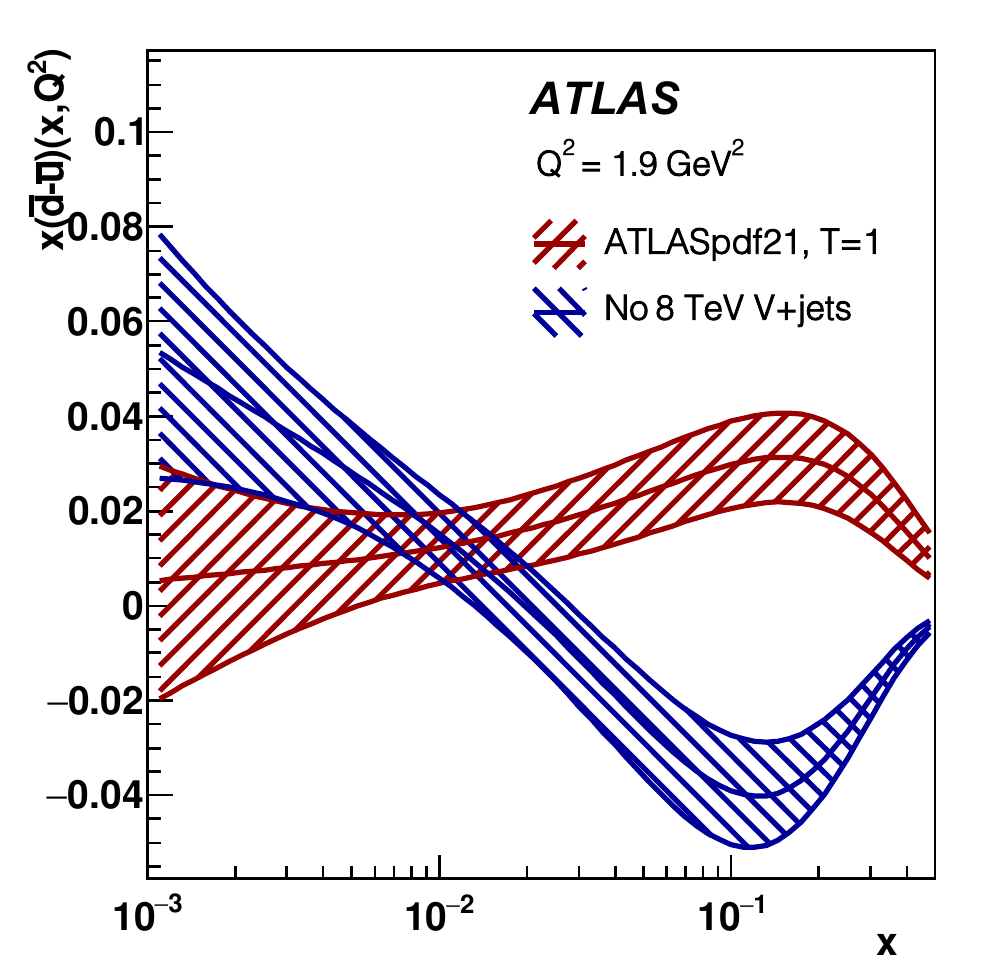}
\caption{ ATLASpdf21 PDFs compared with those from a fit not including $V$+\,jets data. Only experimental uncertainties are shown, evaluated with tolerance $T=1$. Top left: $x\bar{d}$. Top right: $x\bar{s}$. Bottom left: $R_s$. Bottom right: $x(\bar{d}-\bar{u})$.\label{fig:noVjetssbar}
}
\end{centering}
\end{figure*}

\begin{figure*}
\begin{centering}
\includegraphics[width=0.48\textwidth]{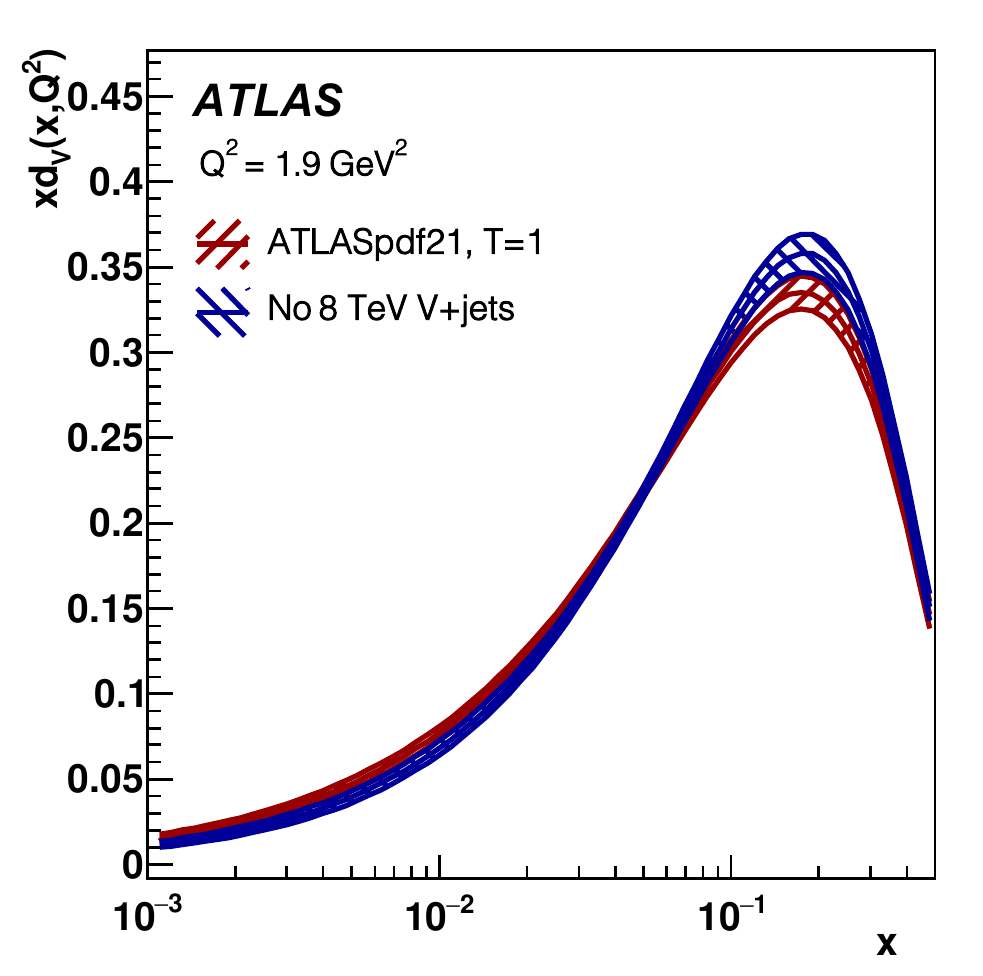}
\includegraphics[width=0.48\textwidth]{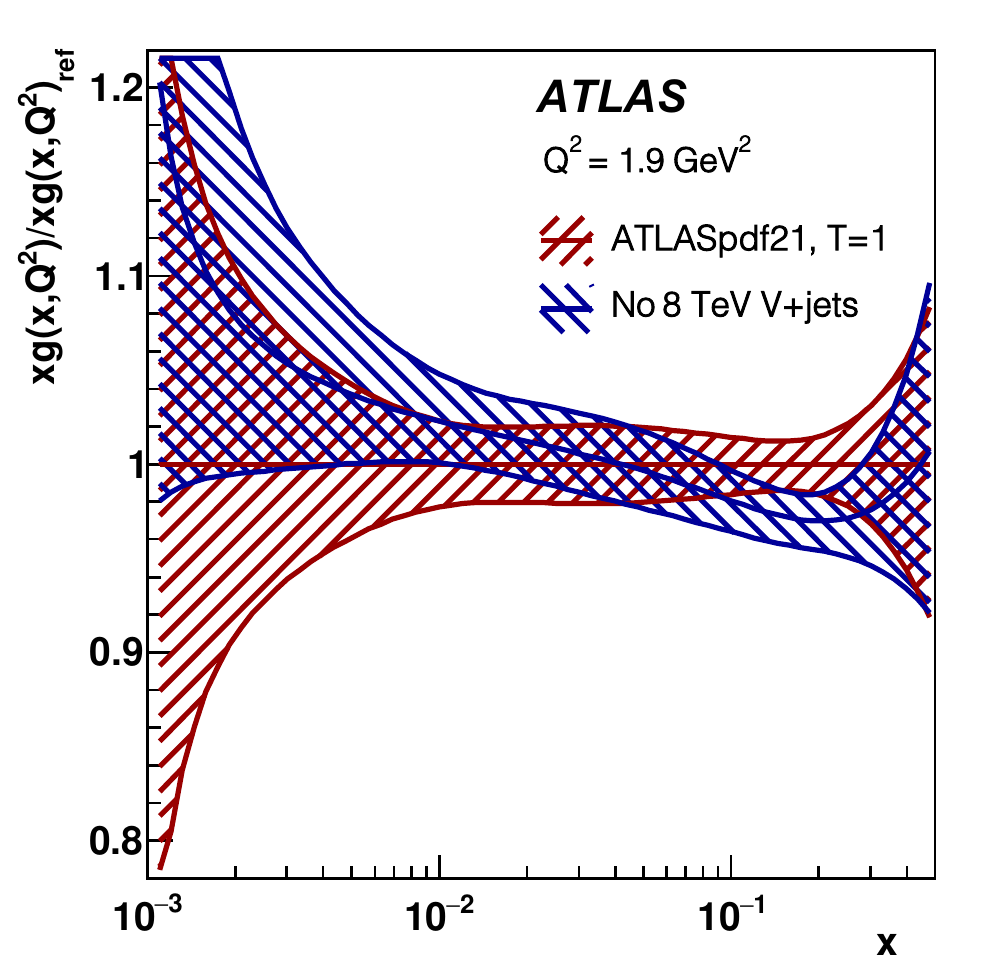}
\caption{ATLASpdf21 PDFs compared with those from a fit not including $V$+\,jets data. Only experimental uncertainties are shown, evaluated with tolerance $T=1$. Left: $xd_v$. Right: the ratio of $xg$ for the two fits.
\label{fig:noVjetsdvg}
}
\end{centering}
\end{figure*}

The impact of the $V$+\,jets data is shown in Figures~\ref{fig:noVjetssbar} and ~\ref{fig:noVjetsdvg}.
There are significant changes in the $x\bar{d}$ and $x\bar{s}$ PDF shapes such that
the high-$x$ $x\bar{s}$ and $x\bar{d}$ PDFs become softer and
harder, respectively, with the input of $V$+\,jets data. Because of the change in the $x\bar{d}$ shape, the difference
$x(\bar{d}-\bar{u})$ is also strongly affected. This is shown in Figure~\ref{fig:noVjetssbar}.
This figure also shows that, without the $V$+\,jets data, there is little information about the ratio $R_s$ at high~$x$.
The changes in Figure~\ref{fig:noVjetssbar} are large because the $V$+\,jets data resolve a double minimum in parameter
space such that the fit now prefers a hard $x\bar{d}$ and soft $x\bar{s}$ at high~$x$, whereas it previously had an additional minimum with
$x\bar{d} \sim x\bar{s}$ at high~$x$, which was marginally preferred. These results are similar to those already seen and fully explained
in the ATLASepWZVjets20 PDF analysis~\cite{Vjets} and the double minimum may be seen in Figure 5 of that paper. A fit with only HERA data, or with HERA plus ATLAS $W,Z$ data at 7~\TeV\ (ATLASepWZ20), prefers the minimum with $x\bar{d} \sim x\bar{s}$ at high~$x$, but once ATLAS $V$ + jets data at 8 TeV are added to the fit the double minimum with a hard $x\bar{d}$ and soft $x\bar{s}$ at high~$x$ is preferred
and the previous minimum disappears. In Figure 3 of Ref.~\cite{Vjets}, it can also be seen that for ATLASepWZ20 (the fit without $V$+\,jets data)  the full uncertainties, including model and parameterisation variations, are very large because they cover both minima, whereas the full
uncertainties of the ATLASepWZVjets20 fit (including $V$ + jets data) are much smaller because there is no double minimum. There is thus no strong inconsistency between
the fits with or without the $V$+\,jets data because of the large uncertainties of the fit without the $V$ + jets data. This also applies in the
present analysis, but in Figure~\ref{fig:noVjetssbar}, only experimental uncertainties are shown. The full uncertainties for the fit without $V$ + jets data in our present analysis are of little interest and are not presented. Model and parametrisation uncertainties for the ATLASpdf21 fit will be discussed in Section~\ref{sec:uncertainties}.
These $V$+\,jets data also effect a modest change in the $xd_v$ shape and the gluon PDF shape as shown in Figure~\ref{fig:noVjetsdvg}. The $u$-type quarks are not strongly affected by these data and thus they are not shown.
There is no tension between the $V$+\,jets data at 8~\TeV\ and other data sets in the fit, since all data are fitted well at the minimum chosen by these $V$+\,jets data.
\clearpage
 
\subsubsection{Impact of $t\bar{t}$ data}
\begin{figure*}[t!]
\begin{centering}
\includegraphics[width=0.48\textwidth]{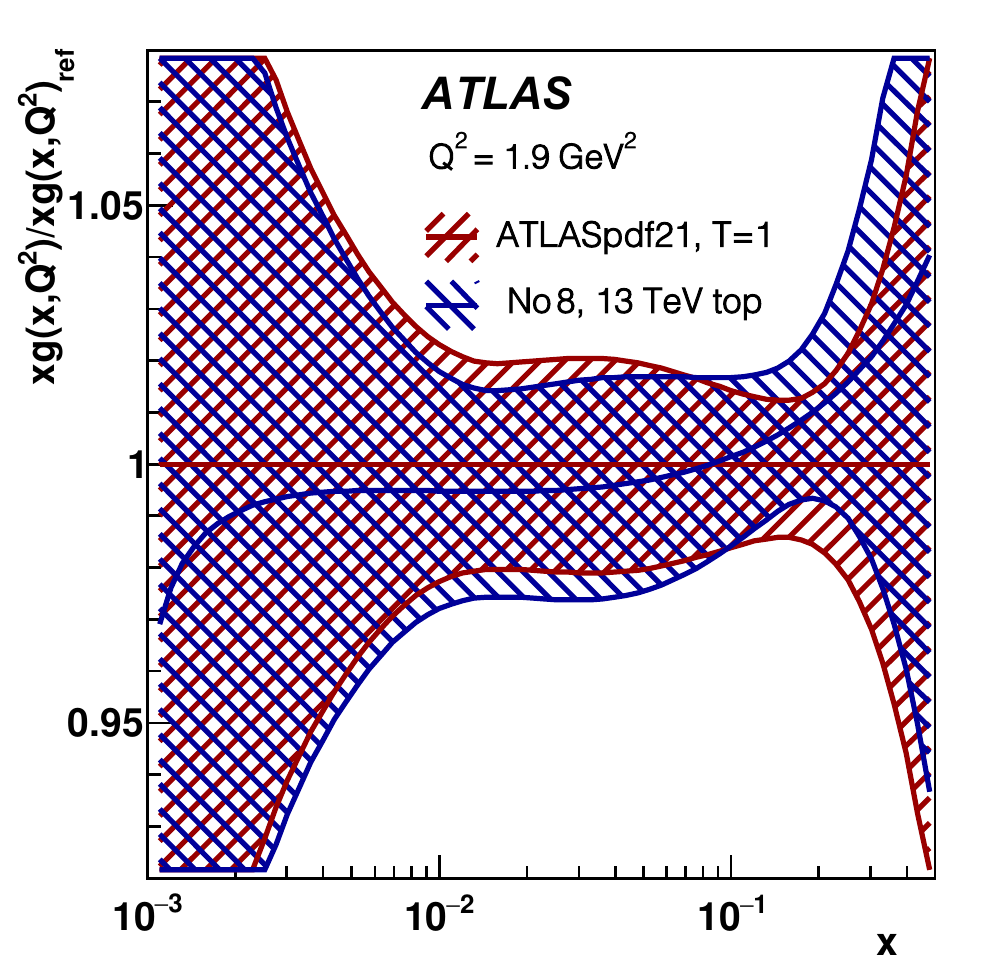}
\includegraphics[width=0.48\textwidth]{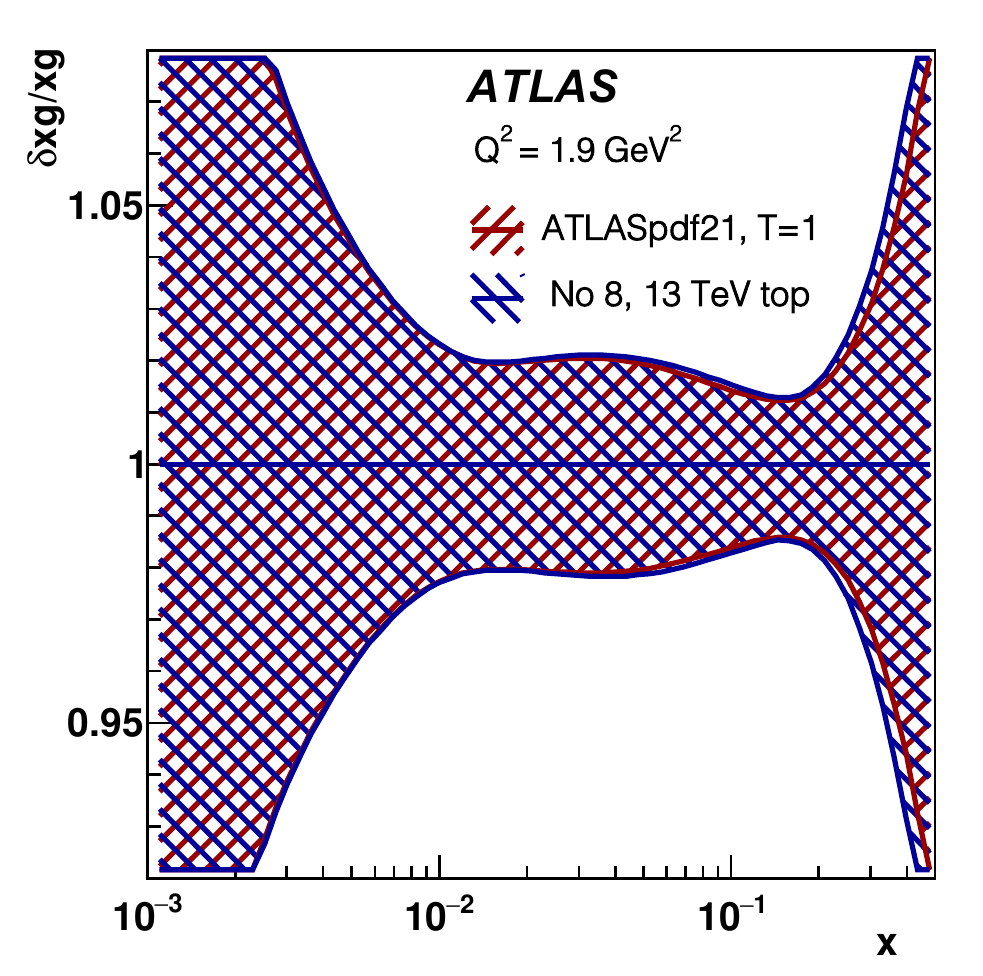}
\includegraphics[width=0.48\textwidth]{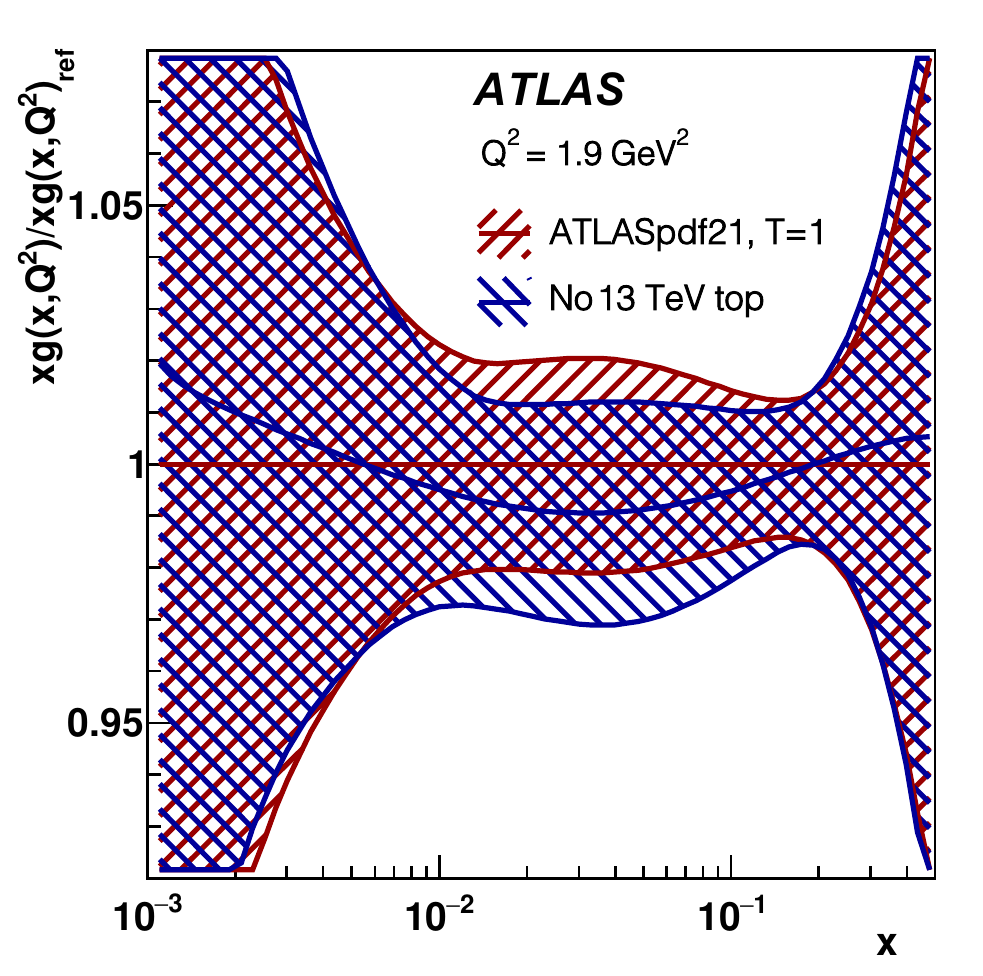}
\includegraphics[width=0.48\textwidth]{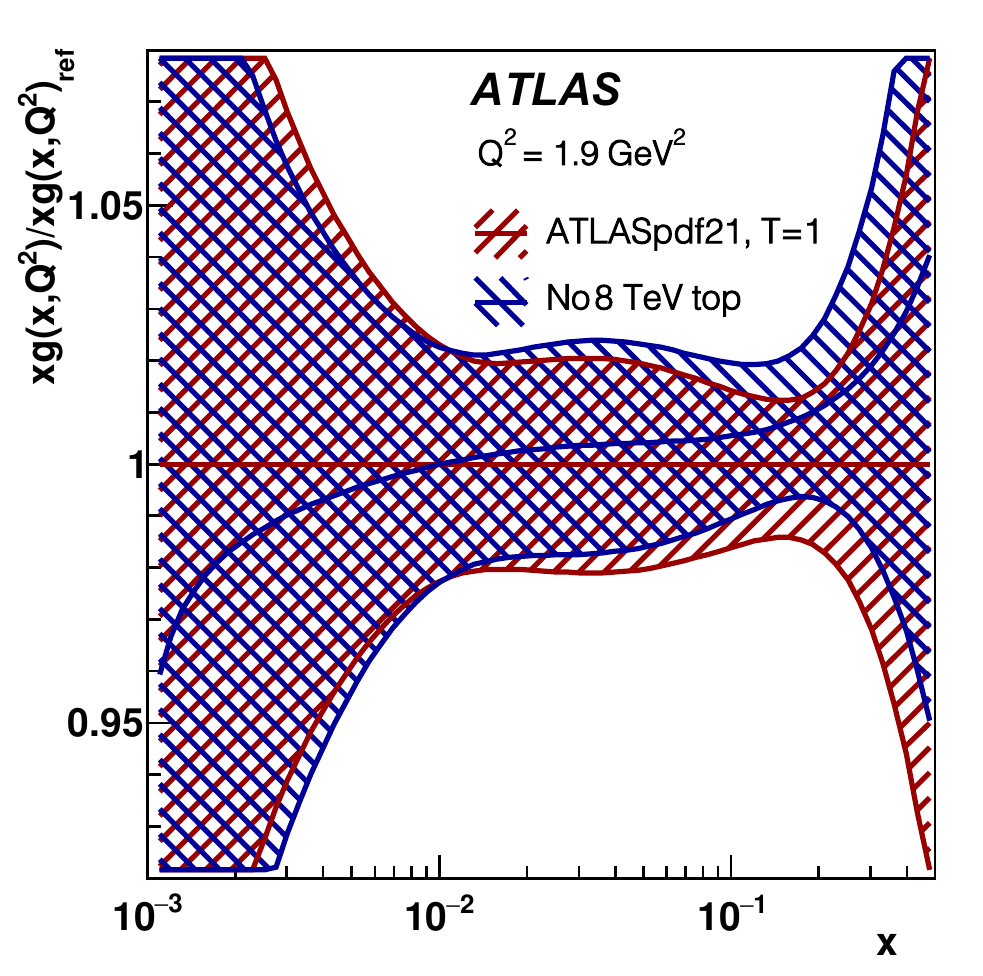}
\caption{ATLASpdf21 $xg$ PDF compared with $xg$ for a fit not including various $t\bar{t}$ data sets. Only experimental uncertainties are shown, evaluated with tolerance $T=1$. Top left: the shape ratio to  a fit not including $t\bar{t}$ data at 8 and 13~\TeV. Top right: the ratio to a fit not
including $t\bar{t}$ data at 8 and 13~\TeV\ for which both distributions are centred on unity. Bottom left: the shape ratio to  a fit not including $t\bar{t}$ data at 13~\TeV. Bottom right: the shape ratio to  a fit not including $t\bar{t}$ data at 8~\TeV.
\label{fig:notop}
}
\end{centering}
\end{figure*}
The impact of the $t\bar{t}$ data is shown in the top half of Figure~\ref{fig:notop}. The high-$x$ gluon distribution is mildly softened when the $t\bar{t}$
data are added to the fit. This effect is opposite to the one observed in the ATLASepWZtop18 fit.
This is because more data which harden the gluon PDF, in particular the $V$+\,jets and inclusive jet
data, are included in the present fit.
The more significant effect is in the uncertainties of the high-$x$ gluon distribution, which are reduced.
There is no significant tension between the $t\bar{t}$ data and the other data in the fit.
Figure~\ref{fig:notop} (bottom half) also shows the impact of removing only the $t\bar{t}$ data at 13~\TeV\ (left) or only the $t\bar{t}$ data at 8~\TeV\ (right). It is clear that the data at 8~\TeV\ have the stronger impact on the shape of the $xg$ PDF but both data sets contribute to a modest reduction in the uncertainties.
 
\subsubsection{Impact of photon data and inclusive jet data}
\begin{figure*}[]
\begin{centering}
\includegraphics[width=0.48\textwidth]{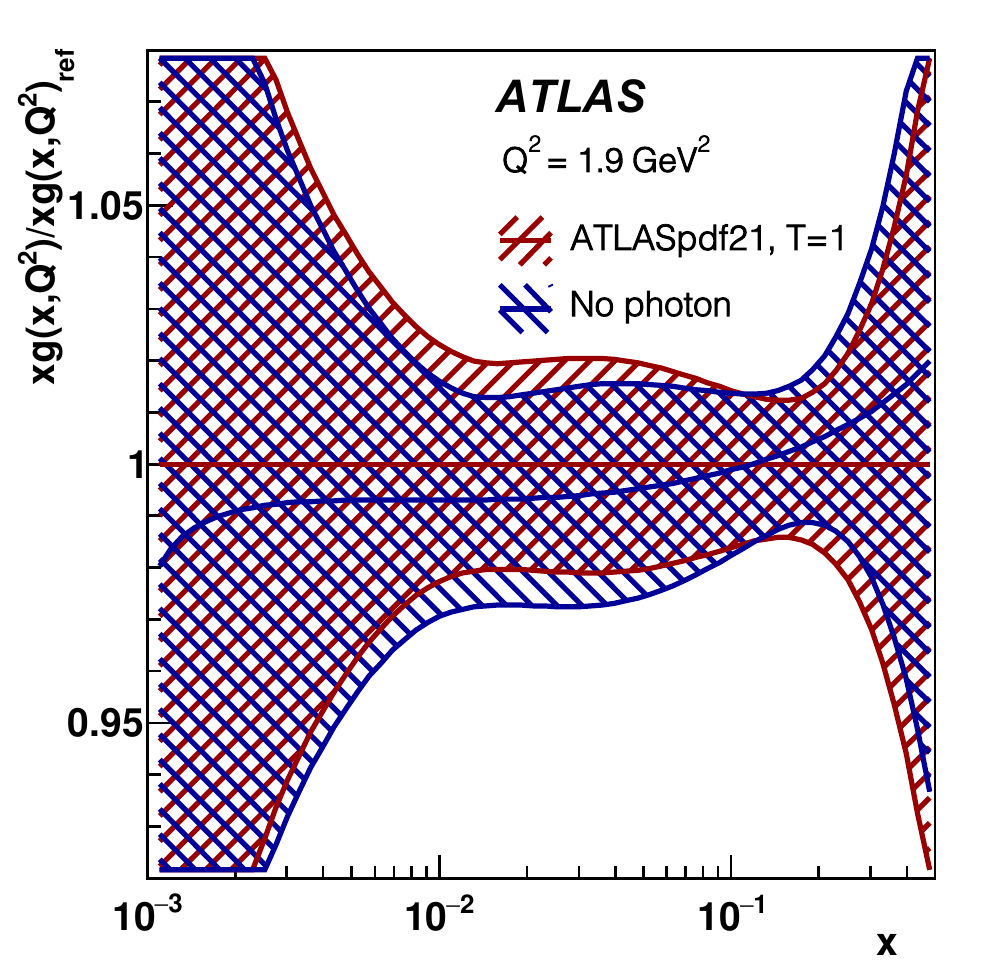}
\includegraphics[width=0.48\textwidth]{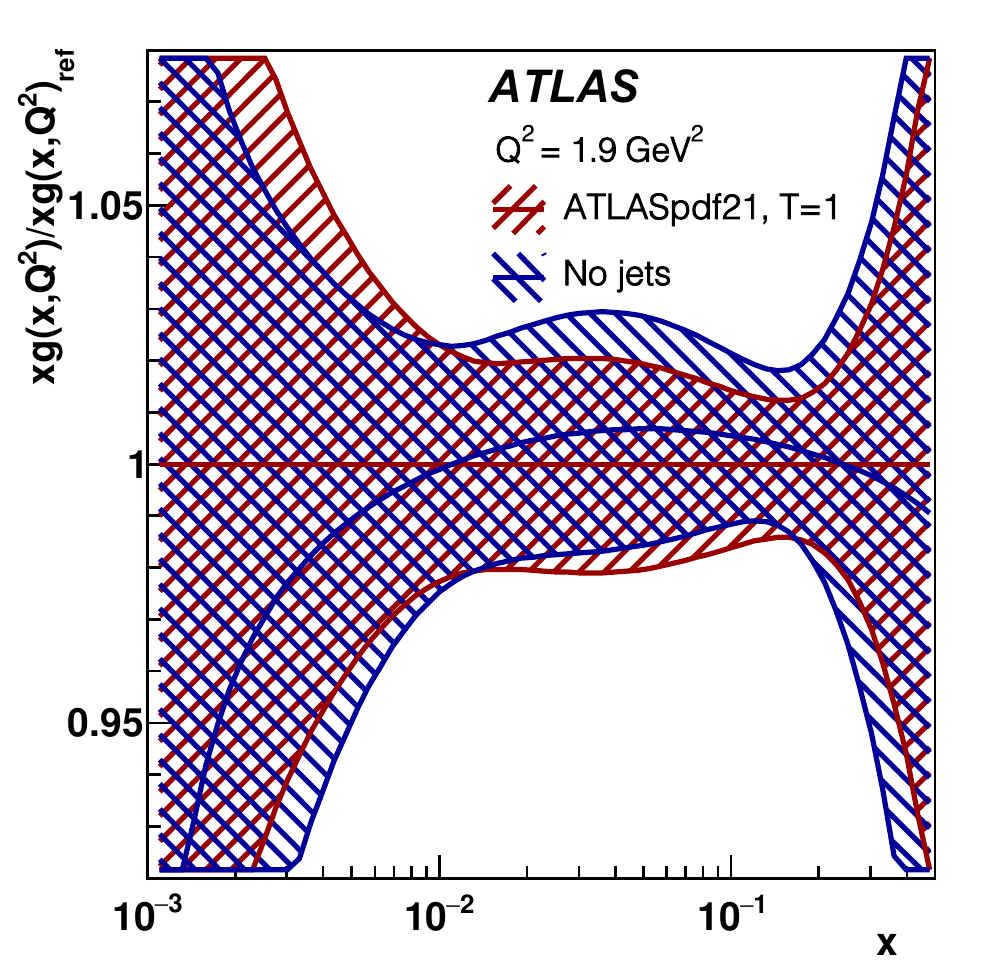}
\caption{ ATLASpdf21 $xg$ PDF compared with $xg$ for fits not including various data sets. Only experimental uncertainties are shown, evaluated with tolerance $T=1$. Left: not including the direct-photon production ratio data taken at 13 and 8~\TeV. Right: not including inclusive jet data at 8~\TeV.
\label{fig:nophotonjets}
}
\end{centering}
\end{figure*}
There is little impact from the addition of the direct-photon production ratio data  apart from a marginal
softening of the high-$x$ gluon distribution as shown in Figure~\ref{fig:nophotonjets} (left). However, it is notable that
these data can now be well fitted at NNLO in QCD, given that they have been excluded from PDF fits
for the last 20~years because of poor fits to lower-energy data~\cite{denterria,1802.03021}.
There is minimal tension with other data sets.
 
The principal impact of the inclusive jet data is on the gluon PDF. The main effect is a considerable decrease in high-$x$
gluon uncertainties, with a mild hardening of the gluon PDF at high~$x$, as shown in Figure~\ref{fig:nophotonjets} (right).
There is minimal tension with other data sets. As explained earlier in Section~\ref{sec:data}, the central ATLASpdf21 fit includes only the inclusive jet data at 8 TeV, with $R=0.6$.
The full uncertainties of the fit, see Section~\ref{sec:model}, include the difference between the choices $R=0.6$ and $R=0.4$. The impact of using inclusive jet data at 7 TeV and at 13 TeV, instead of at 8 TeV, is explored in Appendix~\ref{sec:cofmjets}.

\subsection{Model, theoretical and parameterisation uncertainties}
\label{sec:uncertainties}
The consideration of additional uncertainties affecting the PDFs is
presented in this section. These are classified and labeled here as either
model, theoretical or parameterisation uncertainties.
 
\subsubsection{Model and theoretical uncertainties}
\label{sec:model}
 
\begin{figure*}
\begin{centering}
\includegraphics[width=0.48\textwidth]{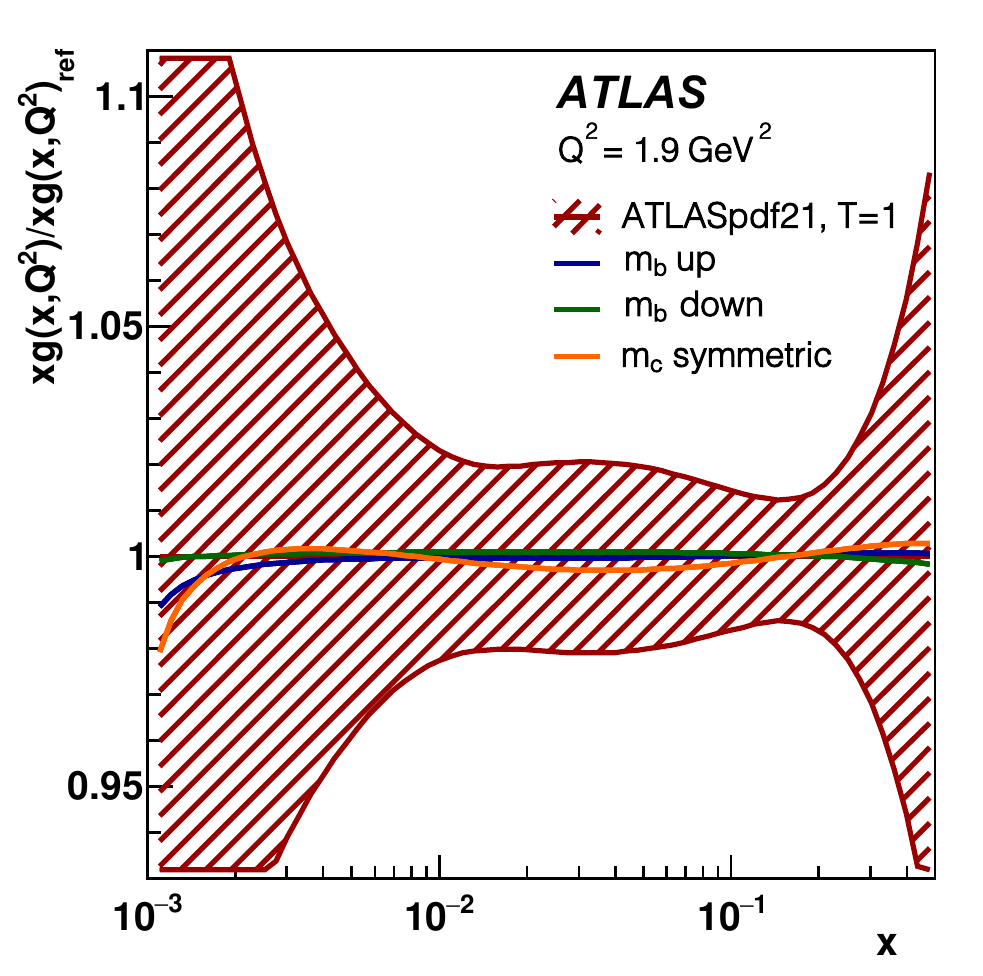}
\includegraphics[width=0.48\textwidth]{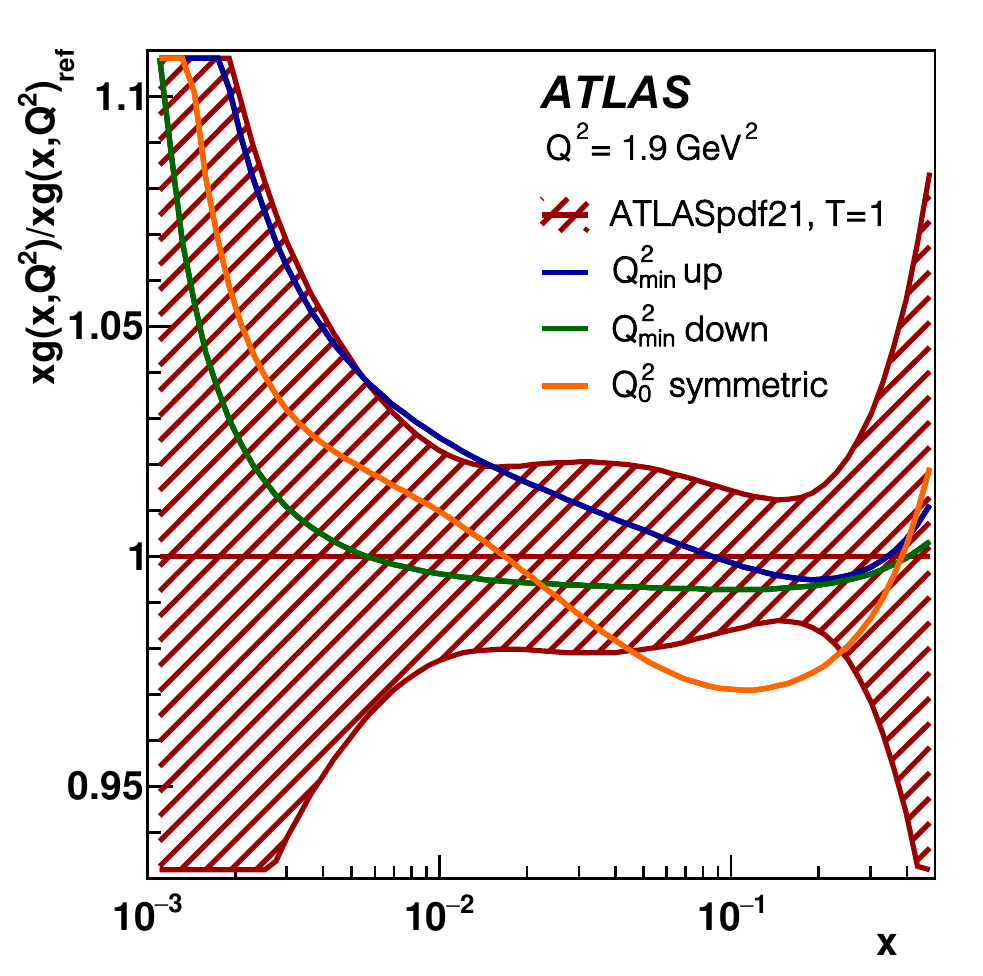}
\includegraphics[width=0.48\textwidth]{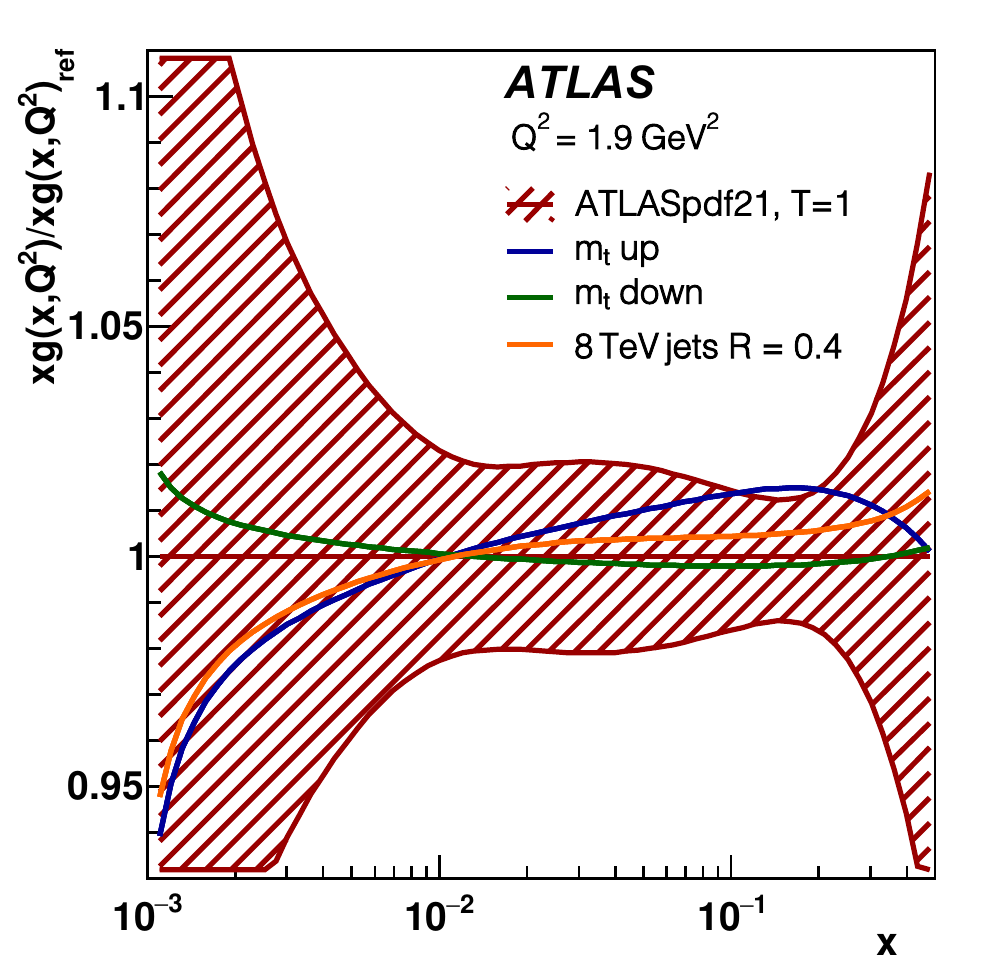}
\caption{Impact of model assumption variations on the ATLASpdf21 gluon PDF illustrated in their ratio to the $xg$ distribution for the central choice. Uncertainties of the central fit are experimental, evaluated with tolerance $T=1$. Top left: variation of $m_c$ and $m_b$ values. The $m_c$ upward variation is shown and this is symmetrised. Top right: impact of  variations of $Q^2_{\mathrm{min}}$ and $Q^2_0$ on the ATLASpdf21 gluon PDF. The $Q^2_0$ downward variation is shown and it is symmetrised. Bottom: impact of the variation of $m_{t}$ and of the choice of jet radius for the inclusive jets on the ATLASpdf21 gluon PDF.
\label{fig:model}
}
\end{centering}
\end{figure*}

The class of model uncertainties includes effects due to variations of the heavy-quark masses input to the TRVFN heavy-quark-mass scheme for the inclusive DIS caluclations,
the minimum $Q^2$ cut on the HERA inclusive DIS data and the value of the starting scale for evolution. The minimum $Q^2$ cut was varied in the range $7.5 < Q^2_{\mathrm{min}} < 12.5$~\GeV$^2$ and the starting scale was varied in the range  $1.6 < Q^2_0 < 2.2$~\GeV$^2$.
In the inclusive DIS calculations, the heavy-quark masses were varied in the ranges
$1.37 < m_c <1.45$~\GeV\ and  $4.1 < m_b< 4.3$~\GeV~\cite{2112.01120}.
The variations of $m_c$ and $Q^2_0$ are coupled since the requirement $Q^2_0 < m_c^2$ must be met. For this reason the upward variation of $m_c$ and the downward variation of $Q^2_0$ are symmetrised.
Figure~\ref{fig:model} illustrates the effect of these variations on the gluon distribution since this is the PDF most sensitive to
these changes. The impact of the choice of $m_c$ and $m_b$ is modest. The effect of the variation of the $Q^2_{\mathrm{min}}$ cut is larger but still within the experimental uncertainties. The largest of these uncertainties is due to the  choice of $Q^2_0$, which gives a gluon PDF differing from the central fit by ${\sim}2\sigma$ for $x \sim 0.1$.
 
An additional model uncertainty comes from the assumed value of the top-quark mass. The interpolation grids for the NNLO
predictions for the $t\bar{t}$ data at 8~\TeV\ are available for pole masses, $m_{t}=172.5, 173.3, 175.0$~\GeV. The smaller and larger values are used to estimate an asymmetric model uncertainty around the central value, shown in Figure~\ref{fig:model}.
For the $t\bar{t}$ 1-D
distributions at 13~\TeV\ the grids are only available for $m_{t}=172.5$~\GeV. The effect of this change in central
value is negligible, as illustrated in Figure~\ref{fig:model}. A cross-check was performed using the double differential $p_{\mathrm{T}}^t$ and $m_{t\bar{t}}$  distributions at 13~\TeV, for which predictions are available for $m_{t}=171.0,172.5,174.0$~\GeV
and the effect was found to be negligible. The 13~\TeV\ $t\bar{t}$ data have a smaller impact in the fit than the 8~\TeV\ $t\bar{t}$ data, so only the $m_{t}$ variations of the 8~\TeV\ data set are input to the final model uncertainty.
 
A further potential model uncertainty comes from the treatment of the jet systematic uncertainties.
Alternative decorrelation scenarios are considered as follows: the alternative option in which the JES Flavour Response is split into only two components rather than three, called ``Decorrelation scenario 1''; complete decorrelation of the Jet
Flavour Response between rapidity bins, called ``FR decorrelated''; and no decorrelation, called ``Fully correlated''.
Table~\ref{tab:chi2jets8R6} gives the total $\chi^2/\mathrm{NDP} $ for the jets
(including all three terms of  Eq.~(\ref{eqn:chi2})) for alternative correlation scenarios,
showing that the difference between full correlations and various choices of decorrelation can have a significant effect
on the  $\chi^2$.
\begin{table}[t]
\caption{$\chi^2$ contributions for the inclusive jet data set at 8~\TeV\ with $R$ = 0.6, for different correlation scenarios, as explained in the text. The $\chi^2$ values given here represent the addition
of all terms in Eq.~(\ref{eqn:chi2}).  \label{tab:chi2jets8R6}}
\begin{center}
\resizebox{\textwidth}{!}{
\begin{tabular}{lcccc}
\hline
Jets 8~\TeV\ $R$ = 0.6 & Fully correlated& FR decorrelated  & Decorrelation scenario 1 &Decorrelation scenario 2 (default)\\
\hline
$\chi^2/\mathrm{NDP}$&289/171&227/171&250/171& 248/171\\
\hline
\hline
\end{tabular}}
\end{center}
\end{table}
However, the difference between the ATLASpdf21 gluon PDFs obtained using jet
data with these different correlation scenarios is
relatively small compared to the model uncertainties considered above, e.g.\ the variation of $Q^2_0$.
There are also no changes in the PDF uncertainties as a result of using different correlation scenarios. Hence, these variations are not considered as a source of significant uncertainty.

A further source of uncertainty for the inclusive jet data comes from the choice of jet radius $R$. The difference between the PDFs due to a different choice of jet radius for the 8~\TeV\ jets is shown in Figure~\ref{fig:model}. The only significant change is in the gluon PDF. This small difference between the PDFs extracted using jet data at 8~\TeV\ for $R = 0.4$ and $R = 0.6$ is considered
as an extra uncertainty because, although it is clear that the choice $R$ = 0.6 is theoretically favoured~\cite{1807.03692},
this paper also inputs $t\bar{t}$ data from the lepton\,+\,jets channel and $V$+\,jets data, and for these data
sets only $R = 0.4$ jets are available.
 
The effect of using jet production data at different centre-of-mass energies of 7, 8 and 13~\TeV\ is shown in Appendix~\ref{sec:cofmjets}. This is not considered as an additional uncertainty but is presented as a cross-check that the choice of centre-of-mass energy for the jet production data does not have a significant influence on the PDFs extracted.
 
All model uncertainties are added in quadrature to form a total model uncertainty.
 
Theoretical uncertainties include the scale uncertainties of the predictions. The largest of these are the
scale uncertainties for the inclusive $W$ and $Z$ data which are
considered in detail in Section~\ref{sec:scaleunc}. Since these are treated as correlated systematic uncertainties in the fit $\chi^2$ they are actually already included in what has been labelled as the experimental uncertainty of the fit.
 
The effect of scale uncertainties for the $V$+\,jets data was studied in Ref.~\cite{Vjets} and was
found to be negligible. The effect of
scale uncertainties for the $t\bar{t}$ lepton\,+\,jets data was studied for the present paper and was also found to be
negligible.
The direct-photon production ratio data are relatively insensitive to scale variations because $100\%$ correlation of the scale
uncertainties is assumed between the 8 and 13~\TeV\ data. Since the fit is relatively insensitive to these data, alternative assumptions for these correlations were not pursued.
 
\begin{figure*}
\begin{centering}
\includegraphics[width=0.48\textwidth]{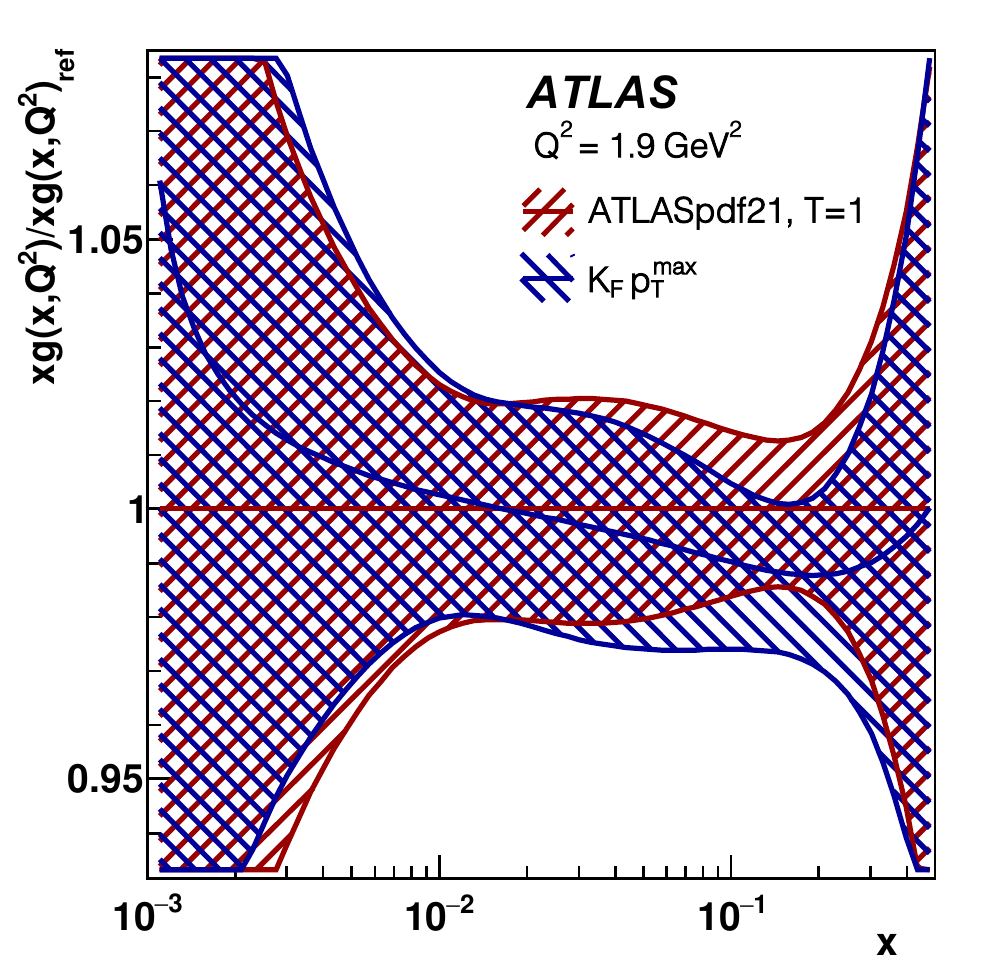}
\includegraphics[width=0.48\textwidth]{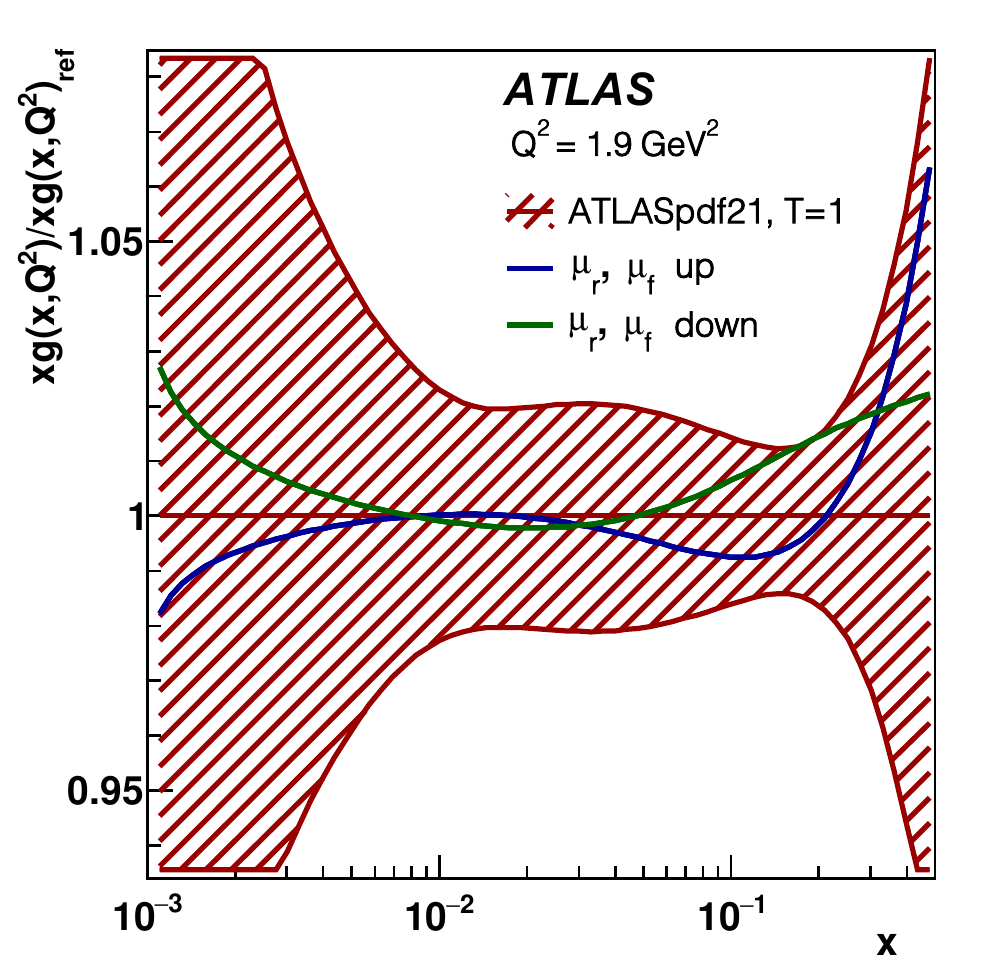}
\caption{Ratios of the ATLASpdf21 gluon PDFs for different inclusive jet scale choices. Left:  $p_{\mathrm{T}}^{\mathrm{max}}$ over $p_{\mathrm{T}}^{\mathrm{jet}}$ (default). Right: $2p_{\mathrm{T}}^{\mathrm{jet}}$ and  $p_{\mathrm{T}}^{\mathrm{jet}}/2$ over $p_{\mathrm{T}}^{\mathrm{jet}}$ (default).  Uncertainties of the central fit are experimental, evaluated with tolerance $T=1$.
\label{fig:jetscale}
}
\end{centering}
\end{figure*}

The scale uncertainty for the inclusive jets at 8~\TeV\ requires further consideration.
These $K$-factors were supplied for two choices of scale:
$\muR=\muF= p_{\mathrm{T}}^{\mathrm{max}}$ or $\muR=\muF= p_{\mathrm{T}}^{\mathrm{jet}}$. A scale related to $p_{\mathrm{T}}^{\mathrm{jet}}$ is favoured over $p_{\mathrm{T}}^{\mathrm{max}}$~\cite{1807.03692}, as already mentioned. However, this difference in scale choice is now investigated.
A comparison of the gluon PDFs for these two scales is shown in Figure~\ref{fig:jetscale} (left), since
this is the PDF
most sensitive to this change. It can be seen that there is no significant difference between the resulting gluon PDFs.
Changes in the nuisance parameter values for the correlated systematic uncertainties absorb the change in the
predictions for the two scales.
The $\chi^2$ values for these fits are given in Table~\ref{tab:chisqjets}.
The only significant change in $\chi^2$ comes from the inclusive jets, not from other data sets.
 
Next, scale variations $\muR=\muF= 2 p_{\mathrm{T}}^{\mathrm{jet}}$ and $\muR=\muF= p_{\mathrm{T}}^{\mathrm{jet}}/2$ are considered.
The effect on the gluon PDF is shown in Figure~\ref{fig:jetscale} (right). The $\chi^2$ values for these fits are given in Table~\ref{tab:chisqjets}.
Again, the only significant change in $\chi^2$ comes from the inclusive jets, not from other data sets.
\begin{table}[h]
\caption{$\chi^2/\mathrm{NDF}$ for the total fit and the contribution $\chi^2/\mathrm{NDP}$ for the inclusive jet data, for various scale choices, and treatment of $K$-factors and their uncertainties. The first row represents the default values used in the central ATLASpdf21 fit.
\label{tab:chisqjets}
}
\begin{center}
\begin{tabular}{lcccc}
\hline
Total $\chi^2/\mathrm{NDF}$&$\chi^2/\mathrm{NDP}$ for jets &Treatment of $K$-factors &Scale choice\\
\hline
2010/1620&248/171&smoothed & $p_{\mathrm{T}}^{\mathrm{jet}}$ scale\\
2019/1620&257/171&smoothed & $p_{\mathrm{T}}^{\mathrm{max}}$ scale\\
2032/1620&272/171&smoothed & $2p_{\mathrm{T}}^{\mathrm{jet}}$ scale\\
1991/1620&228/171&smoothed & $p_{\mathrm{T}}^{\mathrm{jet}}/2$ scale\\
1983/1620&223/171&unsmoothed & $p_{\mathrm{T}}^{\mathrm{jet}}$ scale\\
\hline
\hline
\end{tabular}
\end{center}
\end{table}
\begin{table}[h]
\centering
\caption{Total $\chi^2$/NDF for each model and parameterisation variation considered for the ATLASpdf21 PDF fit. The notation `(sym)' indicates that the upward and downward model variations have been symmetrised.\label{tab:ModParUnc}}
\begin{tabular}{l c }
\hline
\hline
Central $\chi^2$/NDF & 2010/1620\\
\hline
\multicolumn{2}{c}{Model variations}\\
\hline
$Q^{2}_{\mathrm{min}} = 12.5~\GeV^{2}$ &  1947/1571 \\
$Q^{2}_{\mathrm{min}} = 7.5~\GeV^{2}$ & 2076/1660 \\
$m_{c} = 1.45~\GeV$ (sym) & 2025/1620 \\
$Q^{2}_{0} = 1.6~\GeV^{2}$ (sym) & 2018/1620 \\
$m_{b} = 4.3~\GeV$ & 2016/1620 \\
$m_{b} = 4.1~\GeV$ & 2014/1620 \\
$m_{t} = 175.0~\GeV$ & 2063/1620 \\
$m_{t} = 172.5~\GeV$ & 2018/1620 \\
$R=0.4$ & 2080/1620 \\
\hline
\multicolumn{2}{c}{Parameter variations}\\
\hline
$F_{u_{v}}$, $D_{\bar{d}}$ & 2007/1620 \\
\hline
\hline
\end{tabular}
\end{table}
 
Finally, a cross-check is performed in which the $K$-factors are not smoothed but their statistical uncertainties are used as an additional
uncorrelated uncertainty. The $\chi^2$ value for this fit is also given in Table~\ref{tab:chisqjets}.
It is lower than that for smoothed $K$-factors because the statistical
uncertainties of the unsmoothed $K$-factors are ${\sim}1\%$. However, the resulting PDFs are very similar to those obtained using smoothed
$K$-factors. Since none of the scale variation effects produced significant changes in the PDFs, no further theoretical uncertainty is added for this source.
 
Thus the theoretical uncertainties considered so far are either already included in the experimental uncertainties of the fit, or they are
negligible.
It should be noted that the data are also sensitive to the value of $\alphas(\mZ)$, which affects the shape of the gluon PDF.
The correlation between $\alphas(\mZ)$ and the gluon PDF shape is specified by the DGLAP formalism. A determination of $\alphas(\mZ)$ is beyond the scope of the current paper, since a correct NNLO treatment requires variation of the $K$-factor calculations with $\alphas(\mZ)$, or direct NNLO grids for all processes. This is left for future work. The conventional value $\alphas(\mZ) = 0.118$ is used, in line with the value used by the global fitting groups, CT~\cite{Hou:2019efy}, MSHT~\cite{Bailey:2020ooq} and NNPDF~\cite{NNPDF:2017mvq}.
 
\subsubsection{Parameterisation uncertainties}
\label{sec:param}
\begin{figure*}[t!]
\begin{centering}
\includegraphics[width=0.48\textwidth]{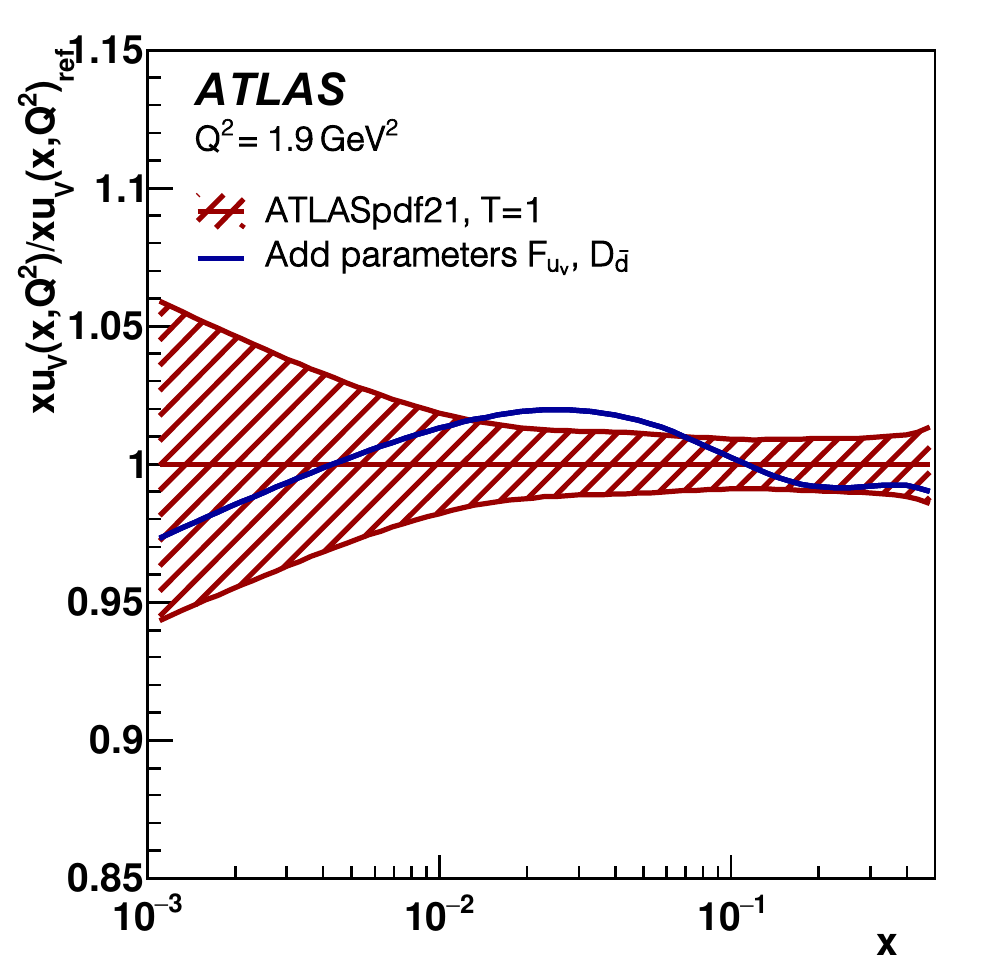}
\includegraphics[width=0.48\textwidth]{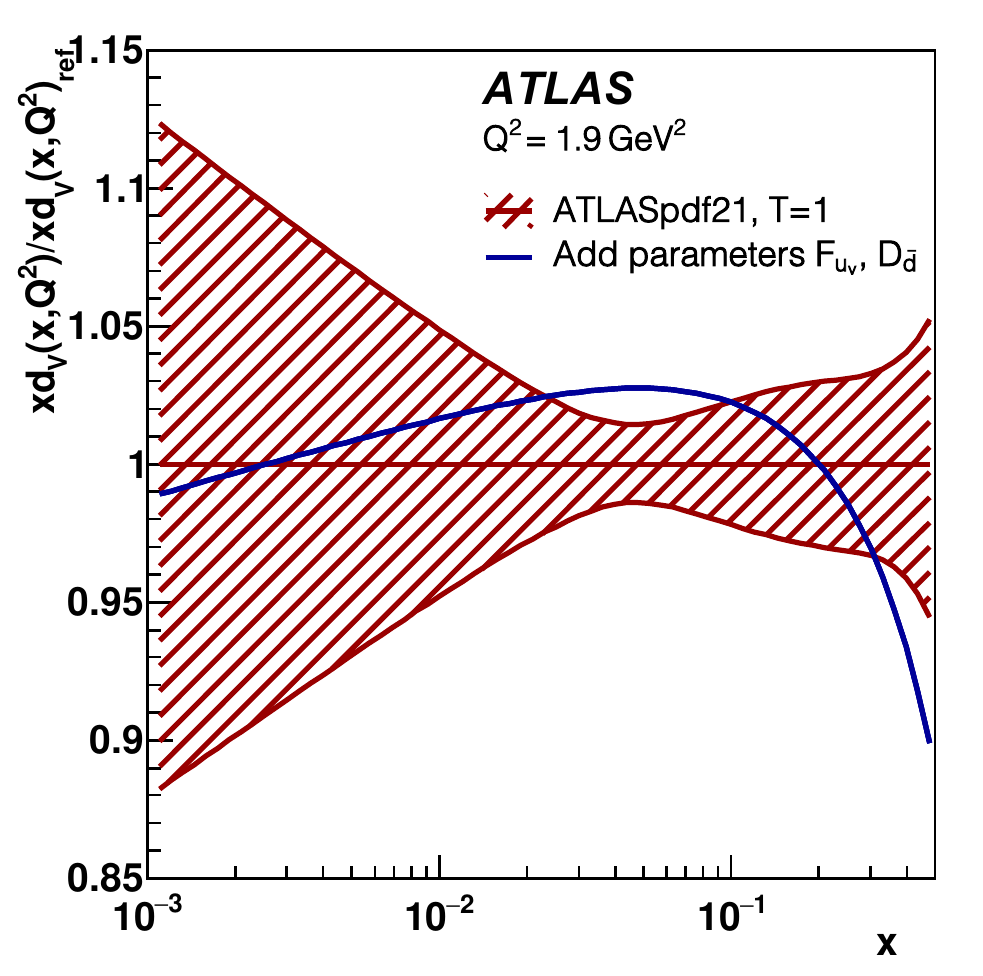}
\includegraphics[width=0.48\textwidth]{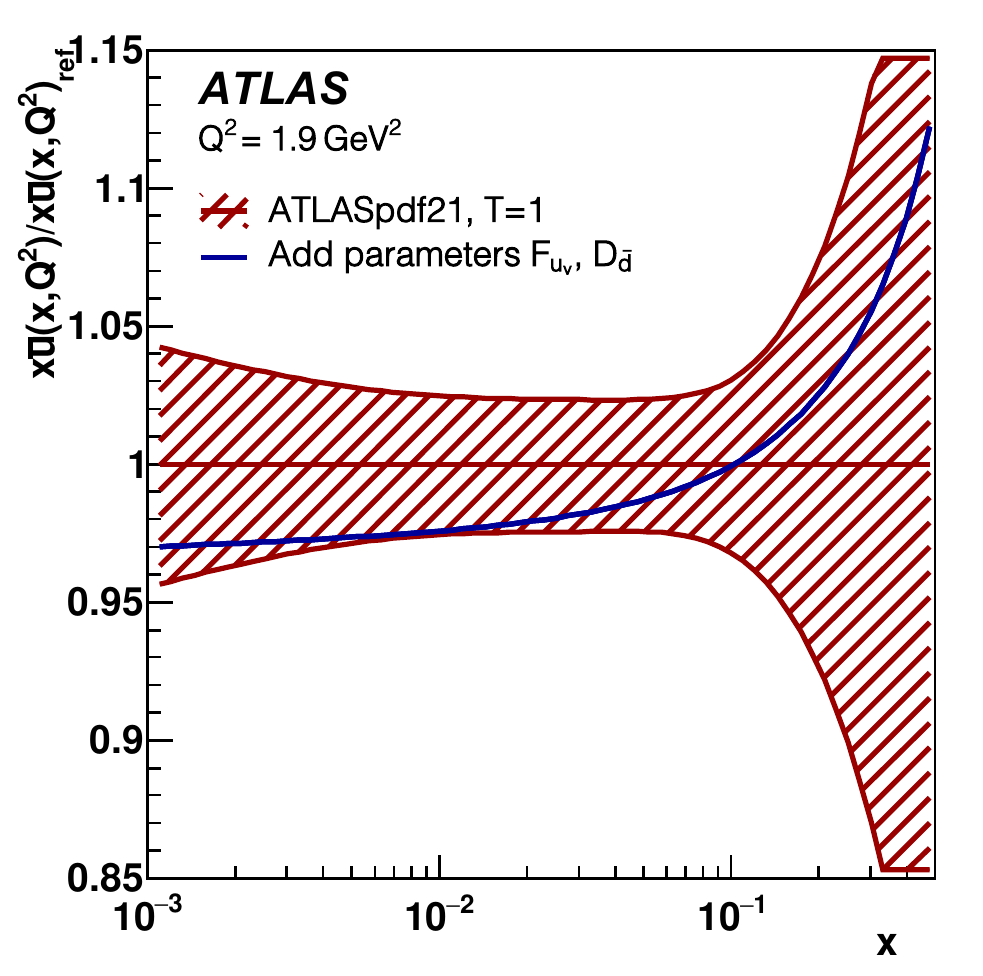}
\includegraphics[width=0.48\textwidth]{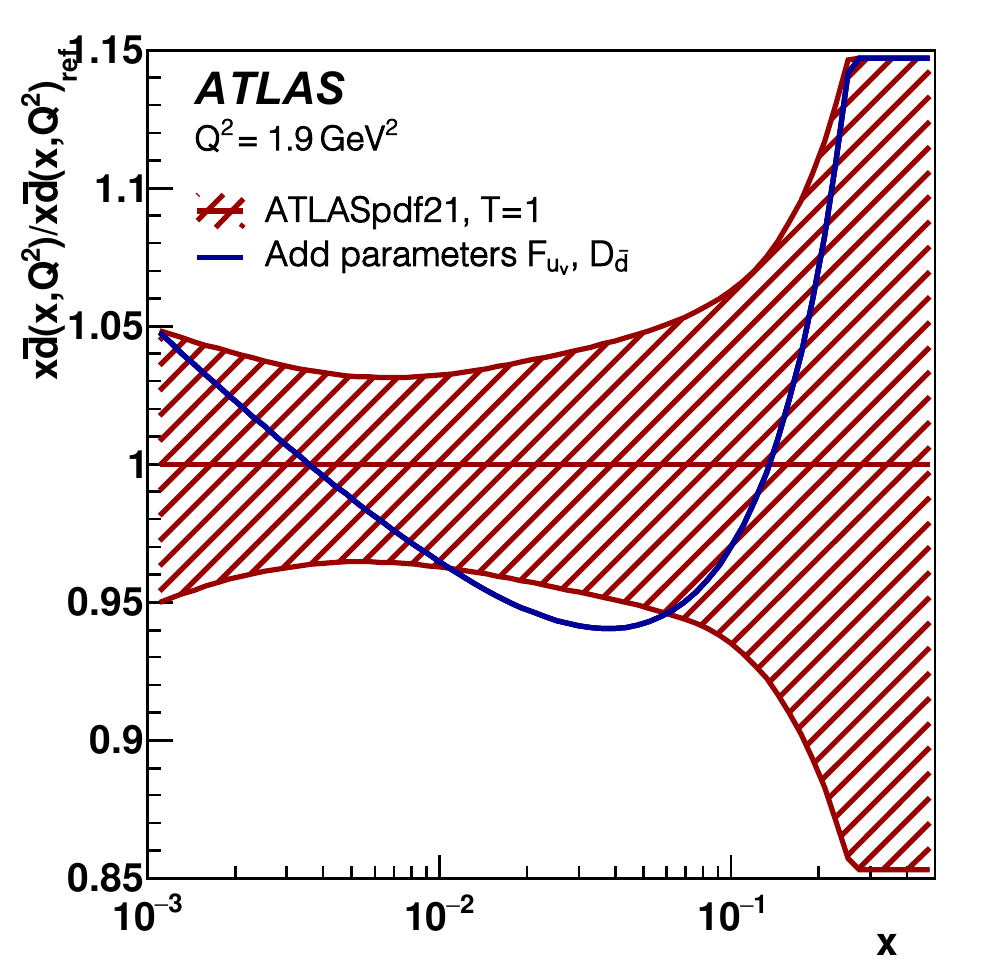}
\caption{Impact of adding $F_{u_{v}}$ and $D_{\bar{d}}$ as free parameters on the valence PDFs in comparison with the central ATLASpdf21 21-parameter fit. Uncertainties of the central fit are experimental, evaluated with tolerance $T=1$.  Top left: $xu_v$.
Top right: $xd_v$. Bottom left: $x\bar{u}$. Bottom right: $x\bar{d}$.
\label{fig:paramuvdv}
}
\end{centering}
\end{figure*}
The optimal number of parameters was determined by `saturation' of the $\chi^2$ as explained in Section~\ref{sec:method}. However, the effect on the
PDFs of adding extra parameters is investigated. Although there is no significant further decrease in $\chi^2$,
some small shape changes are observed when adding
an $F_{u_{v}}$ term to the $u$-valence PDF and/or a $D_{\bar{d}}$ term to the $x\bar{d}$ PDF.
Figure~\ref{fig:paramuvdv} shows the impact of adding both of these as free parameters on the $xu_v$, $xd_v$, $x\bar{u}$ and $x\bar{d}$ PDFs,
which are the PDFs which show the largest variations.
The total parameterisation uncertainty is the envelope of these $D$ and $F$ parameterisation variations.
Model and parameterisation variations, and their effect on the total $\chi^2$ of the fit, are summarised in Table~\ref{tab:ModParUnc}.
\newpage
\begin{figure*}[htb!]
\begin{centering}
\includegraphics[width=0.41\textwidth]{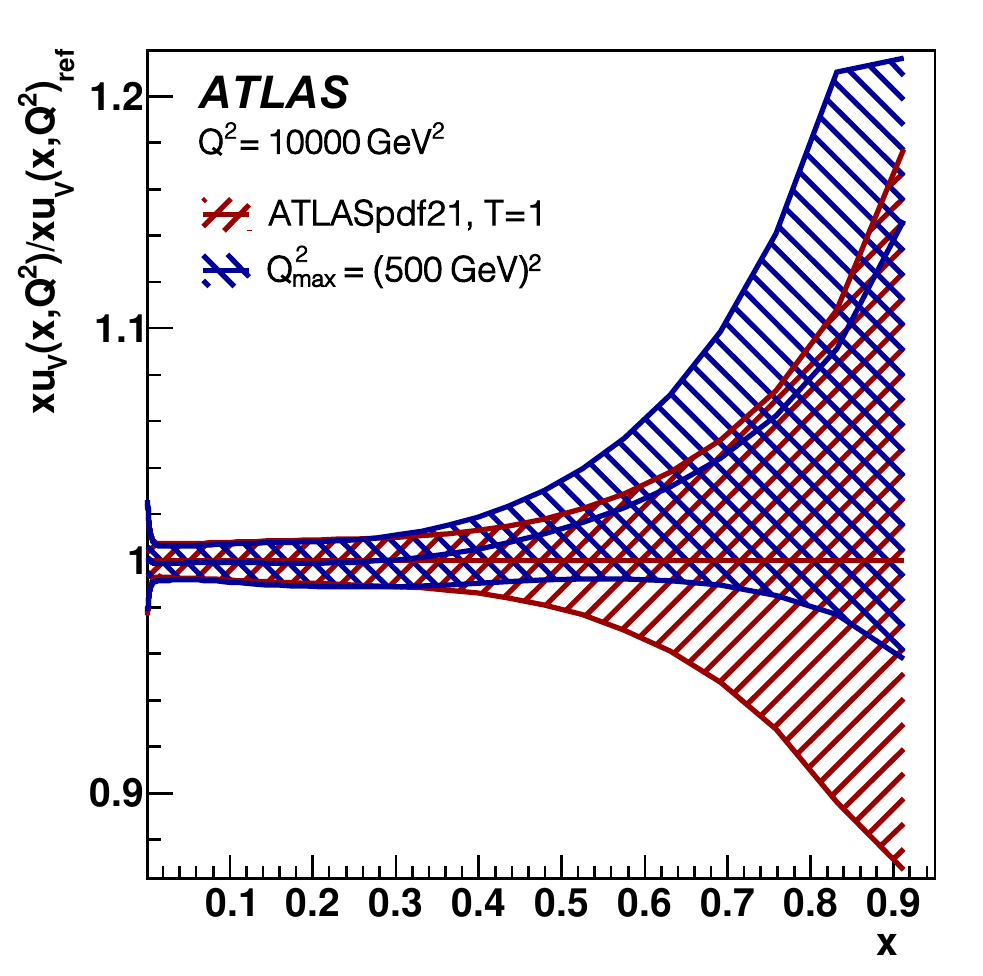}
\includegraphics[width=0.41\textwidth]{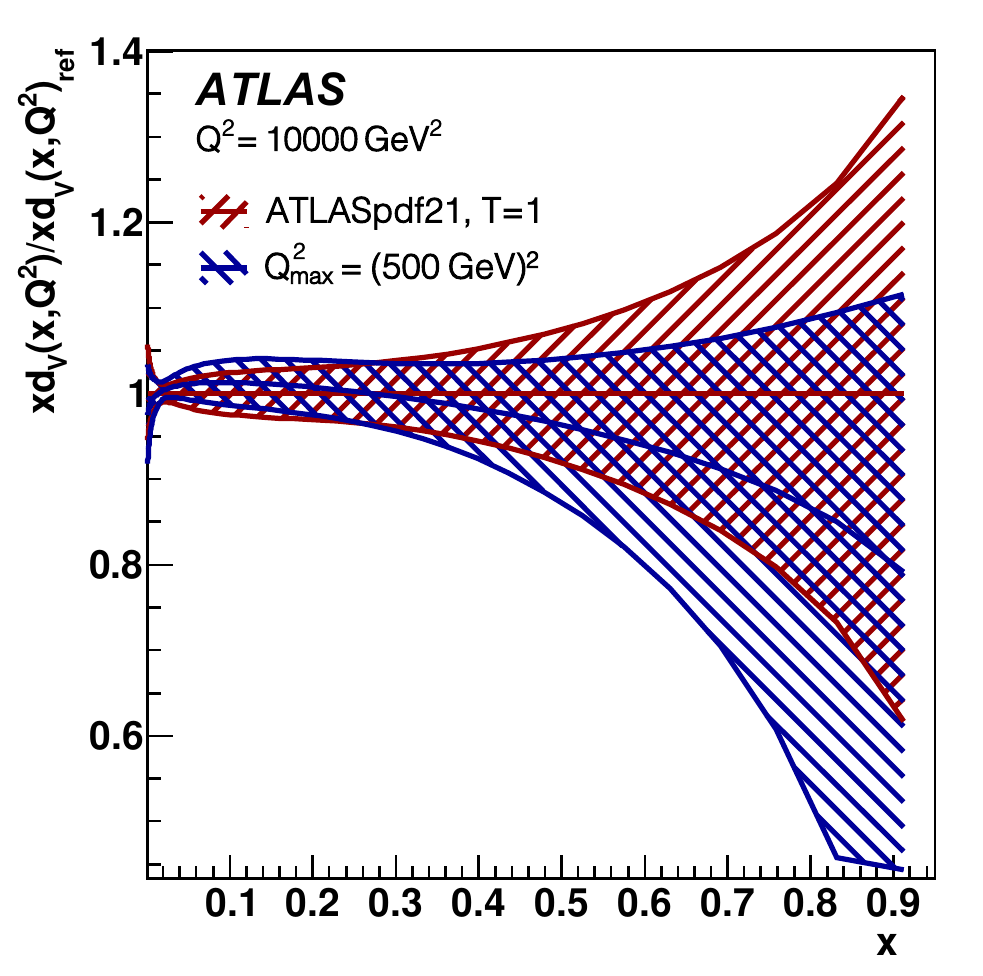}
\includegraphics[width=0.41\textwidth]{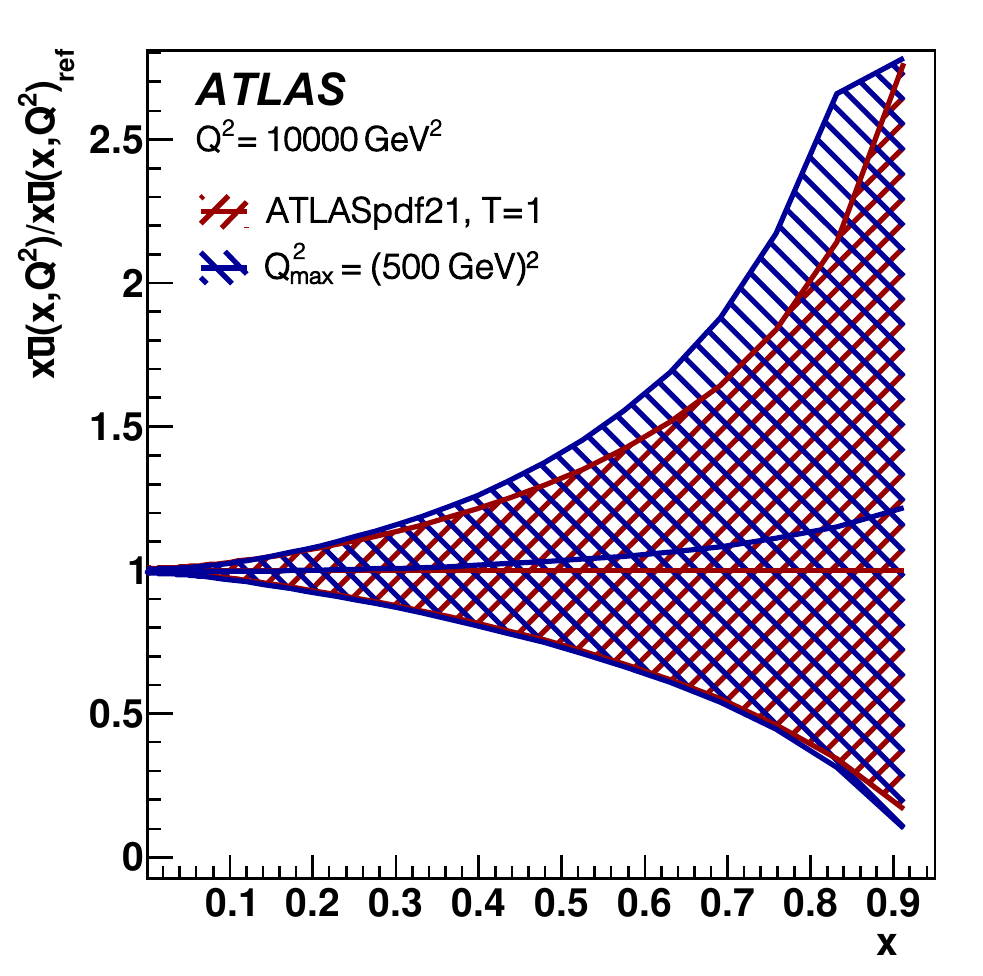}
\includegraphics[width=0.41\textwidth]{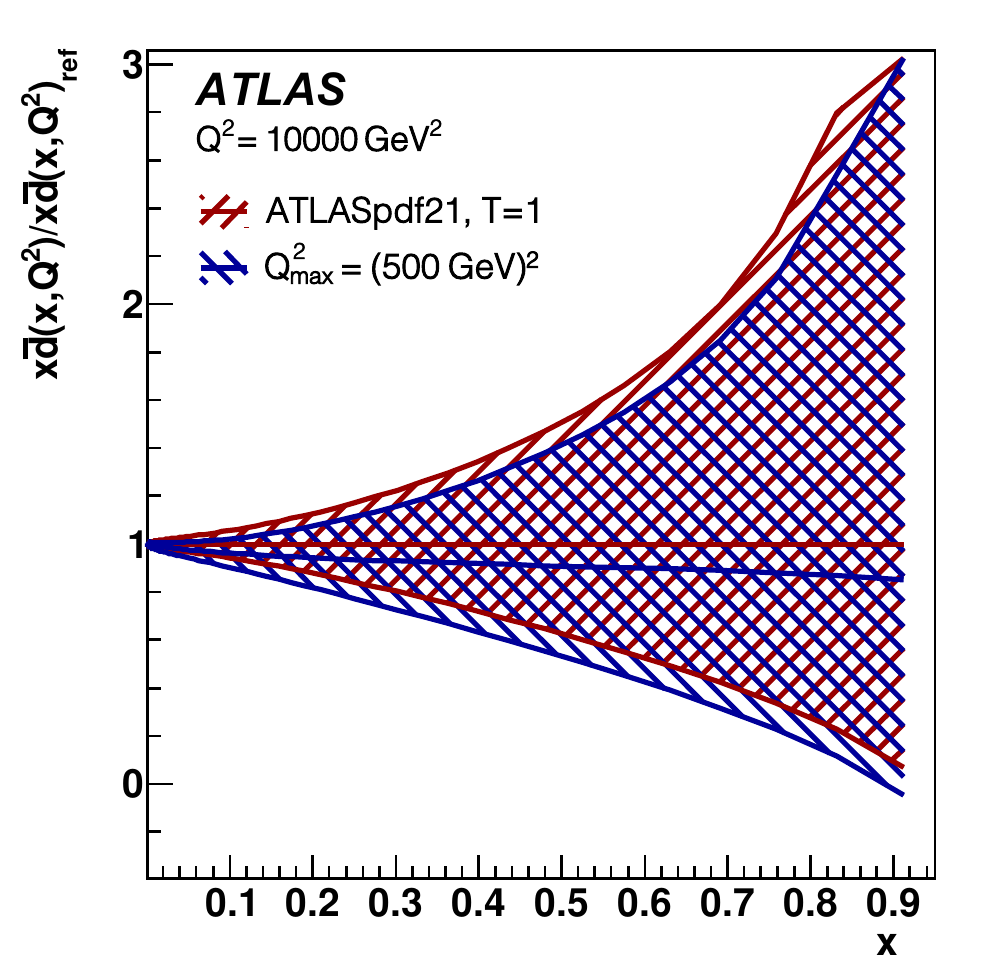}
\includegraphics[width=0.41\textwidth]{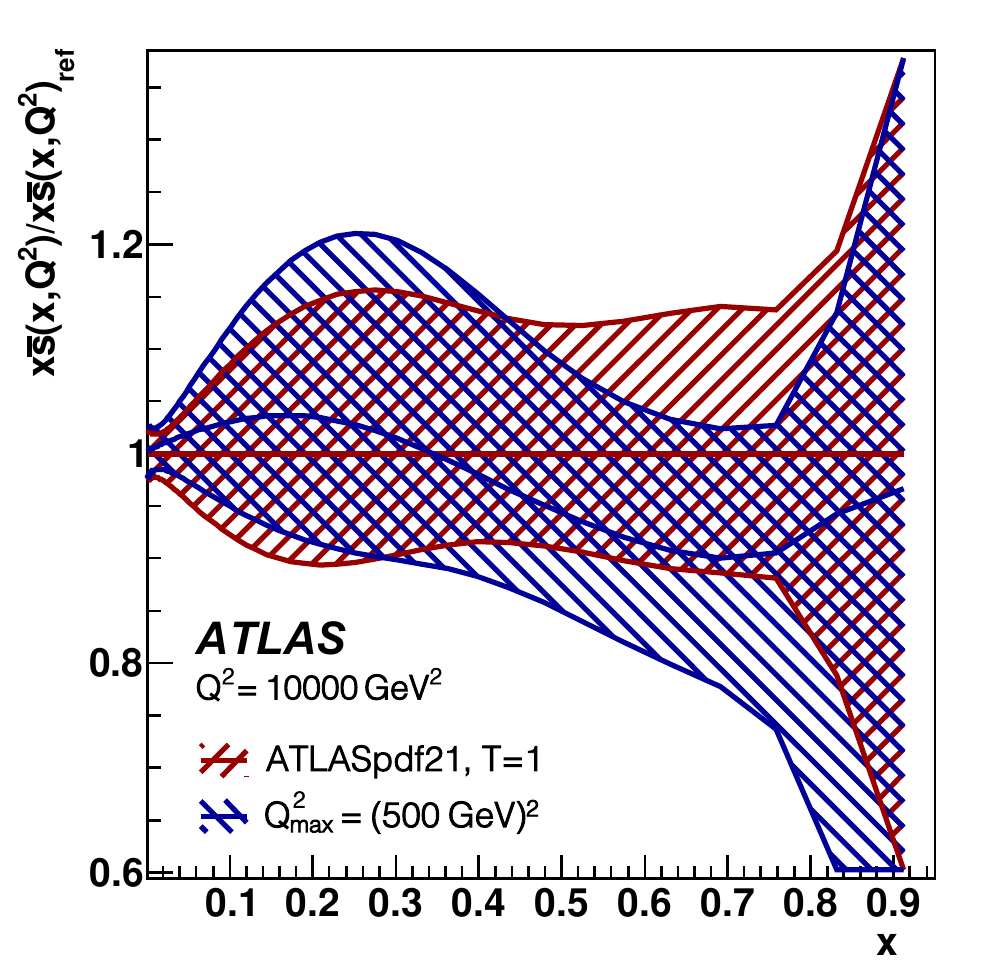}
\includegraphics[width=0.41\textwidth]{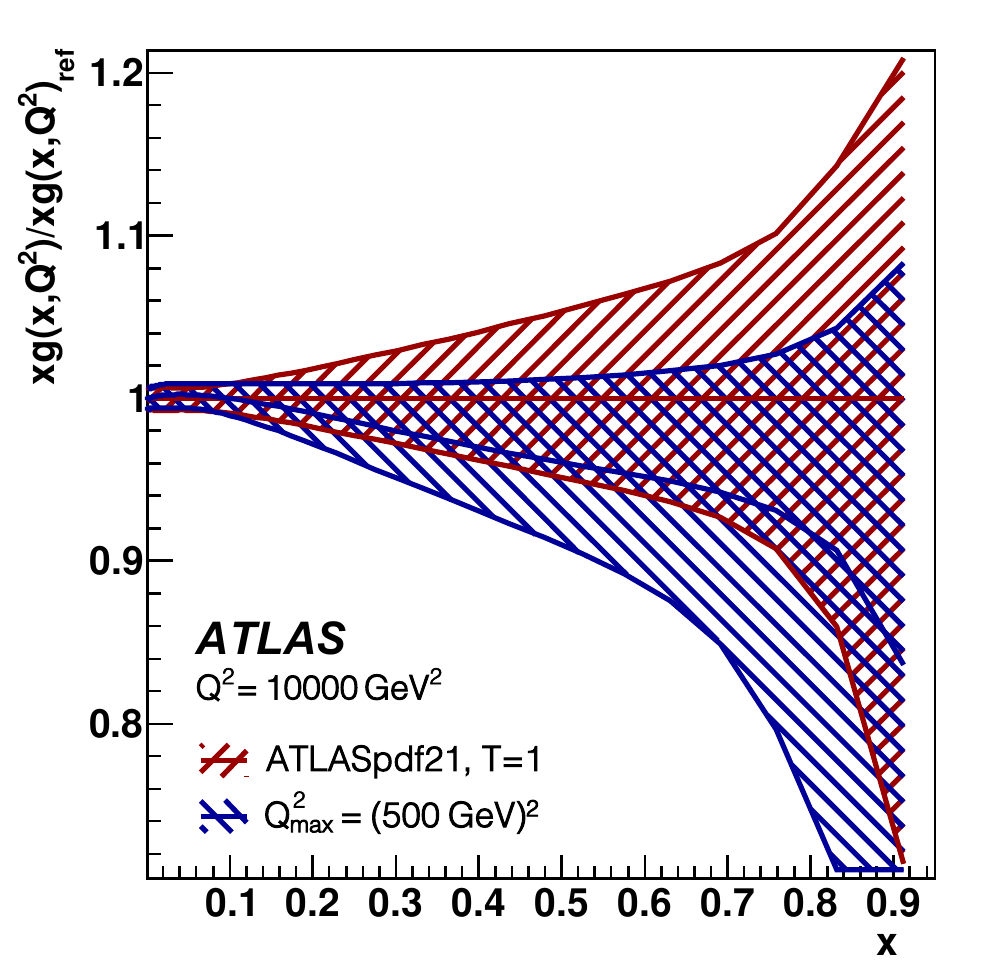}
\caption{Ratios of the PDFs from a fit in which a maximum scale cut of 500~\GeV\ is imposed, to the central ATLASpdf21 at the scale $Q^2 = \num{10000}$~\GeV$^{2}$. Both fits are shown with just experimental uncertainties, evaluated with tolerance $T=1$.
Top left: $xu_v$. Top right: $xd_v$. Middle left: $x\bar{u}$. Middle right: $x\bar{d}$. Bottom left: $xs$. Bottom right: $xg$. Here the $x$-scale is linear.
\label{fig:q2maxevolve}
}
\end{centering}
\end{figure*}
 
\subsubsection{Imposing a maximum $Q^2$ cut on the input data}
A further theoretical uncertainty stems from the possibility that subtle effects of BSM physics may be present in high-scale data. If these are included in a PDF fit, then the estimates of
background to further higher-scale physics from new data could be distorted. A fit is performed in
which the maximum scale for events accepted from each process is 500~\GeV, or $Q^2 =\num{250000}$~\GeV$^2$.
This removes 82 points from the fit: 63
from the inclusive jets data, 6 from the $V$+\,jets data, 11 from the direct-photon ratio data, and one each
from the $t\bar{t}$ $p_{\mathrm{T}}$ spectra at 8 and 13~\TeV. The $\chi^2$/NDF value of the fit improves from
$2010/1620$ to $1870/1538$, with the largest improvement coming from the inclusive jets, for which the $\chi^2$/NDP value improves from $248/171$ to $136/108$.
 
Figure~\ref{fig:q2maxevolve} compares the PDFs at the scale of $Q^2 = \num{10000}$~\GeV$^{2}$, with a linear scale in $x$ so as to emphasise
the differences at high~$x$.
The shapes of the PDFs and their uncertainties are not strongly affected by the cut, and a maximum $Q^2$ cut is not considered further.
 
\subsection{Combination of uncertainties}
 
\begin{figure*}
\begin{centering}
\includegraphics[width=0.49\textwidth]{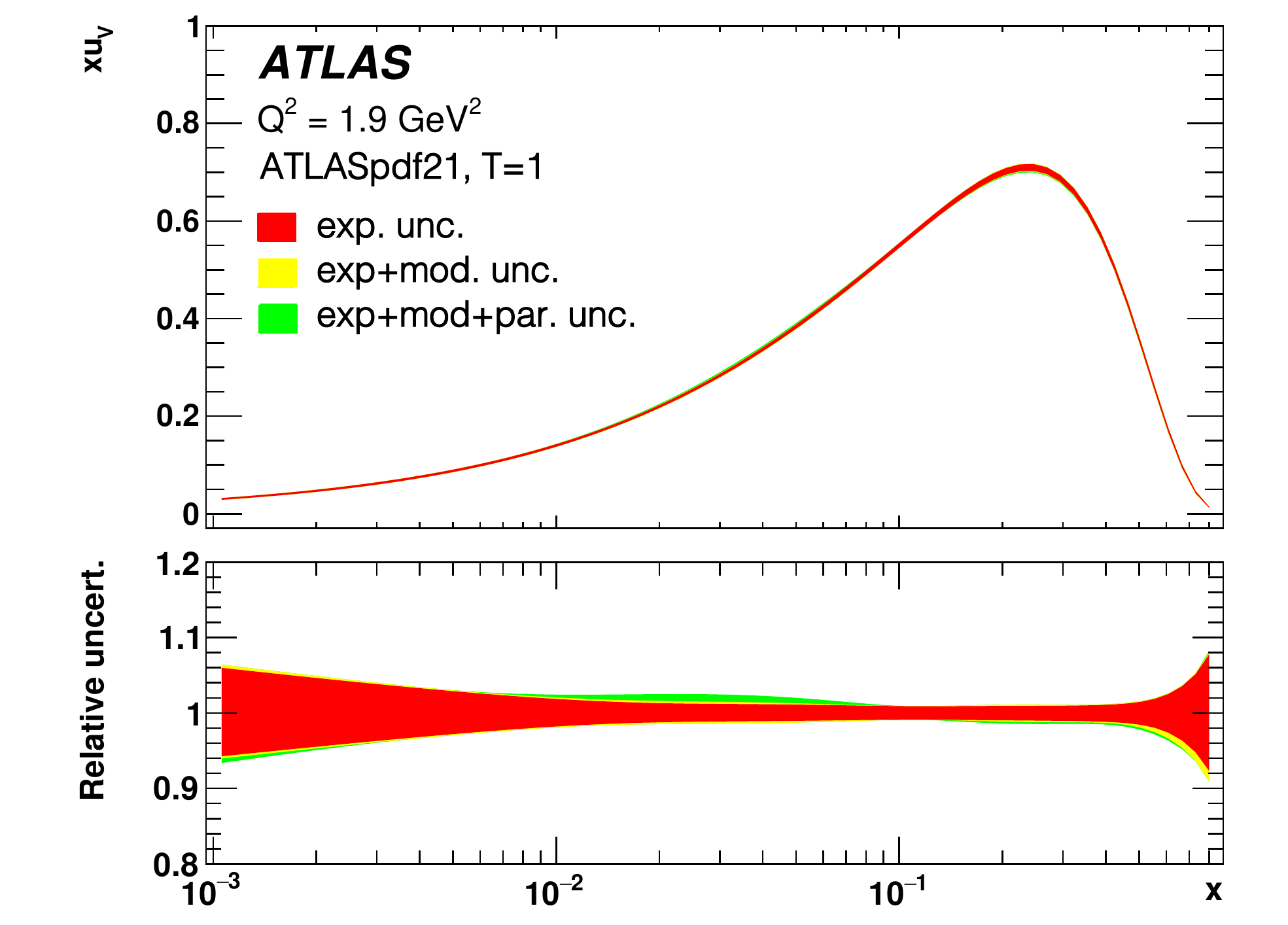}
\includegraphics[width=0.49\textwidth]{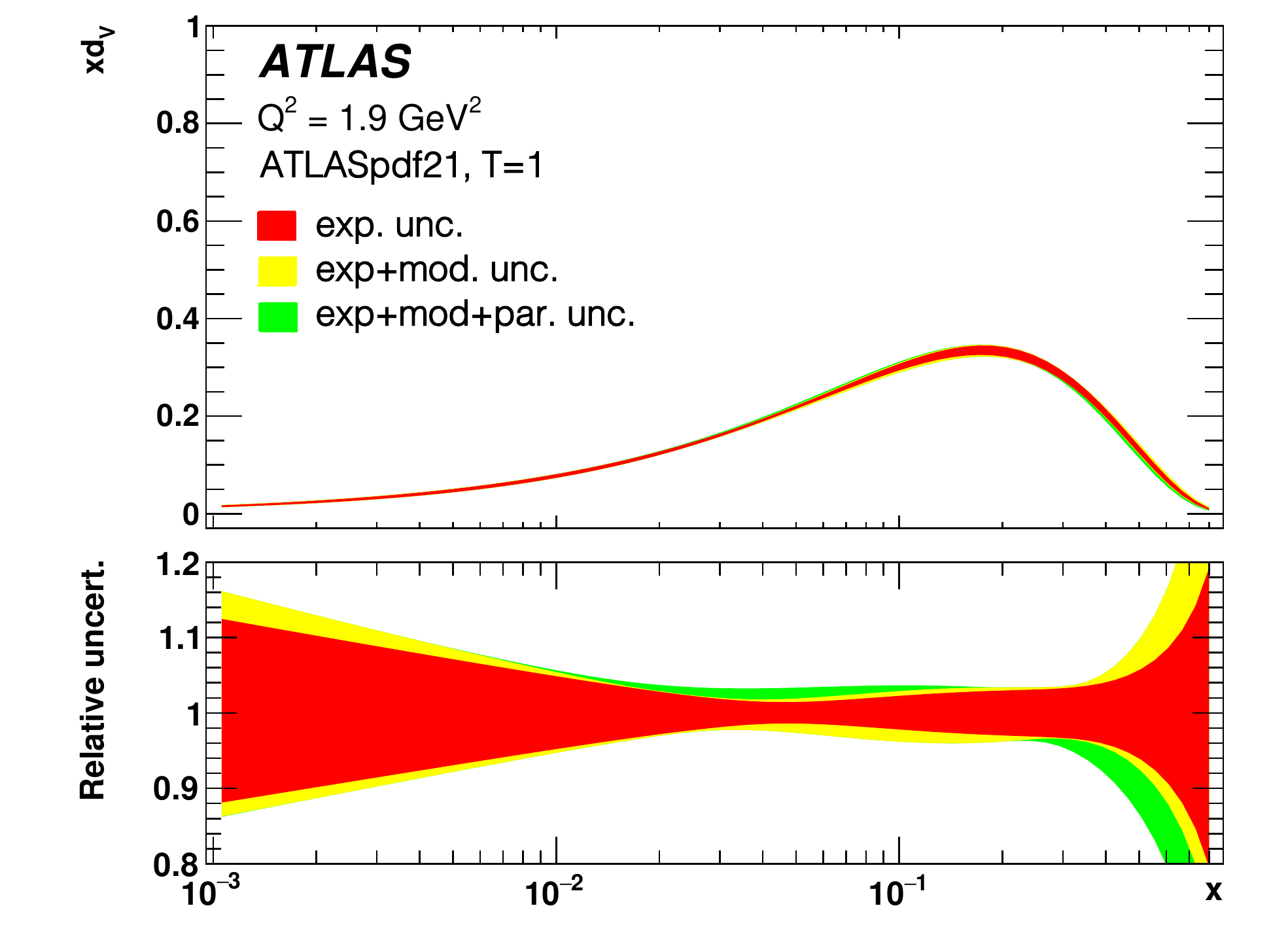}
\includegraphics[width=0.49\textwidth]{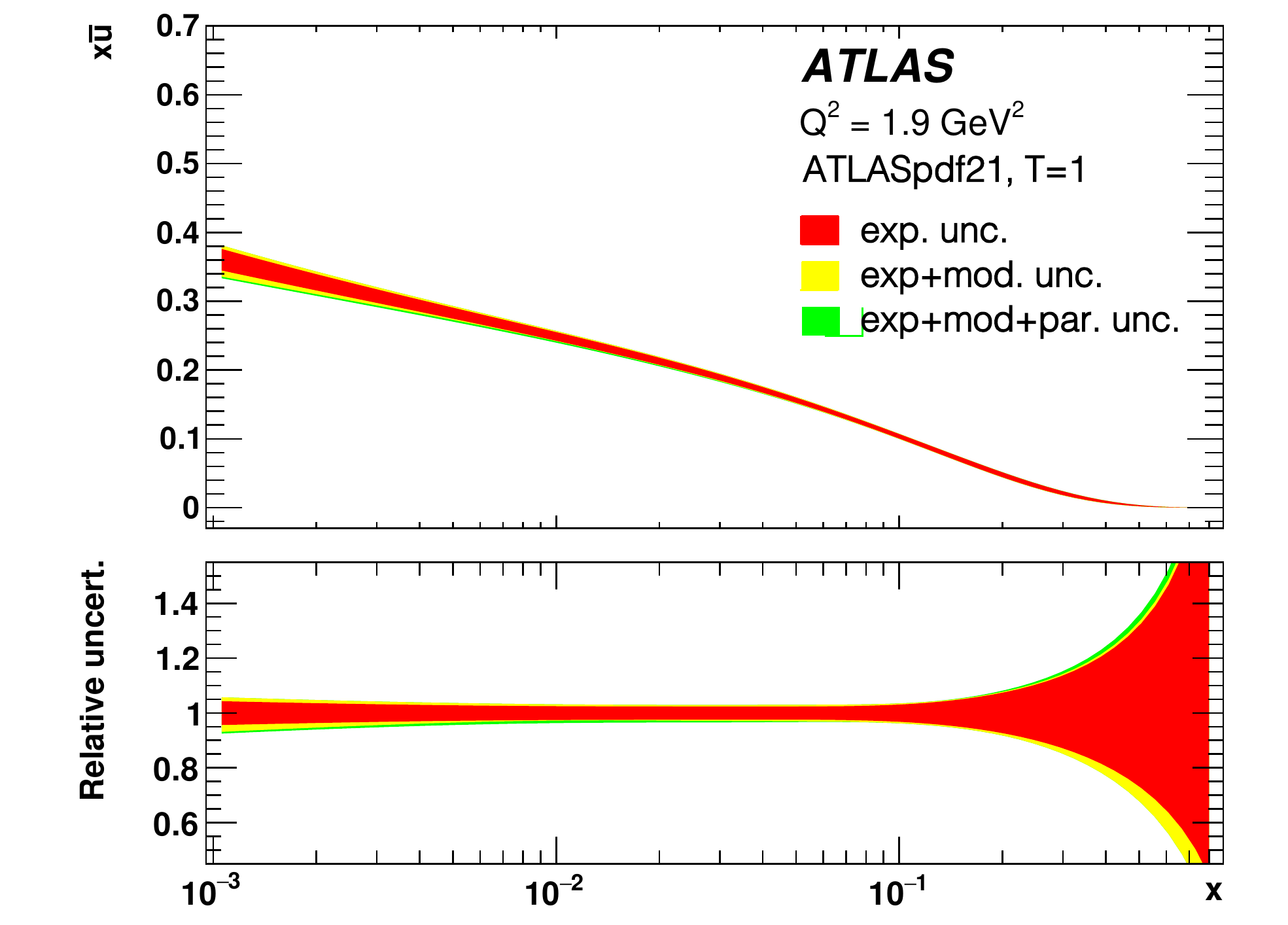}
\includegraphics[width=0.49\textwidth]{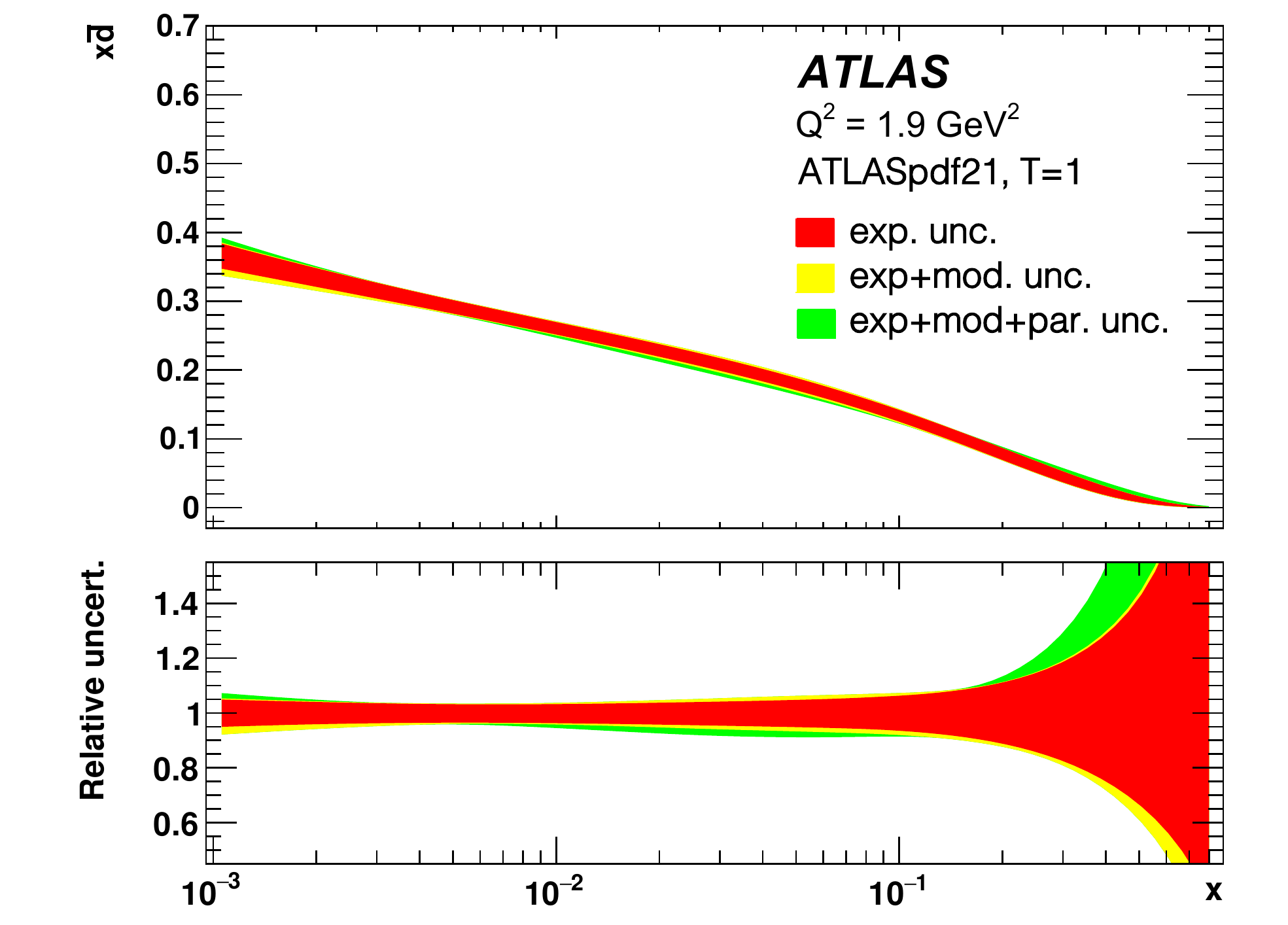}
\caption{ ATLASpdf21 PDFs showing experimental uncertainties evaluated with $T=1$ (red), model (yellow) and parameterisation (green) uncertainties. Experimental, model and parameterisation uncertainties are cumulative. Top left: $xu_v$. Top right: $xd_v$.
Bottom left: $x\bar{u}$. Bottom right: $x\bar{d}$. The lower panels illustrate the fractional uncertainties.
\label{fig:modelparam1}
}
\end{centering}
\end{figure*}

\begin{figure*}
\begin{centering}
\includegraphics[width=0.49\textwidth]{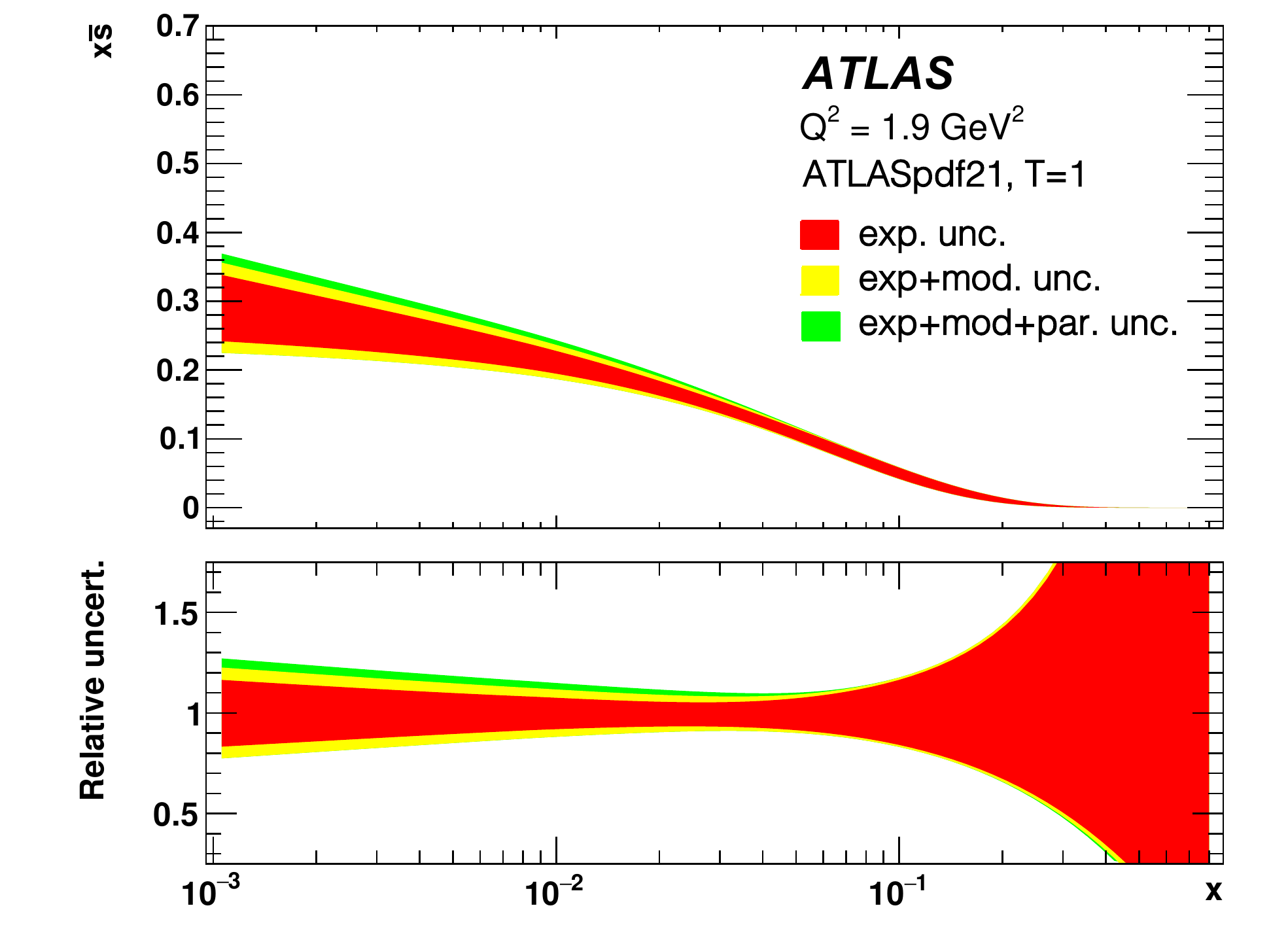}
\includegraphics[width=0.49\textwidth]{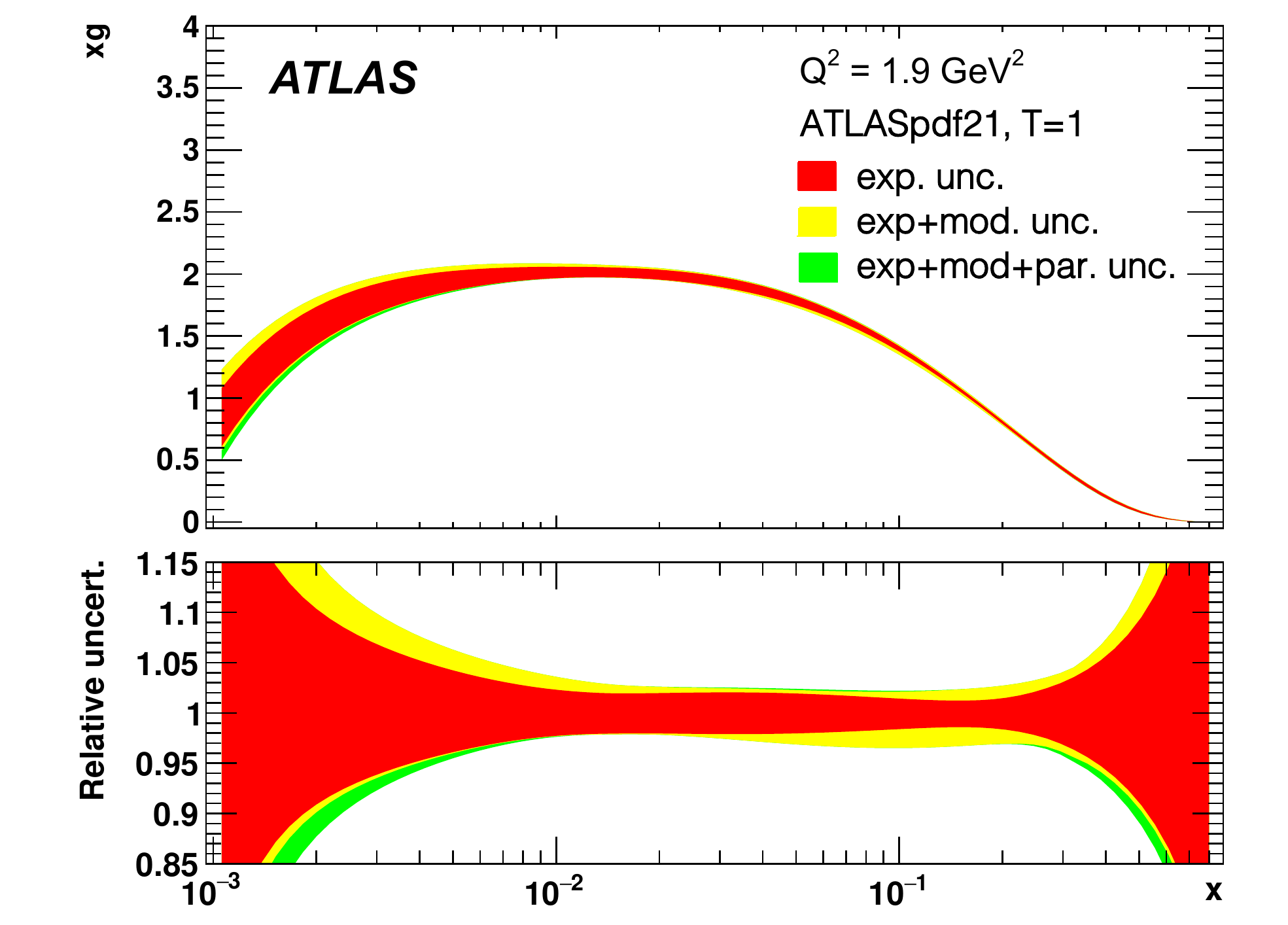}
\includegraphics[width=0.47\textwidth]{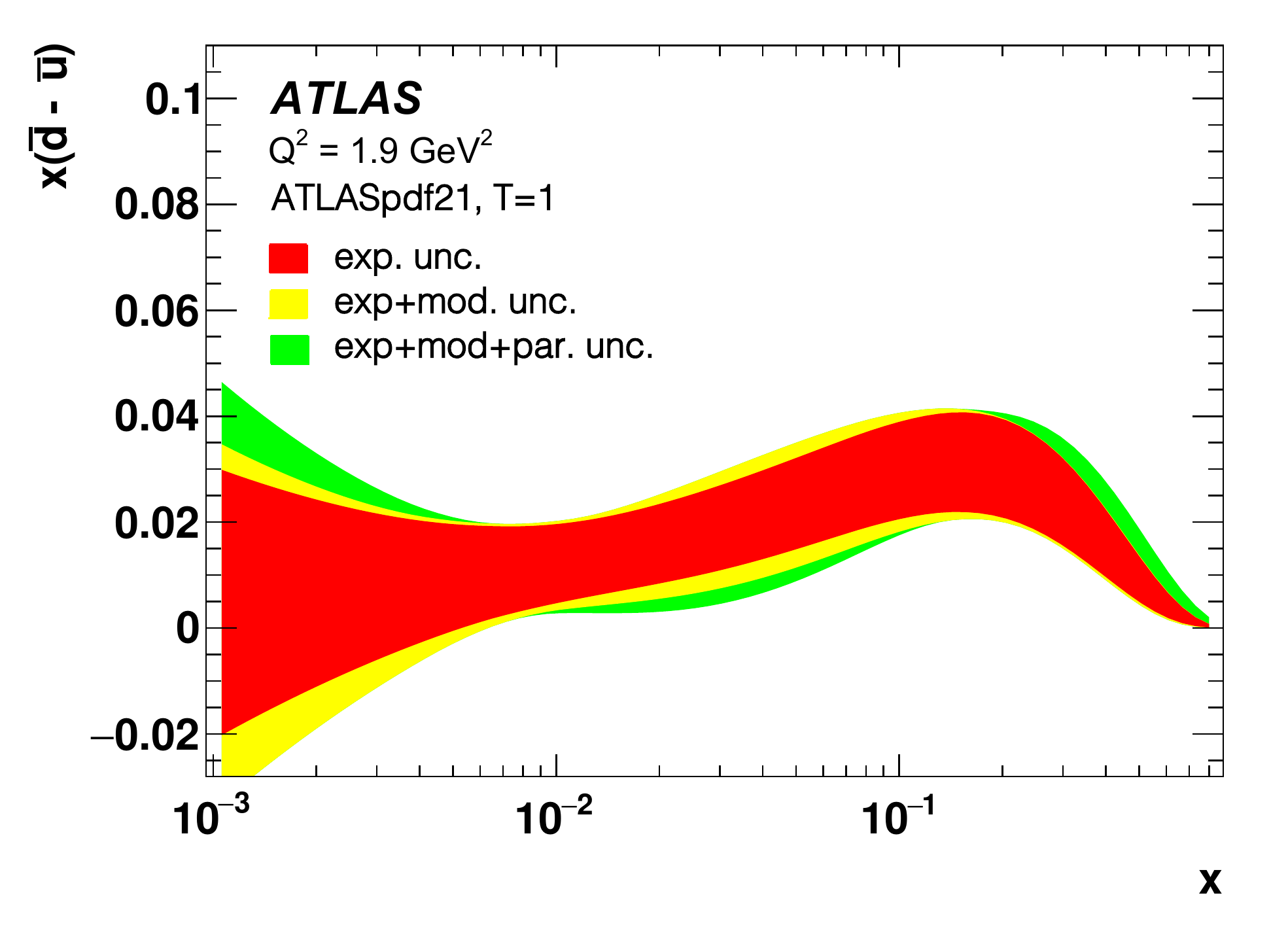}
\caption{ ATLASpdf21 PDFs showing experimental uncertainties evaluated with $T=1$ (red), model (yellow) and parameterisation (green) uncertainties. Experimental, model and parameterisation uncertainties are cumulative. Top left: $x\bar{s}$. Top right: $xg$. Bottom: $x\bar{d}-x\bar{u}$.
The lower panels illustrate the fractional uncertainties.
\label{fig:modelparam2}
}
\end{centering}
\end{figure*}

The PDFs are now presented with experimental uncertainties from the fit and with additional model and parameterisation uncertainties as discussed in this section. These model and parameterisation uncertainties are added in quadrature to the experimental uncertainty. It should be noted that the scale
uncertainties for the inclusive $W,Z$ data appear as part of the experimental uncertainty,
whereas the scale uncertainties for other data sets are negligible.
Figure~\ref{fig:modelparam1} shows the valence distributions and the $x\bar{u}$ and $x\bar{d}$ sea distributions, and Figure~\ref{fig:modelparam2} shows the $x\bar{s}$ and $xg$
distributions and the $x(\bar{d}-\bar{u})$ distribution.
The experimental uncertainties are much larger for the $d$-type sector than for the $u$-type sector. The larger fractional uncertainty of the $x\bar{s}$ PDF in comparison with the $x\bar{d}$ PDF reflects its smaller absolute size as $x\to 1$. The gluon PDF is well determined for $0.01 < x < 0.3$ but not at low or high~$x$.
The model uncertainties in all PDFs are moderate, being largest in the gluon PDF. The parameterisation
uncertainty of $x\bar{d}$ affects both $x\bar{d}$ and $xd_v$ since $xd_{\mathrm{sea}} = x\bar{d}$ and the sea and valence $d$-quarks act together. The $xd_v$ distribution is affected at both low and high~$x$ because of the valence-quark number sum rule. Similarly, the parameterisation uncertainty of $xu_v$ is seen at both low and high~$x$.
The parameterisation uncertainty of the gluon PDF comes from the other PDFs through the momentum sum rule.
\begin{figure*}
\begin{centering}
\includegraphics[width=0.67\textwidth]{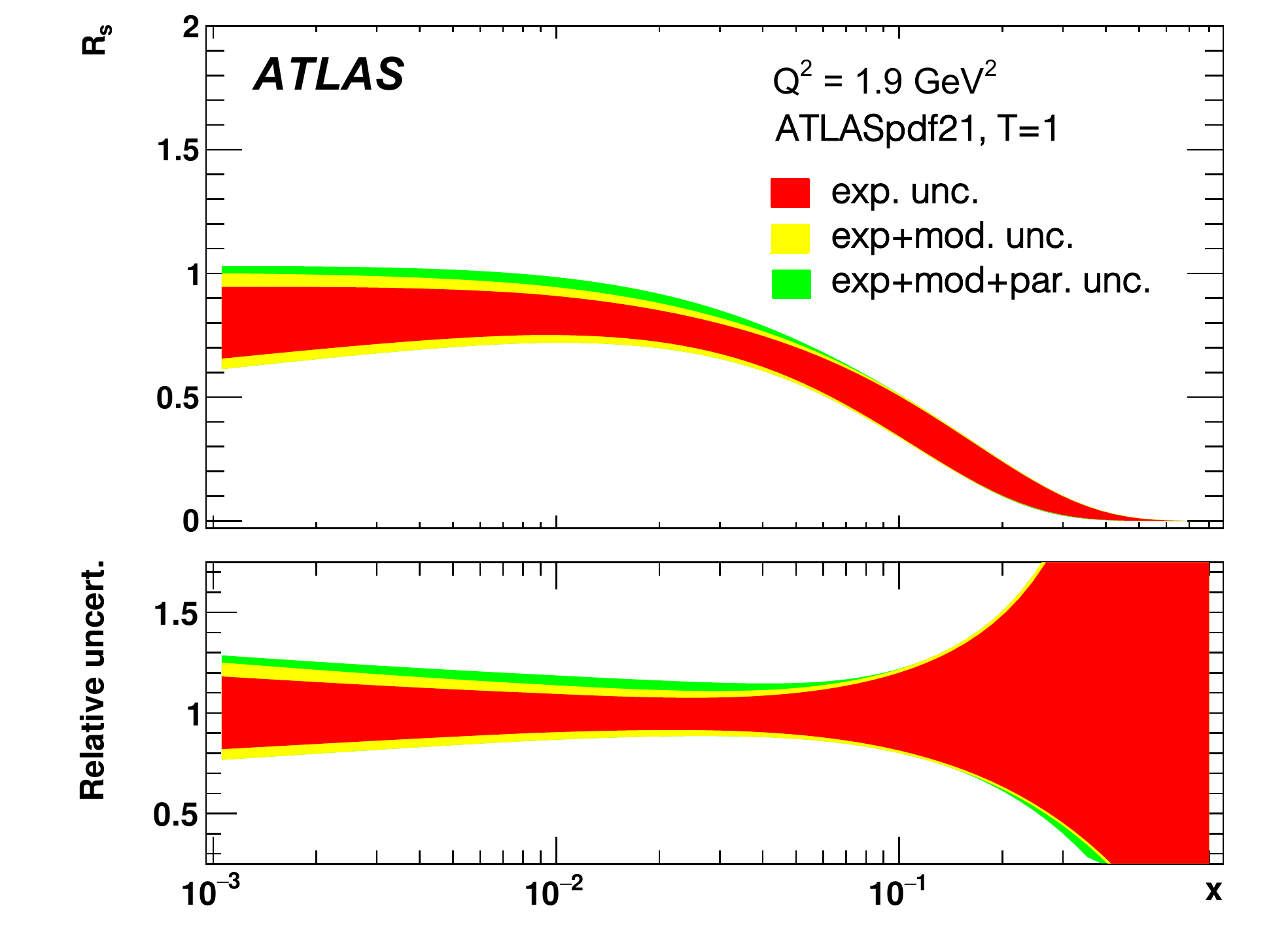}
\caption{ ATLASpdf21 $R_s$ distribution showing experimental uncertainties evaluated with $T=1$ (red), model (yellow) and parameterisation (green) uncertainties. Experimental, model and parameterisation uncertainties are cumulative.
The lower panel illustrates the fractional uncertainties.
\label{fig:modelparam3}
}
\end{centering}
\end{figure*}
 
\begin{figure*}
\begin{centering}
\includegraphics[width=0.49\textwidth]{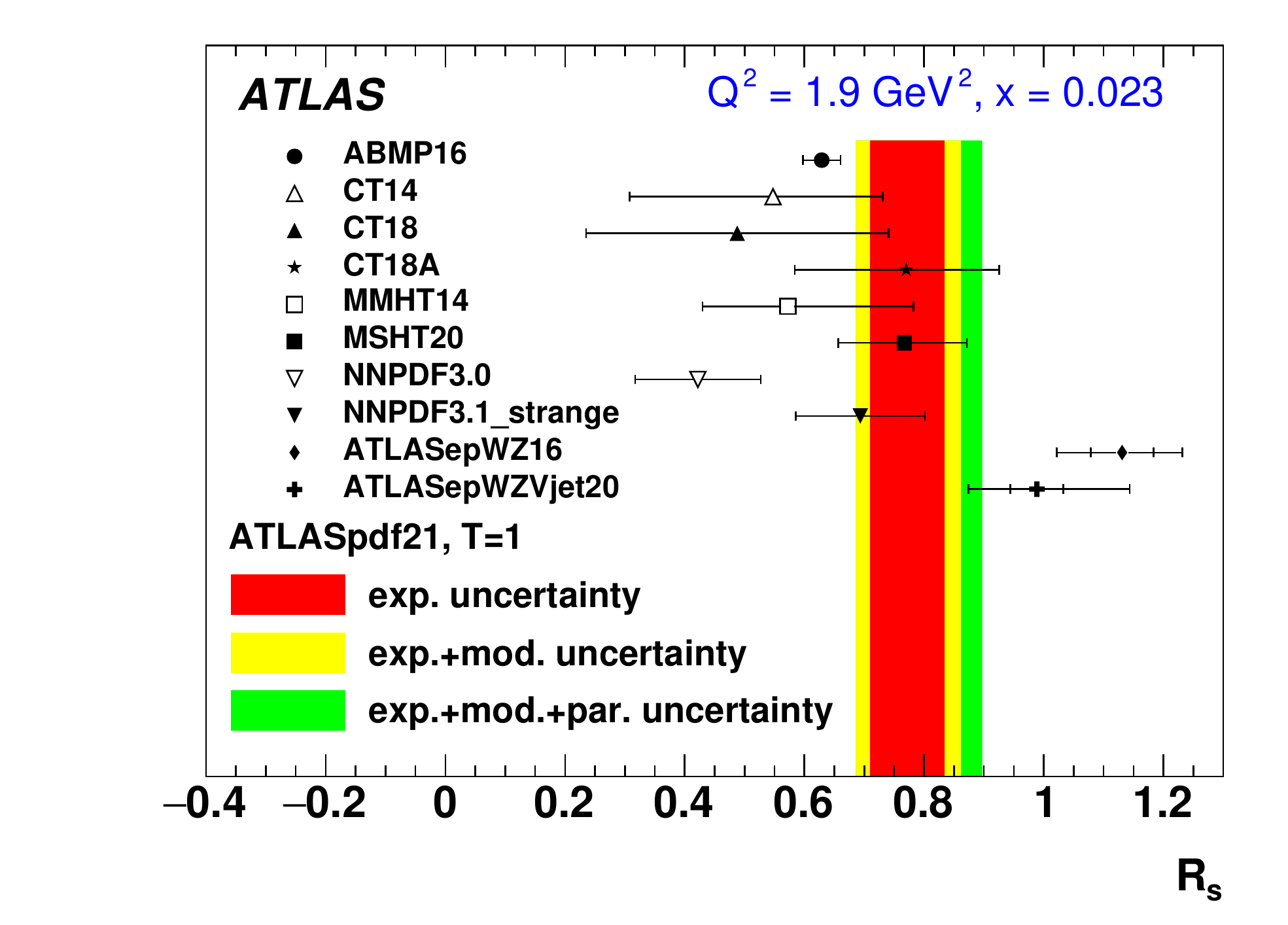}
\includegraphics[width=0.49\textwidth]{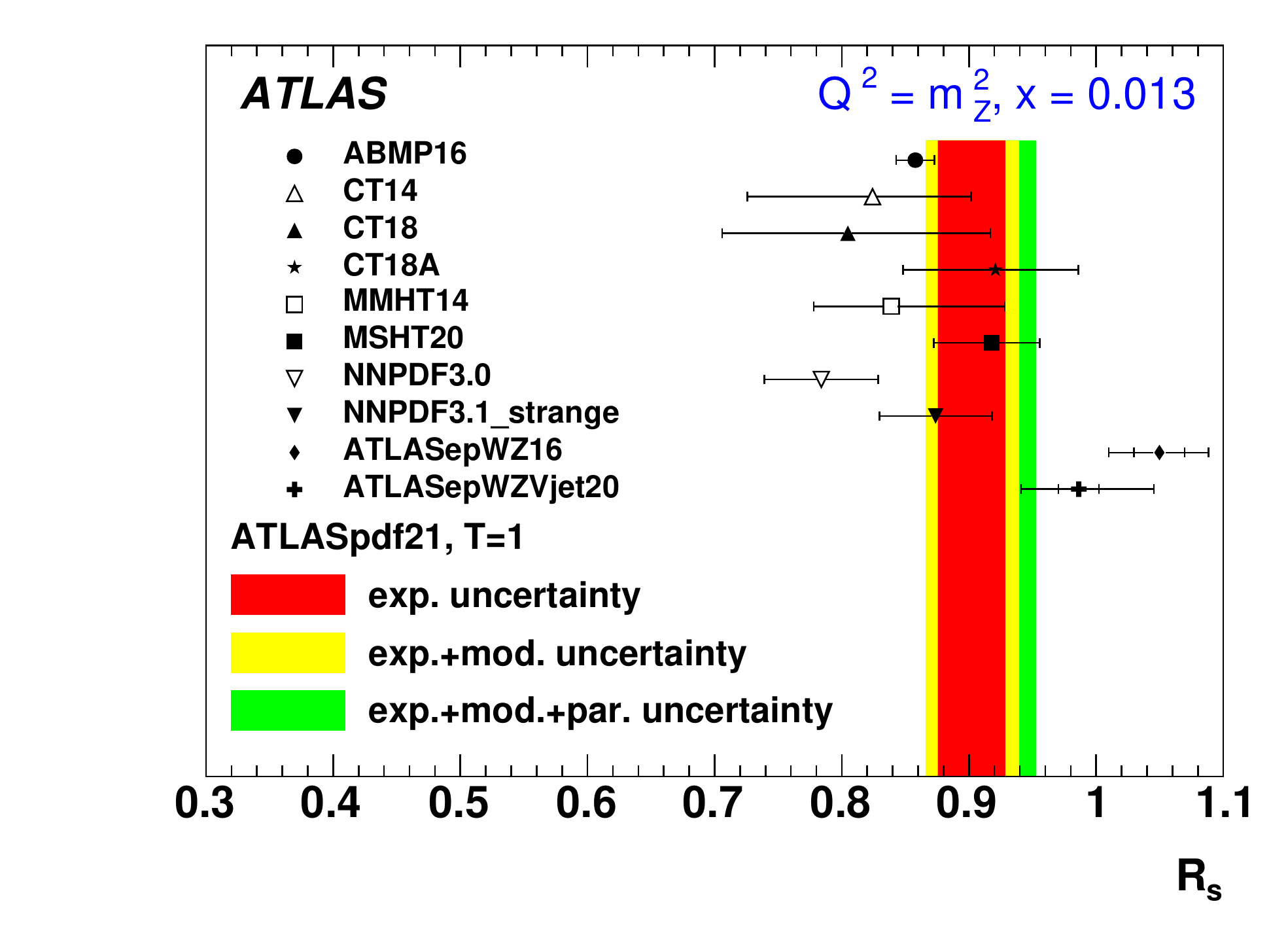}
\caption{ $R_s$ from ATLASpdf21, showing experimental uncertainties evaluated with $T=1$, model and parameterisation uncertainties, compared with other recent PDFs: ABMP16~\cite{Alekhin:2017kpj}, CT14~\cite{Dulat:2015mca}, CT18, CT18A~\cite{Hou:2019efy}, MMHT14~\cite{Harland-Lang:2014zoa}, MSHT20~\cite{Bailey:2020ooq}, NNPDF3.0~\cite{NNPDF:2014otw}, NNPDF3.1{\_}strange~\cite{Faura:2020oom}, ATLASepWZ16~\cite{1612.03016} and ATLASepWZVjets20~\cite{Vjets}. Left: $R_s$ at $Q^{2}=1.9$~\GeV$^2$ and $x=0.023$. Right: $R_{s}$ at $Q^2 = m_{Z}^{2}$ and $x=0.013$.\label{fig:RsT1}}
\end{centering}
\end{figure*}

Figure~\ref{fig:modelparam3} shows the $R_s$ distribution and its uncertainties as a function of $x$, at
$Q^2 = 1.9$~\GeV$^2$.
Figure~\ref{fig:RsT1} shows the value of $R_s$, at $Q^2=1.9$~\GeV$^2$ and $x=0.023$ and at $Q^2 = m_Z^2$ and $x=0.013$, compared with other PDFs. For the ATLASpdf21 fit, the value
of $R_s$ at this low scale has decreased relative to previous ATLAS fits. The ATLASpdf21 central value of $R_s \sim 0.8$ is due firstly to the input of the $V$+\,jets data, as seen in the change from ATLASepWZ16 ($R_s \sim 1.15$) to ATLASepWZVjets20 ($R_s \sim 1.0$), secondly to the new freedom in the low-$x$ parameterisation, whereby the ATLASpdf21 central parameterisation corresponds to the low end of the uncertainty band of the ATLASepWZVjets20 result ($R_s \sim 0.85$), and thirdly to the input of the $W,Z$ data at 8~\TeV.
On the other hand, $R_s$ has increased for some of the other PDFs, e.g.\ MSHT20~\cite{Bailey:2020ooq} compared with MMHT14~\cite{Harland-Lang:2014zoa}, and CT18A~\cite{Hou:2019efy} compared with CT14~\cite{Dulat:2015mca}. This increase is due to the input of the ATLAS precision inclusive $W,Z$ data at 7~\TeV. Thus, the recent evaluations of this ratio of strangeness to light quarks are now in fair agreement. However, the strangeness is still substantially larger at low~$x$ than the older values which estimated $R_s\sim 0.5$. This illustration of the ratio $R_s$ at $x=0.023$ should be supplemented by a comparison of the full shape of the $R_s$ distribution in Section~\ref{sec:globaltol}.
\clearpage
\section{Consideration of $\chi^2$ tolerance and comparison with global fits}
\label{sec:globaltol}
 
\subsection{$\chi^2$ tolerance}
The experimental uncertainties presented for the fitted PDFs were evaluated by the usual criterion of parameter setting, $\Delta\chi^2 = 1$. However, global PDF fitting groups such as CT and MSHT use enhanced $\chi^2$
tolerances,
$\Delta\chi^2 = T^2$, where $T^2$ is a dynamic tolerance ranging from 10 to 100.\footnote{NNPDF uses a completely different method for
estimation of PDF uncertainties, but the magnitude of their uncertainties is comparable to those of MSHT and CT.}
The justification for this comes
from consideration of the consistency, or relative inconsistency,  of input data sets~\cite{MSTW}. In the present case,
whereas the HERA data is a single consistent set of data with common correlated uncertainties, the ATLAS
data are more disparate and a study to find an appropriate tolerance was performed.
 
One way to estimate a suitable tolerance for a fit including the ATLAS data is to consider the dynamic tolerance procedure
of the MSHT group, as first
introduced in the MSTW paper~\cite{MSTW}. Briefly restated, the method is as follows. The inverse of the Hessian matrix of the fit is
diagonalised to give the eigenvector combinations of the parameters of the fit. The eigenvalues are the squares of the uncertainties
in these eigenvector combinations of the parameters. The parameter displacements from the global minimum can then be expanded in a
basis of rescaled eigenvectors, $e_{ik}=\sqrt{\lambda_k} v_{ik}$, where $\lambda_k$ is the $k$-th eigenvalue and $v_{ik}$ is the $i$-th
component of the $k$-th orthonormal eigenvector. Pairs of eigenvector PDFs have parameters given by $a_i(k^{\pm})=a_i^0 \pm t e_{ik}$,
where $a_i^0$ are the central parameters, and $t=1$ should give the usual change in $\chi^2$ of $T^2 = \Delta\chi^2=1$. In the
quadratic approximation, $t=T$, and this is reasonably well obeyed for the present ATLASpdf21 21-parameter fit, at least up to $t=5$.
 
Having determined the eigenvectors, each of these is treated as a different `hypothesis' for a theory which can fit the data, and each
of them is checked to see if it can describe each of the data sets within the hypothesis-testing criterion,
$\chi_n^2 \lesssim N+\sqrt{2N}$, as $t$ is increased and decreased along the eigenvector, where $\chi^2_n$ is the $\chi^2$ for a particular
data set and $N$ is the number of data points for that set. As soon as $\chi^2_n$ goes above this limit, which sets the maximum
value of $t$, and hence of $T$, that can be tolerated for this eigenvector in this direction, for this data set. Each eigenvector
will have a most constraining data set, for each direction, and in principle the values of the tolerances $T$ for the most
constraining data set can be different for each eigenvector and each direction, hence the term `dynamic'.
 
This method is made more sophisticated in a number of ways. Firstly, the hypothesis-testing criterion chosen is not exactly the simple
$\chi^2_n< N + \sqrt{2N}$, but is $\chi^2_n < \xi_{68}$, where $\xi_{68}$ is given by $\int^{\xi_{68}}_0 \dif\chi^2 P_N(\chi^2) =0.68$ and
$P_N(\chi^2)$ is the usual probability density function for the $\chi^2$ distribution with $N$ degrees of freedom. Thus each data set should be described within its $68\%$ confidence level. Secondly, the
criterion imposed is not actually $\chi^2_n < \xi_{68}$, but $\chi^2_n < (\chi^2_{n,0}/\xi_{50})\,\xi_{68}$, to allow for the fact that the central fit $\chi^2_{n,0}$ for the data set may not be perfect, where perfect would imply $\chi^2_{n,0} = \xi_{50}$.
 
For the ATLASpdf21 fit the most constraining data sets for the eigenvectors are the $W,Z$ data at 7~\TeV\ or the $Z/\gamma^{*}$ triple differential cross-section data at
8~\TeV\ (and very occasionally the $W$ data at 8~\TeV). The tolerance values for all the eigenvectors, in both directions, are all in the
range 2.4--4.2. These values are all fairly
similar, so a single value can be used and, since the tolerance should not be set too large whereby a data
set is far outside its $68\%$ CL limit, this suggests the simple choice of $T = 3$.

In the above procedure to evaluate the dynamic tolerance, the fact that the data sets are not fitted perfectly was taken into account.
One may estimate how much of the increased tolerance comes from this part of the procedure
by increasing the uncertainties of each data set by $\sqrt{\chi^2/\mathrm{NDP}}$ for those data, such that $\chi^2/\mathrm{NDP}\sim 1$ is achieved in the fit.
This will clearly increase the uncertainties of the output PDFs.
However, a naive application of this procedure would increase all the experimental uncertainties and this seems
somewhat unnecessary for the statistical uncertainties, which are well known.
Hence, only the systematic uncertainties are increased. Starting with the $\sqrt{\chi^2/\mathrm{NDP}}$ scaling factors, the fit is iterated, adjusting these factors
at each iteration, until a $\chi^2/\mathrm{NDP} \sim 1$ for each data
set is achieved, After three iterations there is no longer any significant change and
the scaling factors range from $1.0$ to $1.6$.
 
It is non-trivial that for these reweightings of the data the PDF shapes hardly differ from those of an unweighted fit. This shows that
the fit is stable for modest reweighting of the input data sets. The uncertainties of the PDFs have
increased, but only by a factor which corresponds to a tolerance of up to $T \sim 1.5$. Thus the greater part of the dynamic tolerance
estimate of $T=3$ comes from the application of the hypothesis-testing criterion to the $\chi^2$ for the eigenvectors of the fit.
 
\begin{figure*}[t!]
\begin{centering}
\includegraphics[width=0.42\textwidth]{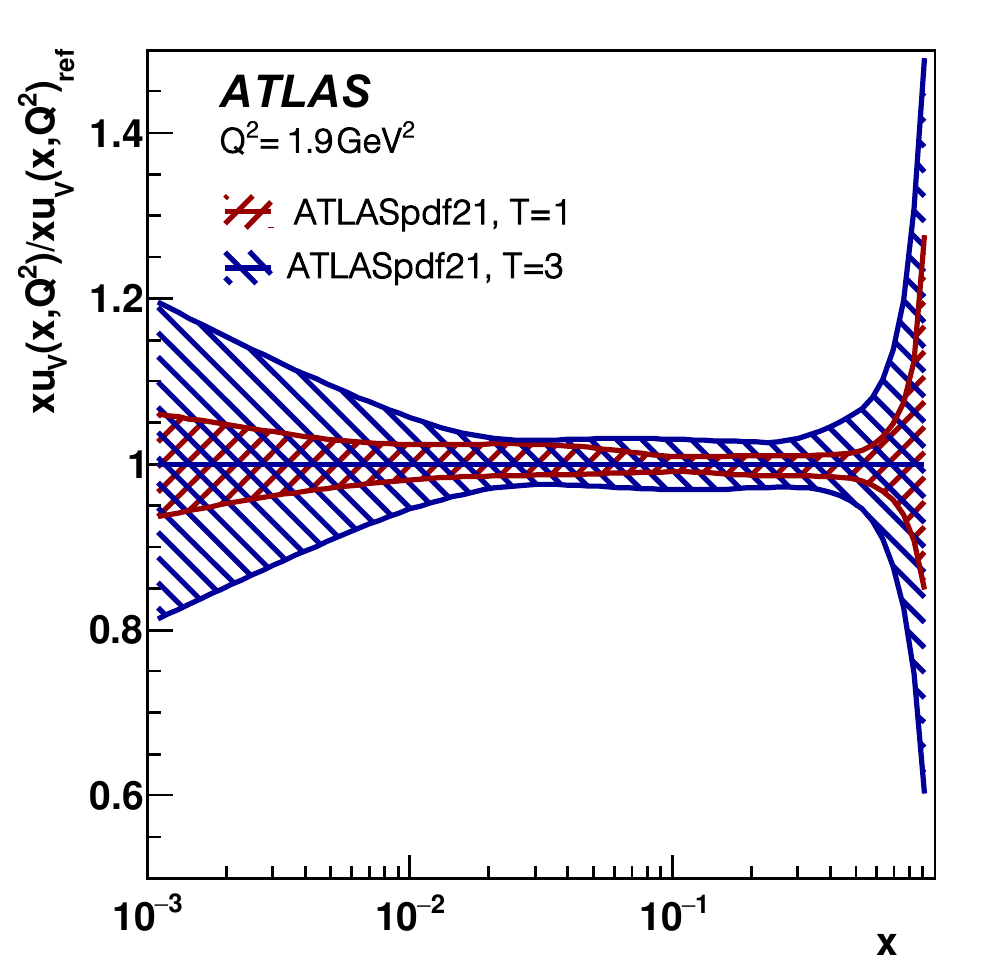}
\includegraphics[width=0.42\textwidth]{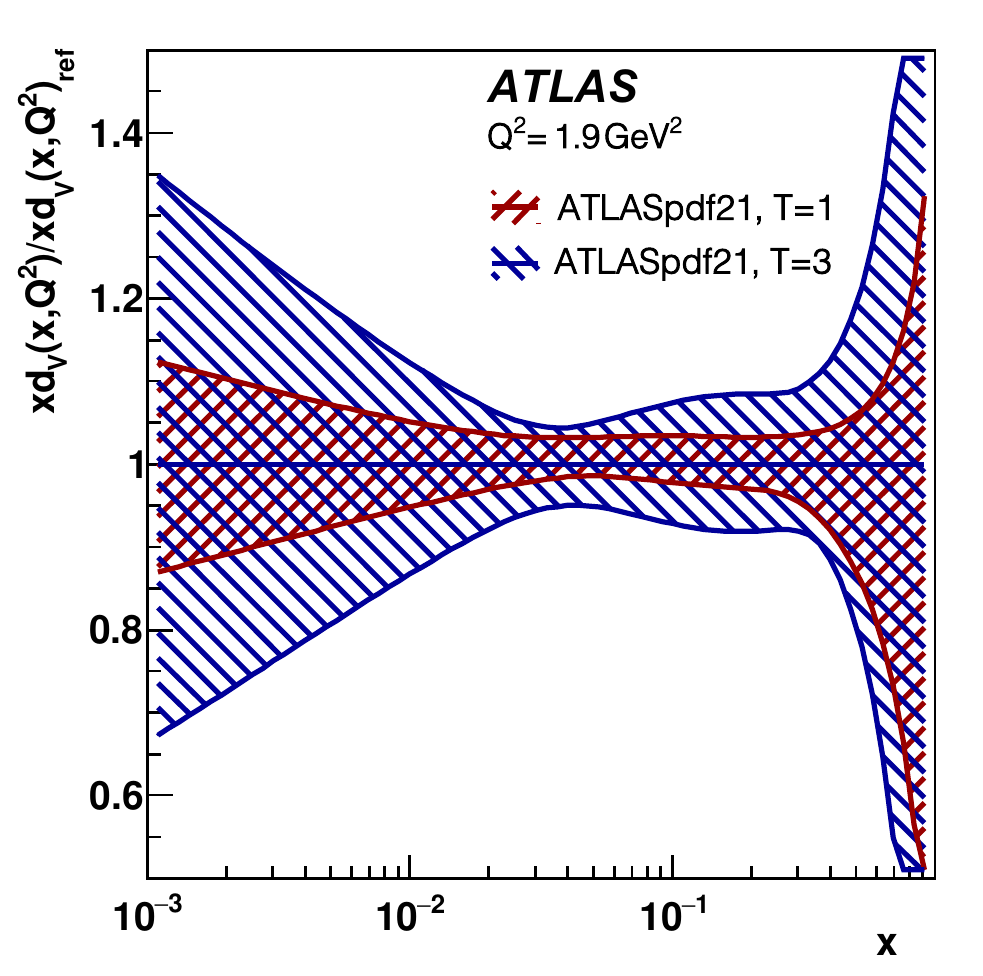}
\includegraphics[width=0.42\textwidth]{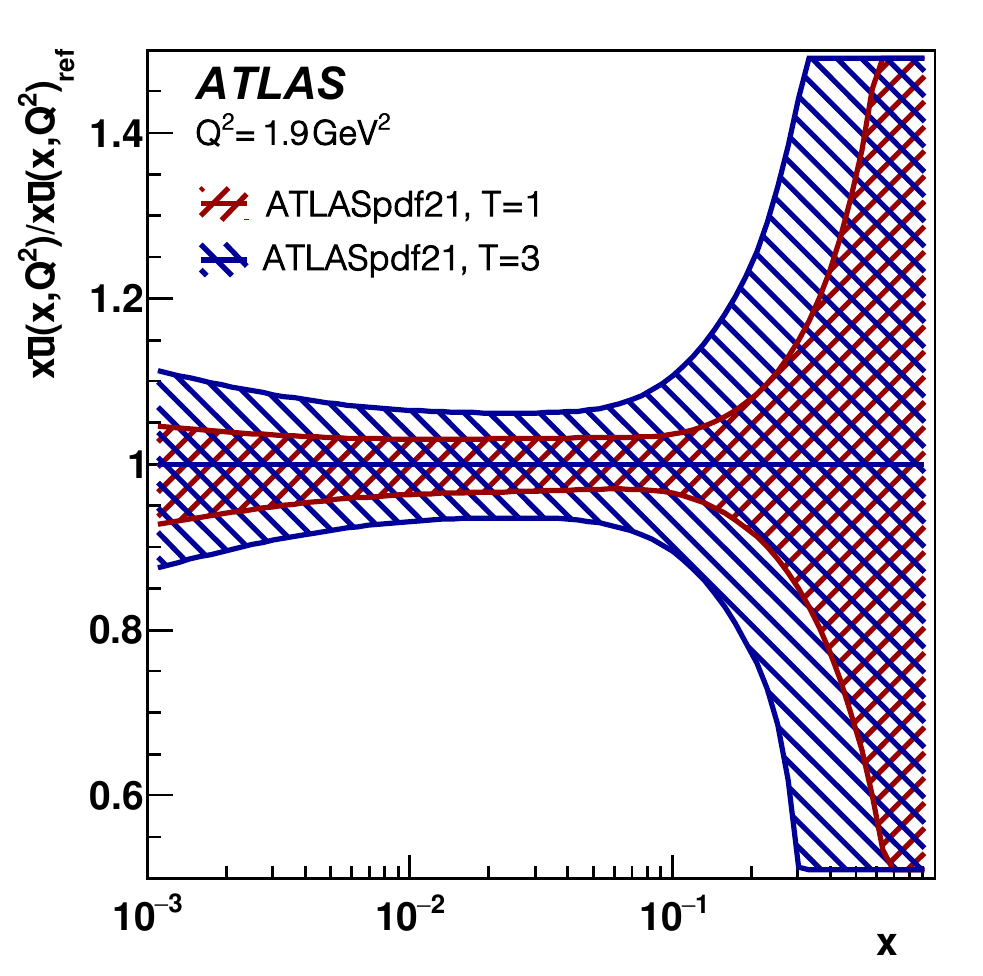}
\includegraphics[width=0.42\textwidth]{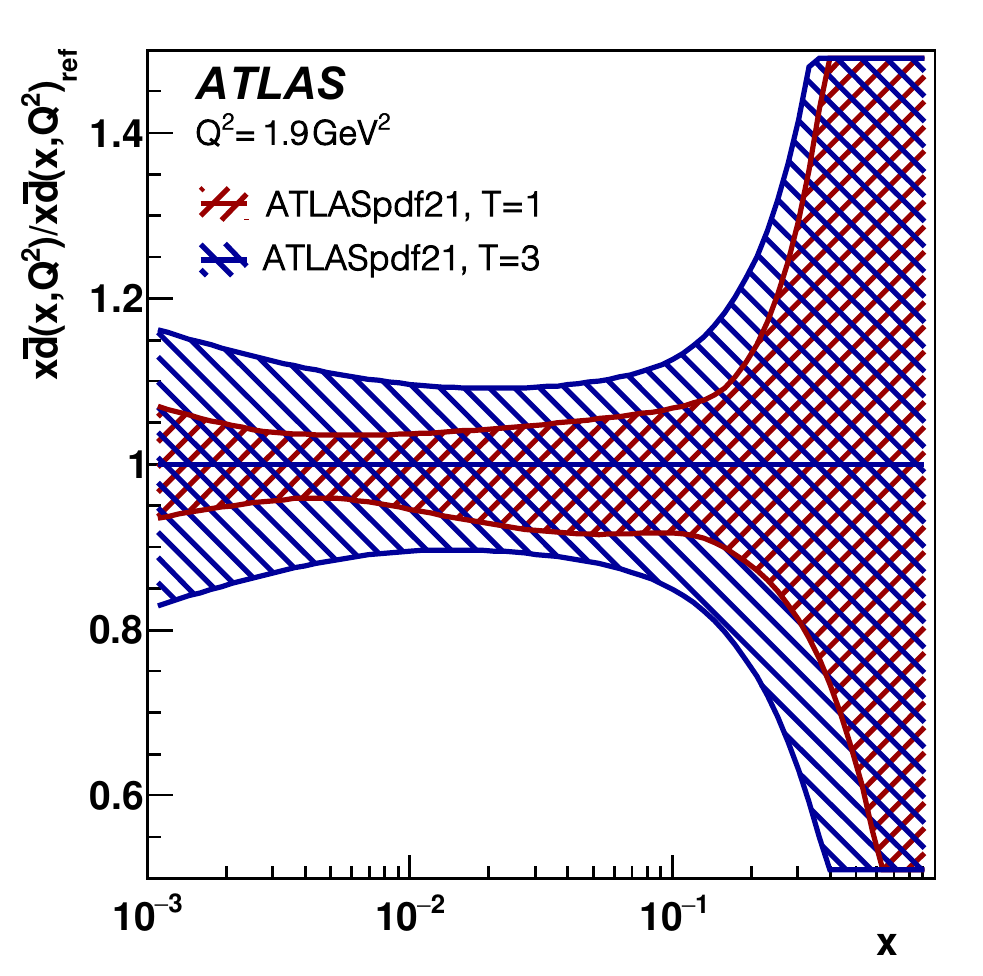}
\includegraphics[width=0.42\textwidth]{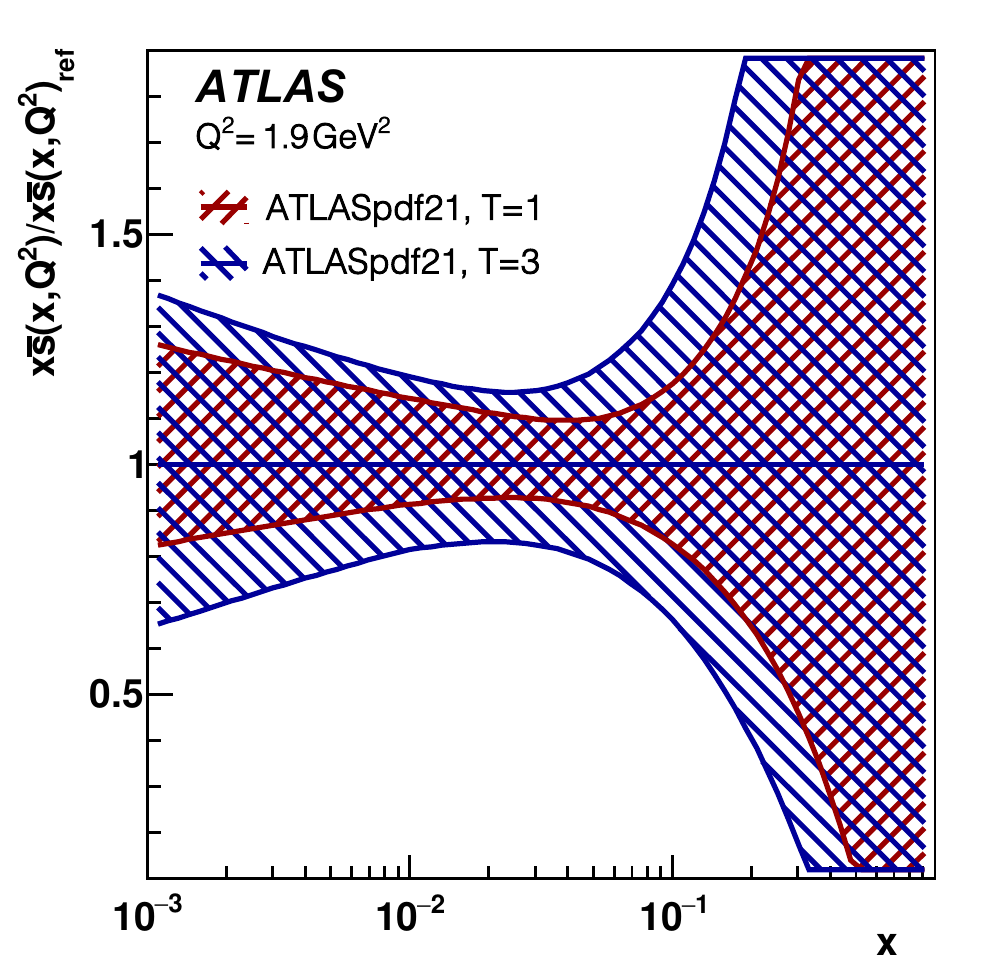}
\includegraphics[width=0.42\textwidth]{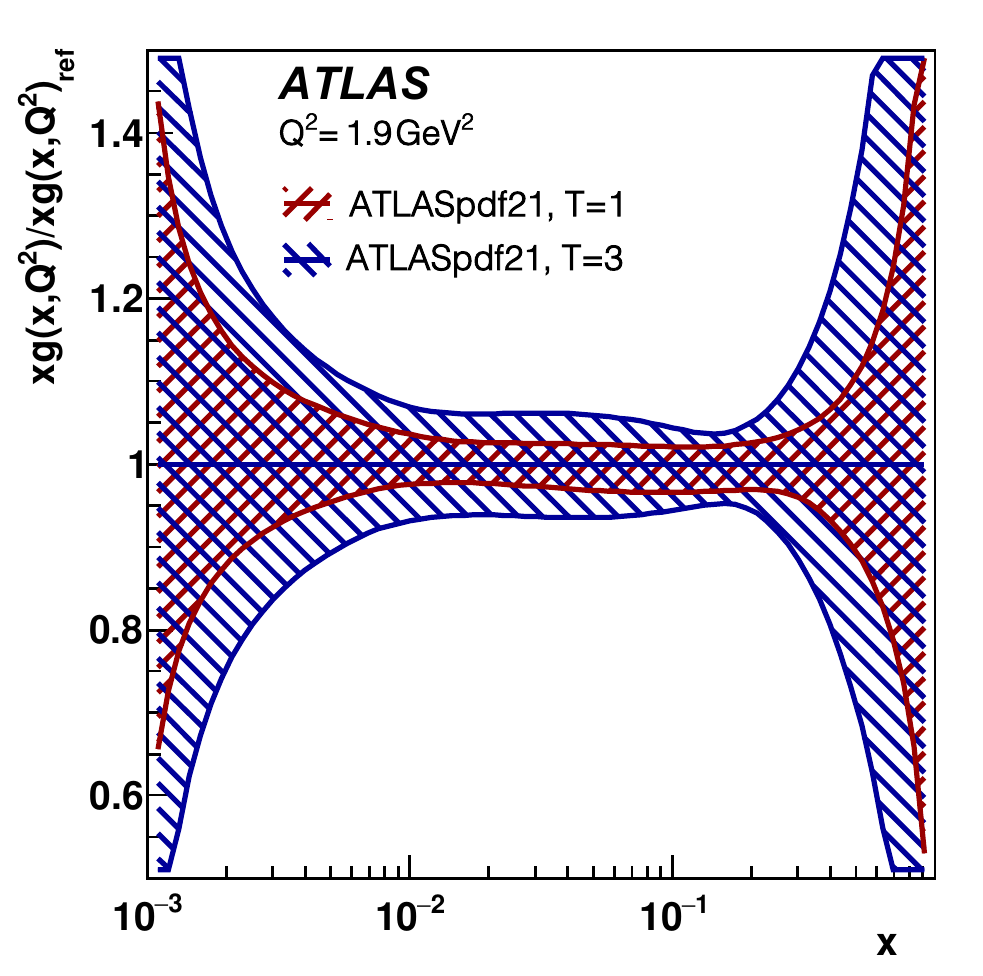}
\caption{ ATLASpdf21 fits comparing the full uncertainties (experimental, model, parameterisation) for tolerance values $T=1$ and $T=3$.
Top left: $xu_v$. Top right: $xd_v$. Middle left: $x\bar{u}$. Middle right: $x\bar{d}$. Bottom left: $x\bar{s}$. Bottom right: $xg$.
\label{fig:3modelparam1}
}
\end{centering}
\end{figure*}

The tolerance $T=3$ is applied for the final estimation of  PDF uncertainty; however, ATLASpdf21 PDFs with both choices of tolerance, $T=1$ and $T=3$, will be made available on the LHAPDF PDF repository~\cite{LHAPDF}.
Having fixed the tolerance for the experimental uncertainties of the PDFs, the full uncertainties,
including experimental, model and parameterisation uncertainties are compared for the choices $T=3$ and $T=1$ in Figure~\ref{fig:3modelparam1}.
When an enhanced tolerance such as this is applied to the experimental uncertainties,
the model and parameterisation uncertainties are far less significant.
 
Comparisons of the differential cross-section measurements used as input to the fit with the predictions of ATLASpdf21 fit, with $T=3$ and full uncertainties, are shown in Appendix~\ref{sec:datafitcomp}.
 
\subsection{ Comparison of ATLASpdf21 with global PDFs}
\begin{figure*}
\begin{centering}
\includegraphics[width=0.48\textwidth]{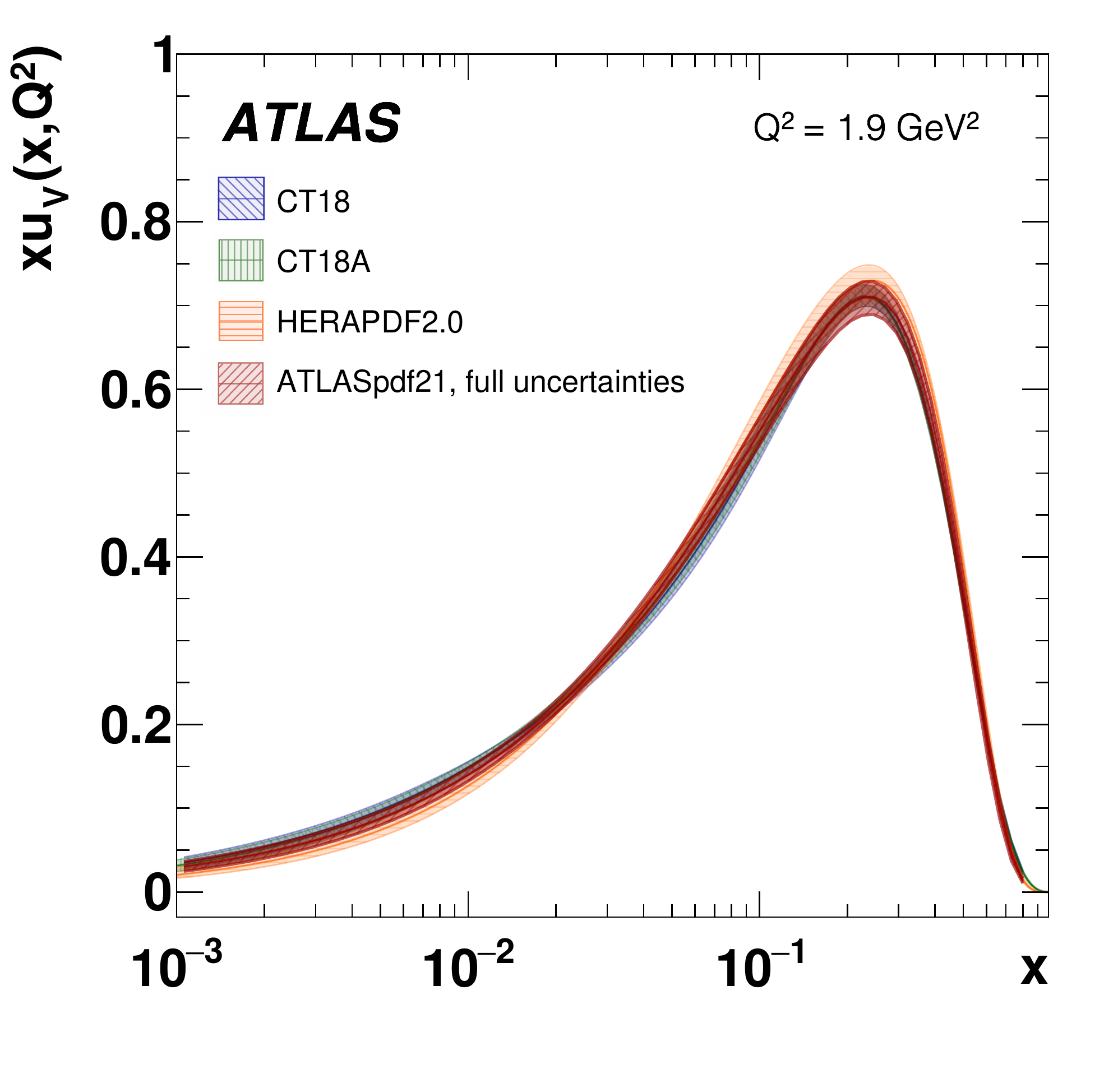}
\includegraphics[width=0.48\textwidth]{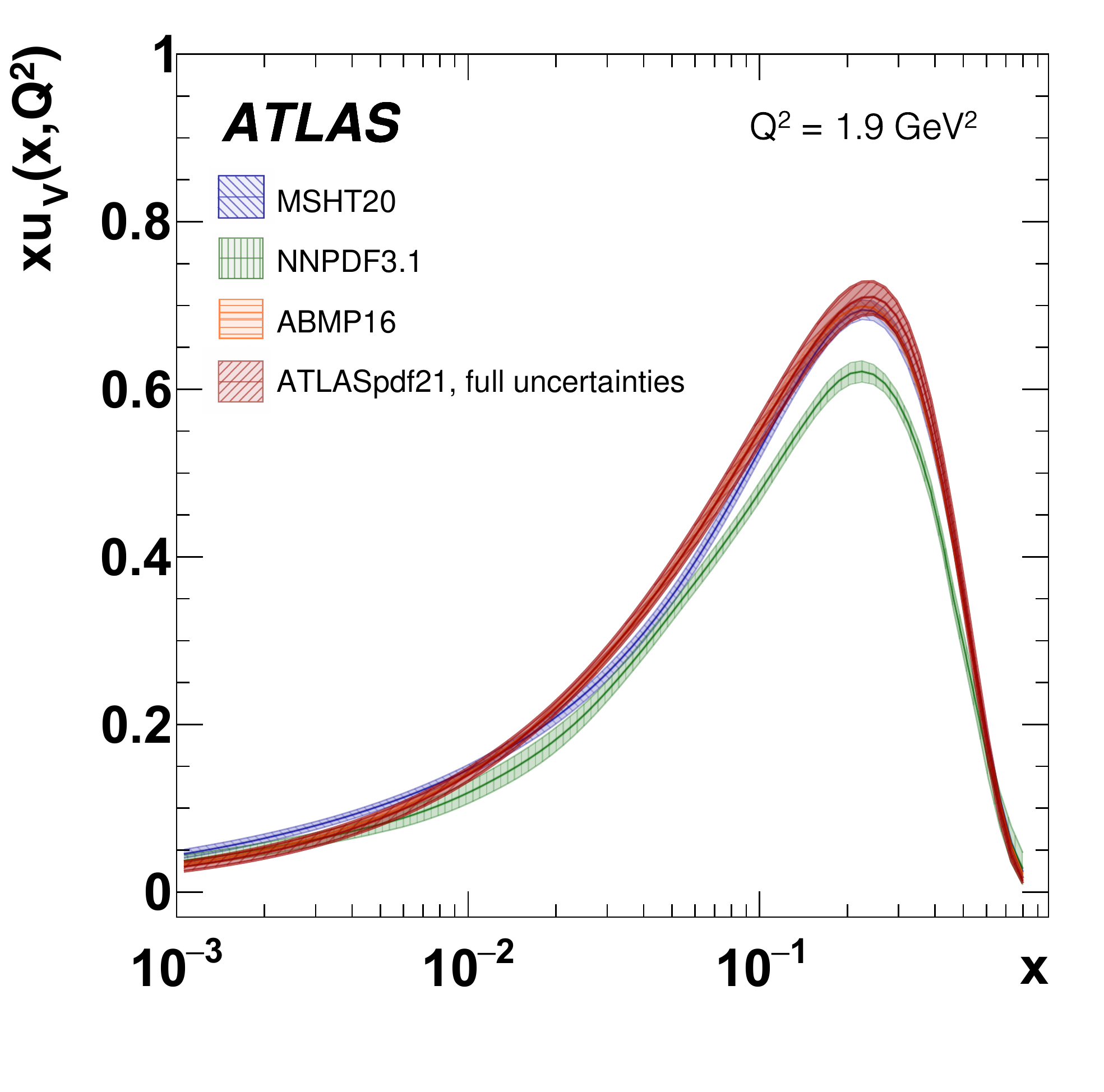}
\includegraphics[width=0.48\textwidth]{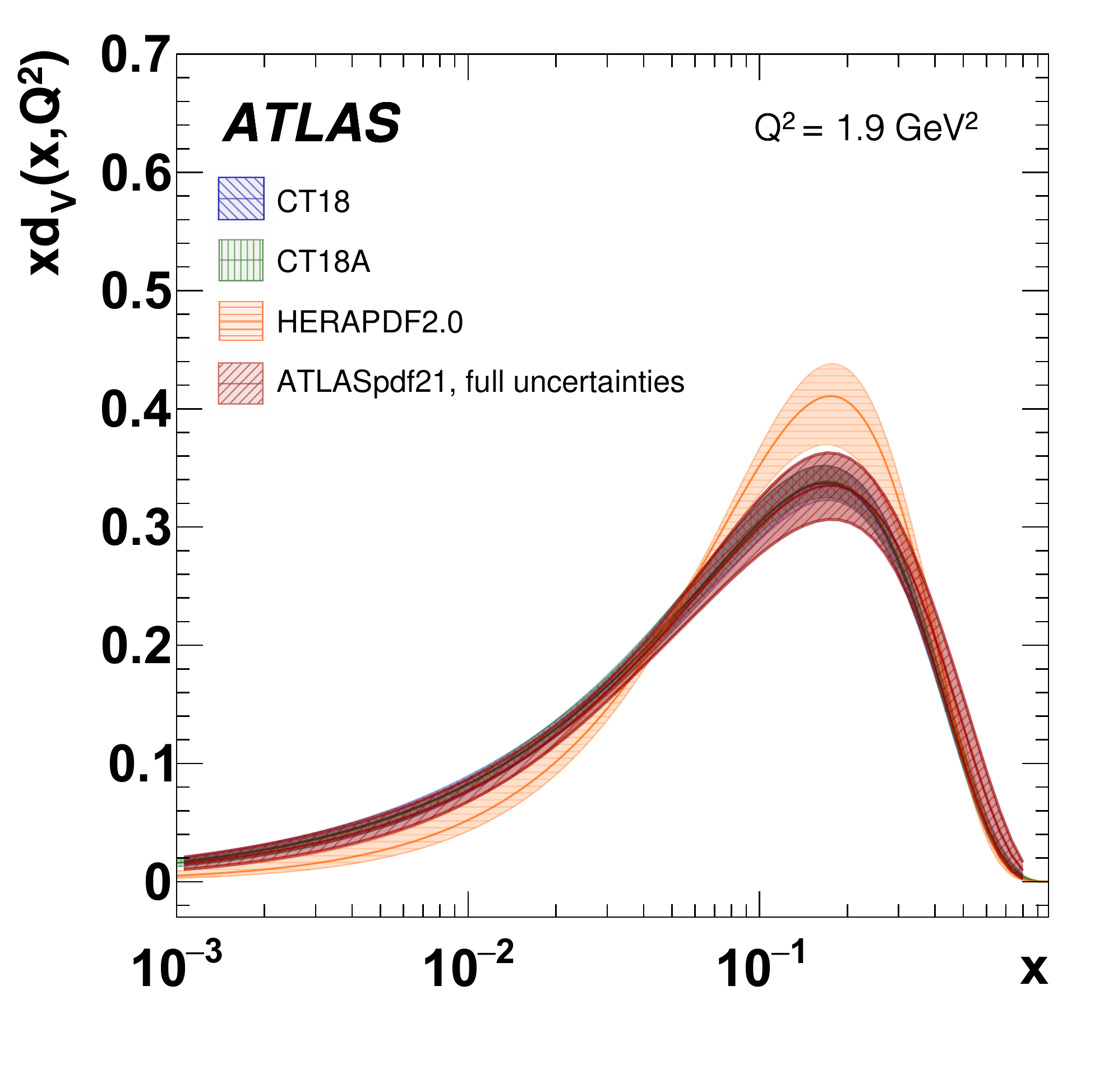}
\includegraphics[width=0.48\textwidth]{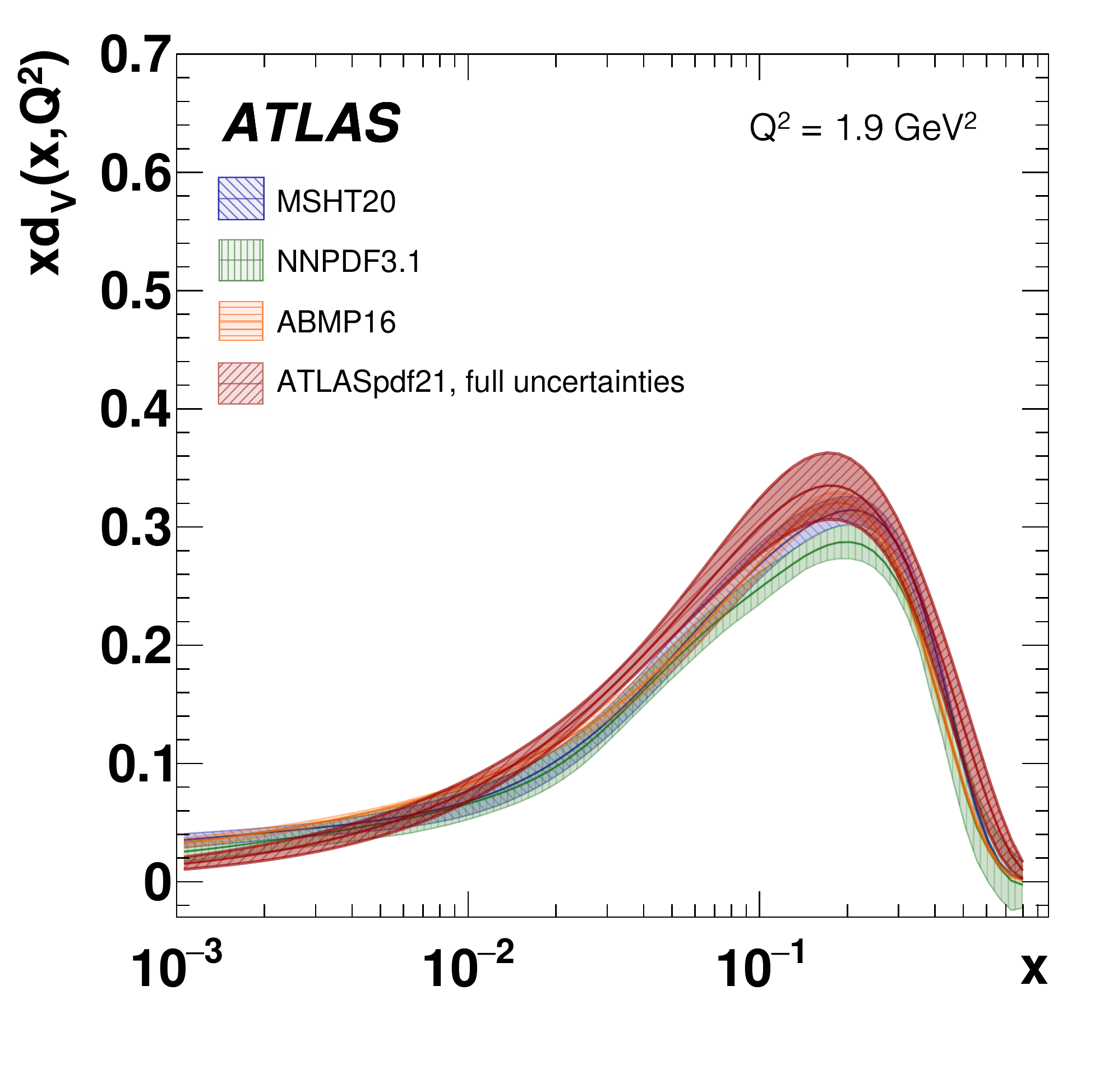}
\caption{ ATLASpdf21 $xu_v$ and $xd_v$ distributions with full uncertainties (experimental $T=3$, model, parameterisation) compared with other PDFs.
Left: CT18, CT18A, HERAPDF2.0. Right: MSHT20, NNPDF3.1, ABMP16.
\label{fig:global1}
}
\end{centering}
\end{figure*}

In Figures~\ref{fig:global1}--\ref{fig:global4}, the ATLASpdf21 PDFs are compared
with the HERAPDF2.0~\cite{herapdf20}, CT18, CT18A~\cite{Hou:2019efy}, NNPDF3.1~\cite{NNPDF:2017mvq}, MSHT20~\cite{Bailey:2020ooq} and ABMP16~\cite{Alekhin:2017kpj} PDFs. All figures in this Section show ATLASpdf21 PDFs with full uncertainties, such that experimental uncertainties are evaluated with $T=3$ and model and parametrisation uncertainties are also included.
It is observed that the addition of the ATLAS
data sets to the HERA data brings the PDFs much closer to those of the global fits than to HERAPDF2.0, particularly for $xd_v$, $x\bar{d}$ and $x\bar{u}$.  The $xu_v$ and $xd_v$ distributions of the ATLASpdf21 fit are in good agreement
with CT18, CT18A, MSHT20 and ABMP16, whereas NNPDF3.1 is an outlier.
The $x\bar{u}$ distribution agrees reasonably well with all the other PDFs.
The $x\bar{d}$ distribution is in fair agreement with CT, MSHT and NNPDF, with
ABMP presenting as somewhat of an outlier.\footnote{Note that ABMP has much smaller uncertainties due to their use of tolerance $T=1$.} However, it should be noted that $x\bar{d}$ from the ATLASpdf21 fit is
somewhat higher at high~$x$. This is further illustrated by the
$x(\bar{d}-\bar{u})$ distribution, which changes from negative at high~$x$ when fitting only HERA data in HERAPDF2.0, to positive at high $x$ for ATLASpdf21, in agreement with the global PDFs.
However, the ATLASpdf21 $x(\bar{d}-\bar{u})$ distribution is more positive than the global PDFs
for $x >0.3$,
in better agreement with the newer E906 Drell--Yan data~\cite{Dove:2021ejl}, rather than the
E866 data which is in the global fits.
This is explored further in Appendix~\ref{sec:extradatafitcomp}.
The central value of the $x\bar{s}$ distribution is larger than those in NNPDF3.1 and CT18 for $x\sim 0.02$, but is in good agreement with CT18A and MSHT20.
The gluon distribution agrees best with NNPDF3.1, but is in reasonable agreement with the other global PDFs.\footnote{The ABMP16 gluon PDF corresponds to $\alphas(\mZ)=0.1145$, a lower value than in the other PDFs illustrated, which all use $\alphas(\mZ)=0.118$.}
The distributions of $R_s$ for all the PDFs illustrated are now in broad agreement but there are differences in detail. The uncertainties of HERAPDF2.0 and CT18 are larger than
the others because they do not include ATLAS inclusive $W,Z$ data at 7 TeV.\footnote{In the case of HERAPDF2.0 the uncertainty in the strange PDF is not measured but estimated.} The uncertainties of ABMP16 are smaller for the usual reason that they use tolerance $T=1$.
The input of ATLAS inclusive $W,Z$ data at 7 TeV is the only difference between the CT18 and CT18A analyses. After input of these data the CT18A $R_s$ PDF ratio
has smaller uncertainties and a larger central value of $R_s$ at low $x$ than CT18. CT18A is in good agreement with ATLASpdf21, for both central value and uncertainty,
over the full $x$ range illustrated.
The PDF analyses of MSHT20 and NNPDF3.1 also use ATLAS inclusive $W,Z$ data at 7 TeV and have a similar size of uncertainty and level of agreement with ATLASpdf21. In addition, the MSHT20
analysis uses ATLAS inclusive $W,Z$ data at 8 TeV, just as the ATLASpdf21 analysis does. Note that NNPDF have updated their study of strangeness recently to NNPDF3.1{\_}strange~\cite{Faura:2020oom}, which has a somewhat larger strangeness at low $x$ than NNPDF3.1.  The modern PDFs which use ATLAS inclusive $W,Z$ data, including ATLASpdf21 itself, all agree that strangeness is not suppressed strongly at low $x$, but there is substantial suppression at high $x$.
 
In summary, the ATLASpdf21 fit now agrees with the global
fits as well as they agree with each other. Thus
ATLAS data seem able to replicate most of the features that the fixed-target
DIS and DY data plus the Tevatron data brought to the global PDFs, see Appendix~\ref{sec:extradatafitcomp}. Using only the
HERA and ATLAS data allows a more rigorous treatment of correlated
systematic uncertainties and, in particular, of correlations between data sets.
\begin{figure*}[bht]
\begin{centering}
\includegraphics[width=0.42\textwidth]{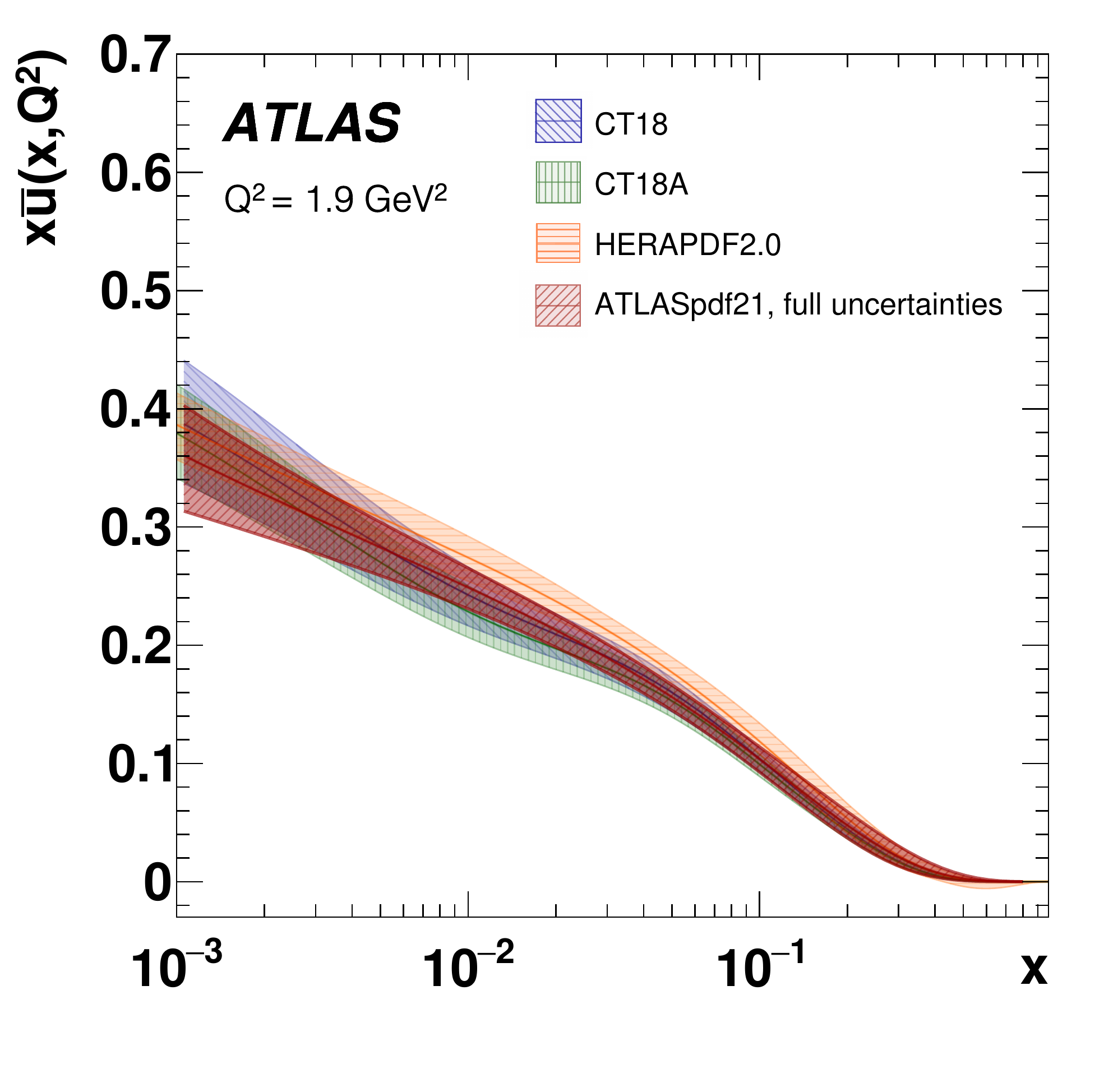}
\includegraphics[width=0.42\textwidth]{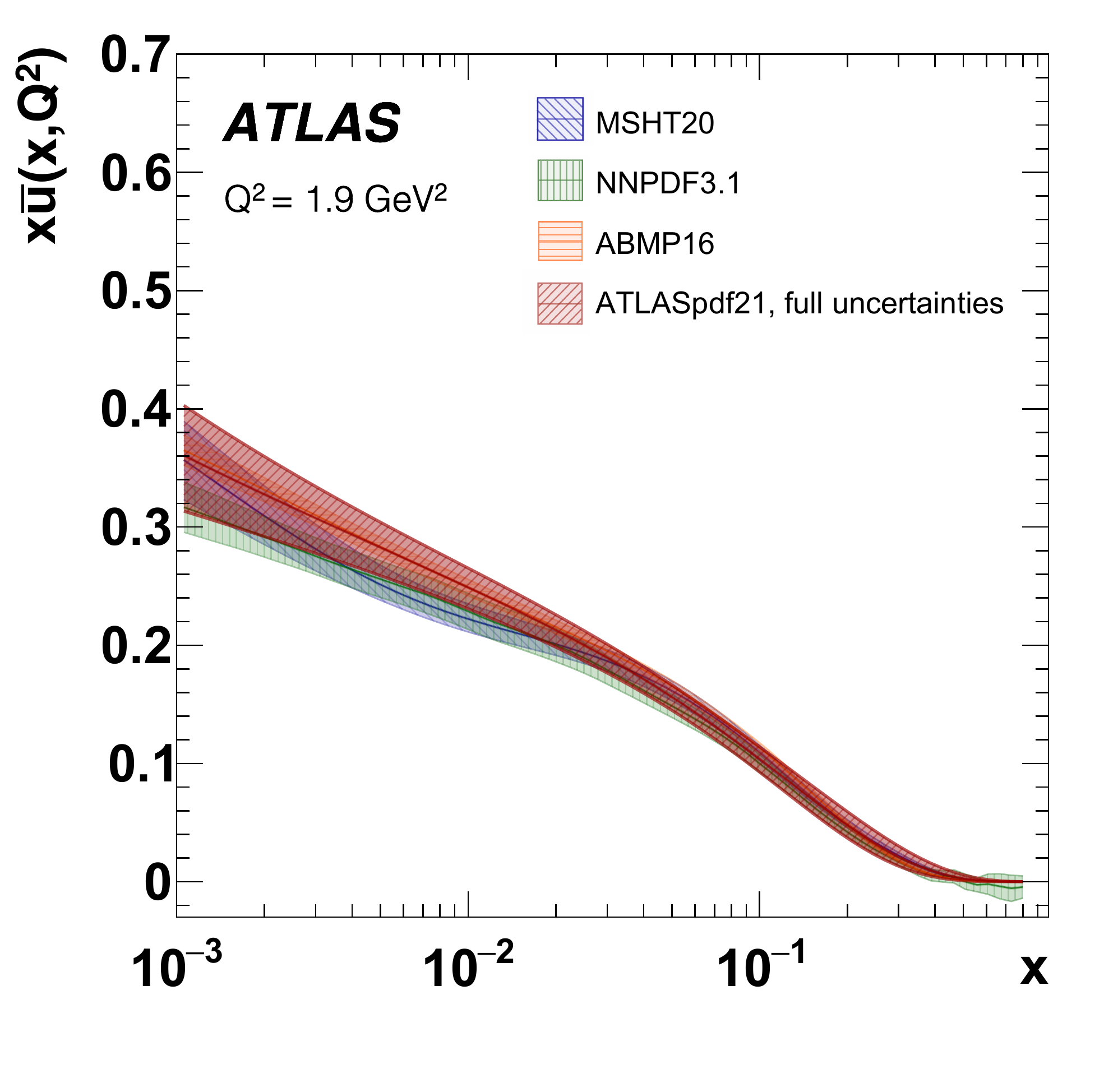}
\includegraphics[width=0.42\textwidth]{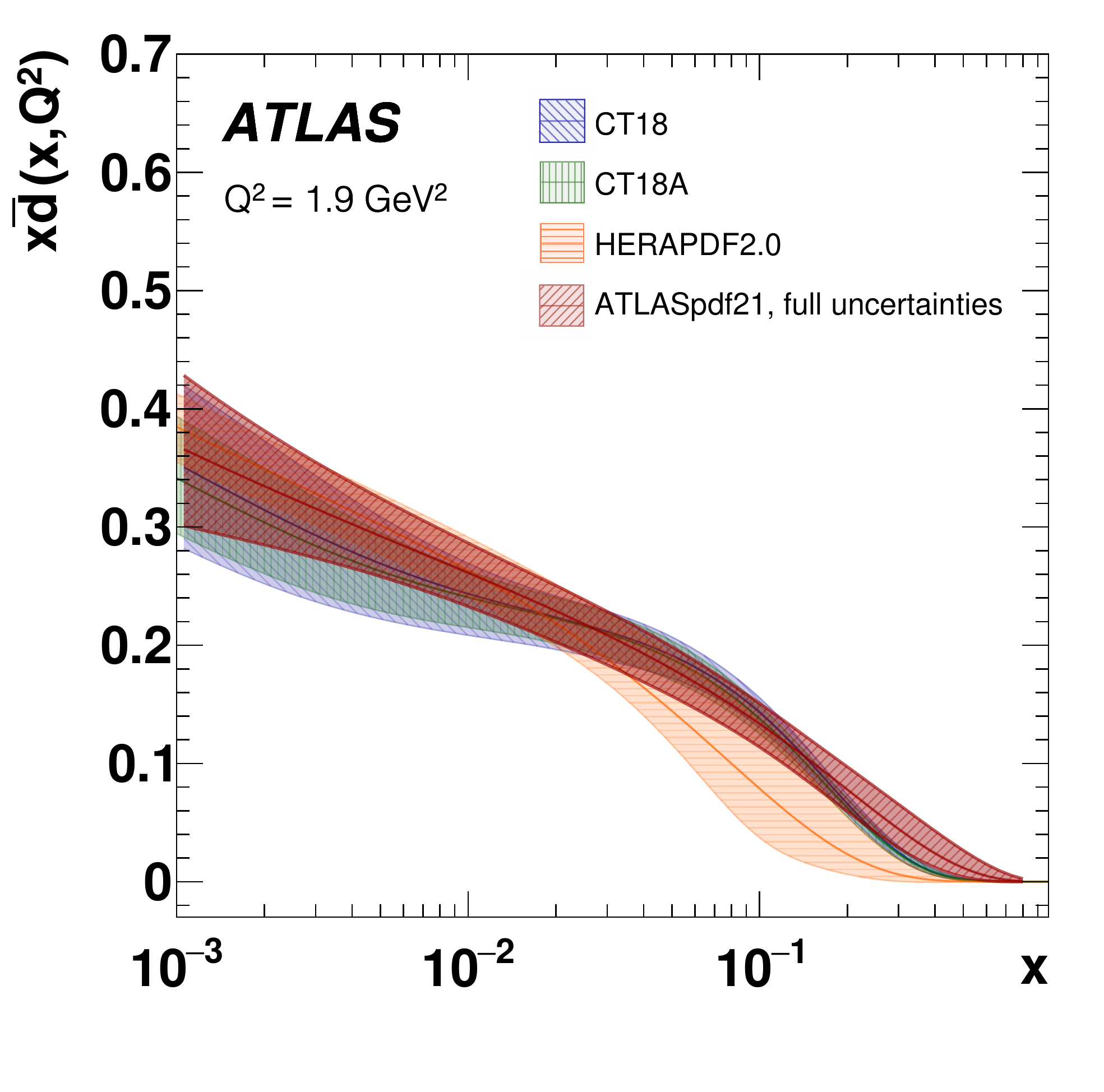}
\includegraphics[width=0.42\textwidth]{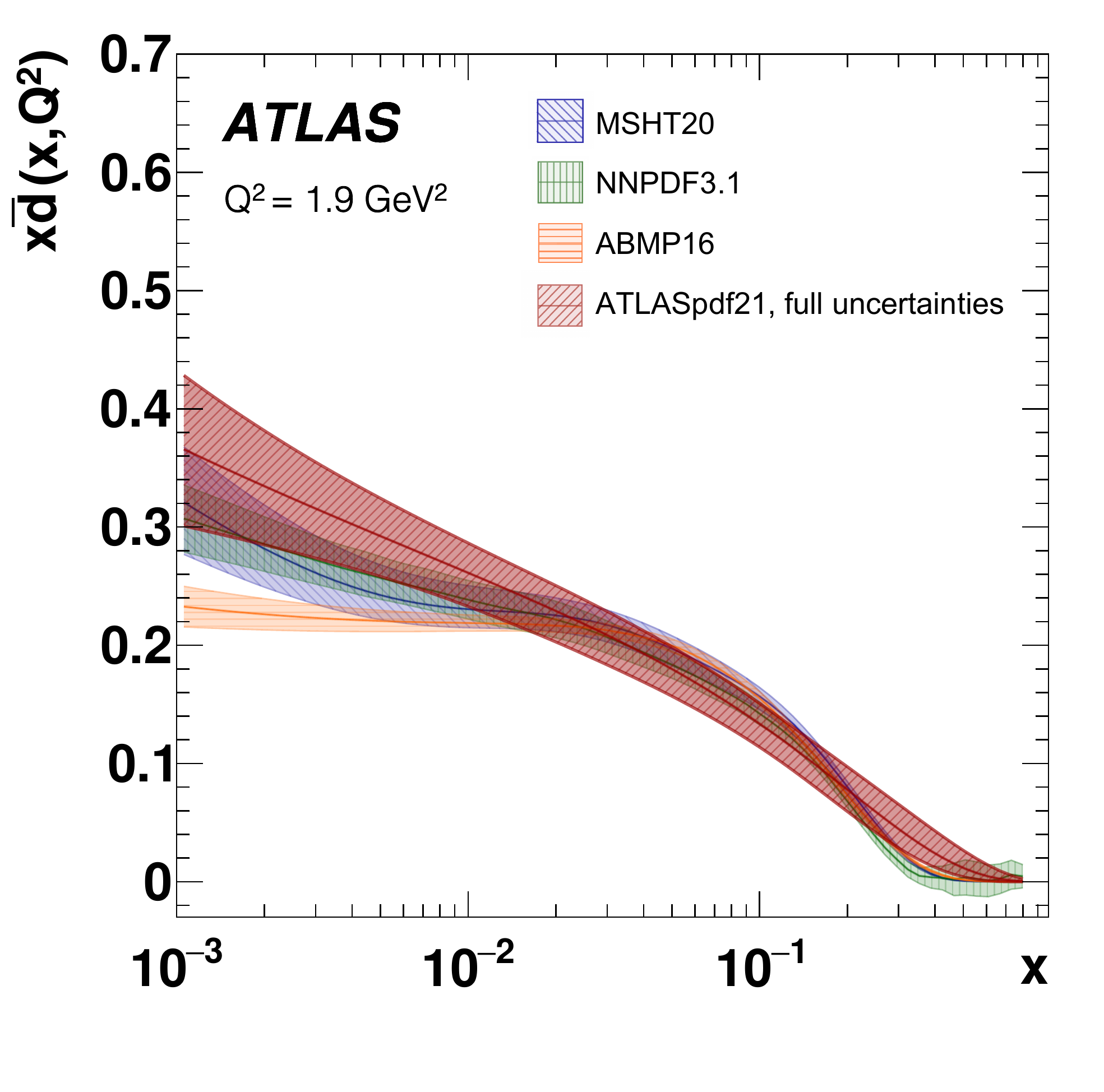}
\includegraphics[width=0.42\textwidth]{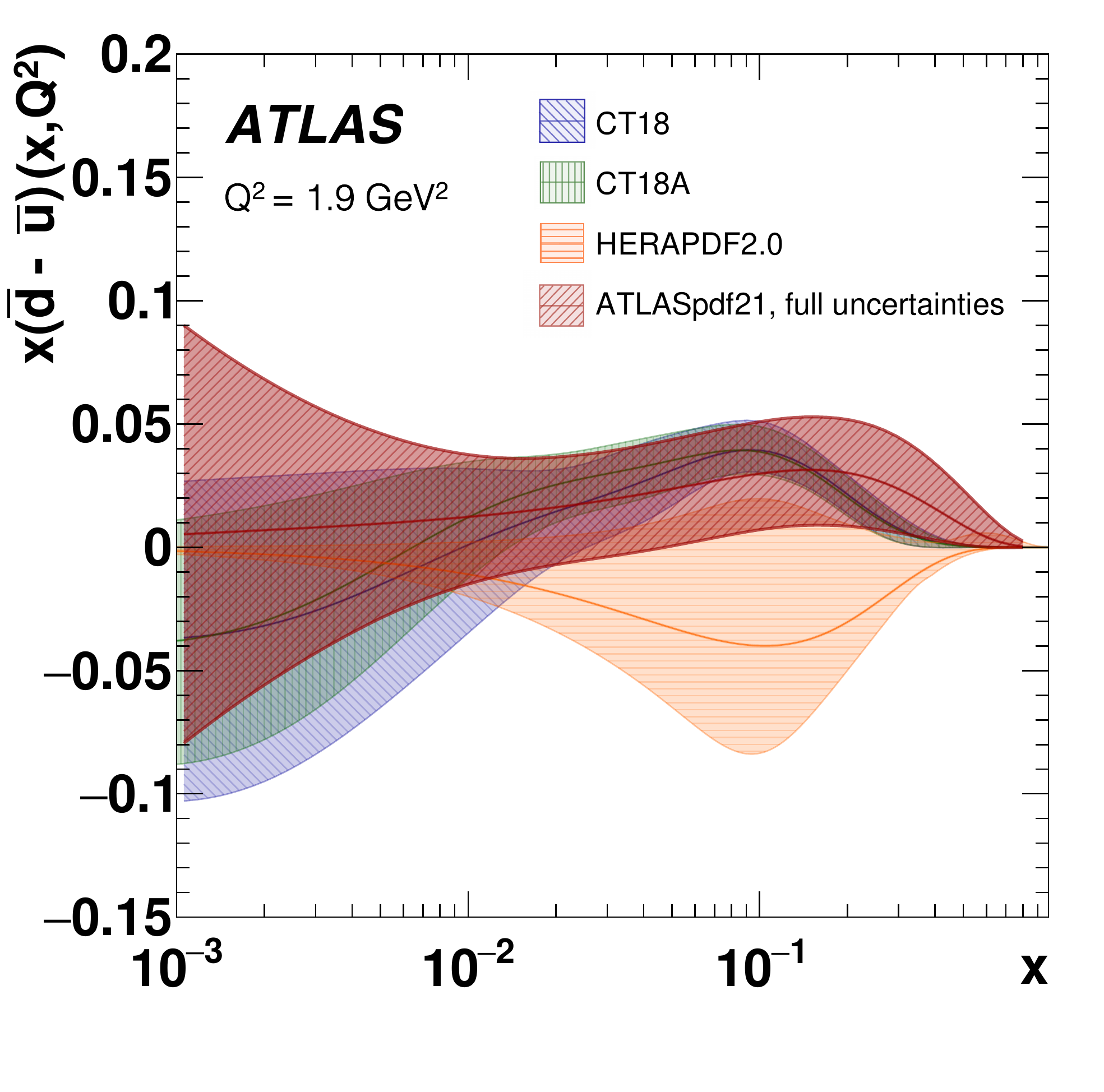}
\includegraphics[width=0.42\textwidth]{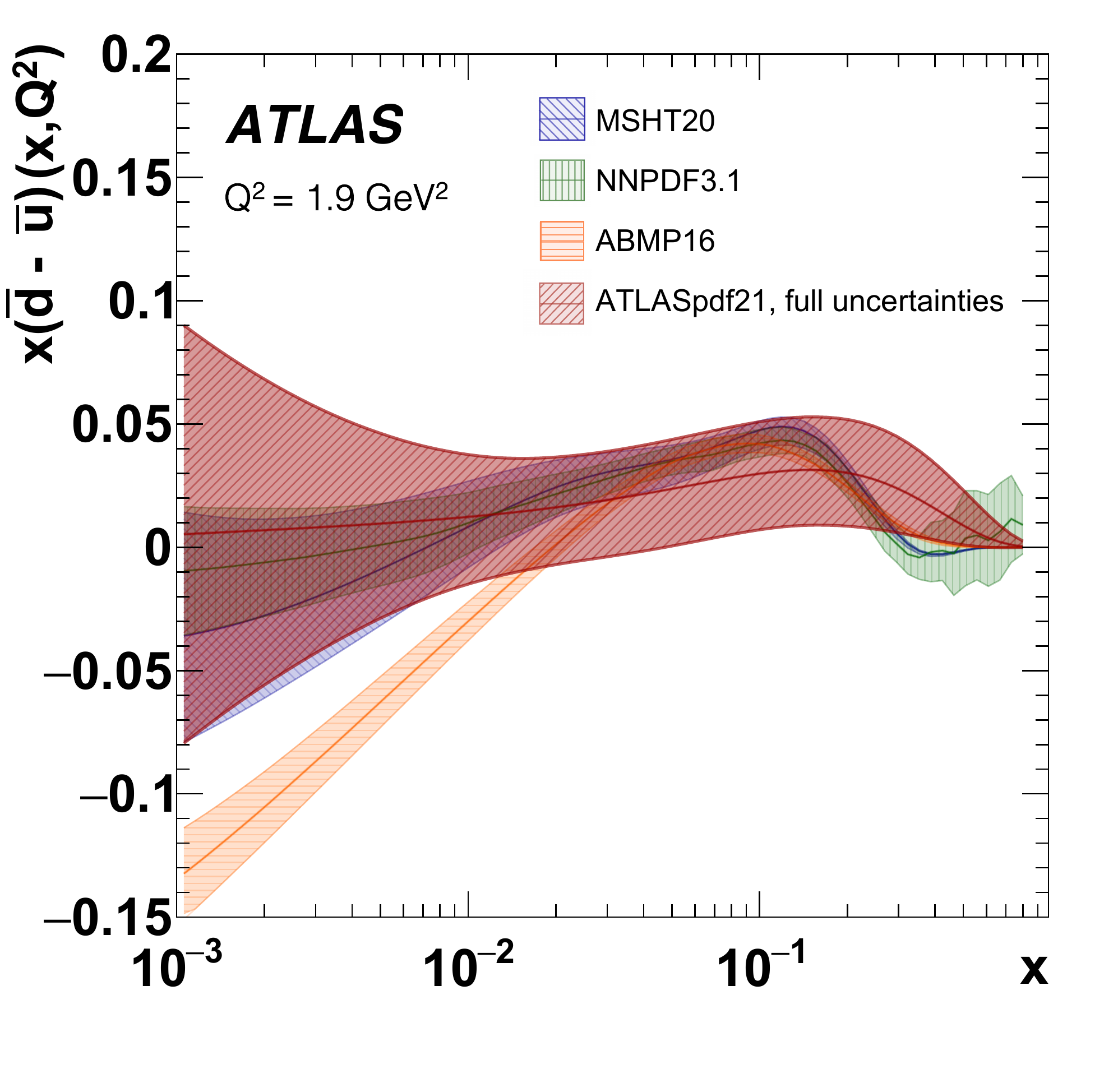}
\caption{ATLASpdf21 $x\bar{u}$, $x\bar{d}$ and $x(\bar{d}-\bar{u})$ distributions with full uncertainties (experimental $T=3$, model, parameterisation) compared with other PDFs.
Left: CT18, CT18A, HERAPDF2.0. Right: MSHT20, NNPDF3.1, ABMP16.
}
\end{centering}
\end{figure*}
 
\begin{figure*}[bht]
\begin{centering}
\includegraphics[width=0.42\textwidth]{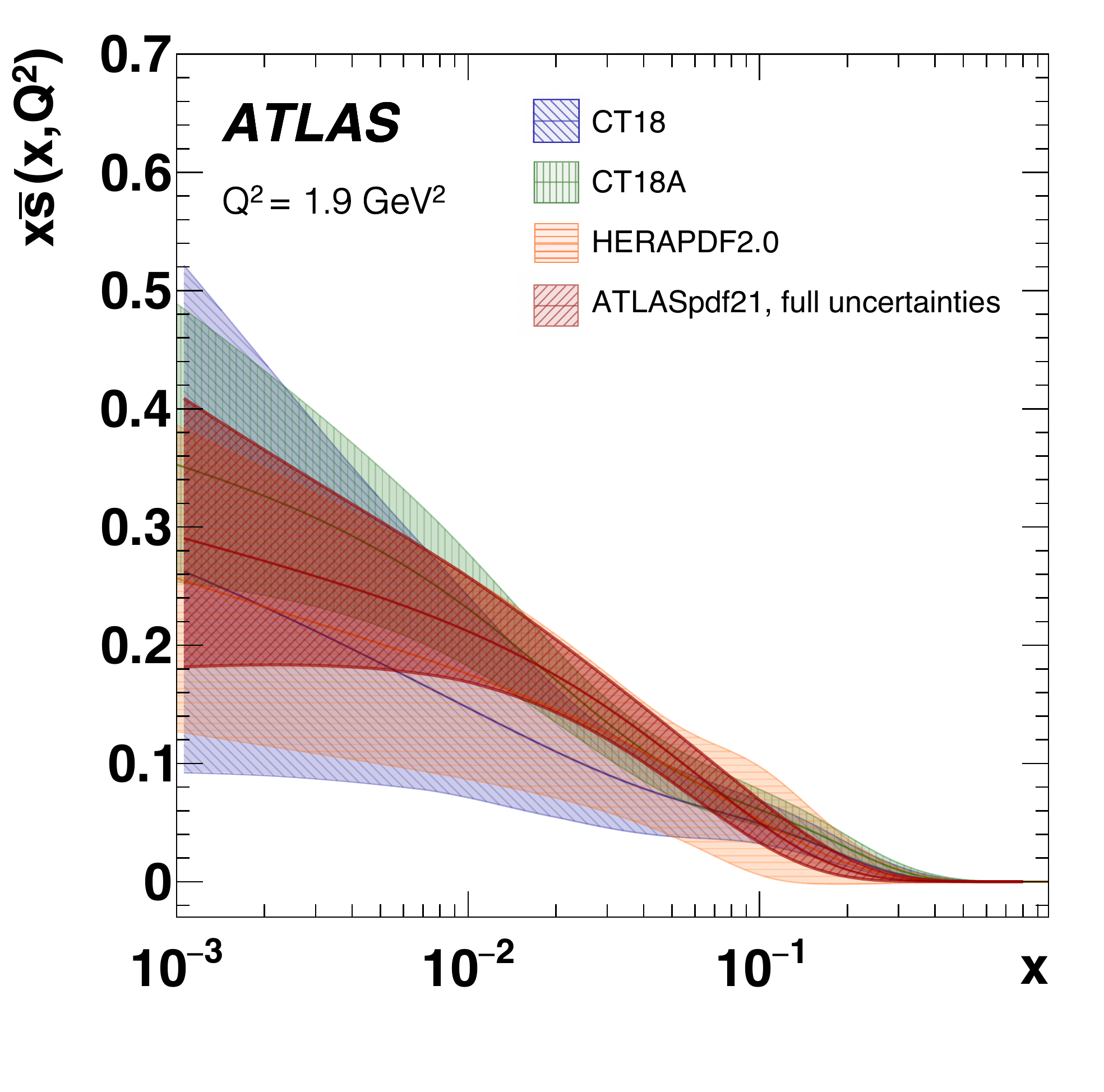}
\includegraphics[width=0.42\textwidth]{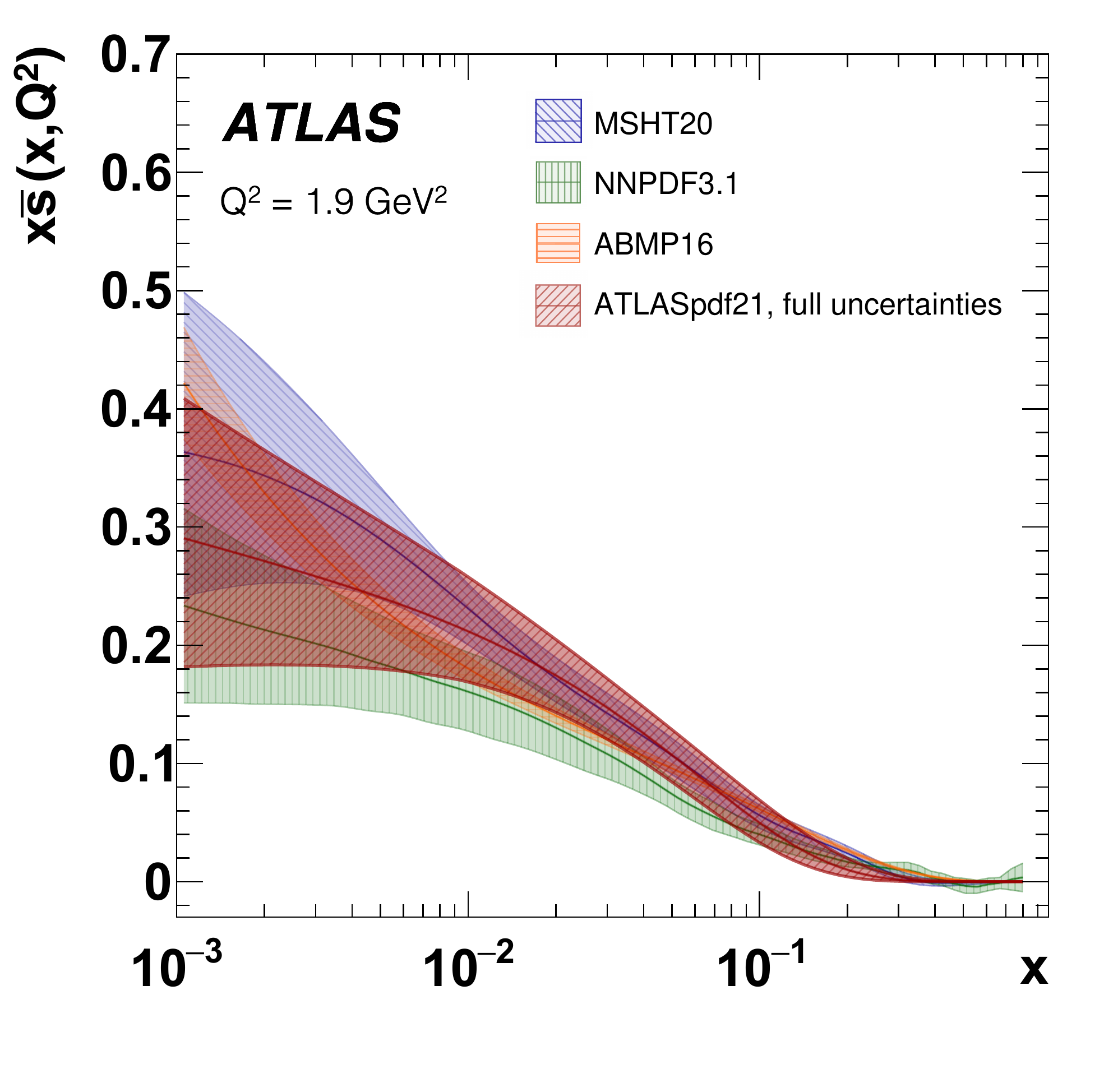}
\includegraphics[width=0.42\textwidth]{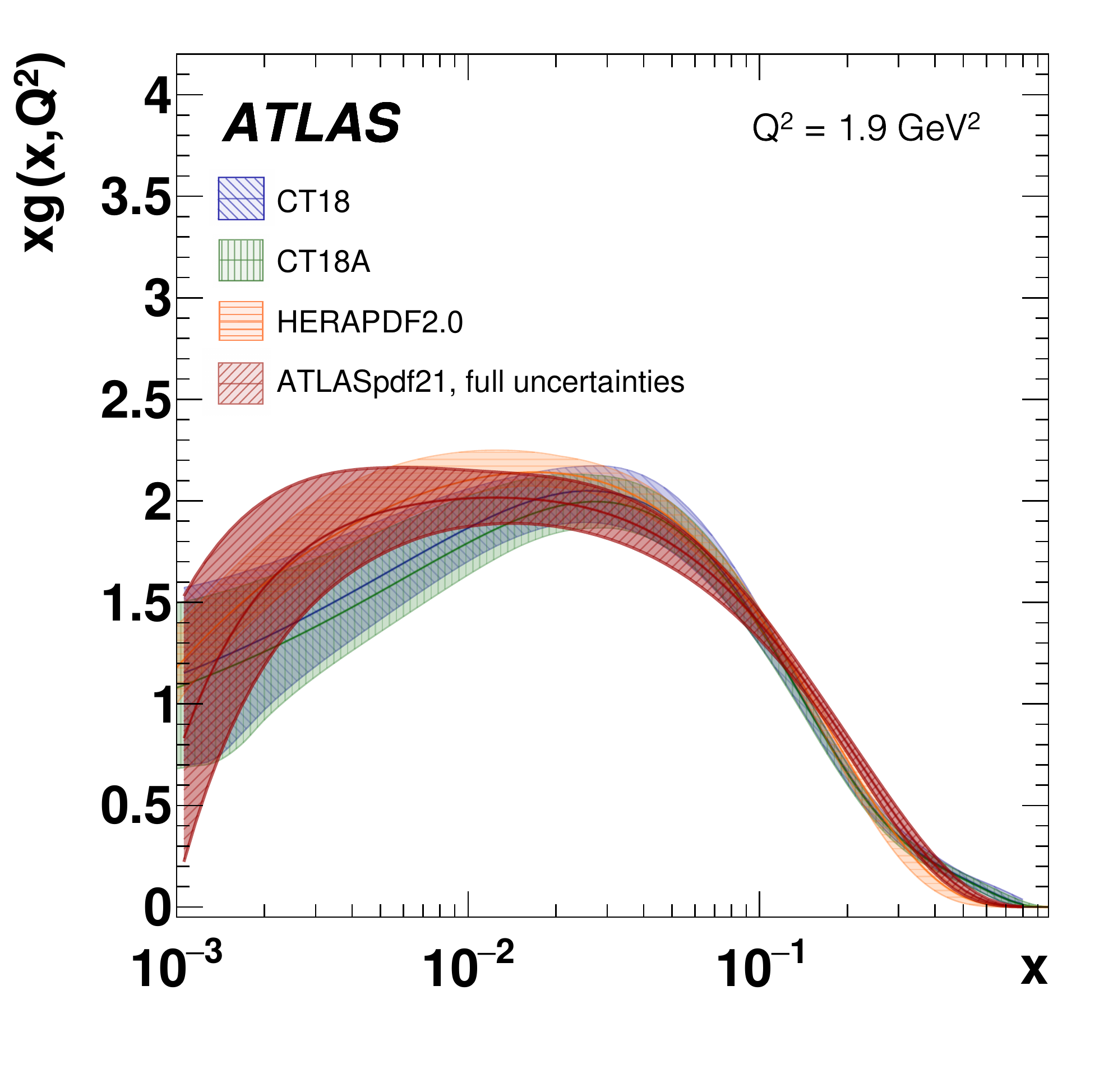}
\includegraphics[width=0.42\textwidth]{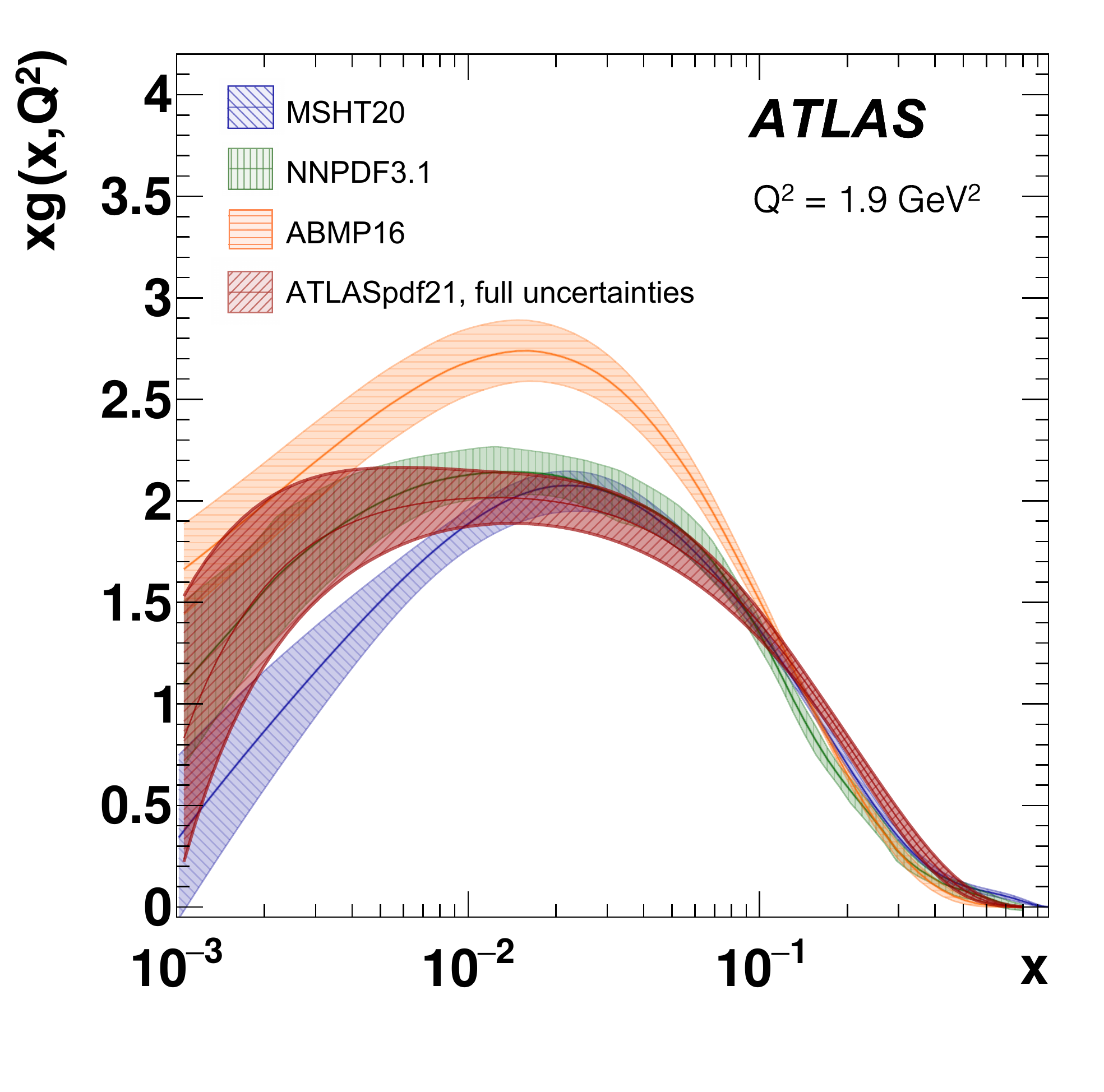}
\includegraphics[width=0.42\textwidth]{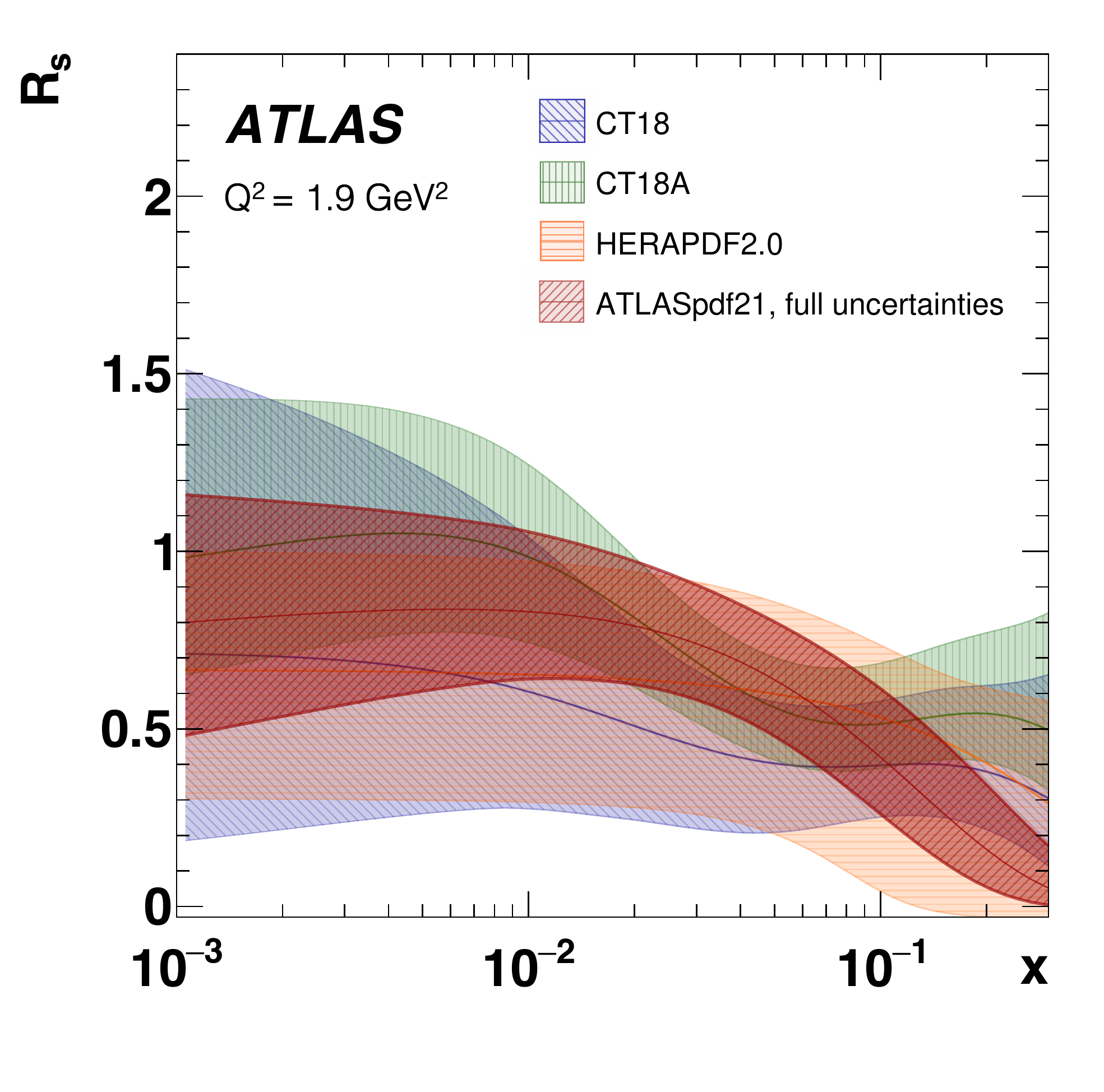}
\includegraphics[width=0.42\textwidth]{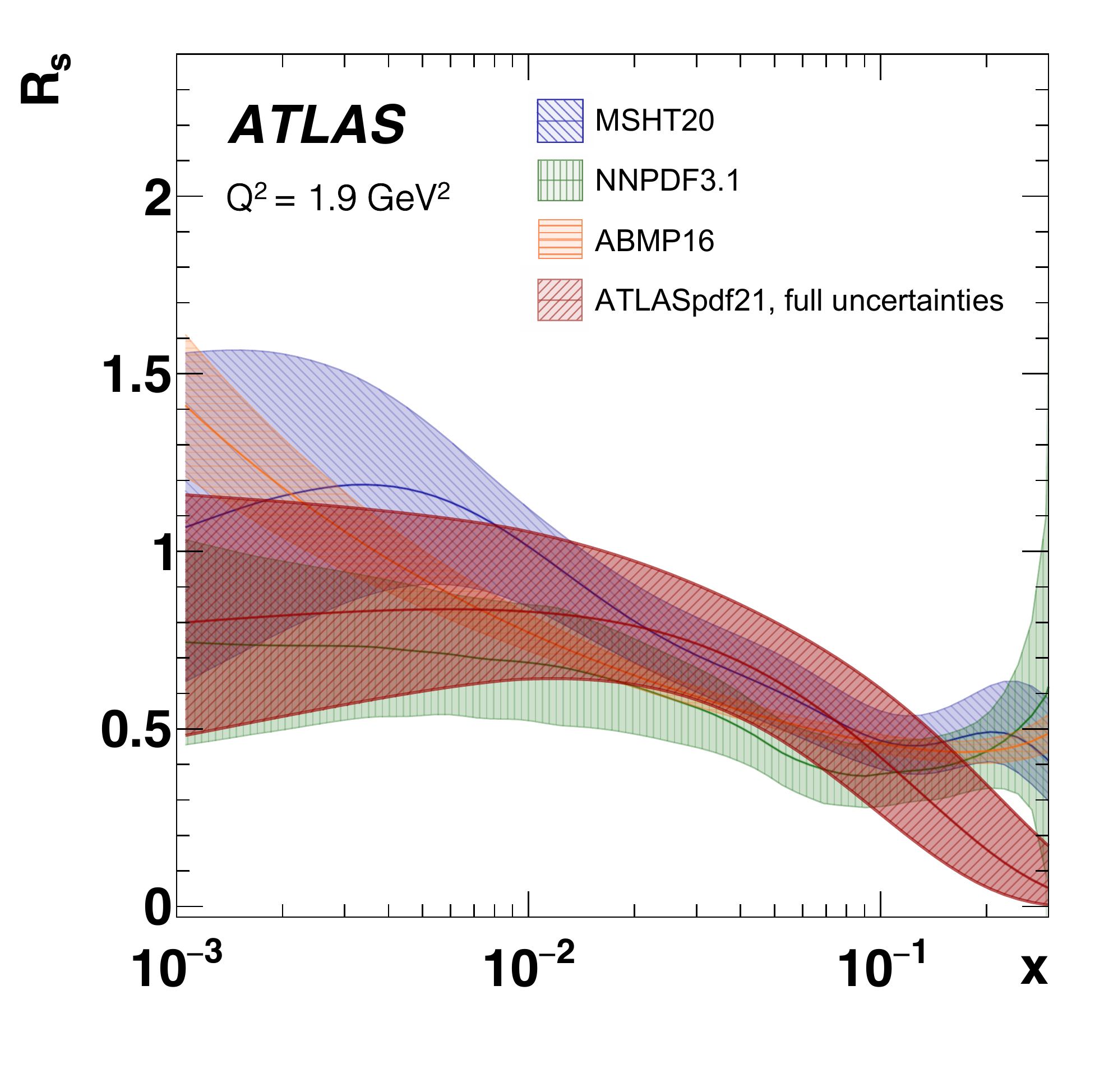}
\caption{ATLASpdf21 $x\bar{s}$, $xg$ and $R_s$ distributions with full uncertainties (experimental $T=3$, model, parameterisation) compared with other PDFs.
Left: CT18, CT18A, HERAPDF2.0. Right: MSHT20, NNPDF3.1, ABMP16.
\label{fig:global4}
}
\end{centering}
\end{figure*}

For the global fits considered here, the $\chi^2$ values for the data sets included in the ATLASpdf21 fit are HERAPDF2.0: 2262, CT18: 2135, CT18A: 2133,
MSHT20: 2218, NNPDF3.1: 2109, compared with ATLASpdf21: 2010.
Although the global PDFs from CT, MSHT and NNPDF have
more flexible parameterisations, the $\chi^{2}$ value of the ATLASpdf21 fit is better for the data sets considered. This is a further indication that the ATLASpdf21 parameterisation is sufficiently flexible.
 
\begin{figure*}[t!]
\begin{centering}
\includegraphics[width=0.42\textwidth]{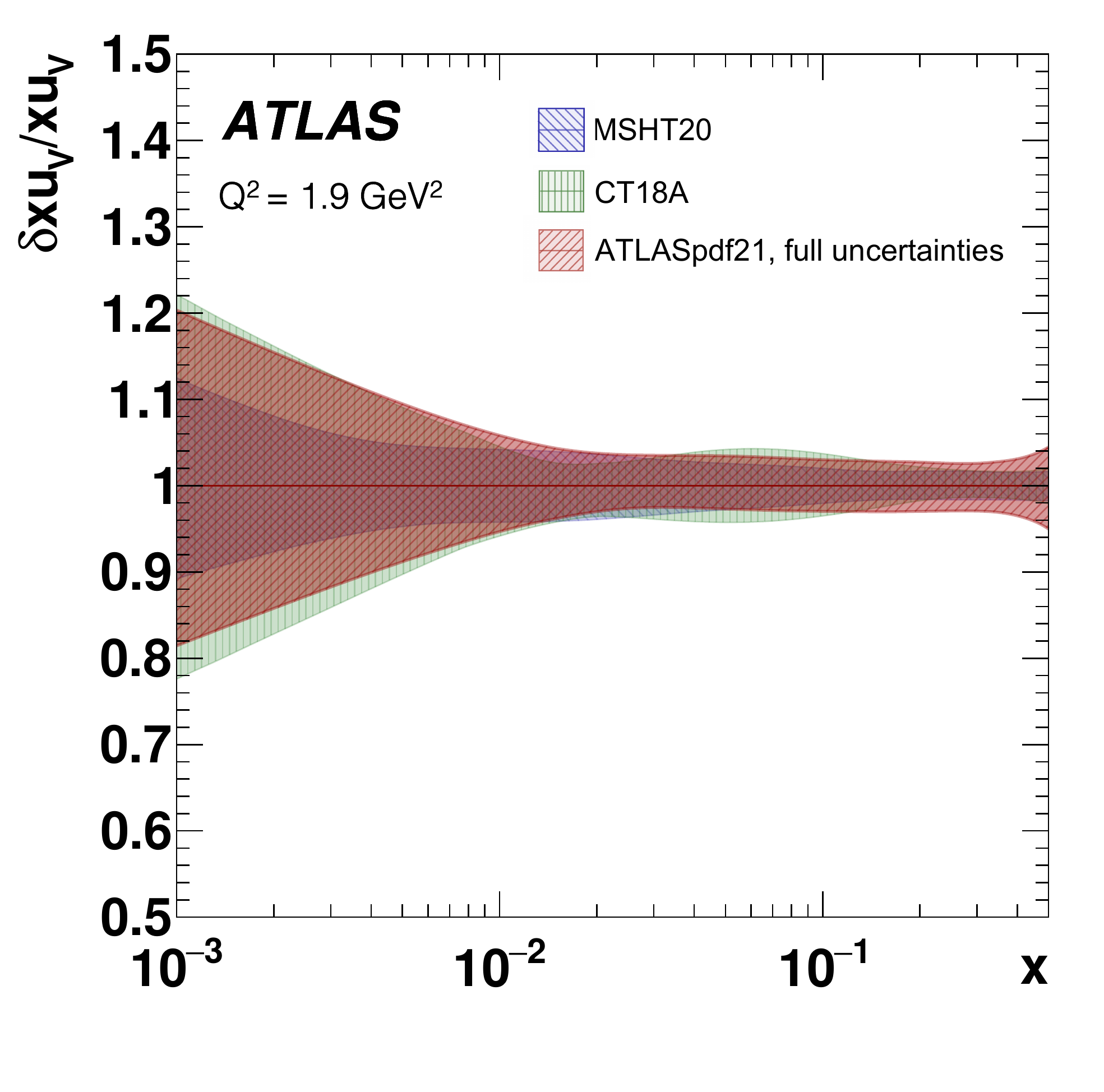}
\includegraphics[width=0.42\textwidth]{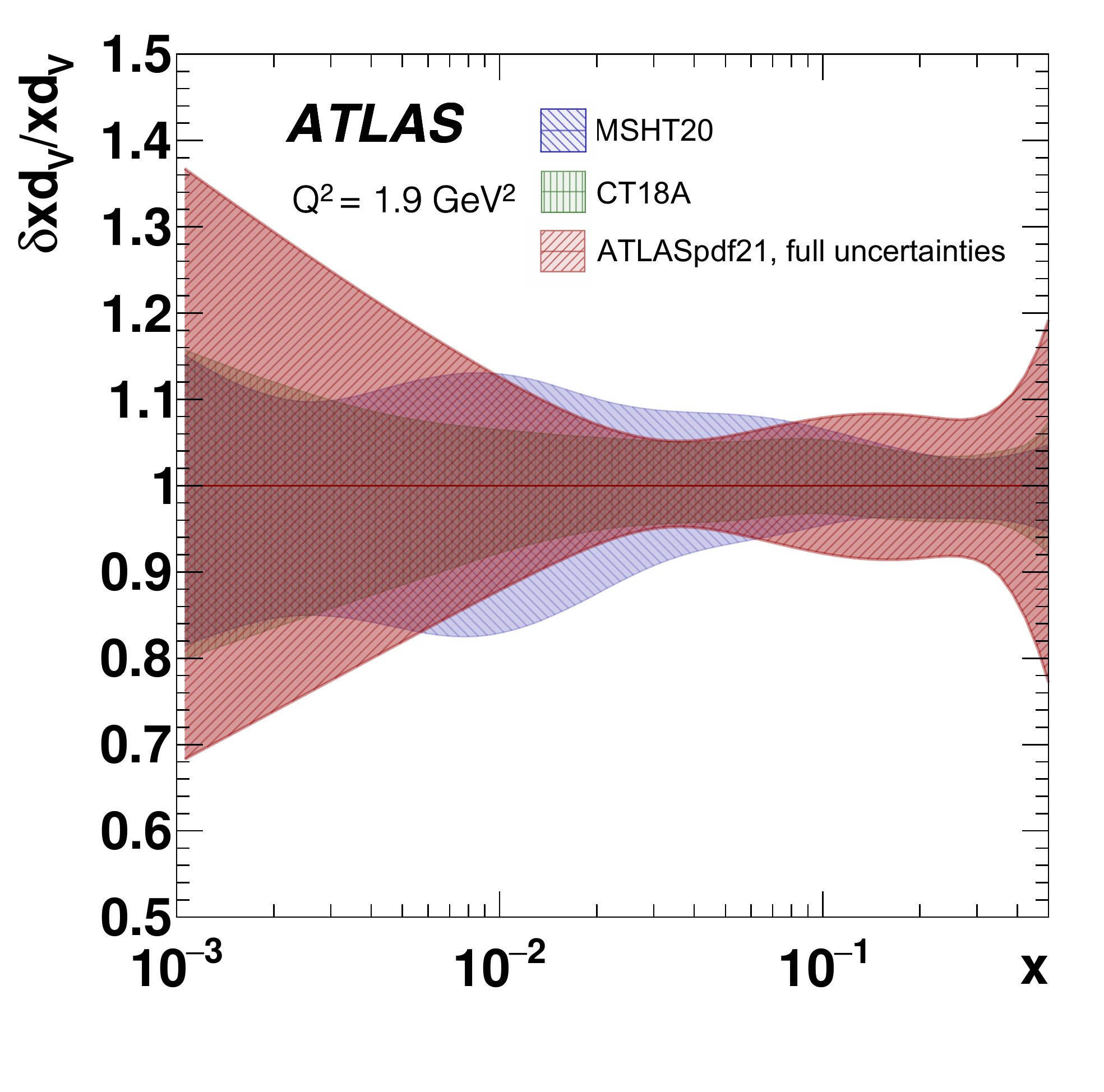}
\includegraphics[width=0.42\textwidth]{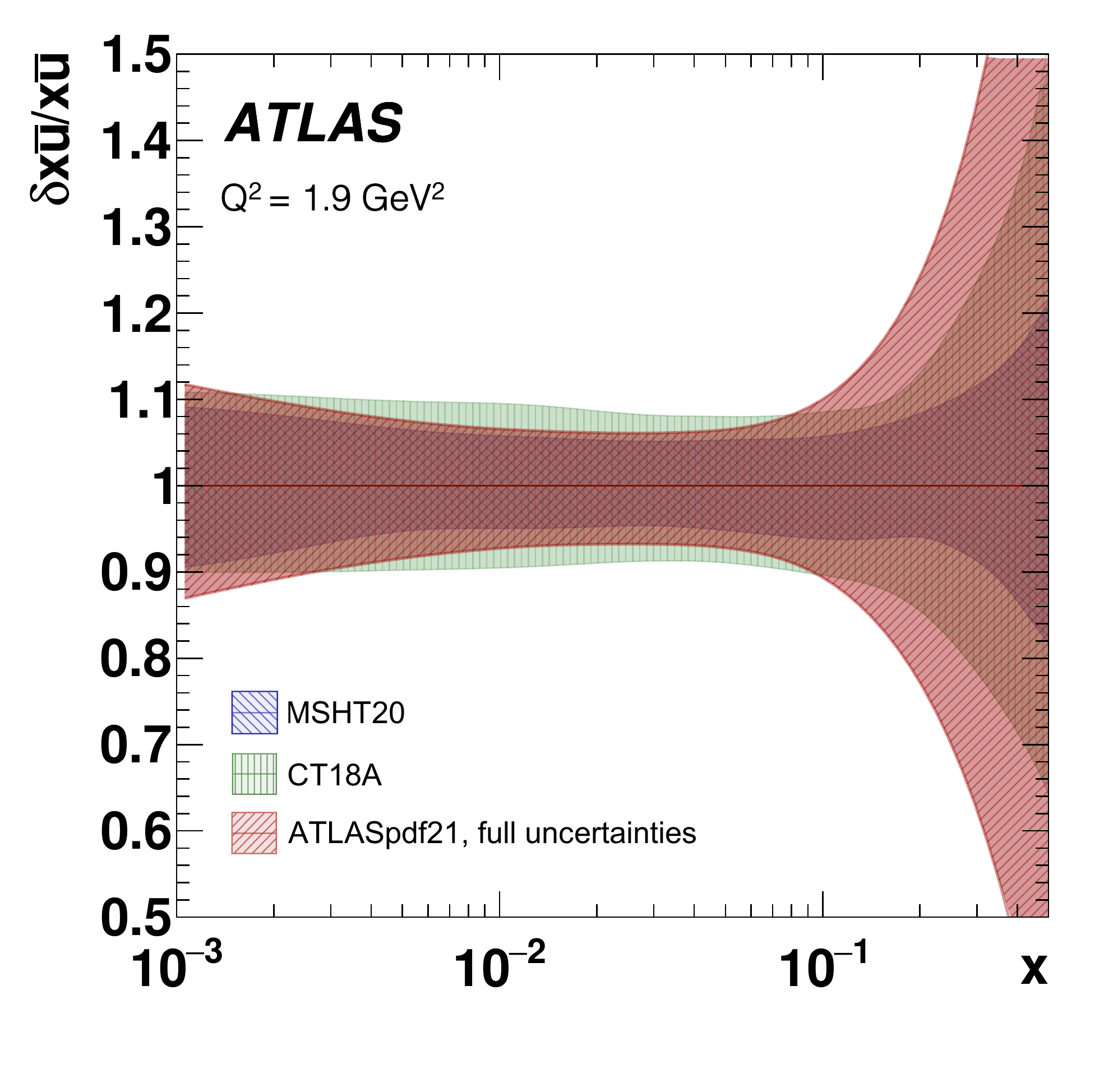}
\includegraphics[width=0.42\textwidth]{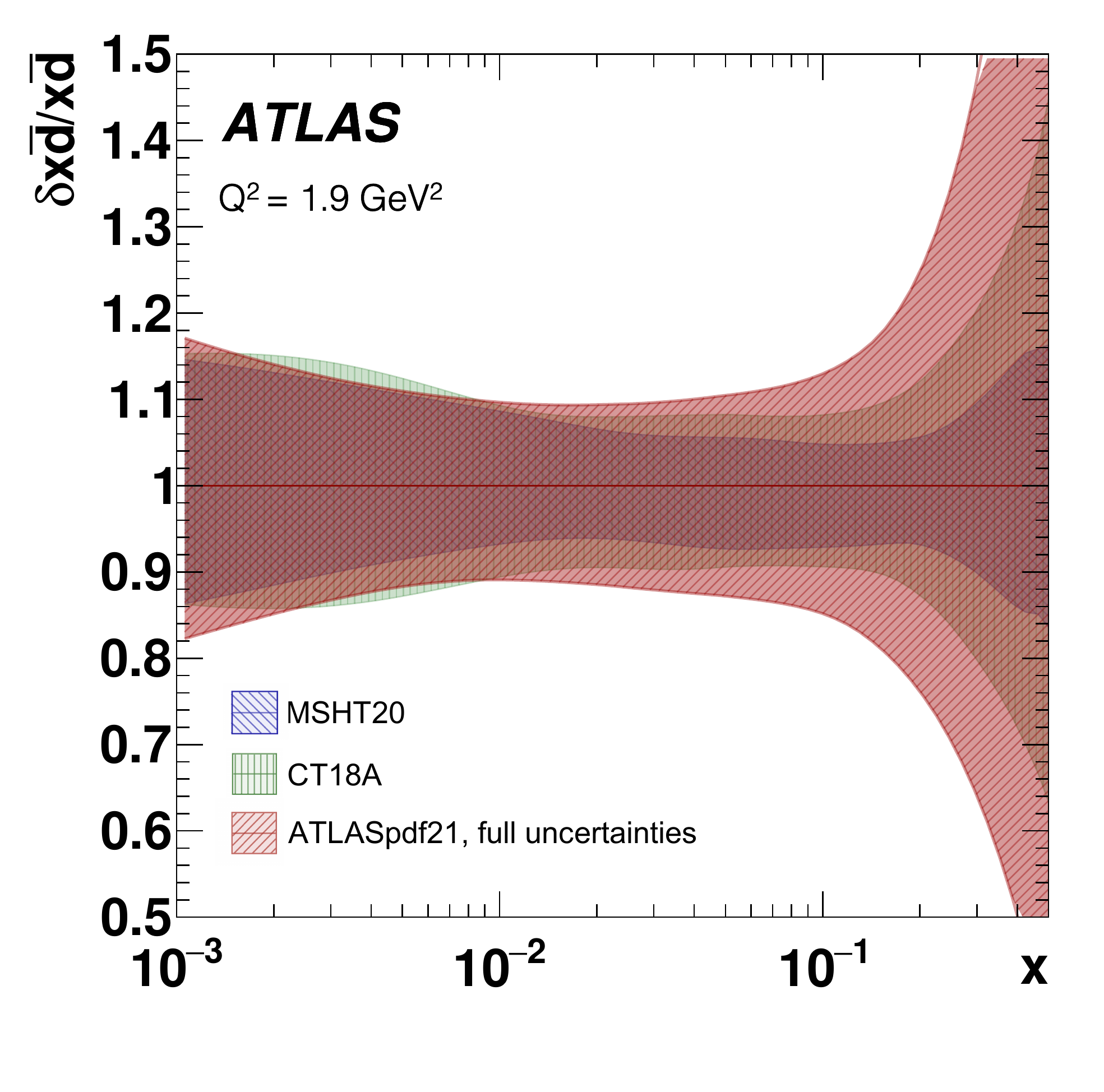}
\includegraphics[width=0.42\textwidth]{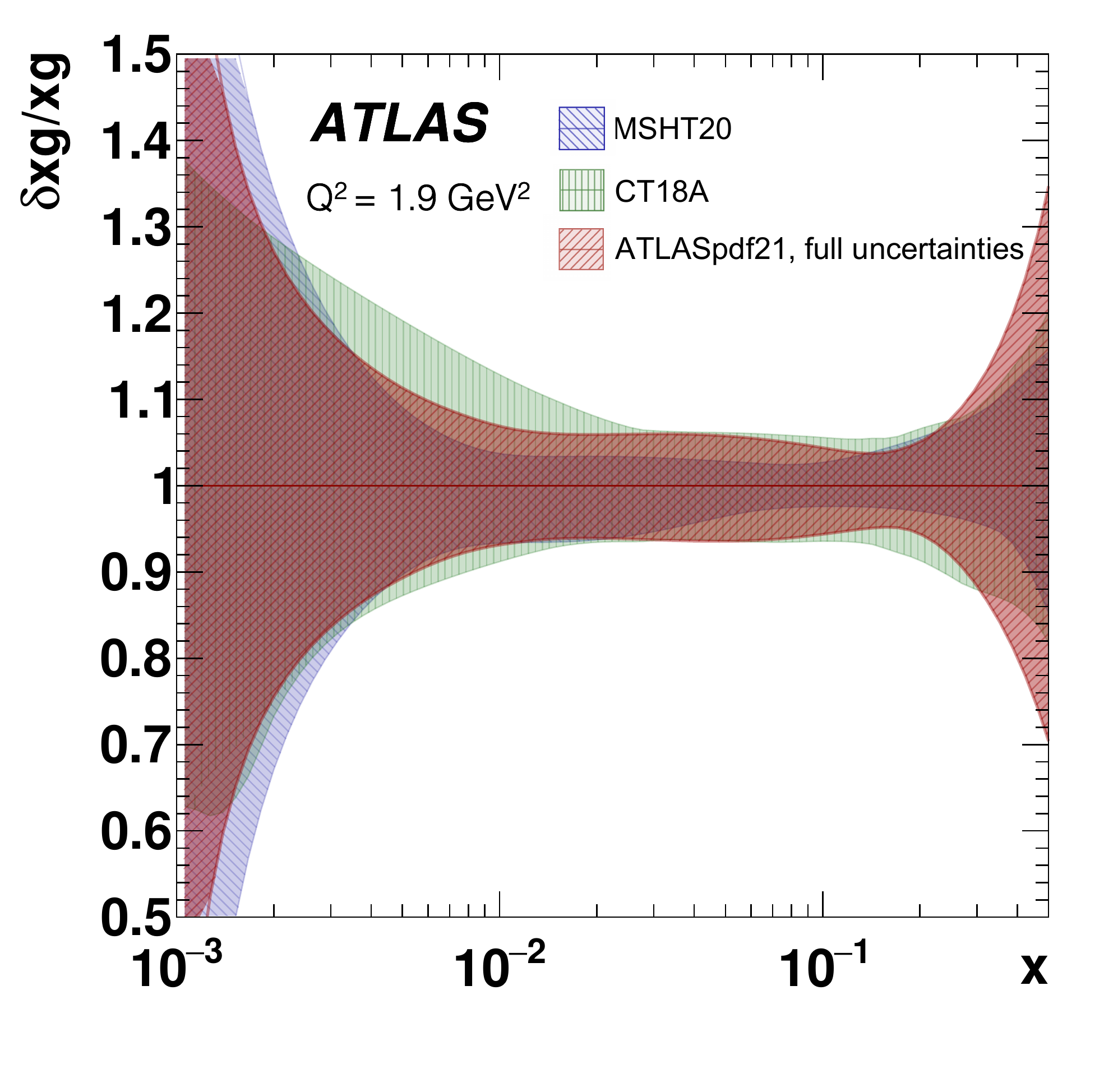}
\includegraphics[width=0.42\textwidth]{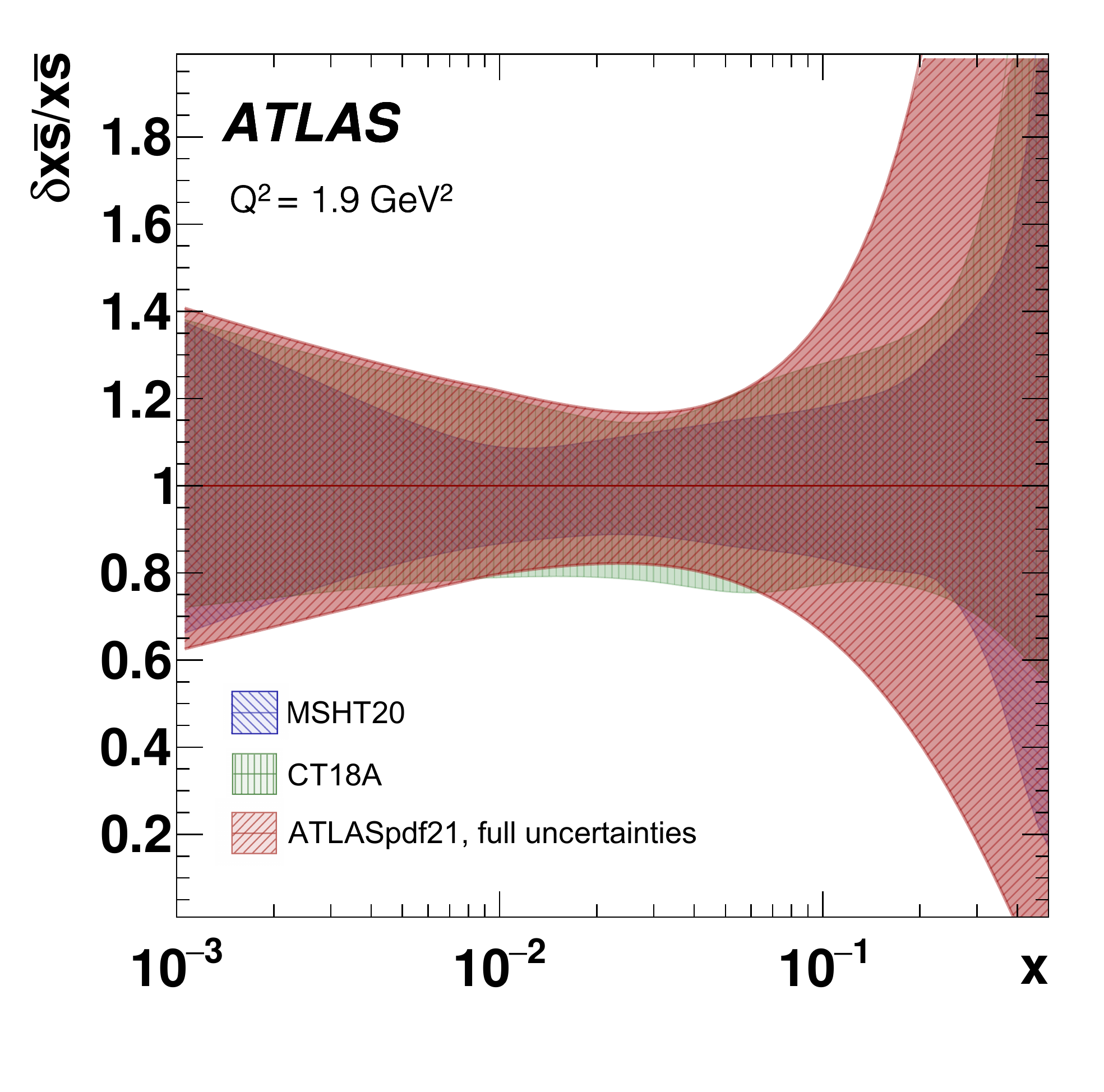}
\caption{ ATLASpdf21 fit with full uncertainties (experimental $T=3$, model, parameterisation) compared with CT18A, MSHT20 shown as a ratio with each distribution centred on unity. Top  left: $xu_v$. Top right: $xd_v$.
Middle left: $x\bar{u}$. Middle right: $x\bar{d}$. Bottom left: $xg$. Bottom right: $x\bar{s}$.
\label{fig:3ratio1}
}
\end{centering}
\end{figure*}
Figure~\ref{fig:3ratio1} shows the uncertainties of
the ATLASpdf21 fit, with the enhanced tolerance $T=3$, compared with CT18A and MSHT20 PDFs. These two PDFs are chosen because they both use the ATLAS $W,Z$ inclusive cross-section data at 7~\TeV\ and they both evaluate uncertainties in a similar way to the present analysis.
In general, the uncertainties of CT18A are somewhat larger than those of MSHT20, even though both represent $68\%$ CL uncertainties. For the $x\bar{u}$ and $x\bar{d}$ sea distributions at low~$x$, the uncertainties of ATLASpdf21 lie
between those of CT18A and MSHT20, becoming larger at high~$x$, $x>0.1$, reflecting the absence of fixed-target
DY data. For the $x\bar{s}$ sea distribution there is a similar pattern, with the low $x$ uncertainties of the ATLAS
PDF being comparable to those of CT18A. The uncertainty of the ATLASpdf21 gluon distribution at $x< 2\times10^{-3}$  is very similar to that of both CT18A and MSHT20, but the high $x$
gluon uncertainty is somewhat larger for the ATLASpdf21 fit, reflecting the absence of Tevatron jet data.
In the valence sector, the $xu_v$ uncertainties of the ATLASpdf21 fit are comparable to those of CT18A at low~$x$,
but larger at high~$x$, reflecting the absence of Tevatron $W,Z$ data. For $xd_v$, the uncertainties
are larger for all $x$ except $x\sim 0.003$--0.008, reflecting the fact that there is less information about the
$d$-quark than the $u$-quark in the absence of deuterium target data.
Appendix~\ref{sec:extradatafitcomp} shows comparisons between the ATLASpdf21 predictions and some of these older data sets which, while not fitted, are generally well described.\\
 
\begin{figure*}[t!]
\begin{centering}
\includegraphics[width=0.42\textwidth]{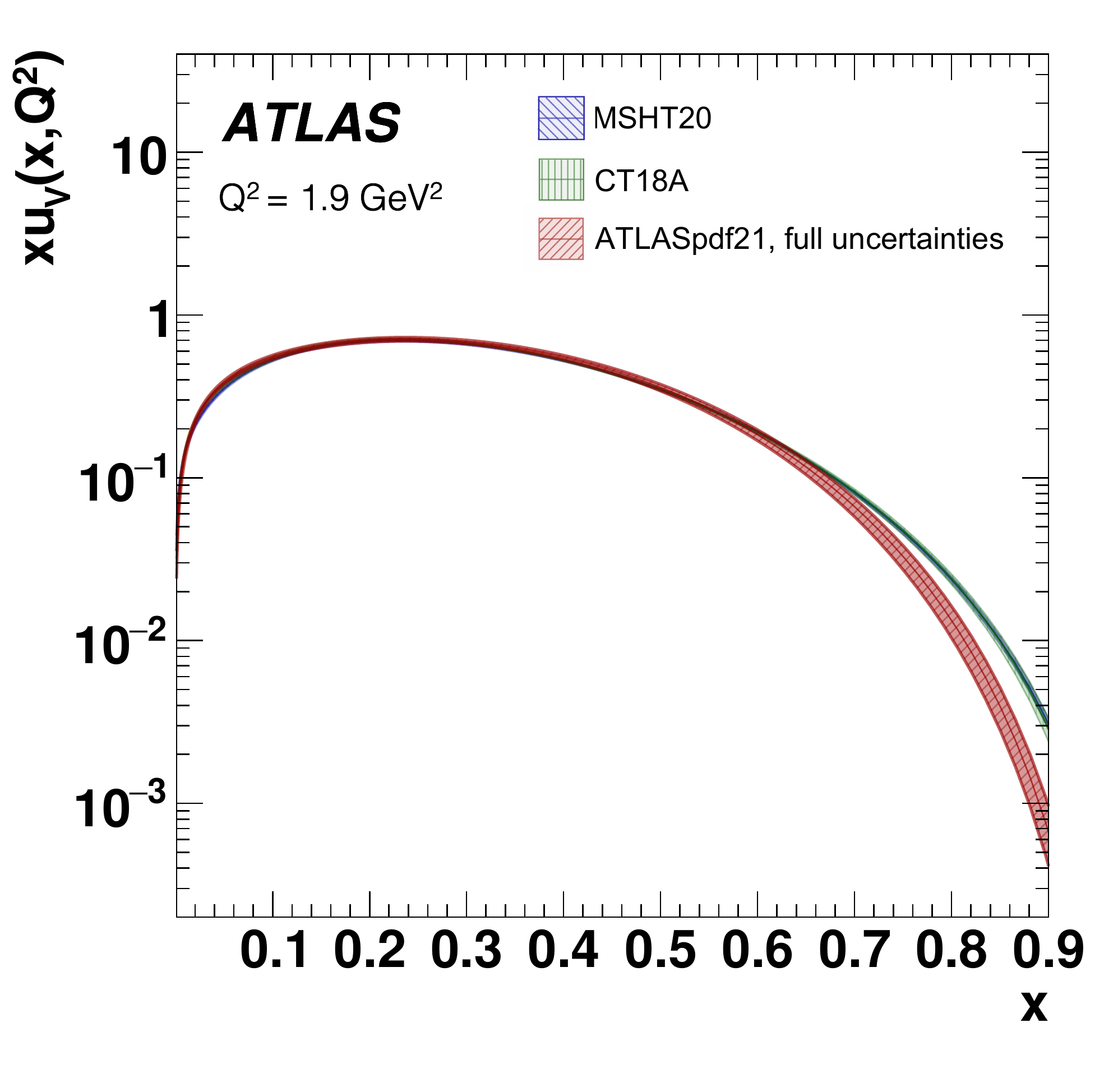}
\includegraphics[width=0.42\textwidth]{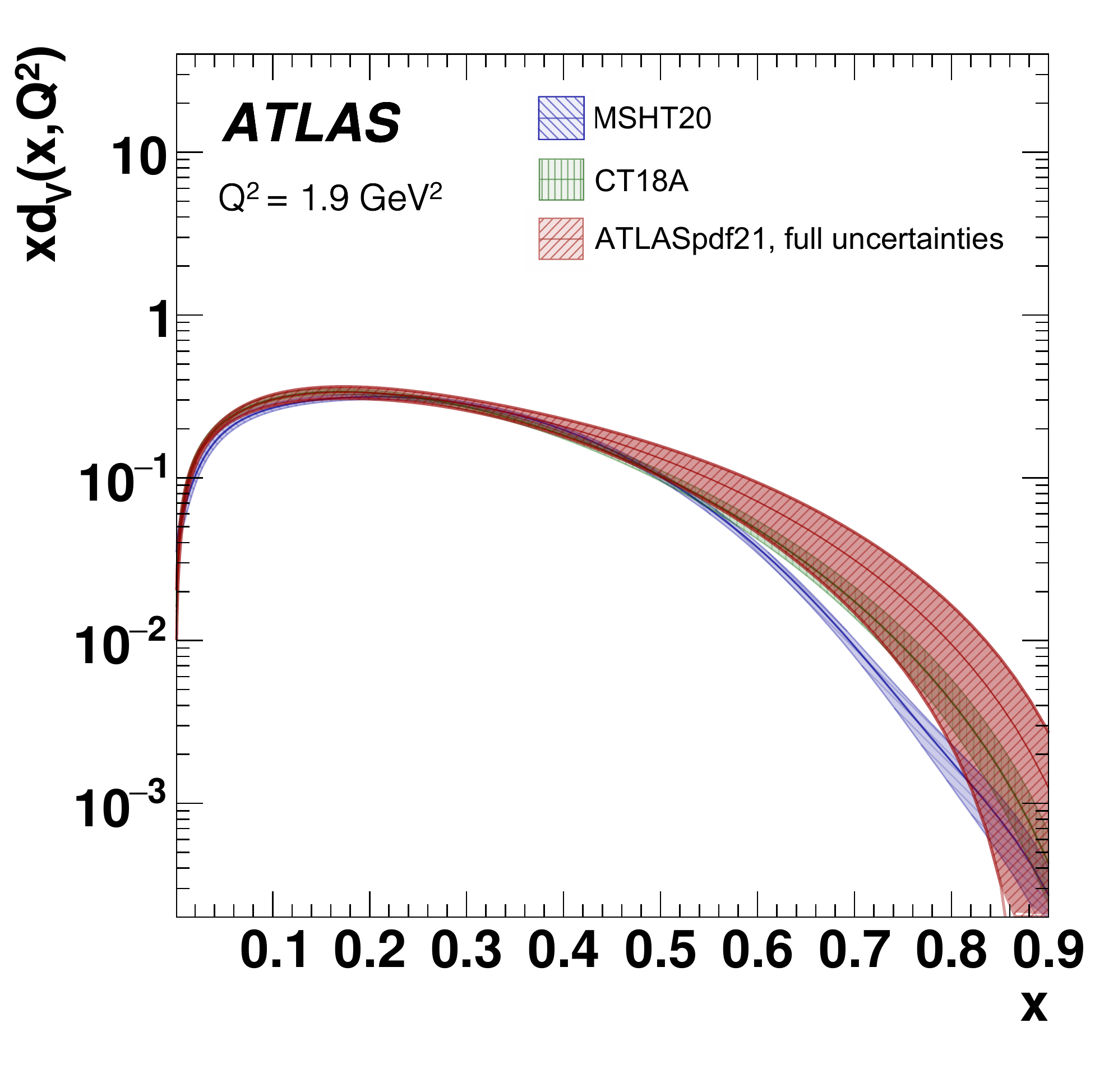}
\includegraphics[width=0.42\textwidth]{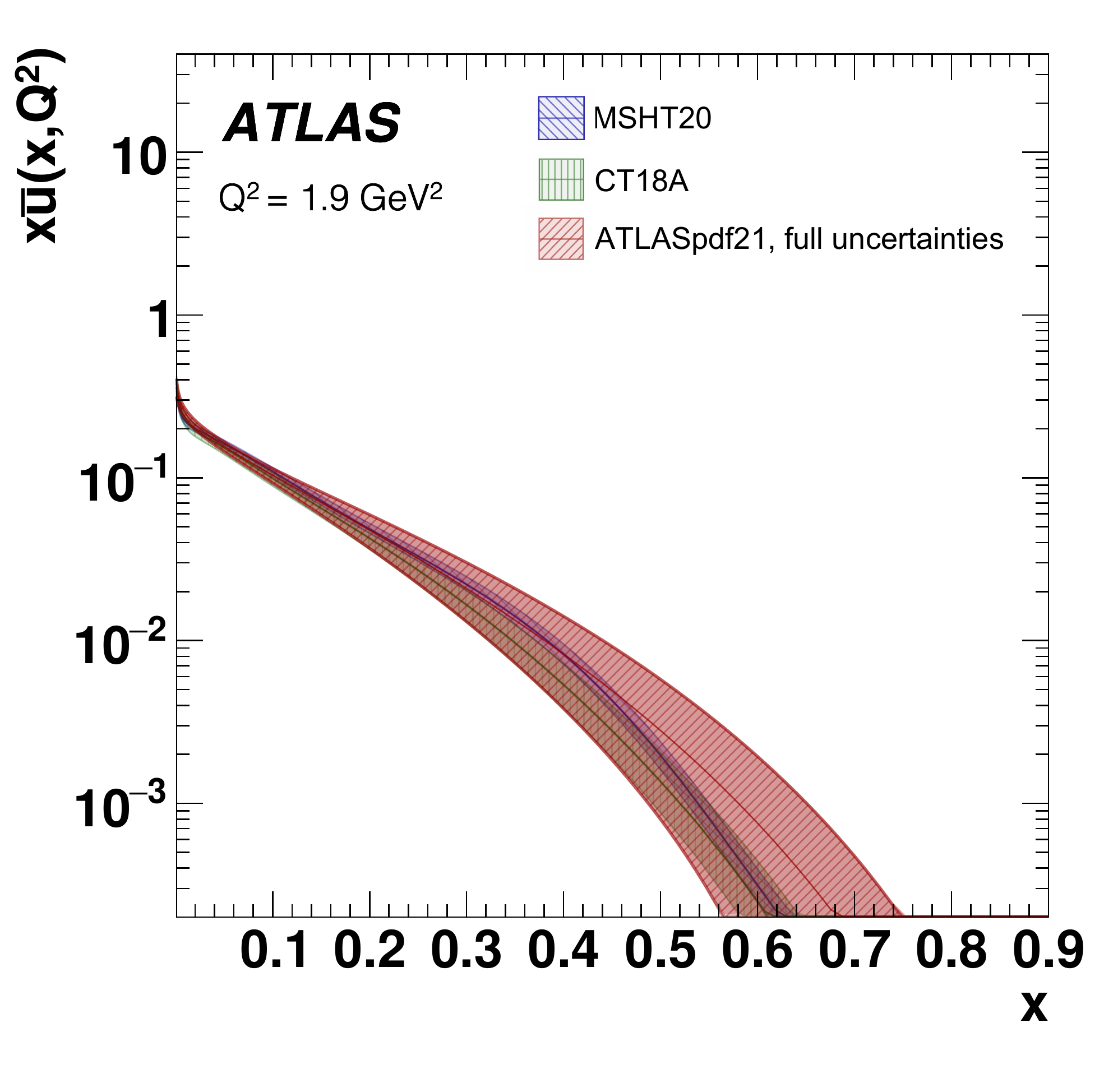}
\includegraphics[width=0.42\textwidth]{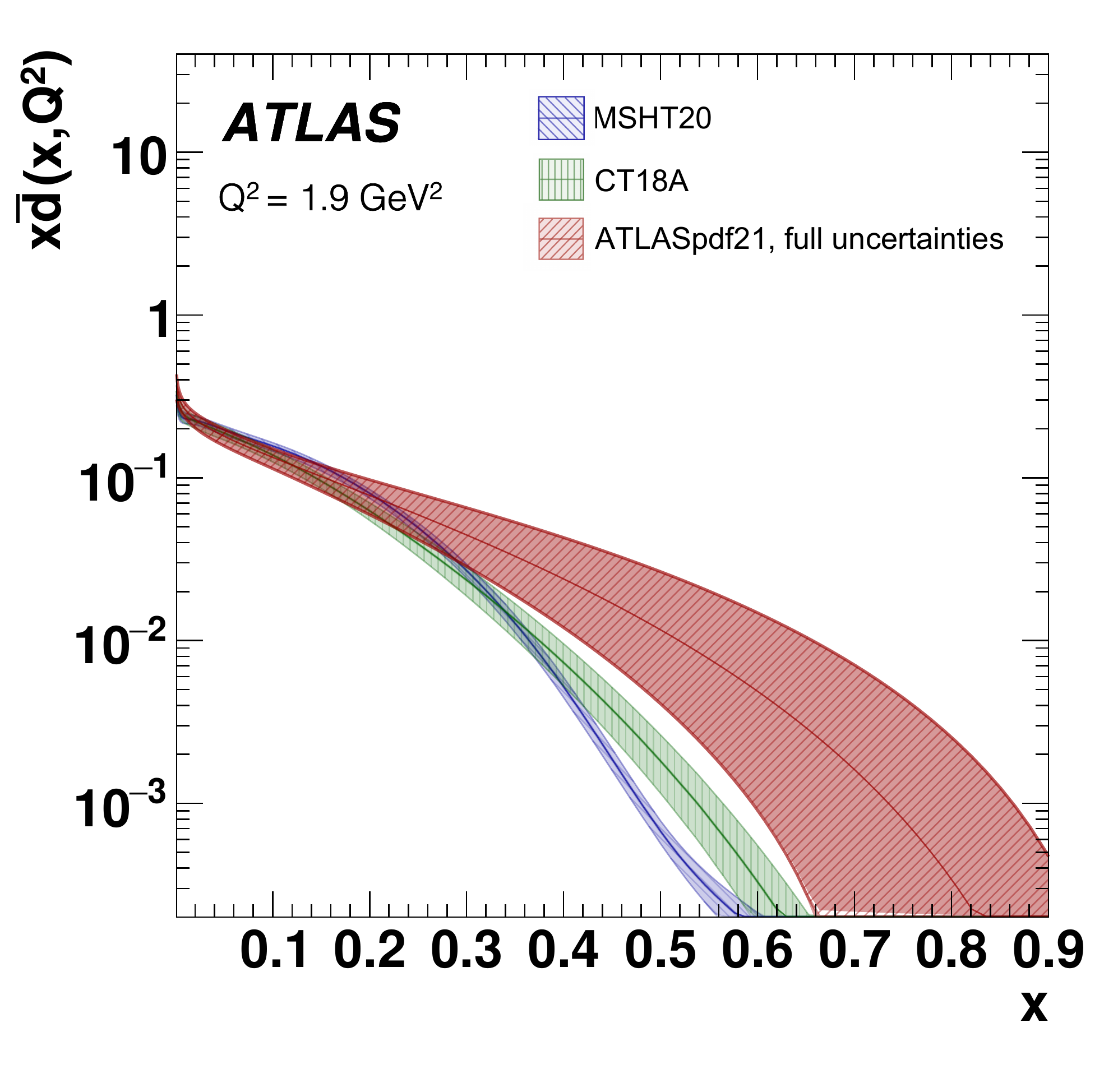}
\includegraphics[width=0.42\textwidth]{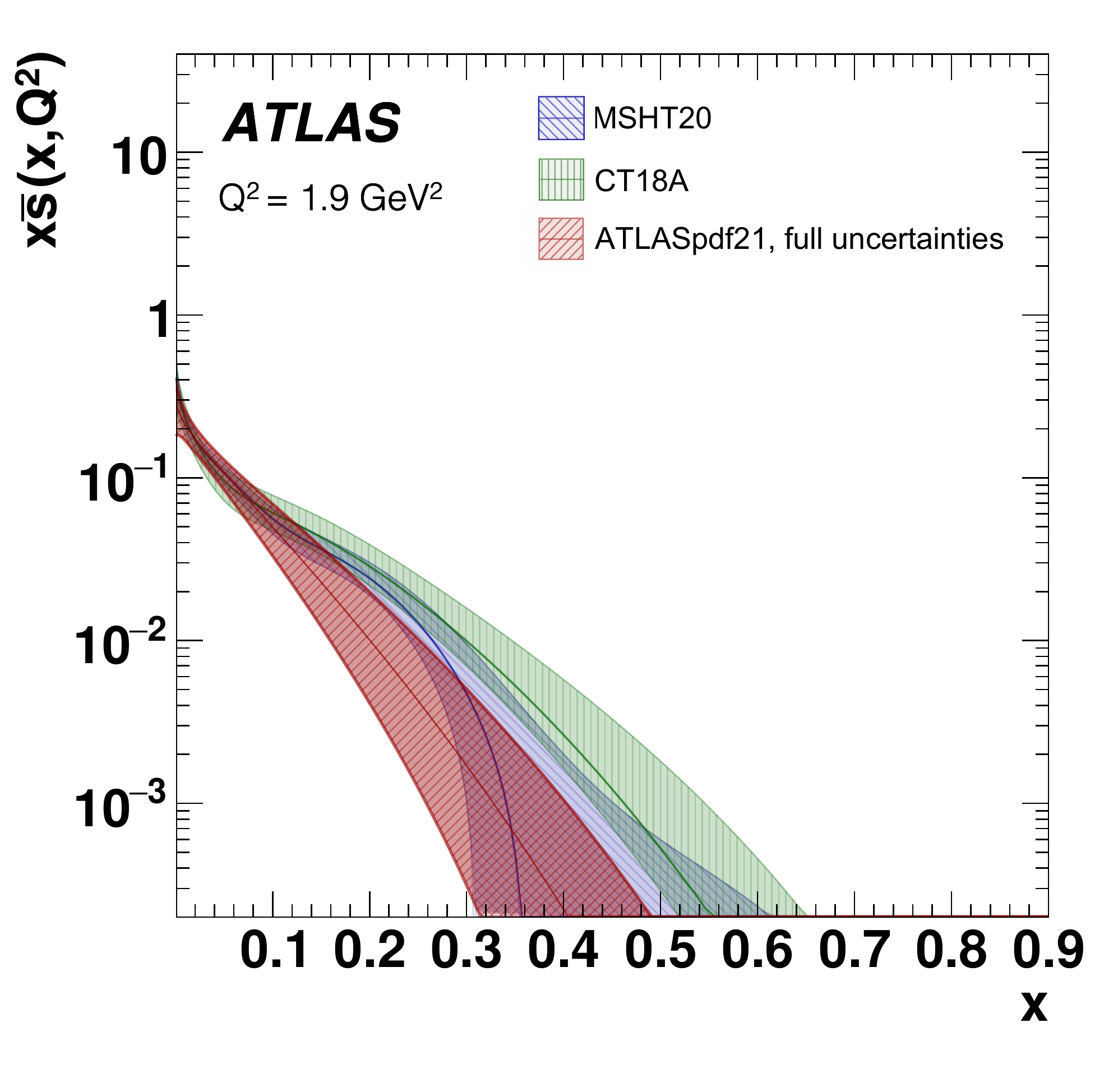}
\includegraphics[width=0.42\textwidth]{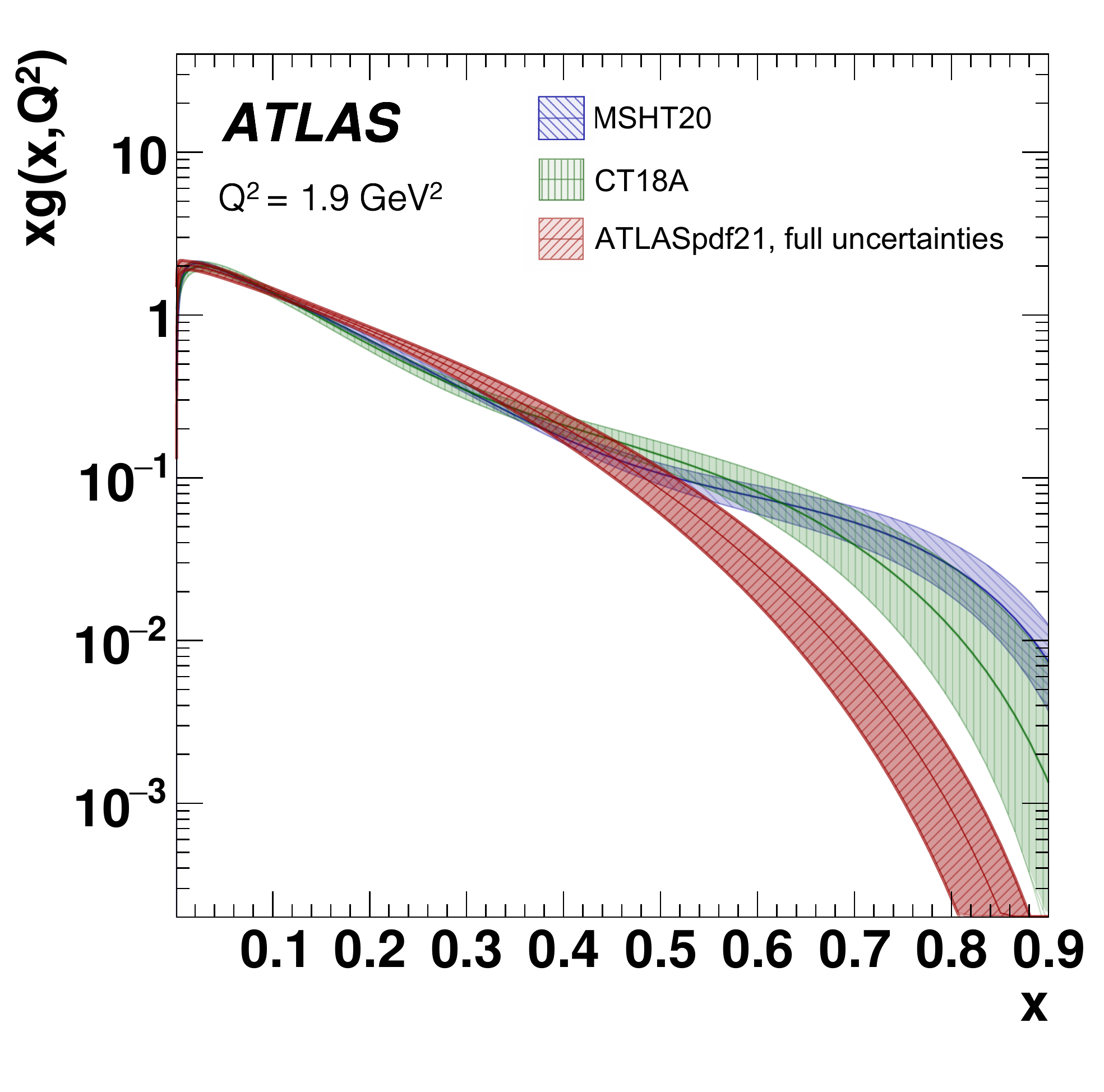}
\caption{ATLASpdf21 fit with full uncertainties (experimental $T=3$, model, parameterisation) compared with global PDFs: CT18A and MSHT20, focusing on high $x$ behaviour.
Top left: $xu_v$. Top right: $xd_v$. Middle left: $x\bar{u}$. Middle right: $x\bar{d}$. Bottom left: $x\bar{s}$. Bottom right: $xg$.
\label{fig:global_LinXLogY_val}
}
\end{centering}
\end{figure*}
In order to focus more on the high $x$ regime, which is important in searches for new physics, the comparisons between the ATLASpdf21
fit, with the  enhanced tolerance of $T=3$, and the  CT18A and MSHT20 PDFs, are repeated with a linear-$x$ and log-$y$ scale in
Figure~\ref{fig:global_LinXLogY_val}. Some discrepancies, not evident before, appear for $x \geq 0.7$. In this high $x$ region there is no data to determine any of the PDFs. Nevertheless, there is agreement within ${\sim}2\sigma$.
 
\clearpage
 
\section{Conclusion}
\label{sec:conclusion}

A PDF fit is presented using selected proton--proton collision data sets recorded by the ATLAS experiment at the LHC together with the final combined inclusive HERA data in order to assess the impact of the ATLAS data with a minimum of other input data. The resulting PDF set is called ATLASpdf21.
The addition of ATLAS inclusive $W$ and $Z$ data to HERA data allows a fit to be made without any constraints imposed
on the relationship between $x\bar{u}$, $x\bar{d}$ and $x\bar{s}$, nevertheless, $x\bar{d}\sim x\bar{u}$ is observed at low~$x$. The strangeness at low $x$
is found not to be as suppressed as found in earlier PDFs such as CT14, MMHT14 and NNPDF3.0, but to be in agreement with modern PDFs such as CT18A, MSHT20 and NNPDF3.1{\_}strange which use these ATLAS inclusive $W, Z$ data. This increase in the low $x$ strange was first seen in the ATLASepWZ12 and ATLASepWZ16 PDF analyses, but is now
somewhat moderated by further ATLAS data, and the freedom of the parametrisation at low $x$.
These inclusive $W,Z$ data also exhibit sensitivity to the valence quark distributions and the gluon distribution, considerably reducing their uncertainty.
The ATLAS $V$+\,jets data constrain the
light-quark sea at higher~$x$, such that strangeness is strongly suppressed at high $x$, and impact the high $x$ gluon distribution.
The $t\bar{t}$ data serve to moderately reduce the uncertainty in the high $x$ gluon distribution. Ratios of direct photon production at different proton--proton collision energies have only a mild
impact on the gluon PDF, but it is notable that they may now be reliably fitted to NNLO in QCD.
The jet production data are sensitive to the gluon distribution at medium to high~$x$ and they reduce
its uncertainty considerably.
 
The role of scale uncertainties is considered for all data sets
and these theoretical uncertainties are implemented in
the fit for the inclusive $W,Z$ data, where they are most impactful. The effects of such scale uncertainties are small.
 
A study is made of excluding high-scale data (with a scale above 500~\GeV),  which may contain subtle effects of physics beyond the Standard Model, from the fit. The resulting PDFs are not strongly affected by this restriction of the fitted kinematic region.
 
It is observed that the addition of the ATLAS
data sets to the HERA data brings the PDFs much closer to the global PDFs of MSHT, CT and NNPDF
than to HERAPDF2.0. The ATLASpdf21 PDFs agree with these global
fits as well as they agree with each other. Thus,
ATLAS data seem to be able to replicate many of the features that the fixed-target
deep inelastic scattering and Drell--Yan data plus the Tevatron Drell--Yan data bring to the global PDFs. Using only the
HERA and ATLAS data allows a detailed treatment of correlated systematic uncertainties.
 
Correlations of systematic uncertainties within and between ATLAS data sets are considered, with particular emphasis on the larger
systematic uncertainties appertaining to the jet energy scale. Information about correlations between data
sets is provided such that the major correlations could be considered by the global fitting groups CT, MSHT and NNPDF in future
fits. The effects of these
correlations are relatively small, $O({\sim}1\%$), at the scale $Q^2\sim \num{10000}$~GeV$^2$ considered for
precision physics at the LHC,  but large enough to be considered in future precision data analyses such as the measurement of
\mW and  $\sin^2\!\theta_{\text{W}}$, given that such a level of accuracy is now desired to resolve effects of physics
beyond the Standard Model in the measurements of Standard Model parameters.


\section*{Acknowledgements}
 

We thank CERN for the very successful operation of the LHC, as well as the
support staff from our institutions without whom ATLAS could not be
operated efficiently.
 
We acknowledge the support of
ANPCyT, Argentina;
YerPhI, Armenia;
ARC, Australia;
BMWFW and FWF, Austria;
ANAS, Azerbaijan;
SSTC, Belarus;
CNPq and FAPESP, Brazil;
NSERC, NRC and CFI, Canada;
CERN;
ANID, Chile;
CAS, MOST and NSFC, China;
Minciencias, Colombia;
MSMT CR, MPO CR and VSC CR, Czech Republic;
DNRF and DNSRC, Denmark;
IN2P3-CNRS and CEA-DRF/IRFU, France;
SRNSFG, Georgia;
BMBF, HGF and MPG, Germany;
GSRI, Greece;
RGC and Hong Kong SAR, China;
ISF and Benoziyo Center, Israel;
INFN, Italy;
MEXT and JSPS, Japan;
CNRST, Morocco;
NWO, Netherlands;
RCN, Norway;
MEiN, Poland;
FCT, Portugal;
MNE/IFA, Romania;
JINR;
MES of Russia and NRC KI, Russian Federation;
MESTD, Serbia;
MSSR, Slovakia;
ARRS and MIZ\v{S}, Slovenia;
DSI/NRF, South Africa;
MICINN, Spain;
SRC and Wallenberg Foundation, Sweden;
SERI, SNSF and Cantons of Bern and Geneva, Switzerland;
MOST, Taiwan;
TAEK, Turkey;
STFC, United Kingdom;
DOE and NSF, United States of America.
In addition, individual groups and members have received support from
BCKDF, CANARIE, Compute Canada and CRC, Canada;
COST, ERC, ERDF, Horizon 2020 and Marie Sk{\l}odowska-Curie Actions, European Union;
Investissements d'Avenir Labex, Investissements d'Avenir Idex and ANR, France;
DFG and AvH Foundation, Germany;
Herakleitos, Thales and Aristeia programmes co-financed by EU-ESF and the Greek NSRF, Greece;
BSF-NSF and GIF, Israel;
Norwegian Financial Mechanism 2014-2021, Norway;
NCN and NAWA, Poland;
La Caixa Banking Foundation, CERCA Programme Generalitat de Catalunya and PROMETEO and GenT Programmes Generalitat Valenciana, Spain;
G\"{o}ran Gustafssons Stiftelse, Sweden;
The Royal Society and Leverhulme Trust, United Kingdom.
 
The crucial computing support from all WLCG partners is acknowledged gratefully, in particular from CERN, the ATLAS Tier-1 facilities at TRIUMF (Canada), NDGF (Denmark, Norway, Sweden), CC-IN2P3 (France), KIT/GridKA (Germany), INFN-CNAF (Italy), NL-T1 (Netherlands), PIC (Spain), ASGC (Taiwan), RAL (UK) and BNL (USA), the Tier-2 facilities worldwide and large non-WLCG resource providers. Major contributors of computing resources are listed in Ref.~\cite{ATL-SOFT-PUB-2021-003}.
 

\clearpage
\appendix
\part*{Appendix}
\addcontentsline{toc}{part}{Appendix}
\section{Scale uncertainties in $W,Z$ inclusive data}
\label{sec:scale}
\begin{figure*}[th!]
\begin{centering}
\includegraphics[width=0.42\textwidth]{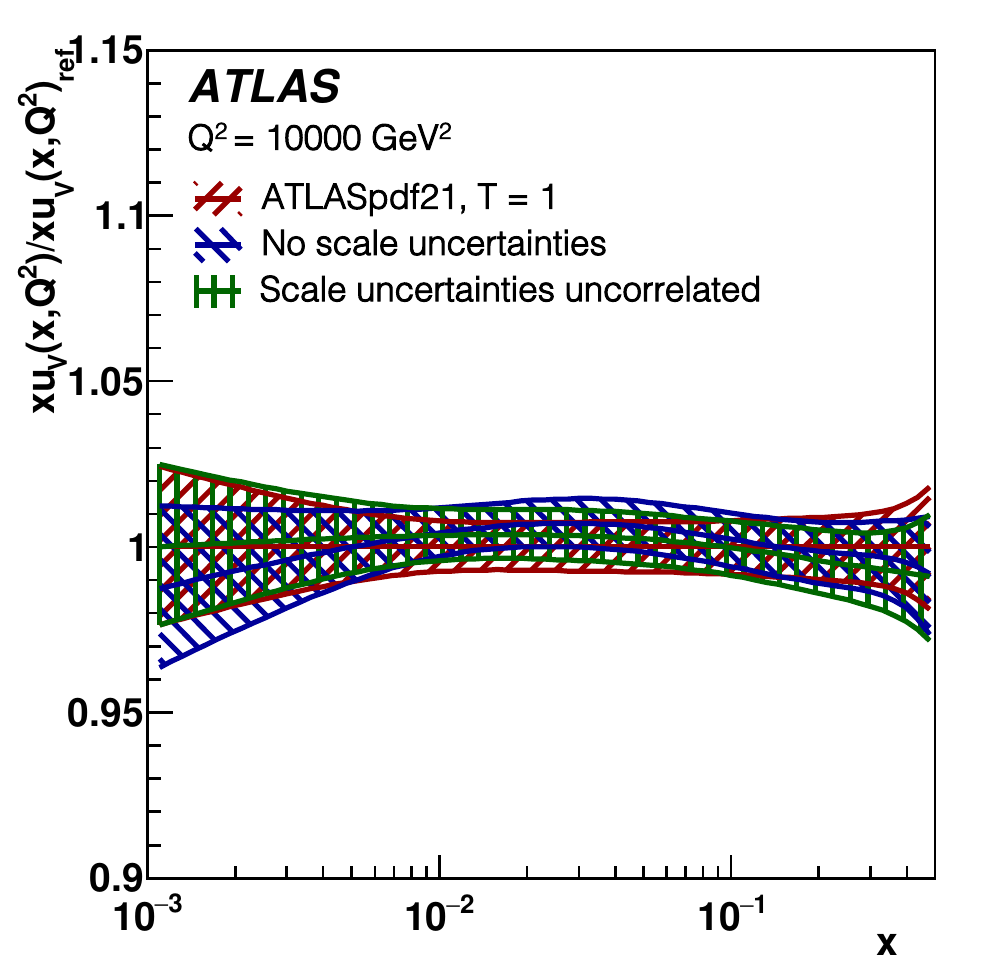}
\includegraphics[width=0.42\textwidth]{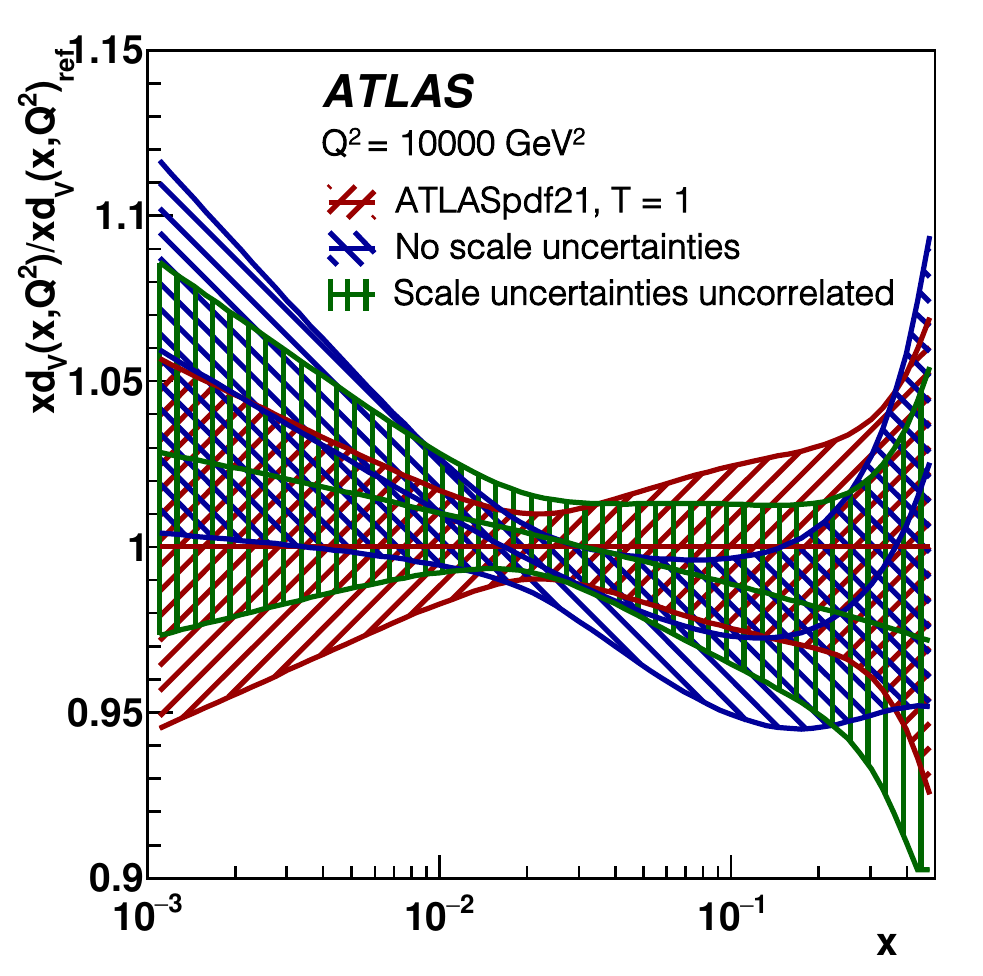}
\includegraphics[width=0.42\textwidth]{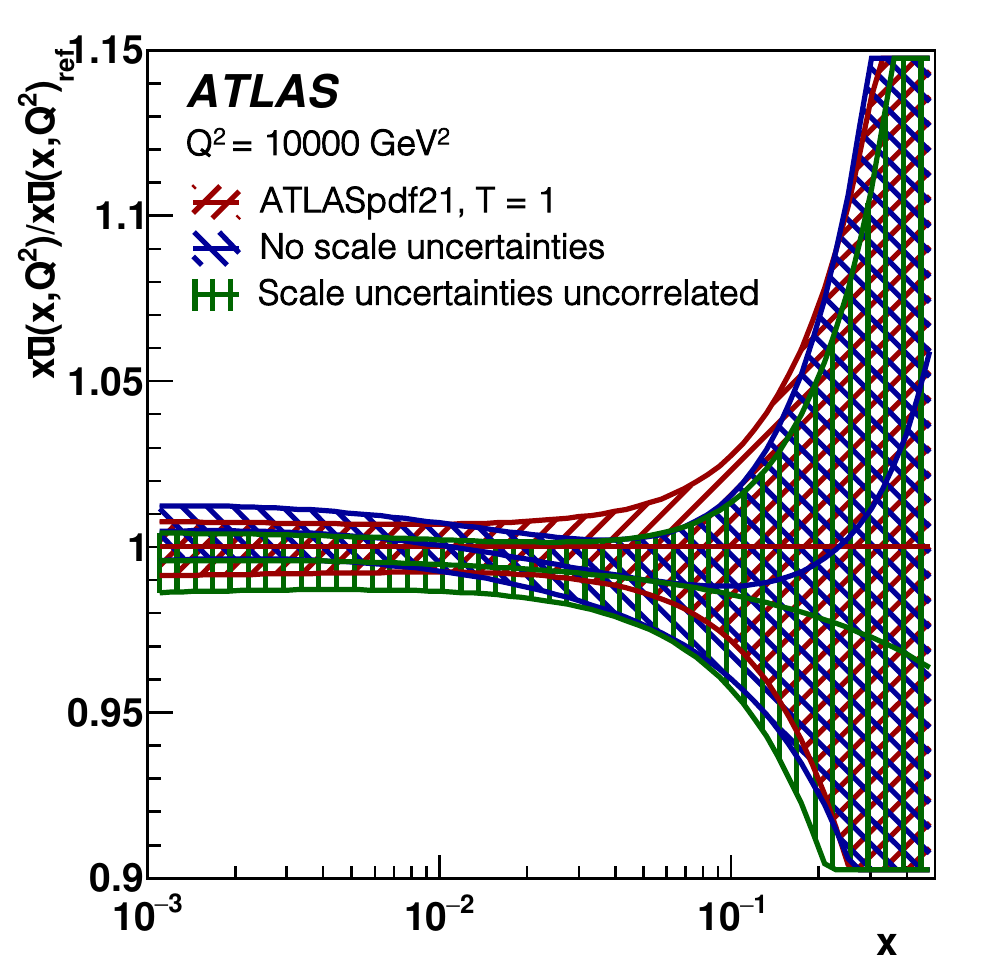}
\includegraphics[width=0.42\textwidth]{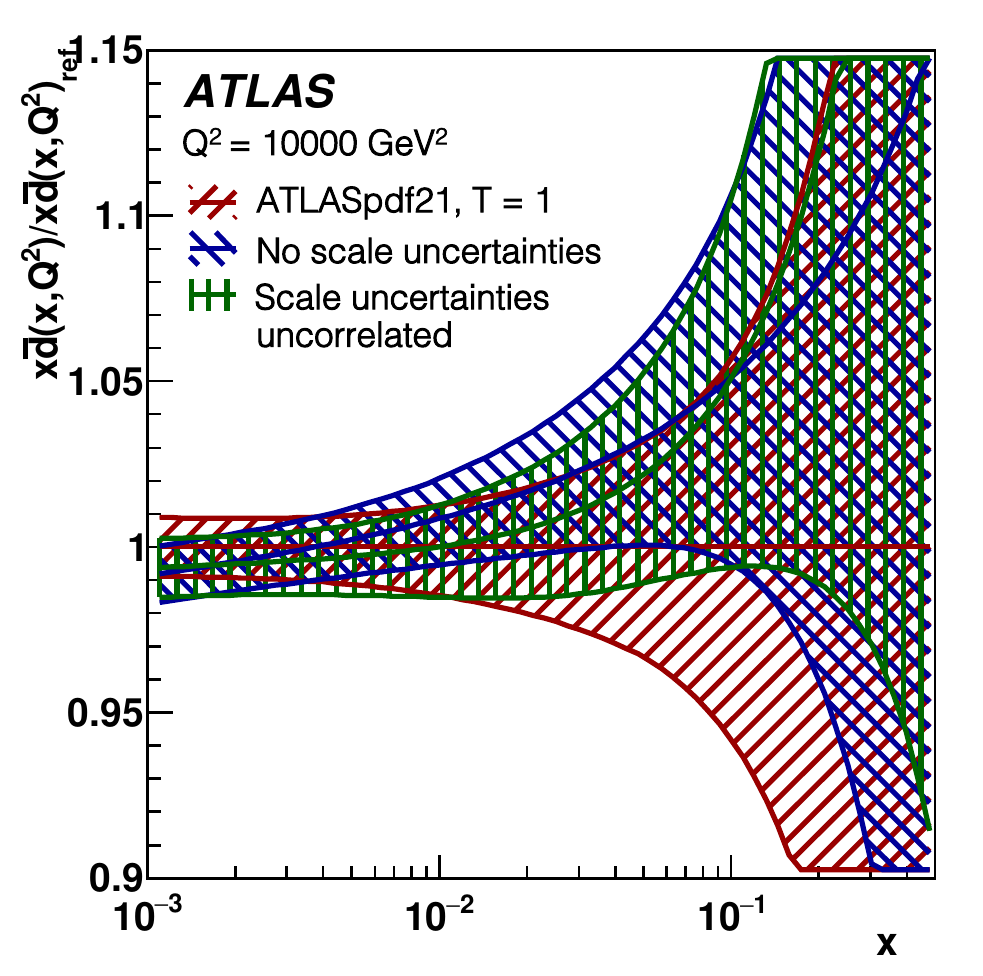}
\includegraphics[width=0.42\textwidth]{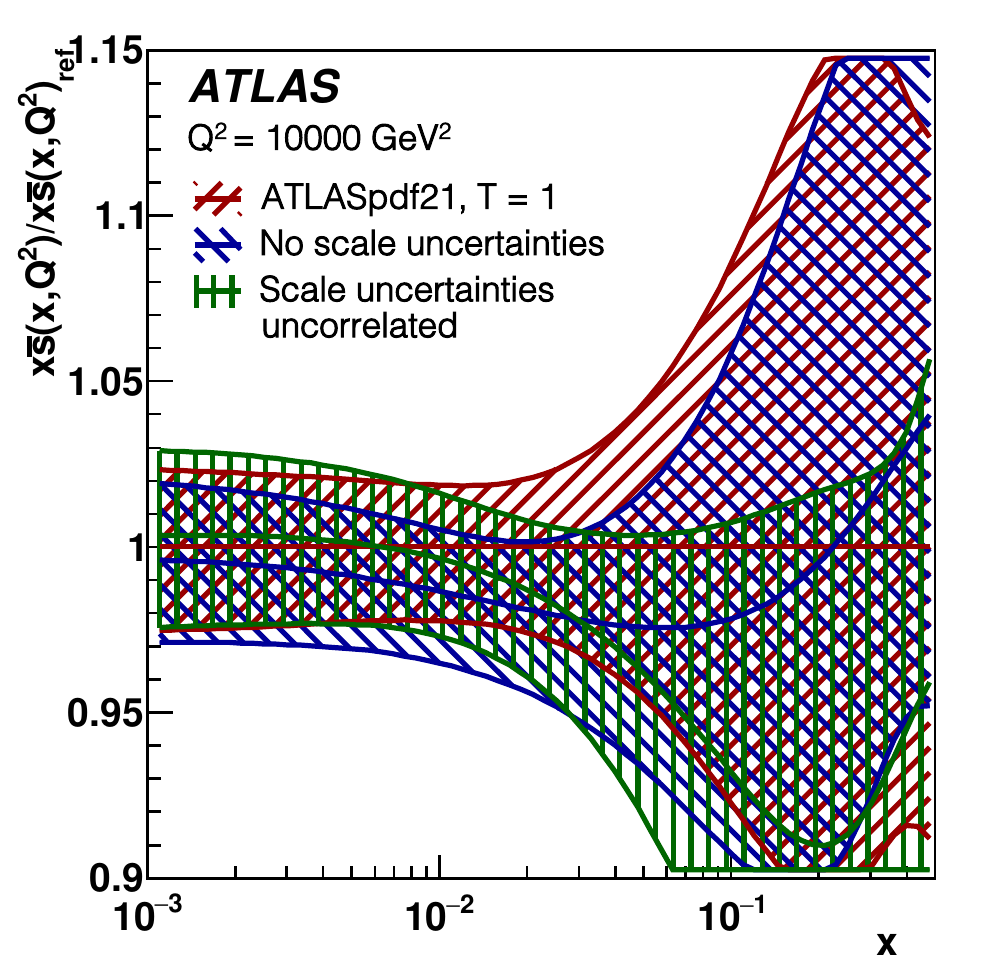}
\includegraphics[width=0.42\textwidth]{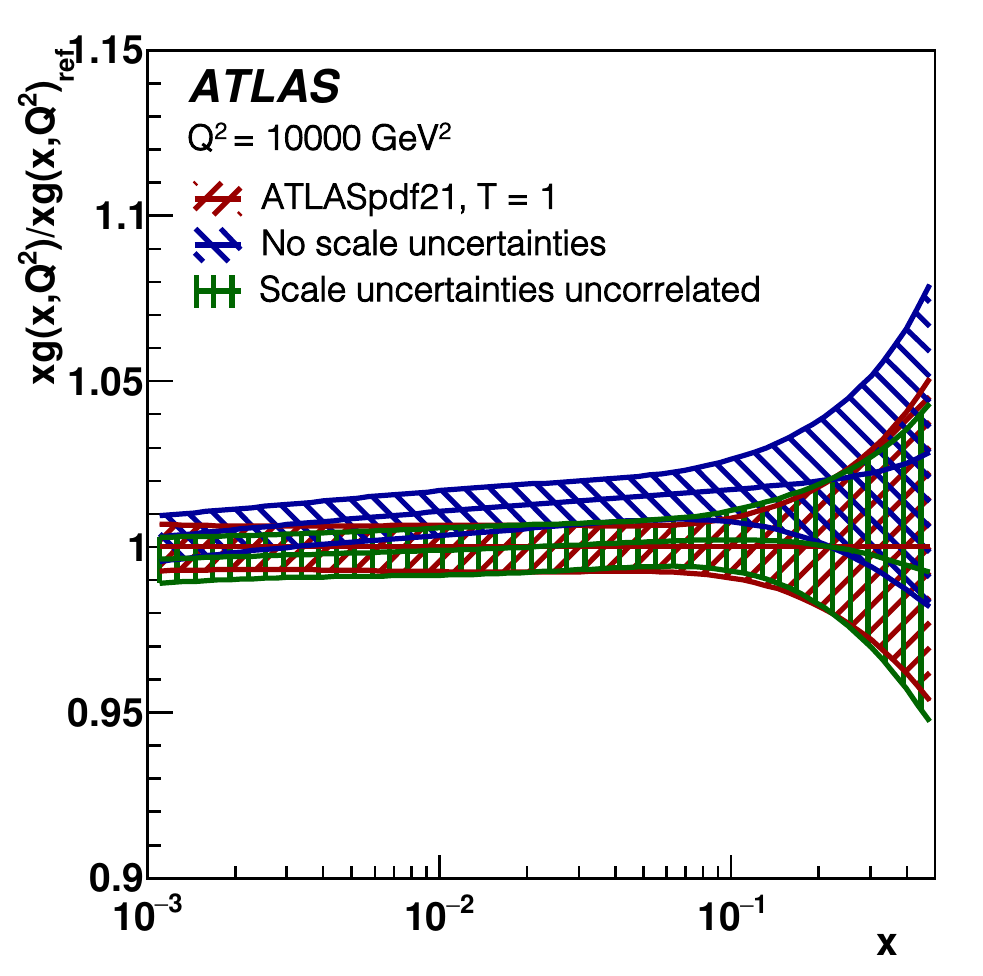}
\caption{ ATLASpdf21, showing the ratios of a fit not including theoret- ical scale uncertainties in the inclusive W, Z data to the central fit which does include these uncertainties, at the scale $Q^2 = \num{10000}$~\GeV$^2$. Also shown is the result of a fit in which the scale uncertainties are applied but not correlated between the 7 and 8 TeV data. All three fits are shown with just experimental uncertainties, evaluated with tolerance $T=1$. Top left: $xu_v$. Top right: $xd_v$. Middle left: $x\bar{u}$. Middle right: $x\bar{d}$. Bottom left: $xs$. Bottom right: $xg$.
\label{fig:ratioscaleunc}
}
\end{centering}
\end{figure*}
 
In this appendix, different choices for the treatment of the scale uncertainties in inclusive $W,Z$ data are considered.
For these data sets the experimental uncertainties are comparable to the scale uncertainties of NNLO predictions (${\sim}0.5\%$).  These theoretical uncertainties can therefore have some significant impact on the PDF fit.
For the central fit, these theoretical uncertainties are applied
in the same way as correlated systematic uncertainties.
Changes in renormalisation and factorisation scales are taken to be correlated
between $W$ and $Z$ data and to be correlated between the 7 and 8~\TeV\ data
sets. Two alternatives are considered here: i) the scale uncertainties are not
correlated between the 7 and 8~\TeV\ data sets and ii) scale uncertainties are not applied
at all.
 
The results of fits for these two cases, compared with the central fit, are shown as a ratio at a scale $Q^2 = \num{10000}$~GeV$^2$,
relevant for LHC physics, in Figure~\ref{fig:ratioscaleunc}. If should be noted that the central fit is shown with just experimental uncertainties, evaluated with tolerance $T=1$, so that its uncertainties are directly comparable with those of the fit without scale uncertainties. The uncertainties are very similar in size. The differences between the shapes of the PDFs are not large, but they can be important if the desired accuracy of the PDFs is $O({\sim}1\%)$.
The difference between the cases where the scale uncertainties are applied as being correlated or uncorrelated between the 7 and 8~\TeV\ inclusive $W,Z$ data sets is shown by the green line in these figures and it can be seen that it is
generally a smaller effect.
\clearpage
\section{Comparison of the impact of inclusive jet data at different centre-of-mass energies}
\label{sec:cofmjets}
 
\begin{figure*}[t!]
\begin{centering}
\includegraphics[width=0.45\textwidth]{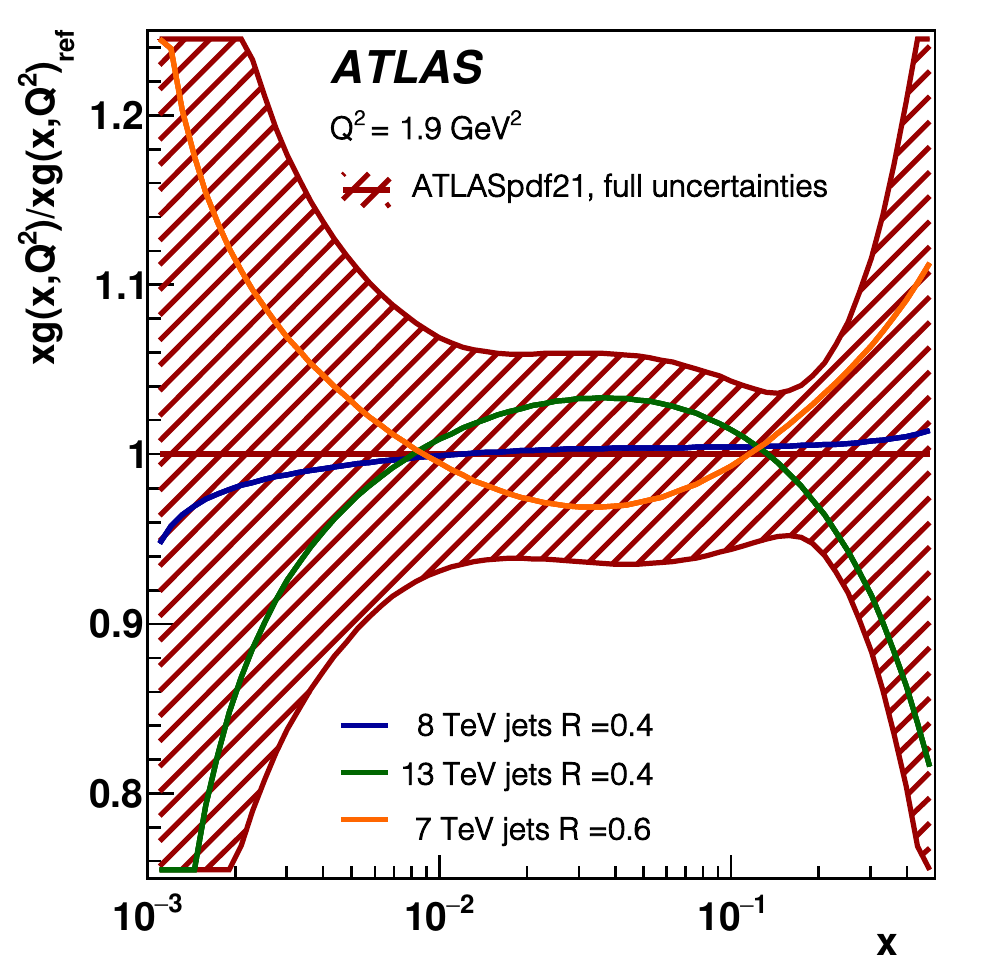}
\includegraphics[width=0.45\textwidth]{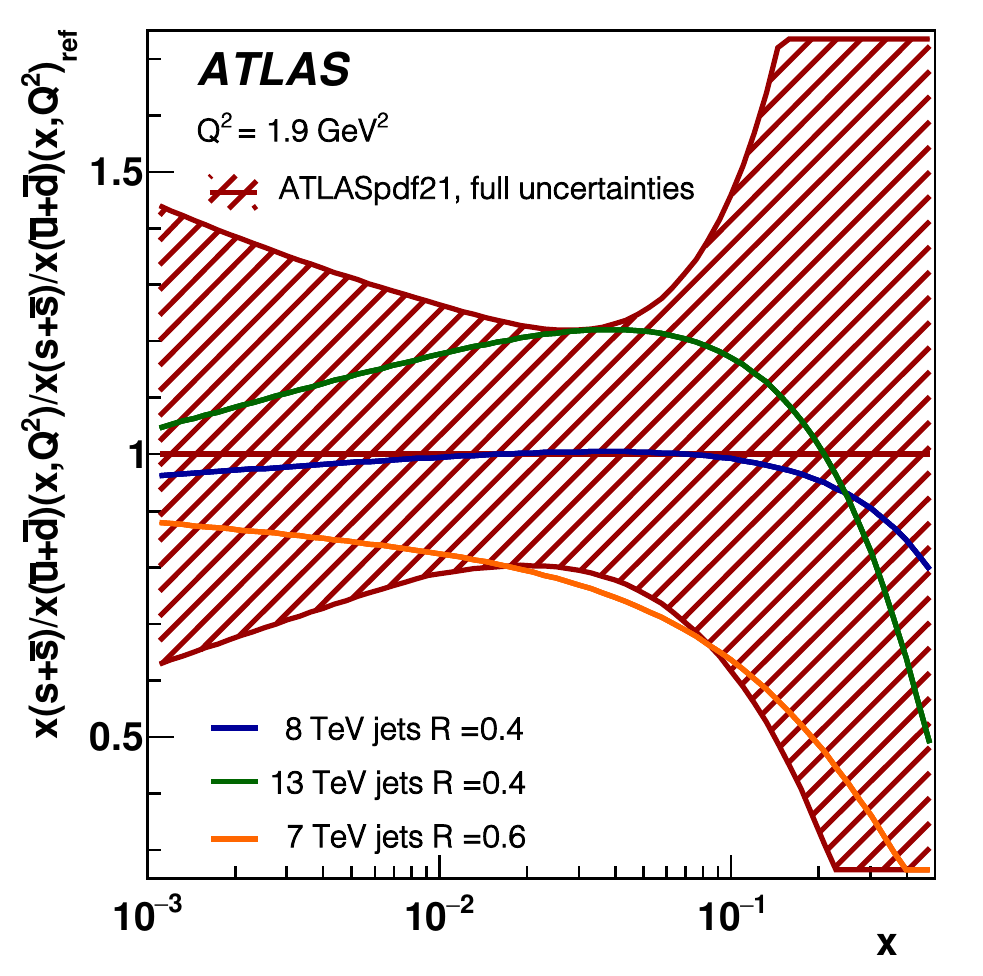}
\caption{Comparison showing the ratio of the ATLASpdf21 gluon PDF and the $R_s$ PDF ratio, using inclusive jet data at 8~\TeV\ with $R = 0.6$, to the gluon PDF and $R_s$ PDF ratio for fits using various different
jet production data sets at 7, 8 and 13~\TeV, with differing choices of jet radius.
The data sets in these plots are for scale choice $p_{\mathrm{T}}^{\mathrm{jet}}$.
Uncertainties of the central fit are  full uncertainties: experimental, evaluated with tolerance $T=3$, plus model and parameterisation uncertainties.
\label{fig:jets82}
}
\end{centering}
\end{figure*}

It is not possible to fit inclusive jet production data at different centre-of-mass energies simultaneously, because the
full experimental systematic uncertainty correlations between these data sets are not known. The inclusive jet production data at 8~\TeV\ with
$R = 0.6$ were selected for input to the central fit. In this appendix, fits using the inclusive jet production data at 7 or 13~\TeV\ instead of the data at 8~\TeV\ are compared with the central fit. The data at 7~\TeV\ shown here were extracted for $R = 0.6$ whereas the data at 7~\TeV\
were extracted only for $R =0.4$. The gluon PDF and $R_s$ PDF ratio using these jet production data sets at 7 and 13~\TeV\ are shown in their ratio to the central fit results in Figure~\ref{fig:jets82}. The scale choice was $p_{\mathrm{T}}^{\mathrm{jet}}$ for all the jet  production
data sets included in this figure. Since the effect of using $R = 0.4$ or $R = 0.6$ is very similar,
as illustrated for the jet production data at 8~\TeV, the differences between the PDFs are dominated by the change in centre-of-mass energy. The PDFs using the jet production data at 8~\TeV\ lie between those of the jet production data at 7~\TeV\ and the jet production data at 13~\TeV.
However, these differences are not significant compared to the full uncertainties of the PDFs evaluated with $T = 3$.
\clearpage
\section{Goodness of the fit: comparison with data sets included in the fit}
\label{sec:datafitcomp}
 
\begin{figure*}[h!]
\begin{centering}
\includegraphics[width=0.48\textwidth]{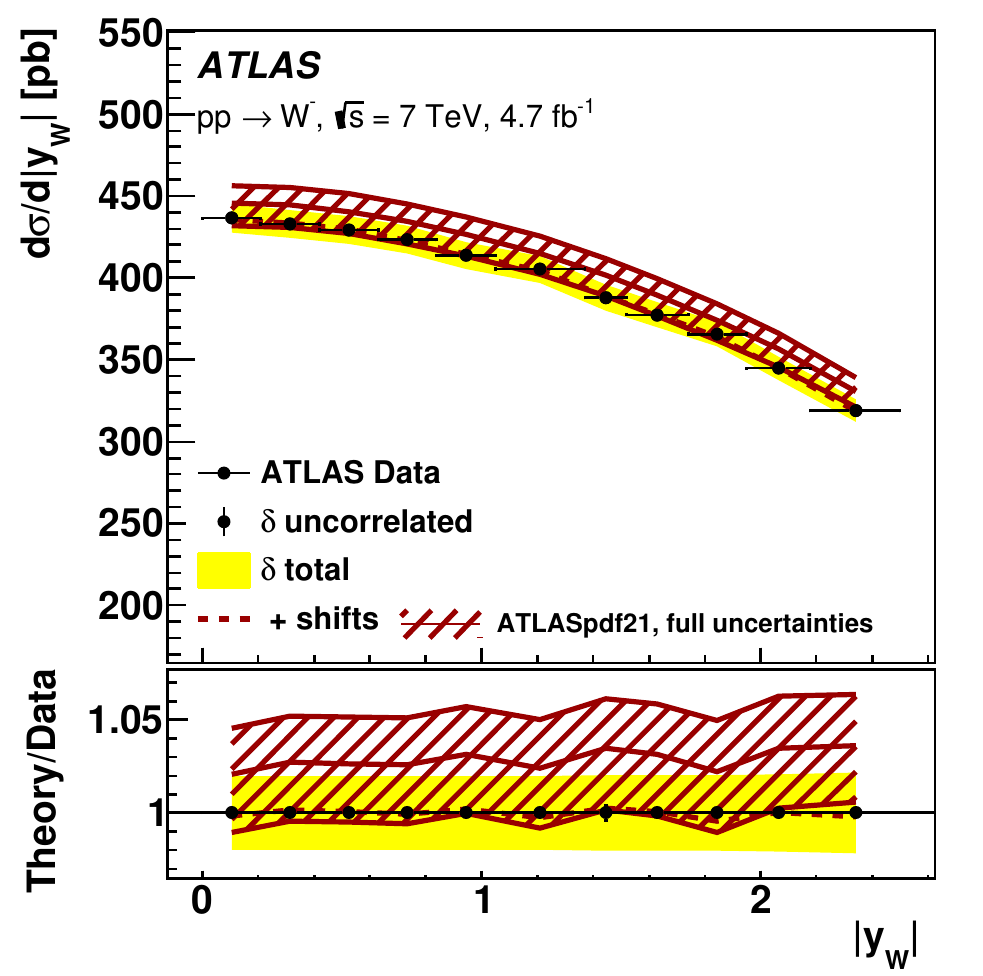}
\includegraphics[width=0.48\textwidth]{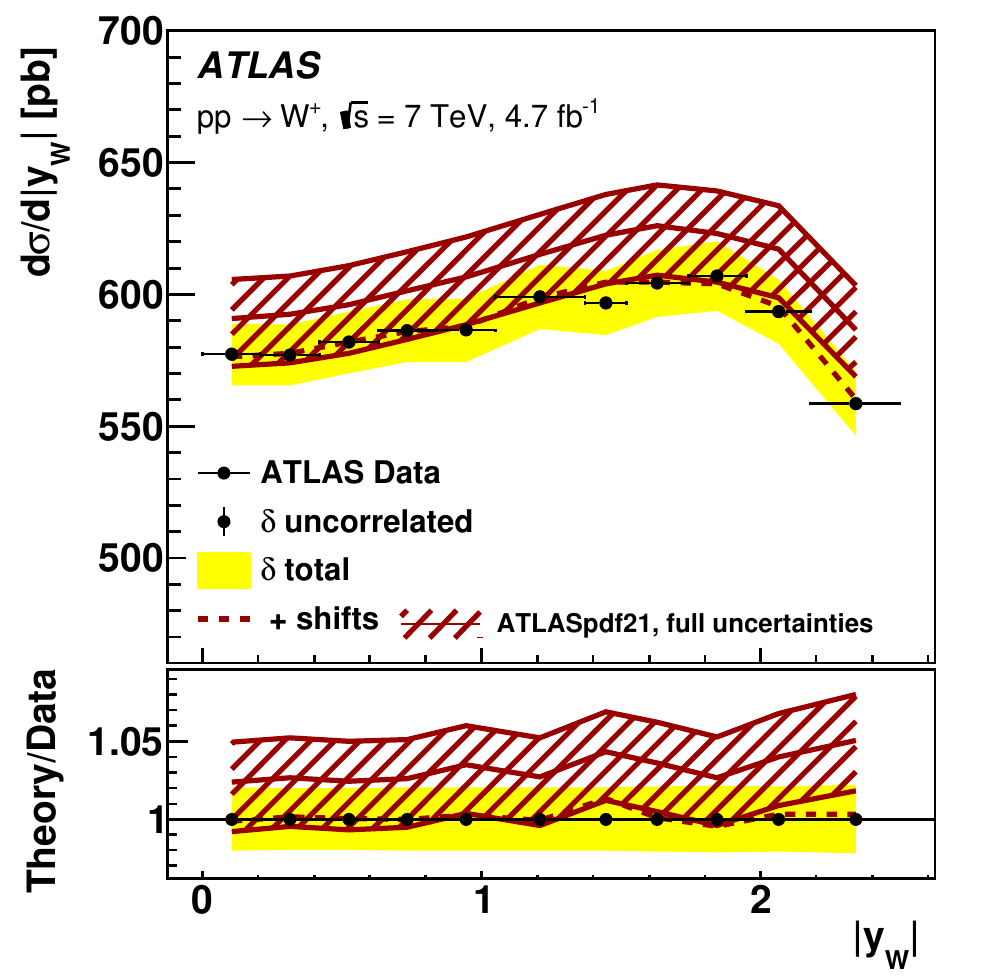}
\caption{ The differential cross-section measurements of (left) $W^{-}$ and (right) $W^{+}$ bosons in Ref.~\cite{1612.03016} (black points) as a function of their absolute rapidity, $|y_{W}|$. The bin-to-bin uncorrelated part of the data uncertainties is shown as black error bars, while the total uncertainties are shown as a yellow band. The cross sections are compared with the predictions computed with the PDFs resulting from the ATLASpdf21 fit. The solid line shows the predictions without shifts of the systematic uncertainties, while for the dashed line the $b_j$ parameters associated with the experimental systematic uncertainties as shown in Eq.~(\ref{eqn:chi2}) are allowed to vary to minimise the $\chi^{2}$. The red band represents the full uncertainty (experimental (evaluated with $T=3$) +~model +~parameterisation) of the fit prediction.
\label{fig:WZ7TeV_W}
}
\end{centering}
\end{figure*}

Figures~\ref{fig:WZ7TeV_W}--\ref{fig:IncJets_8TeVptjet} show comparisons of the various ATLAS differential cross-section measurements used in the ATLASpdf21 fit, together with the predictions of this fit. Further details are provided in the figure captions.
\begin{figure*}[t!]
\begin{centering}
\includegraphics[width=0.48\textwidth]{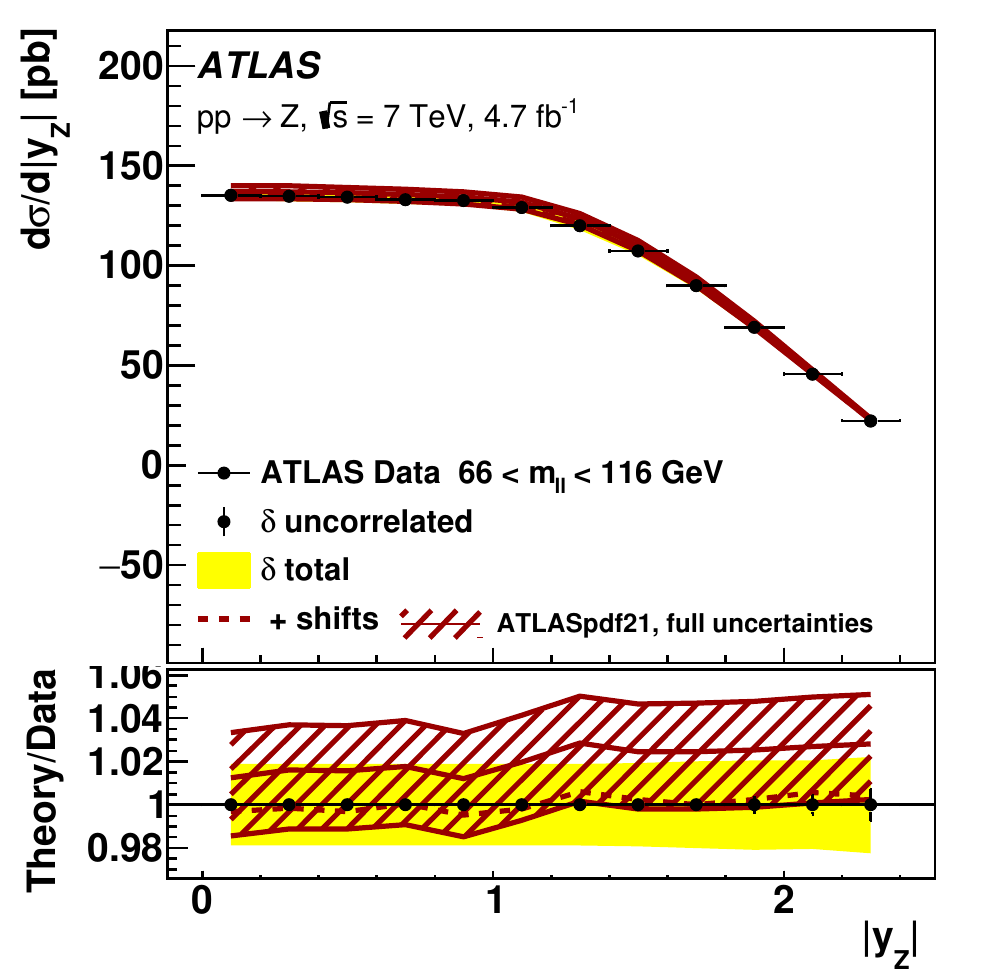}
\includegraphics[width=0.48\textwidth]{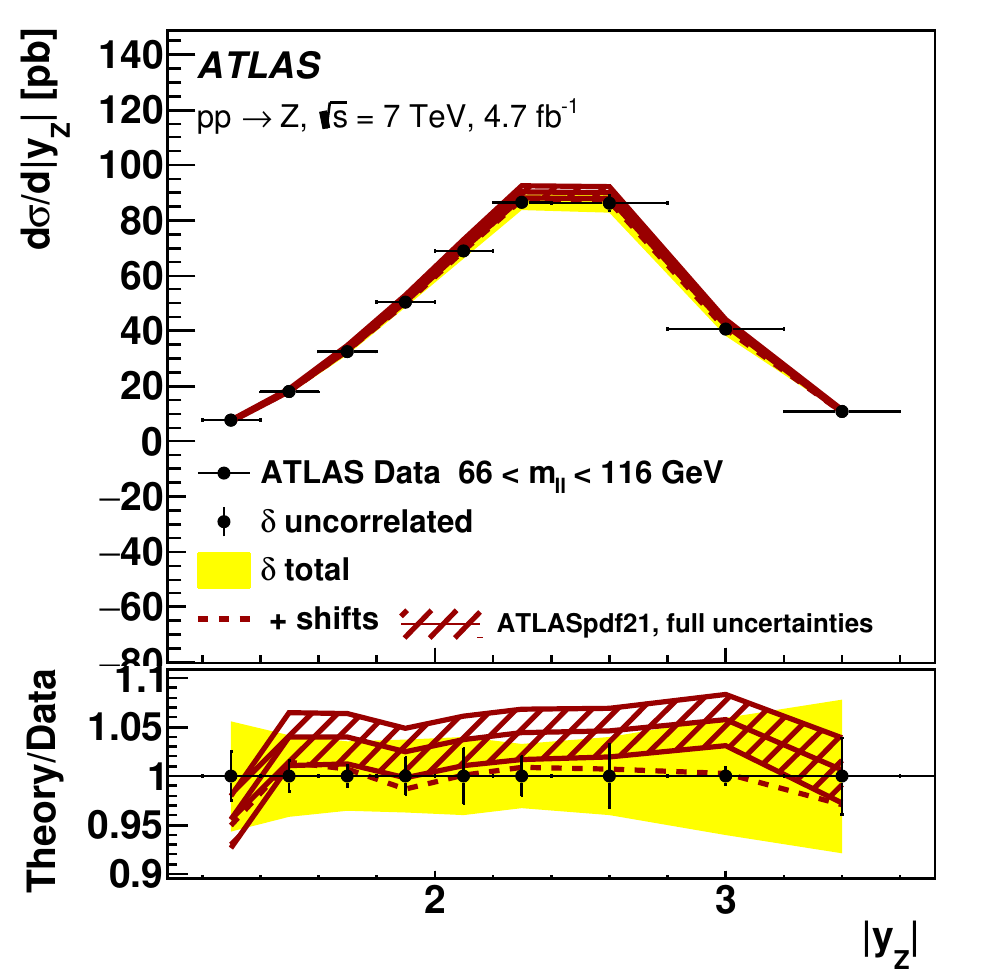}
\includegraphics[width=0.48\textwidth]{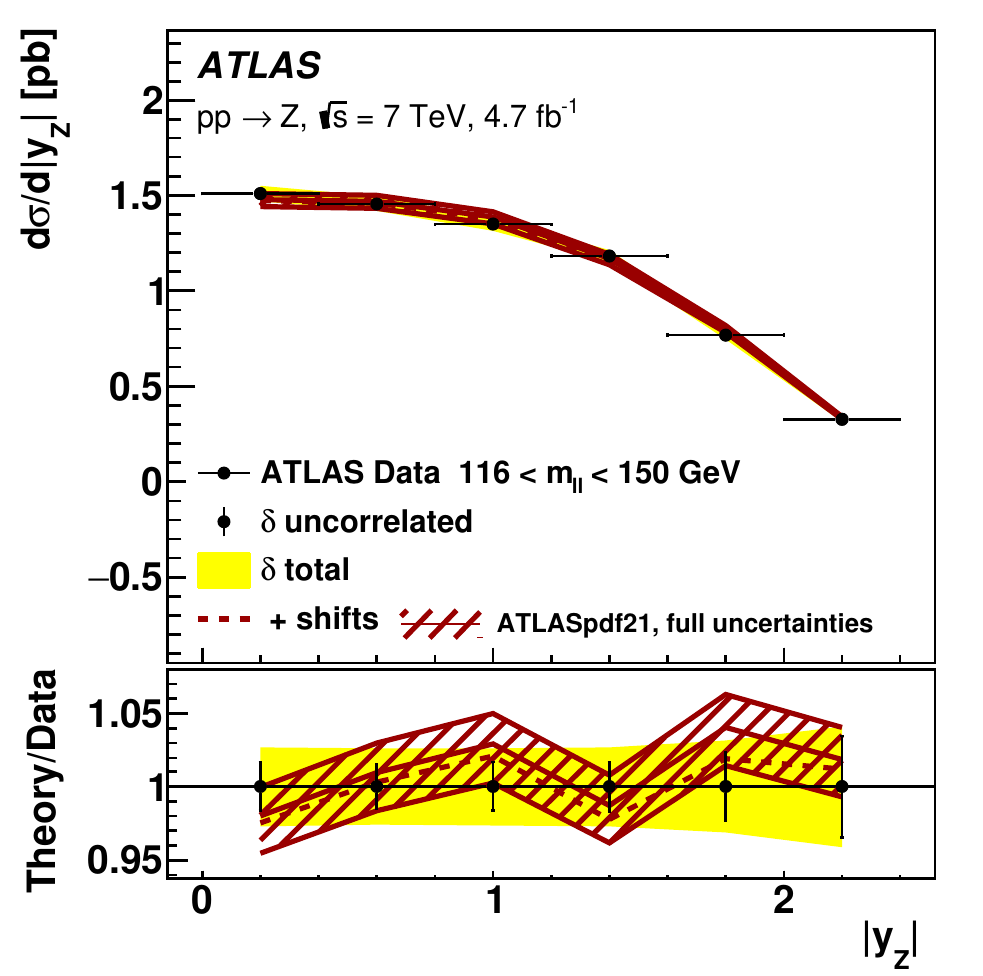}
\includegraphics[width=0.48\textwidth]{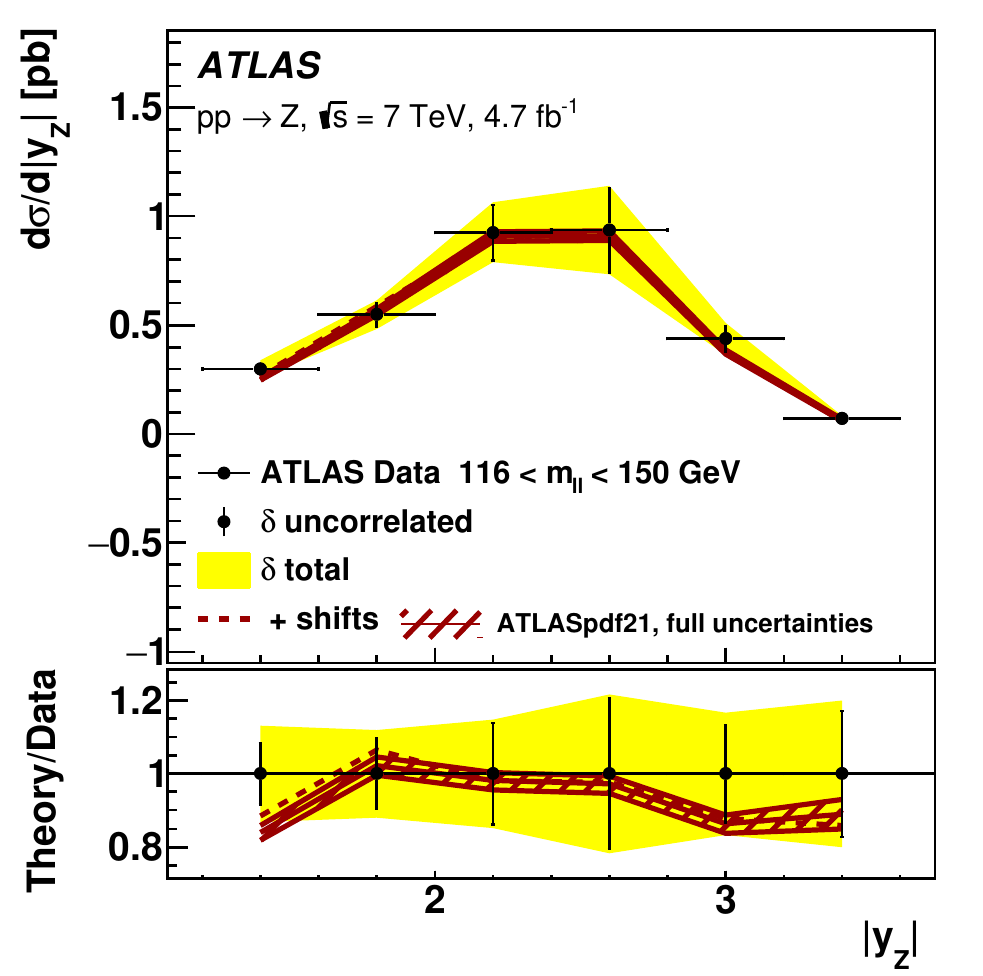}
\caption{ The differential cross section measurements of $Z$ bosons for (left) central-central and (right) central-forward events in Ref.~\cite{1612.03016} (black points) as a function of their absolute rapidity, $|y_{Z}|$. The top row shows differential cross sections in the $66<m_{\ell\ell}<116$~\GeV\ mass range, while the bottom row shows the $116<m_{\ell\ell}<150$~\GeV\ mass range. The bin-to-bin uncorrelated part of the data uncertainties is shown as black error bars, while the total uncertainties are shown as a yellow band. The cross sections are compared with the predictions computed with the PDFs resulting from the ATLASpdf21 fit. The solid line shows the predictions without shifts of the systematic uncertainties, while for the dashed line the $b_j$ parameters associated with the experimental systematic uncertainties as shown in Eq.~(\ref{eqn:chi2}) are allowed to vary to minimise the $\chi^{2}$. The red band represents the full uncertainty (experimental (evaluated with $T=3$) +~model +~parameterisation) of the fit prediction. It is interesting to observe that the N$^{3}$LO cross section for $Z$ bosons is expected to be ${\sim}2\%$ lower than the NNLO cross section~\cite{2107.09085}, which would bring the data into agreement with theory without need of shifts of systematic uncertainties.
}
\end{centering}
\end{figure*}
\begin{figure*}[t!]
\begin{centering}
\includegraphics[width=0.35\textwidth]{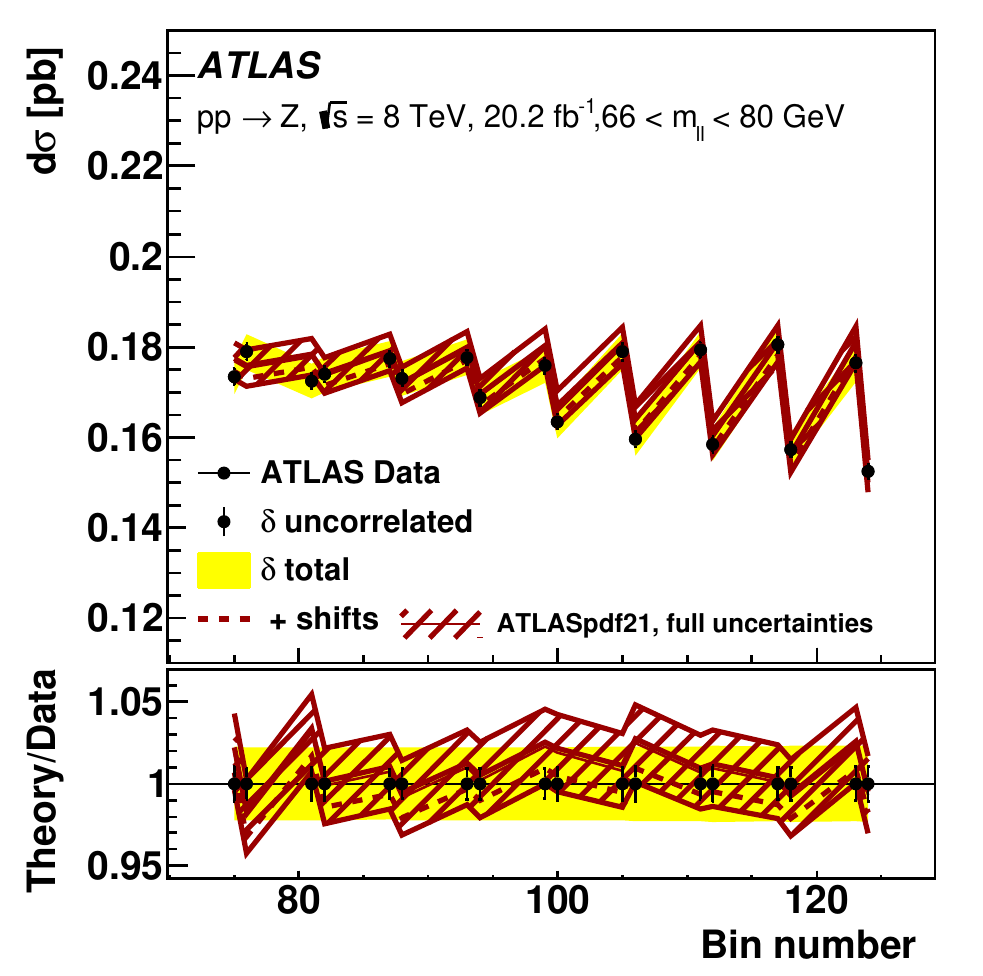}
\includegraphics[width=0.35\textwidth]{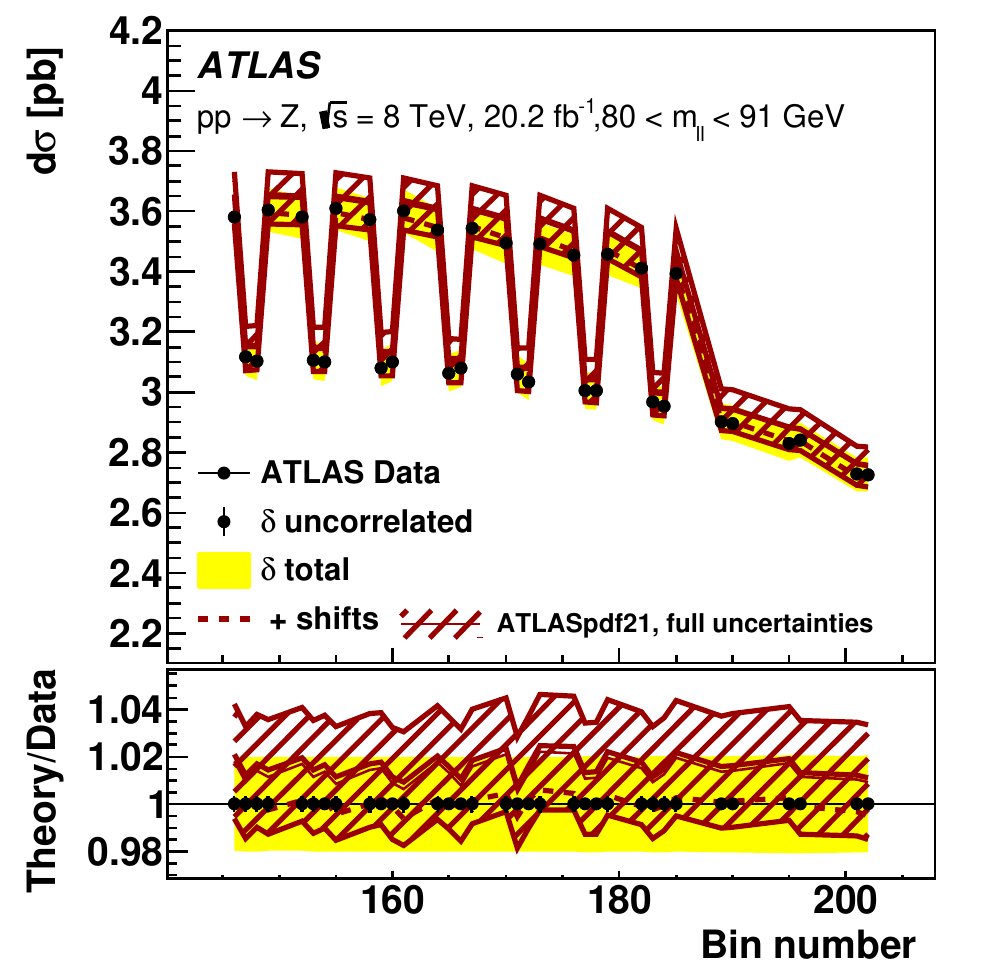}
\includegraphics[width=0.35\textwidth]{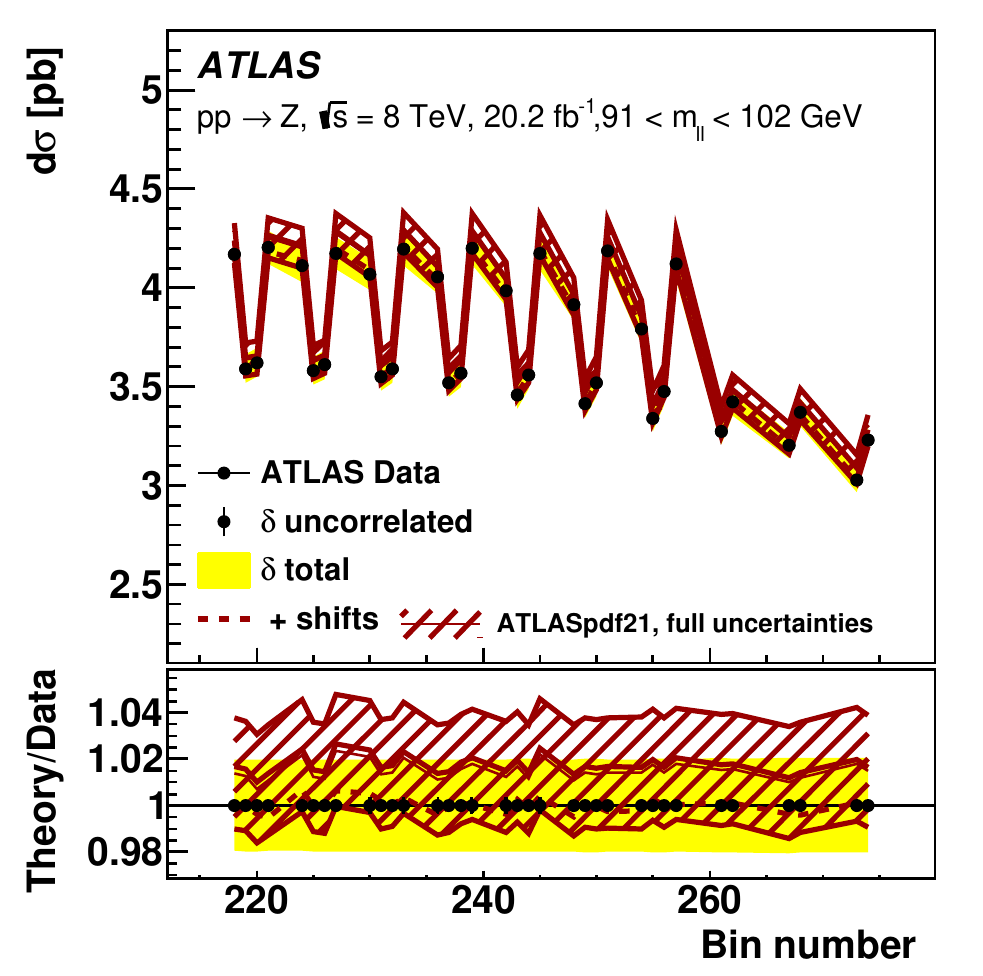}
\includegraphics[width=0.35\textwidth]{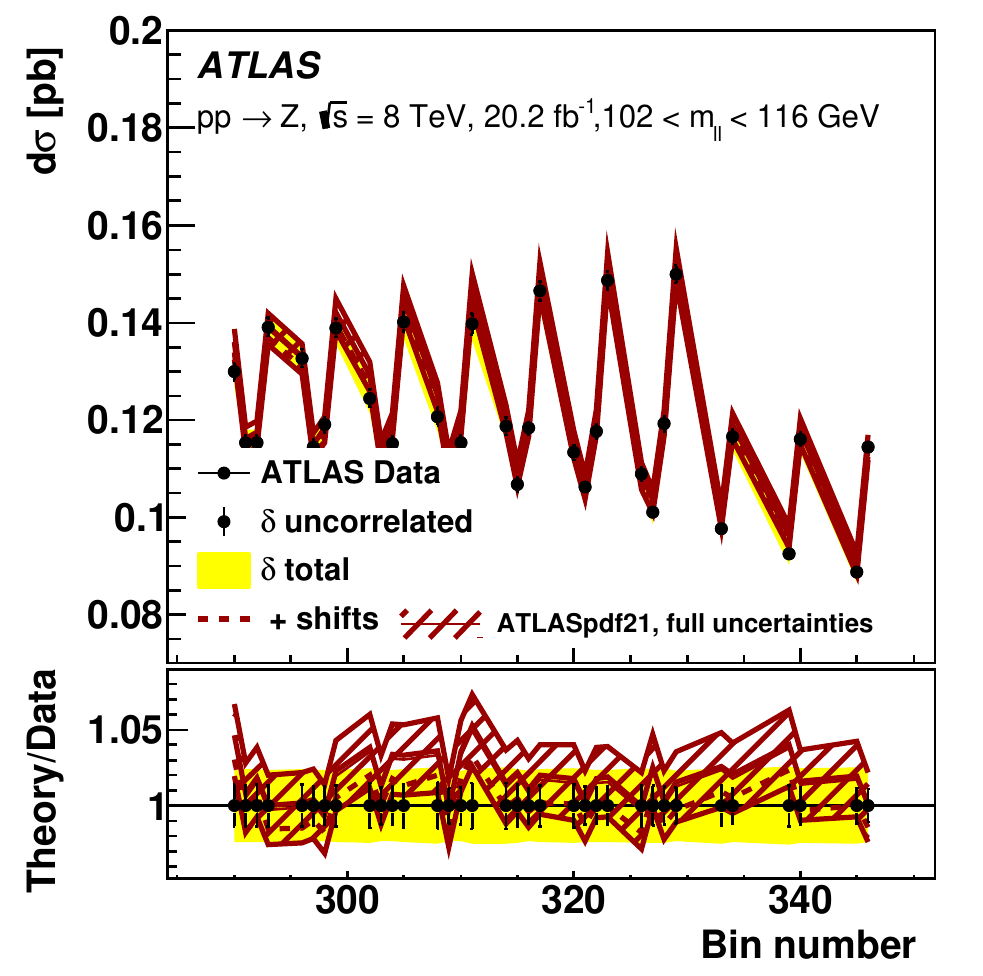}
\includegraphics[width=0.35\textwidth]{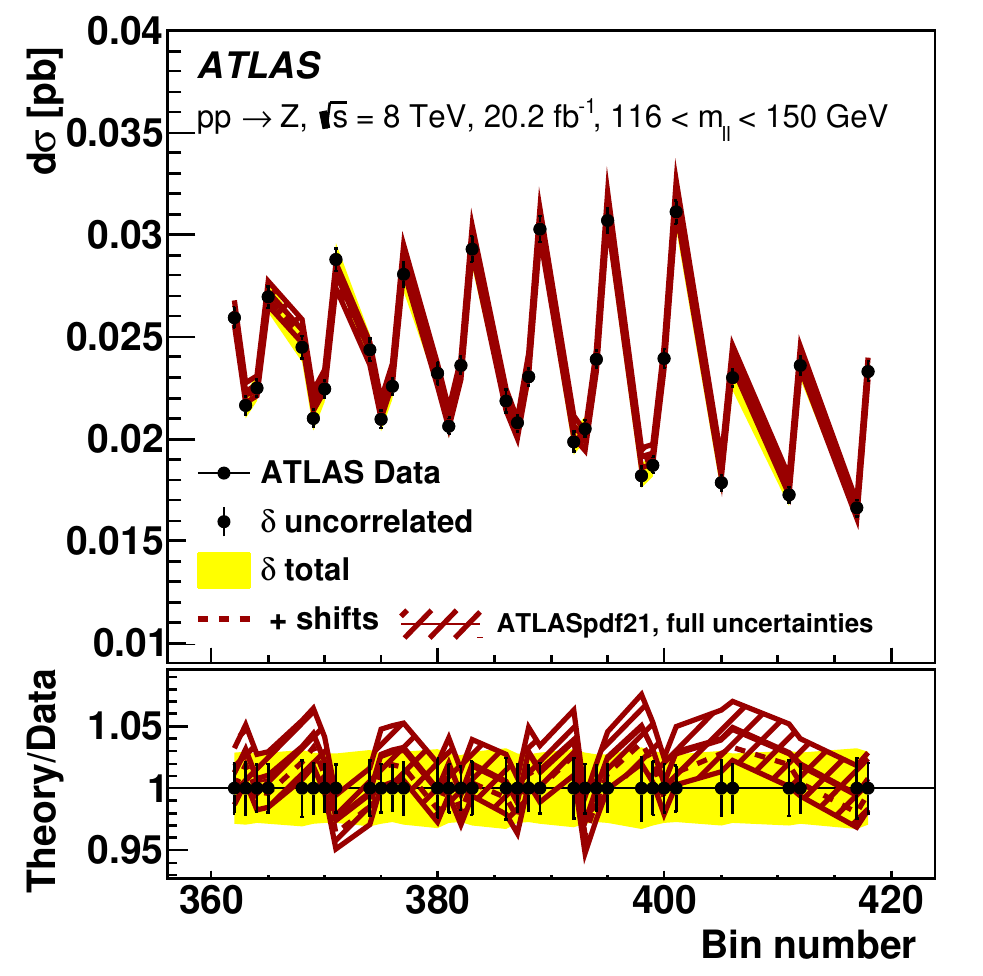}
\includegraphics[width=0.35\textwidth]{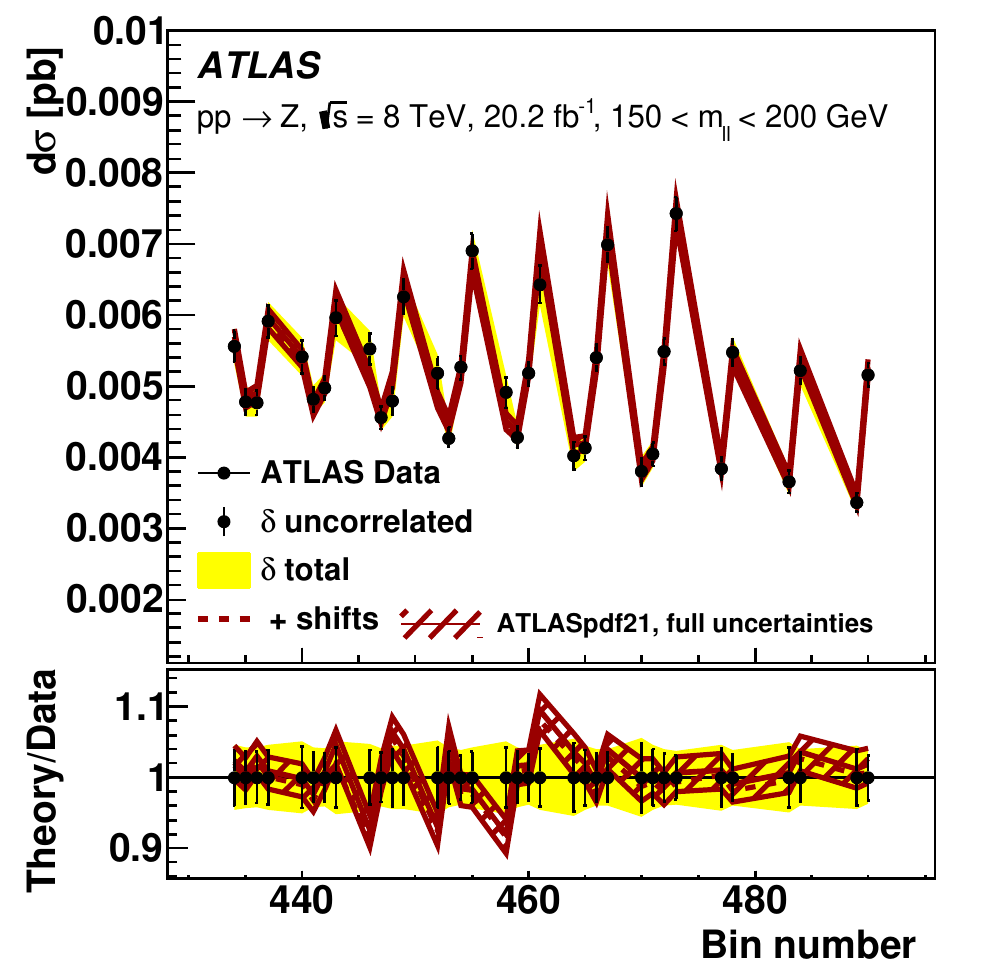}
\caption{ The triple differential cross-section measurements of $Z/\gamma^{*}$ at 8~\TeV\ in Ref.~\cite{z3d} (black points), in
the central-central rapidity region of the dilepton pair. The six plots show data in bins of the dilepton mass $m_{\ell\ell}$
(66--80--91--102--116--150--200~\GeV). Within each plot the absolute rapidity increases from left to right in 12 steps of 0.2, ranging from 0.0
to 2.4, although the full rapidity range is not accessed for every mass region. Nearby data points show variation with
the Collins--Soper angle at the same rapidity. The bin-to-bin uncorrelated part of the data uncertainties is shown as
black error bars, while the total uncertainties are shown as a yellow band. The cross sections are compared with the predictions
computed with the PDFs resulting from the ATLASpdf21 fit. The solid line shows the predictions without shifts
of the systematic uncertainties, while for the dashed line the $b_j$ parameters associated with the experimental systematic
uncertainties as shown in Eq.~(\ref{eqn:chi2}) are allowed to vary to minimise the $\chi^{2}$. The red band represents the full
uncertainty (experimental (evaluated with $T=3$) +~model +~parameterisation) of the fit prediction. It is interesting to observe that the N$^{3}$LO cross section for $Z$ bosons is expected to be ${\sim}2\%$ lower than the NNLO cross section~\cite{2107.09085}, which would bring data into agreement with theory without need of shifts of systematic uncertainties. The $Z$ mass-peak data are in the top right and middle left plots.
}
\end{centering}
\end{figure*}
\begin{figure*}
\begin{centering}
\includegraphics[width=0.48\textwidth]{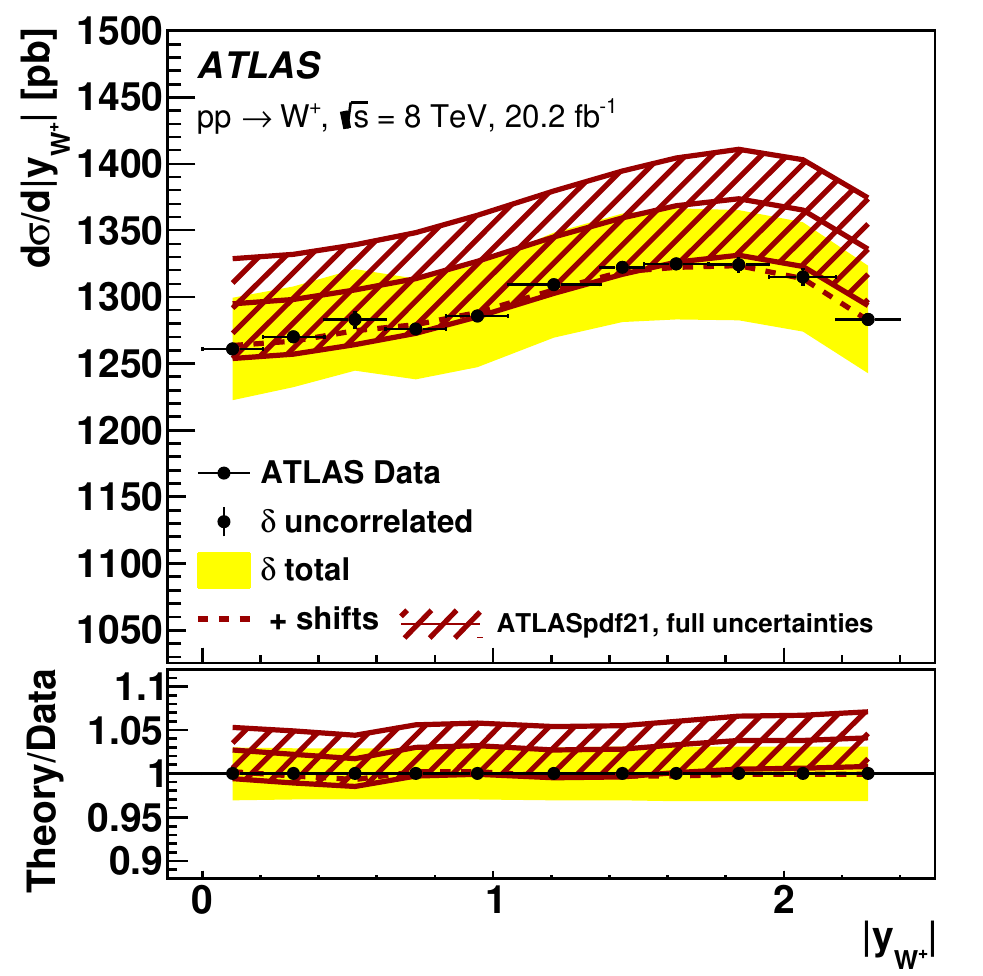}
\includegraphics[width=0.48\textwidth]{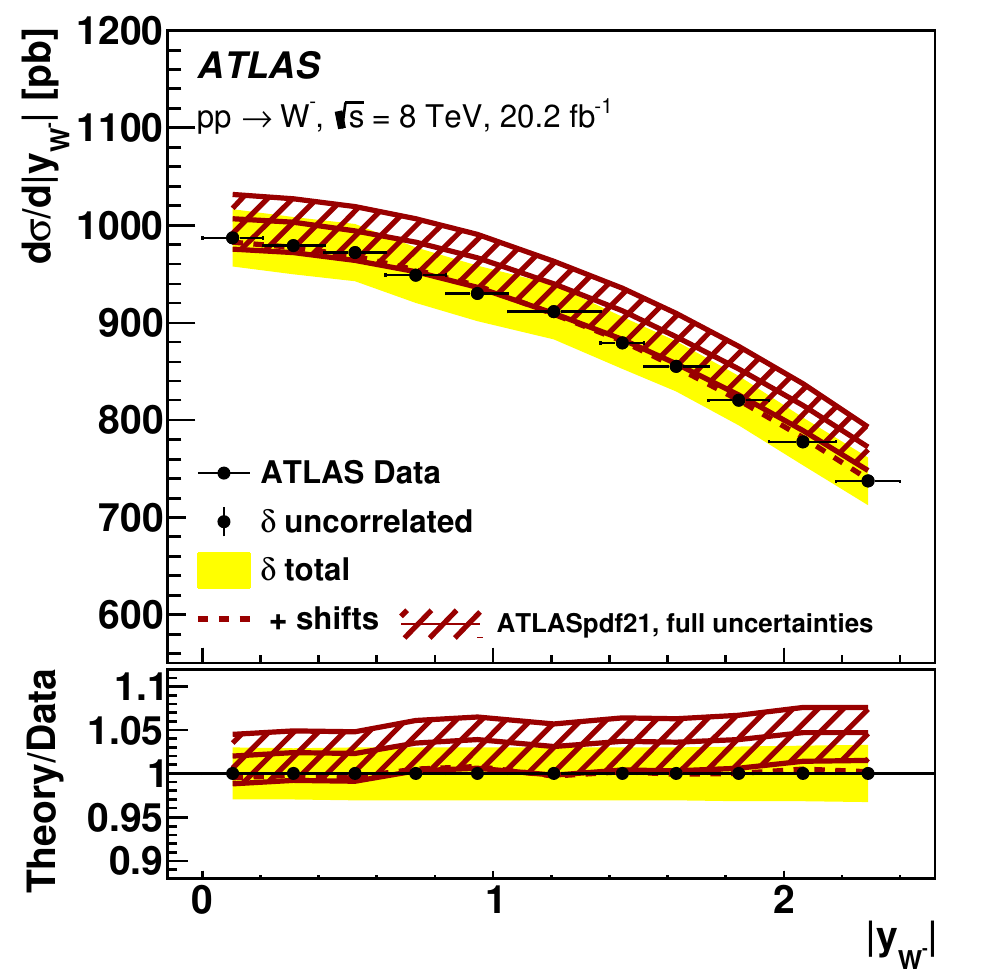}
\caption{ The differential cross-section measurements of (left) $W^{-}$ and (right) $W^{+}$ bosons at 8~\TeV\ in Ref.~\cite{W8} (black points) as a function of their absolute rapidity, $|y_{W}|$. The bin-to-bin uncorrelated part of the data uncertainties is shown as black error bars, while the total uncertainties are shown as a yellow band. The cross sections are compared with the predictions computed with the PDFs resulting from the ATLASpdf21 fit. The solid line shows the predictions without shifts of the systematic uncertainties, while for the dashed line the $b_j$ parameters associated with the experimental systematic uncertainties as shown in Eq.~(\ref{eqn:chi2}) are allowed to vary to minimise the $\chi^{2}$. The red band represents the full uncertainty (experimental (evaluated with $T=3$) +~model +~parameterisation) of the fit prediction.
}
\end{centering}
\end{figure*}
\begin{figure*}[t!]
\begin{centering}
\includegraphics[width=0.48\textwidth]{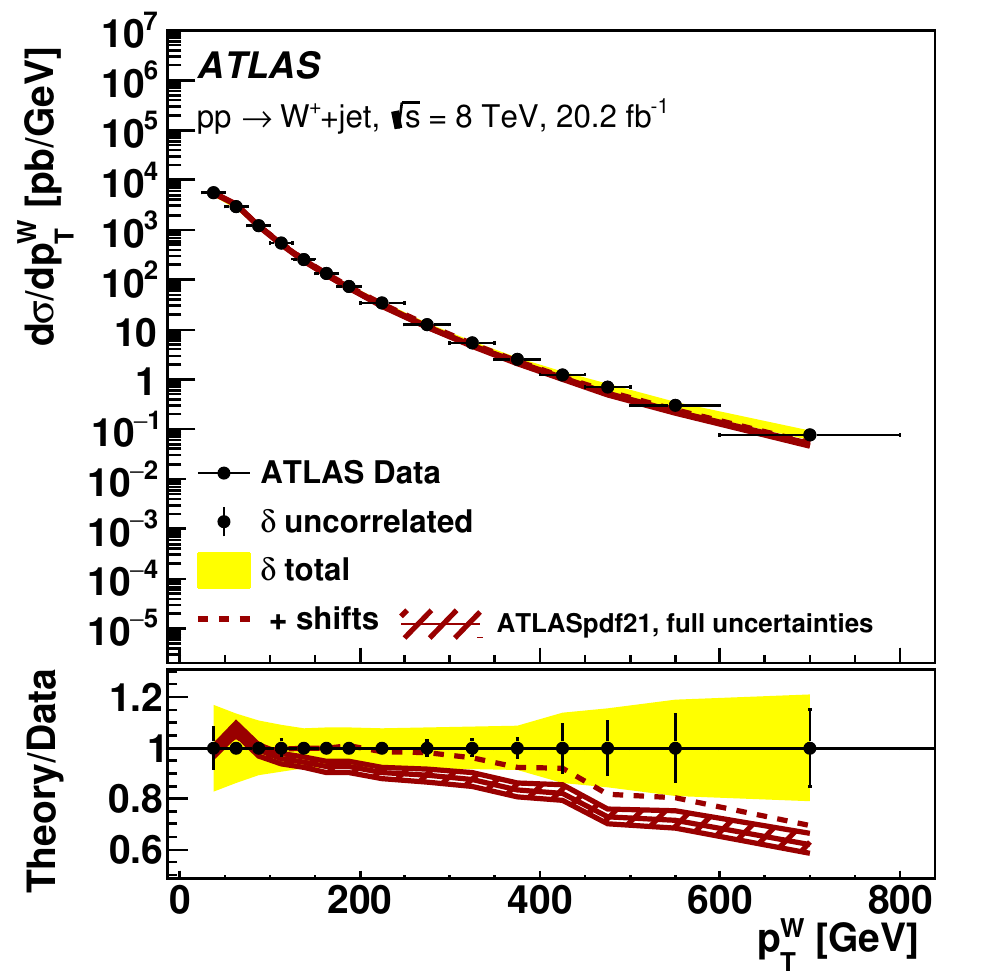}
\includegraphics[width=0.48\textwidth]{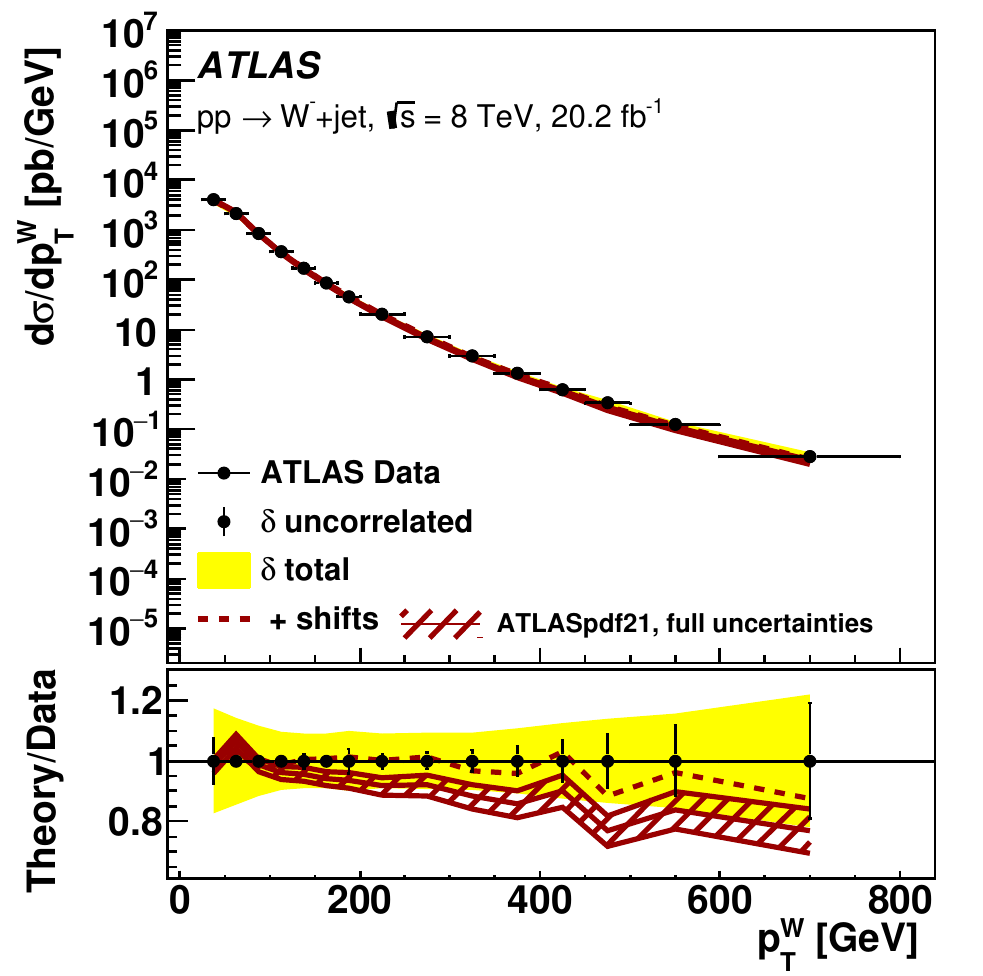}
\caption{ The differential cross-section measurements of (left) $W^{-}$\,+\,jets and (right) $W^{+}$\,+\,jets at 8~\TeV\ in Ref.~\cite{wjetspaper} (black points) as a function of the transverse momentum of the $W$ boson, $p_{\mathrm{T}}^{W}$ . The bin-to-bin uncorrelated part of the data uncertainties is shown as black error bars, while the total uncertainties are shown as a yellow band. The cross sections are compared with the predictions computed with the PDFs resulting from the ATLASpdf21 fit. The solid line shows the predictions without shifts of the systematic uncertainties, while for the dashed line the $b_j$ parameters associated with the experimental systematic uncertainties as shown in Eq.~(\ref{eqn:chi2}) are allowed to vary to minimise the $\chi^{2}$. The red band represents the full uncertainty (experimental (evaluated with $T=3$) +~model +~parameterisation) of the fit prediction.
}
\end{centering}
\end{figure*}
\begin{figure*}[t!]
\begin{centering}
\includegraphics[width=0.38\textwidth]{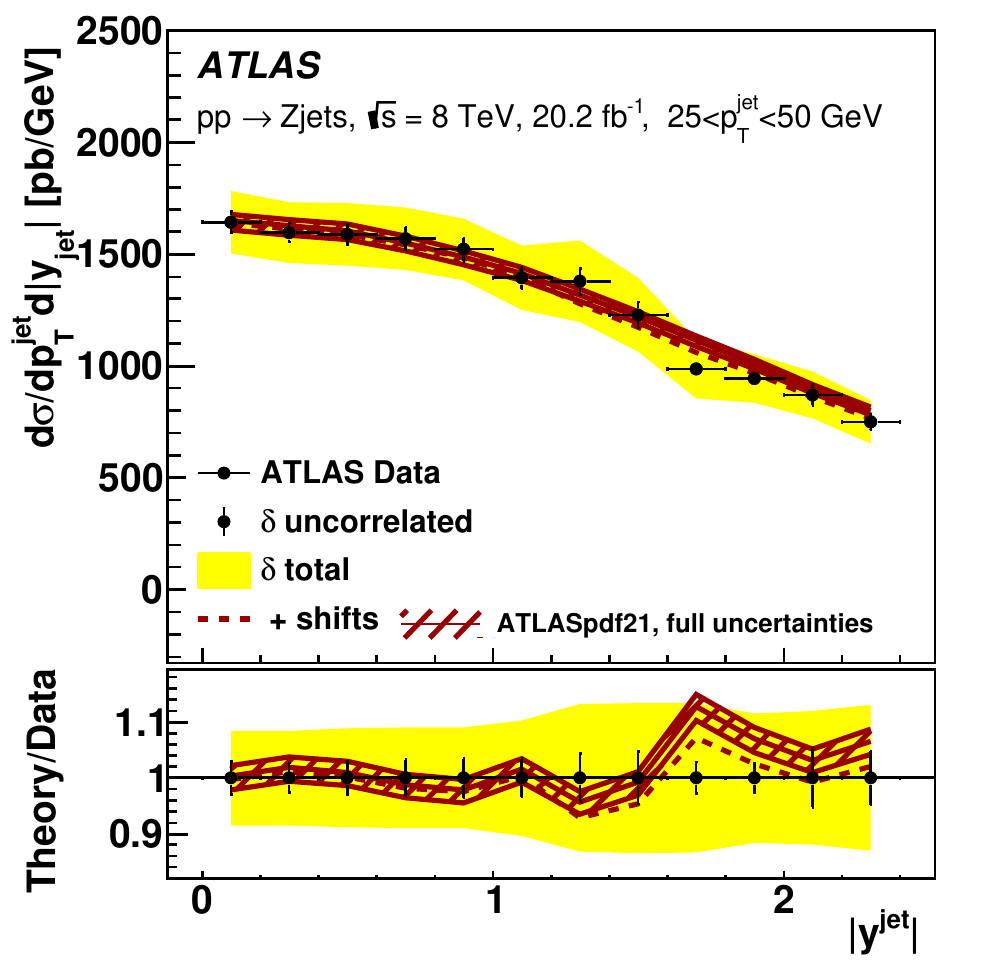}
\includegraphics[width=0.38\textwidth]{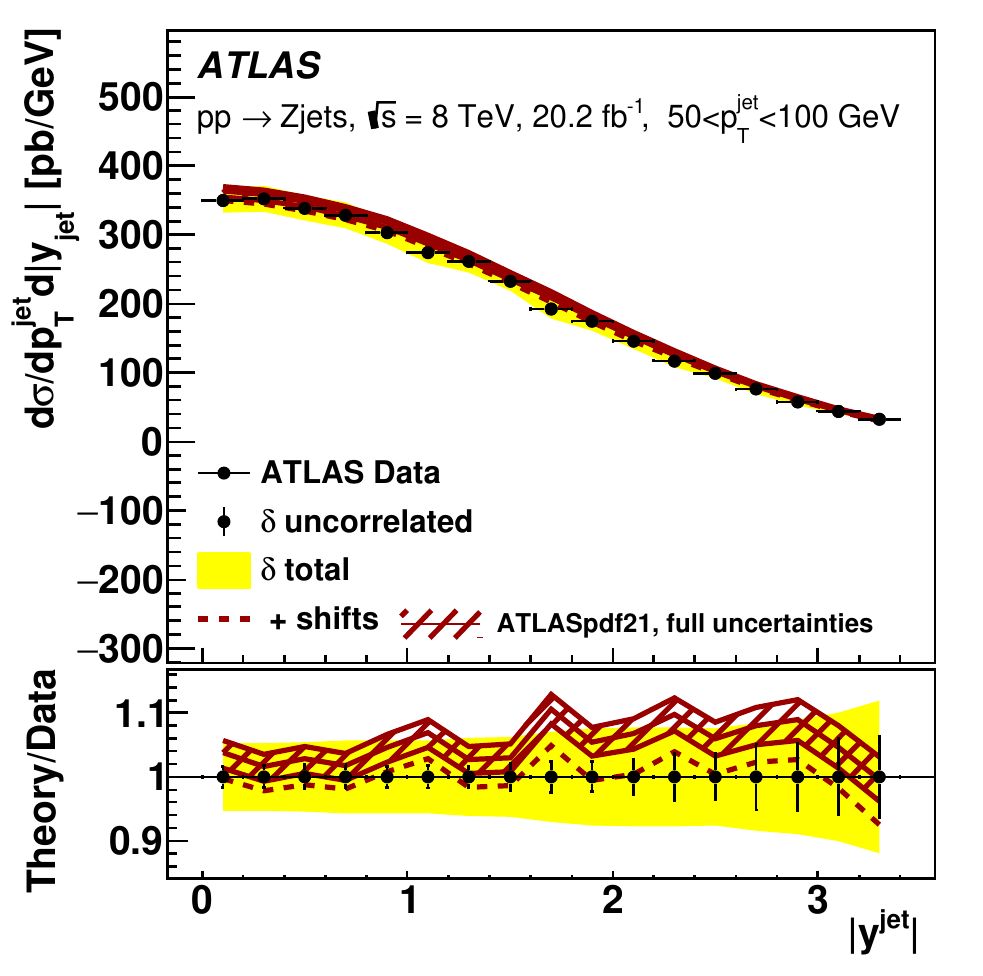}
\includegraphics[width=0.38\textwidth]{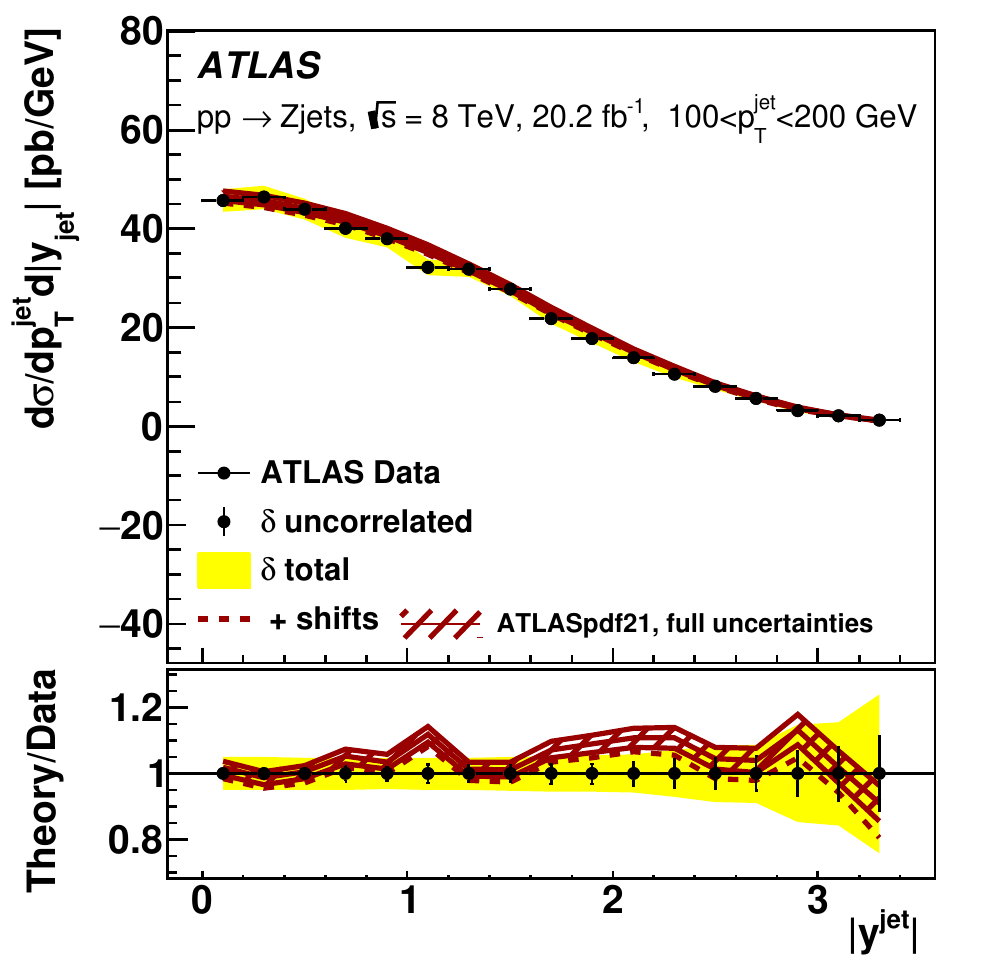}
\includegraphics[width=0.38\textwidth]{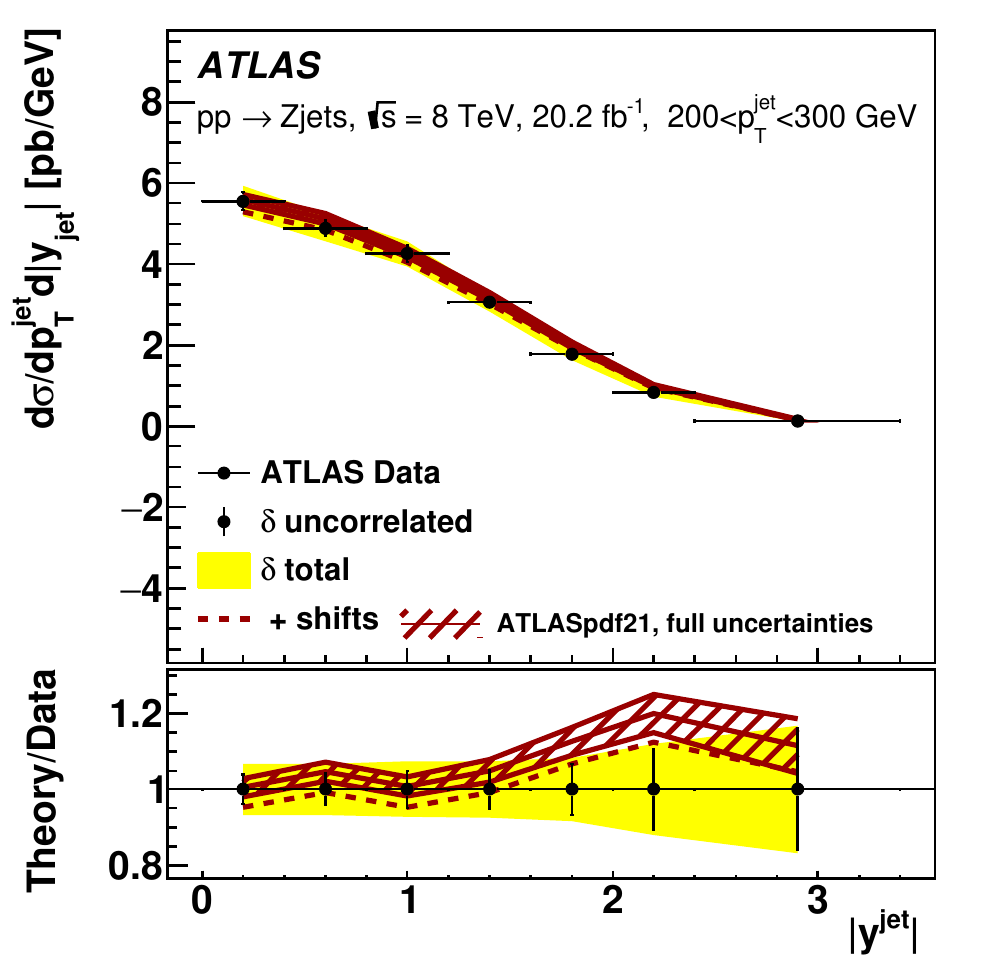}
\includegraphics[width=0.38\textwidth]{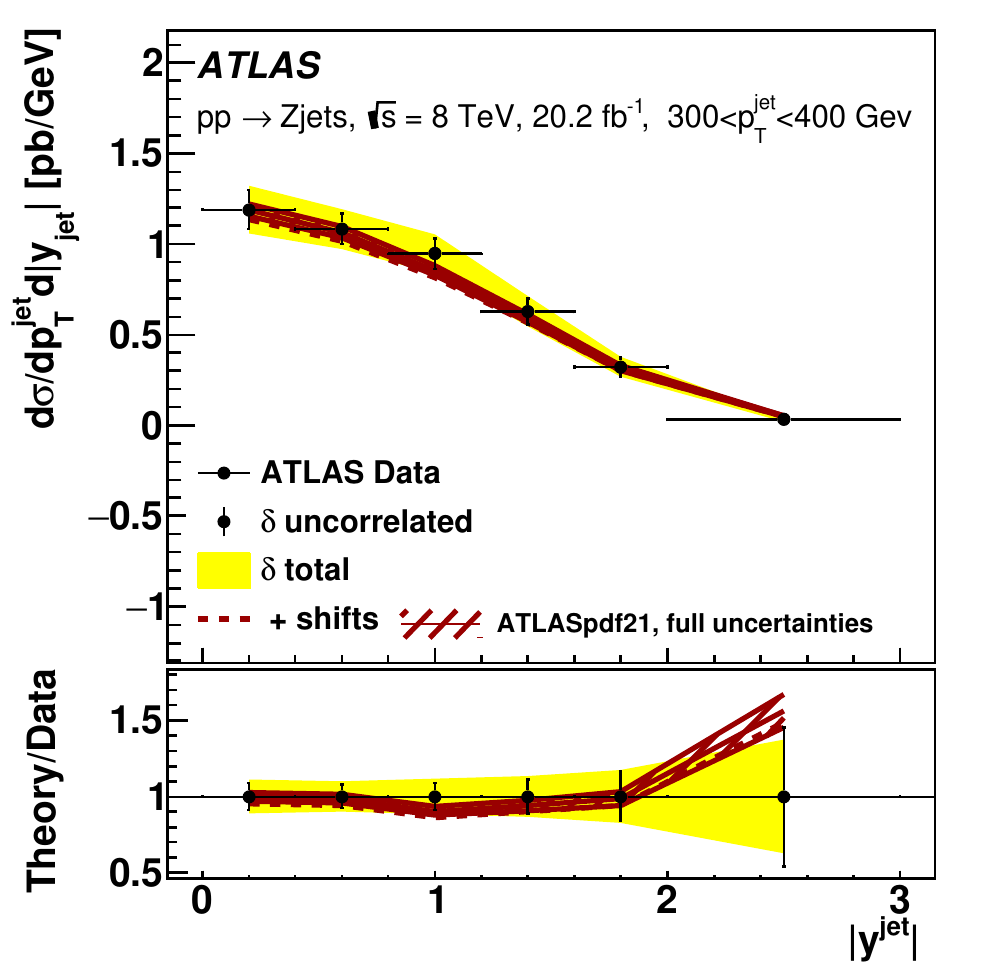}
\includegraphics[width=0.38\textwidth]{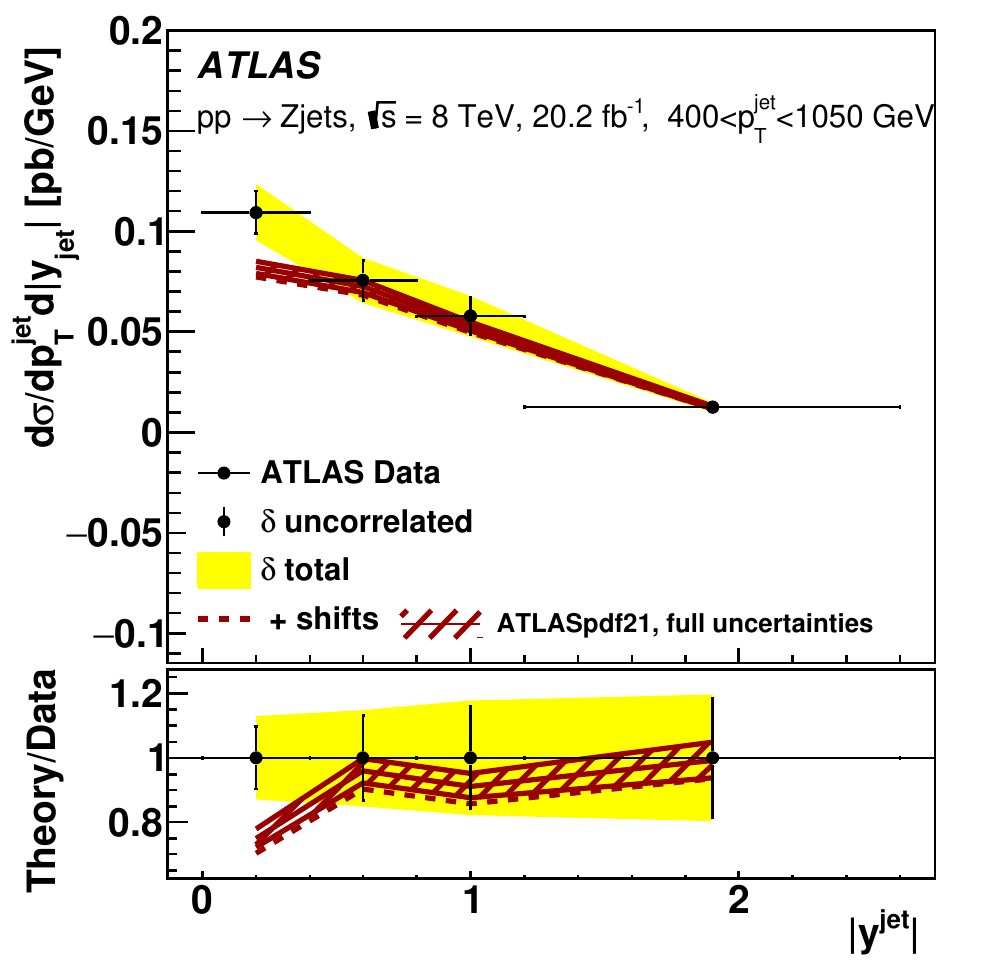}
\caption{ The differential cross-section measurements of $Z$\,+\,jets at 8~\TeV\ in Ref.~\cite{zjetspaper} (black points) as a function of the absolute rapidity of inclusive jets, $|y^{\mathrm{jet}}|$, in bins of $p_{\mathrm{T}}^{\mathrm{jet}}$, the transverse momentum of the inclusive jets. Top left: $25 < p_{\mathrm{T}}^{\mathrm{jet}} < 50$~\GeV. Top right: $50 < p_{\mathrm{T}}^{\mathrm{jet}} < 100$~\GeV. Middle left: $100 < p_{\mathrm{T}}^{\mathrm{jet}} < 200$~\GeV. Middle right: $200 < p_{\mathrm{T}}^{\mathrm{jet}} < 300$~\GeV. Bottom left: $300 < p_{\mathrm{T}}^{\mathrm{jet}} < 400$~\GeV. Bottom right: $400 < p_{\mathrm{T}}^{\mathrm{jet}} < 1050$~\GeV. The bin-to-bin uncorrelated part of the data uncertainties is shown as black error bars, while the total uncertainties are shown as a yellow band. The cross sections are compared with the predictions computed with the PDFs resulting from the ATLASpdf21 fit. The solid line shows the predictions without shifts of the systematic uncertainties, while for the dashed line the $b_j$ parameters associated with the experimental systematic uncertainties as shown in Eq.~(\ref{eqn:chi2}) are allowed to vary to minimise the $\chi^{2}$. The red band represents the full uncertainty (experimental (evaluated with $T=3$) +~model+~parameterisation) of the fit prediction.
}
\end{centering}
\end{figure*}
\begin{figure*}[t!]
\begin{centering}
\includegraphics[width=0.48\textwidth]{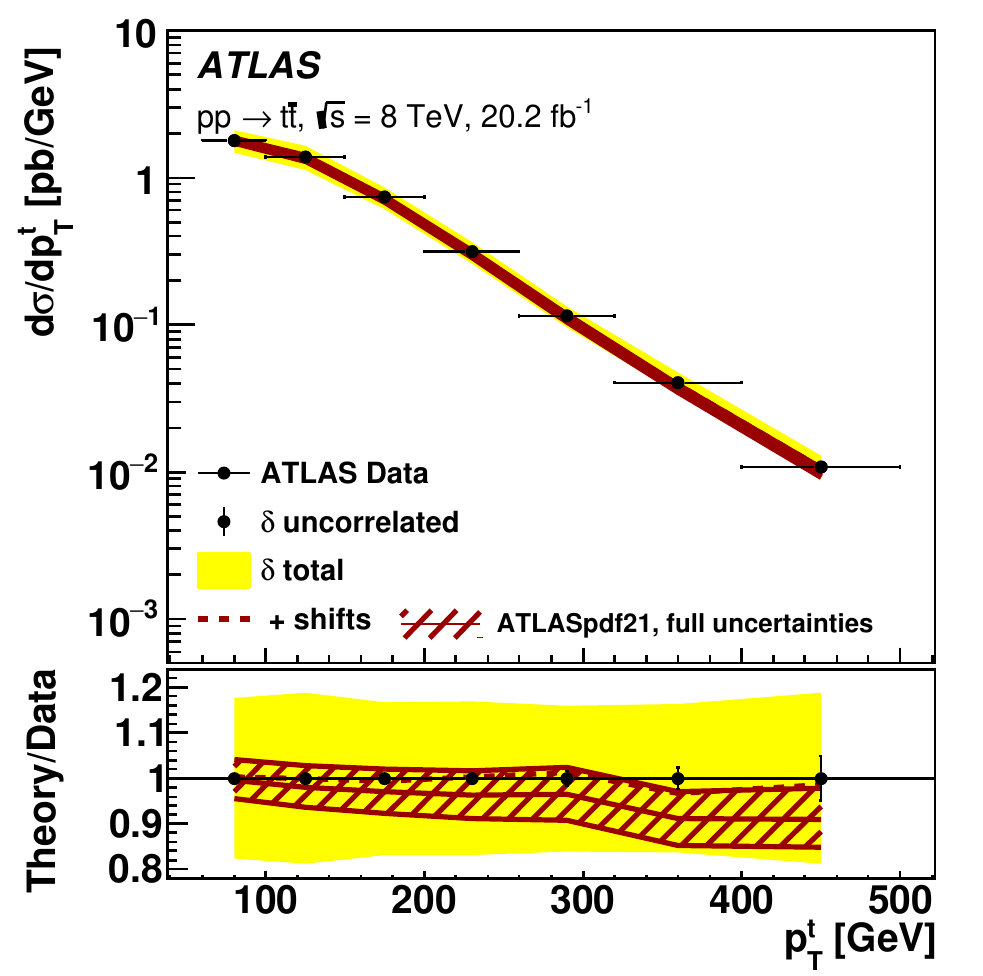}
\includegraphics[width=0.48\textwidth]{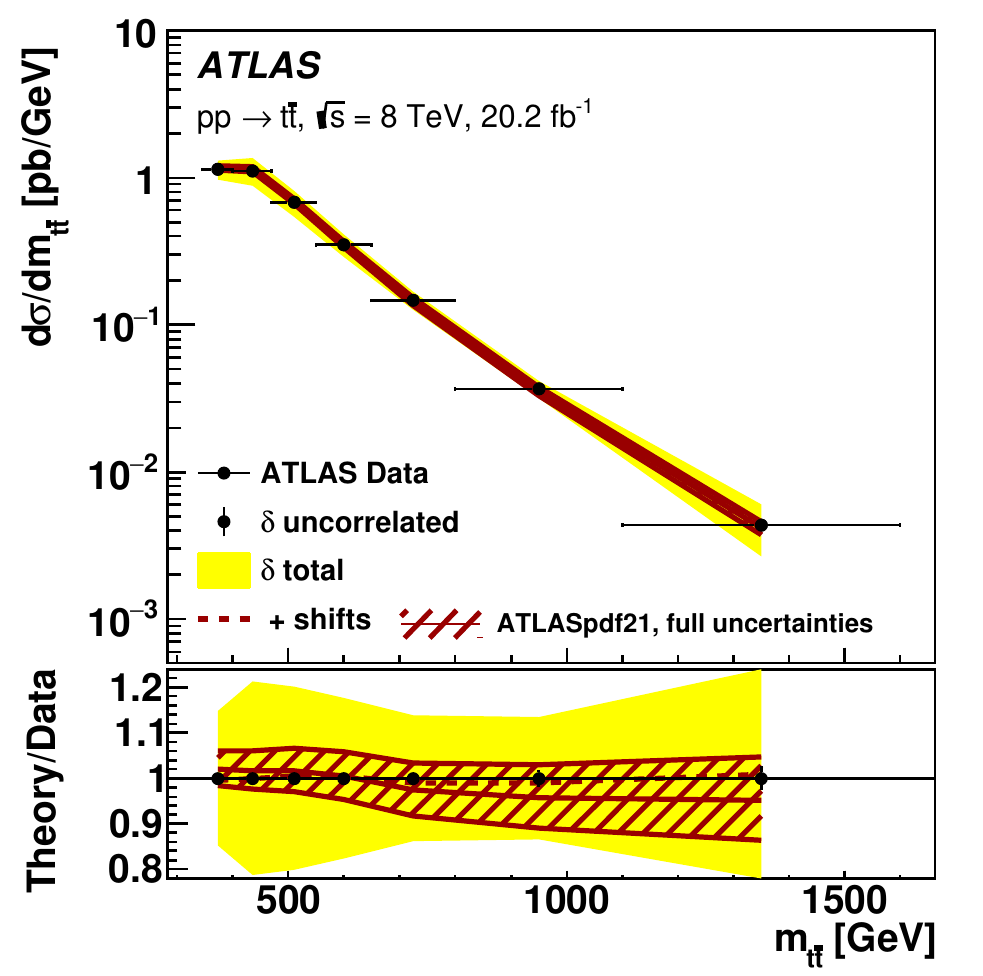}
\includegraphics[width=0.48\textwidth]{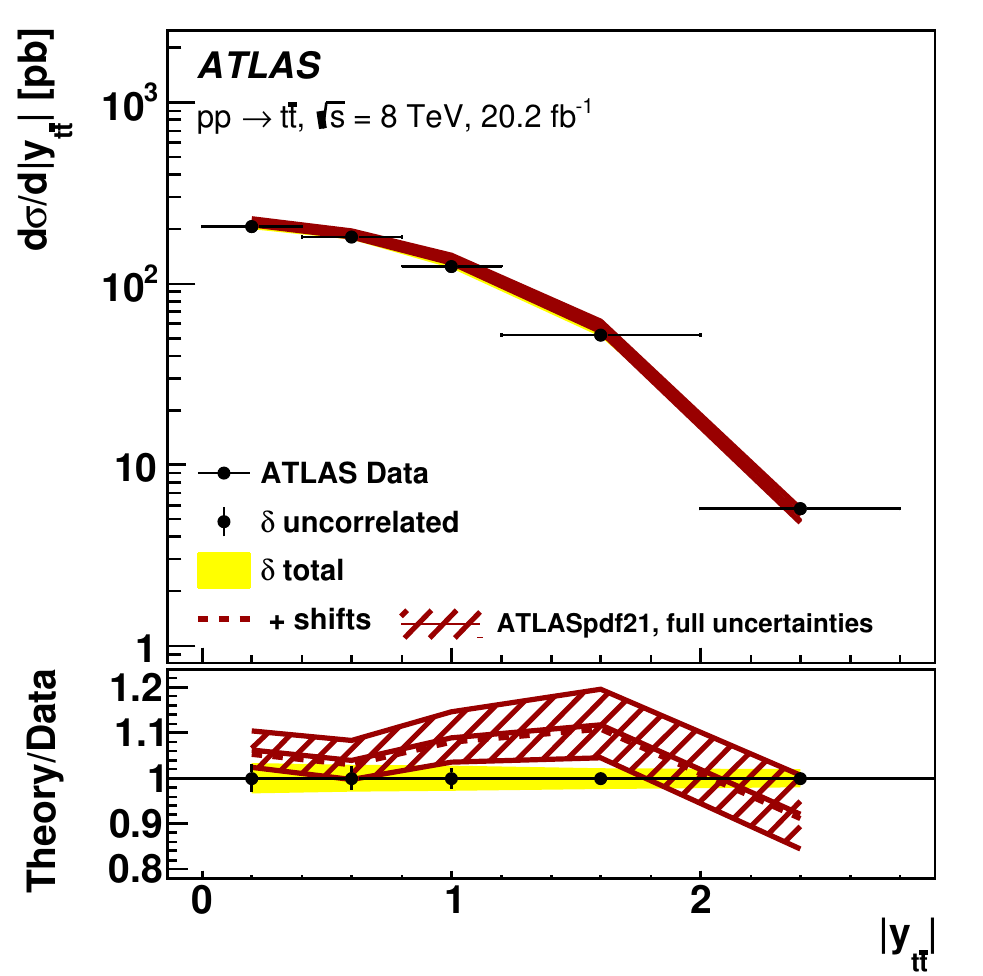}
\caption{ Top: the differential cross-section measurements of $t\bar{t}$ at 8~\TeV\ as a function of (left) the average top momentum, $p_{\mathrm{T}}^{t}$, and (right) the invariant mass of the $t\bar{t}$ system, $m_{t\bar{t}}$, in Ref.~\cite{1511.04716} (black points) in the lepton\,+\,jets decay channel. Bottom: the differential cross-section measurements of $t\bar{t}$ at 8~\TeV\ as a function of the absolute rapidity of the $t\bar{t}$ pair, $|y_{t\bar{t}}|$, in Ref.~\cite{1607.07281} (black points) in the dilepton decay channel. The bin-to-bin uncorrelated part of the data uncertainties is shown as black error bars, while the total uncertainties are shown as a yellow band. The cross sections are compared with the predictions computed with the PDFs resulting from the ATLASpdf21 fit. The solid line shows the predictions without shifts of the systematic uncertainties, while for the dashed line the $b_j$ parameters associated with the experimental systematic uncertainties as shown in Eq.~(\ref{eqn:chi2}) are allowed to vary to minimise the $\chi^{2}$. The red band represents the full uncertainty (experimental (evaluated with $T=3$) +~model +~parameterisation) of the fit prediction.
}
\end{centering}
\end{figure*}
\begin{figure*}
\begin{centering}
\includegraphics[width=0.48\textwidth]{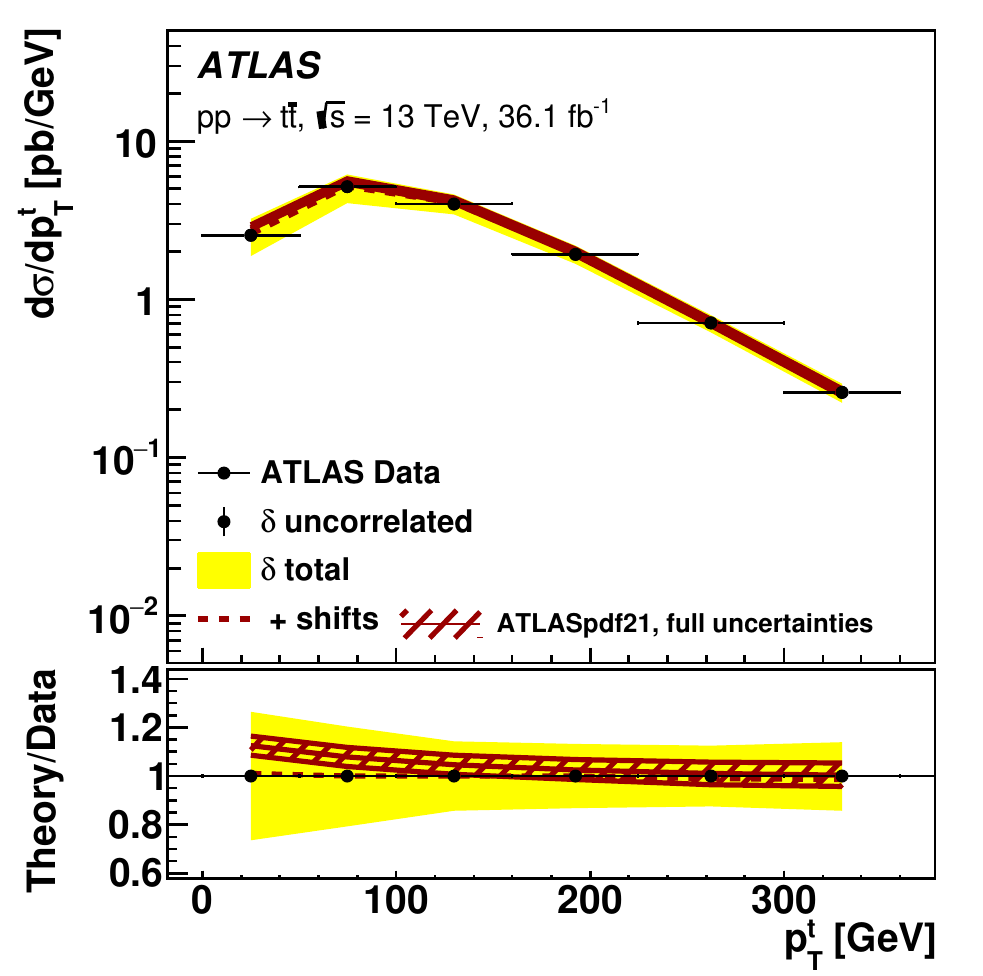}
\includegraphics[width=0.48\textwidth]{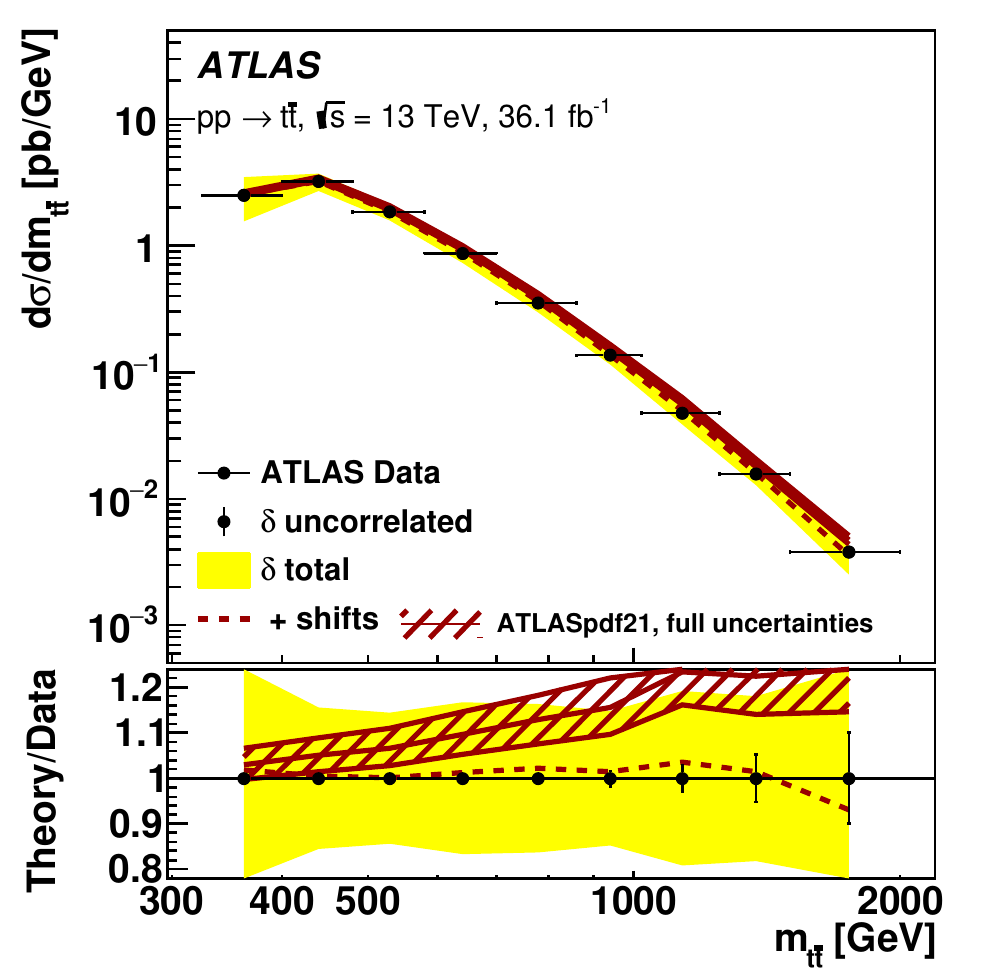}
\includegraphics[width=0.48\textwidth]{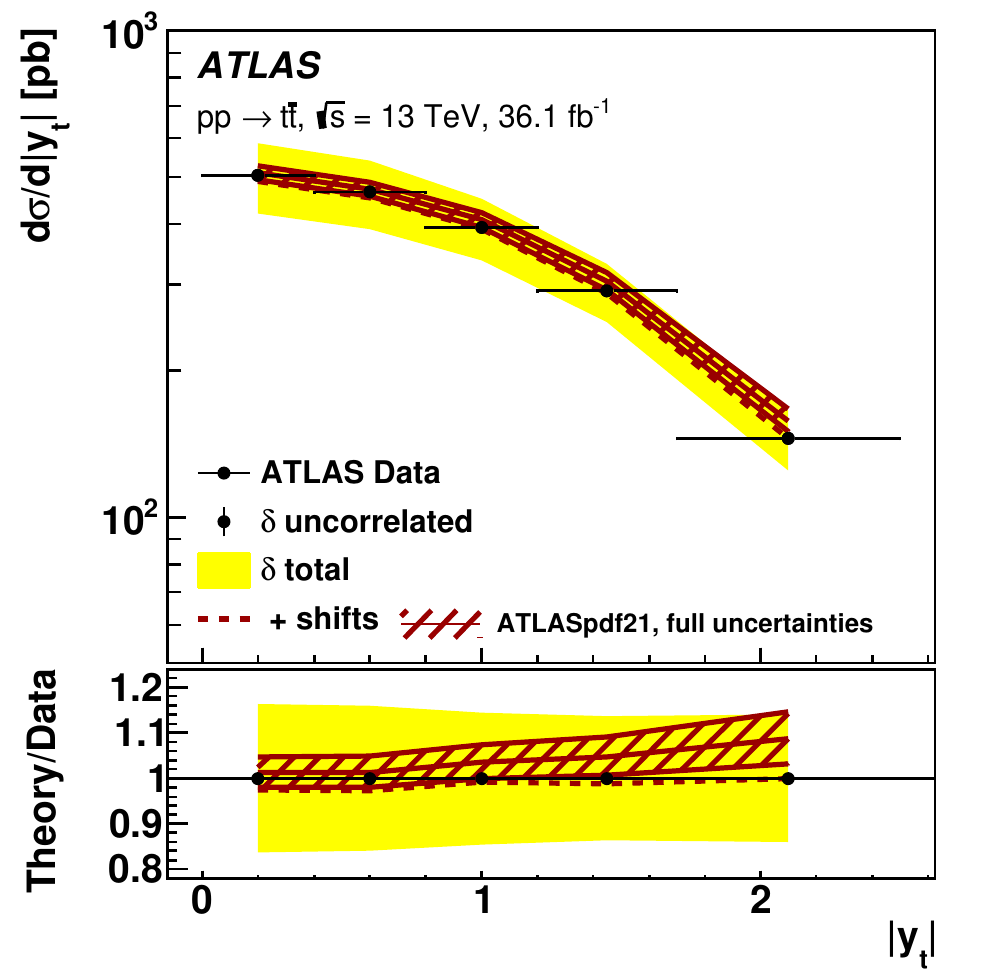}
\includegraphics[width=0.48\textwidth]{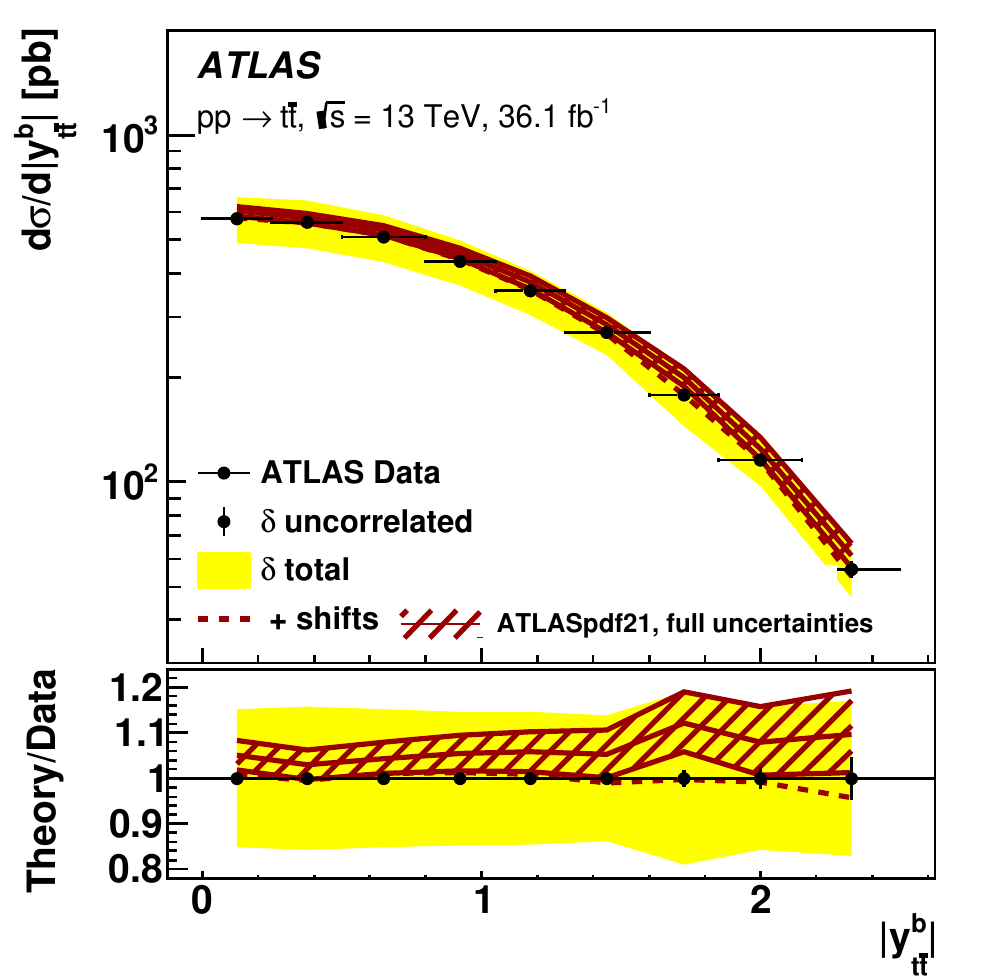}
\caption{The differential cross-section measurements of $t\bar{t}$ at 13~\TeV\ in Ref.~\cite{1908.07305} (black points) as functions of the (top left) average top transverse momentum $p_{\mathrm{T}}^t$, (top right) invariant mass of the $t\bar{t}$ pair $m_{t\bar{t}}$, (bottom left) average top absolute rapidity $|y_t|$ and (bottom right) absolute boosted  rapidity of the $t\bar{t}$ pair, $|y^{\mathrm{b}}_{t\bar{t}}|$. The bin-to-bin uncorrelated part of the data uncertainties is shown as black error bars, while the total uncertainties are shown as a yellow band. The cross sections are compared with the predictions computed with the PDFs resulting from the ATLASpdf21 fit. The solid line shows the predictions without shifts of the systematic uncertainties, while for the dashed line the $b_j$ parameters associated with the experimental systematic uncertainties as shown in Eq.~(\ref{eqn:chi2}) are allowed to vary to minimise the $\chi^{2}$. The red band represents the full uncertainty (experimental (evaluated with $T=3$) +~model +~parameterisation) of the fit prediction.
}
\end{centering}
\end{figure*}
\begin{figure*}
\begin{centering}
\includegraphics[width=0.48\textwidth]{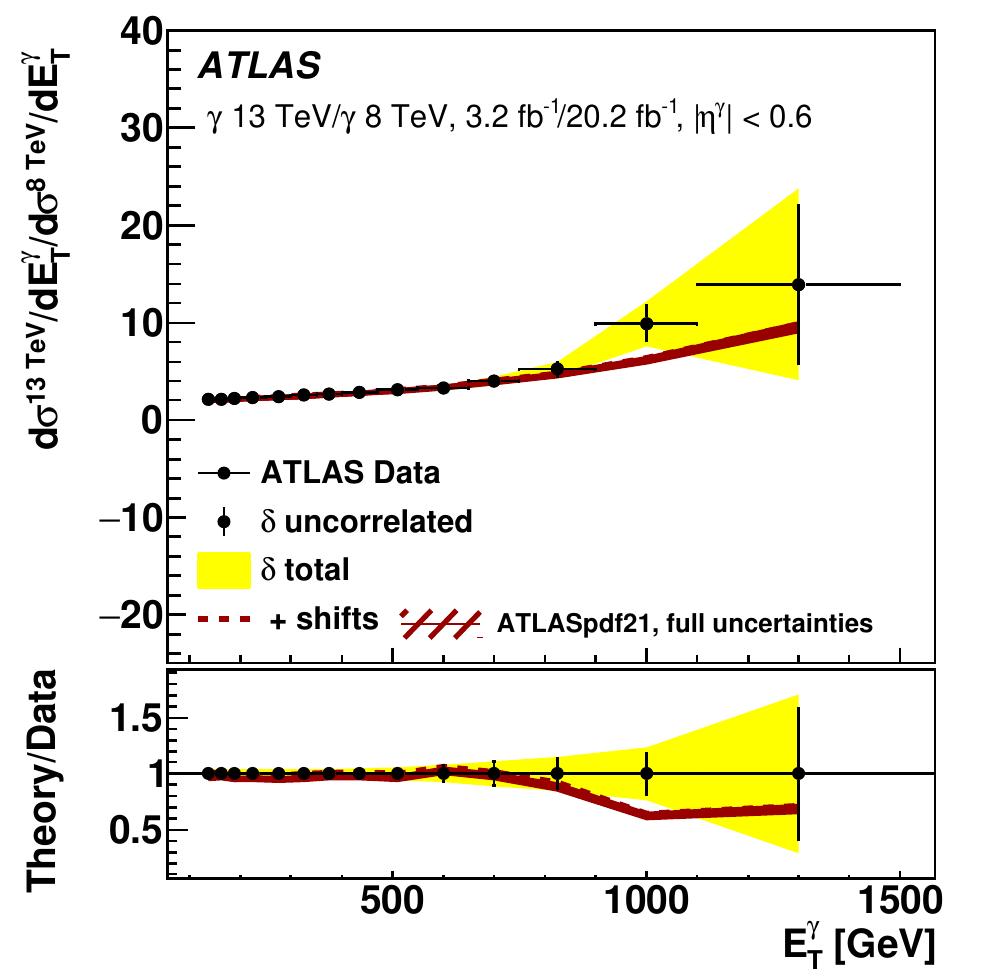}
\includegraphics[width=0.48\textwidth]{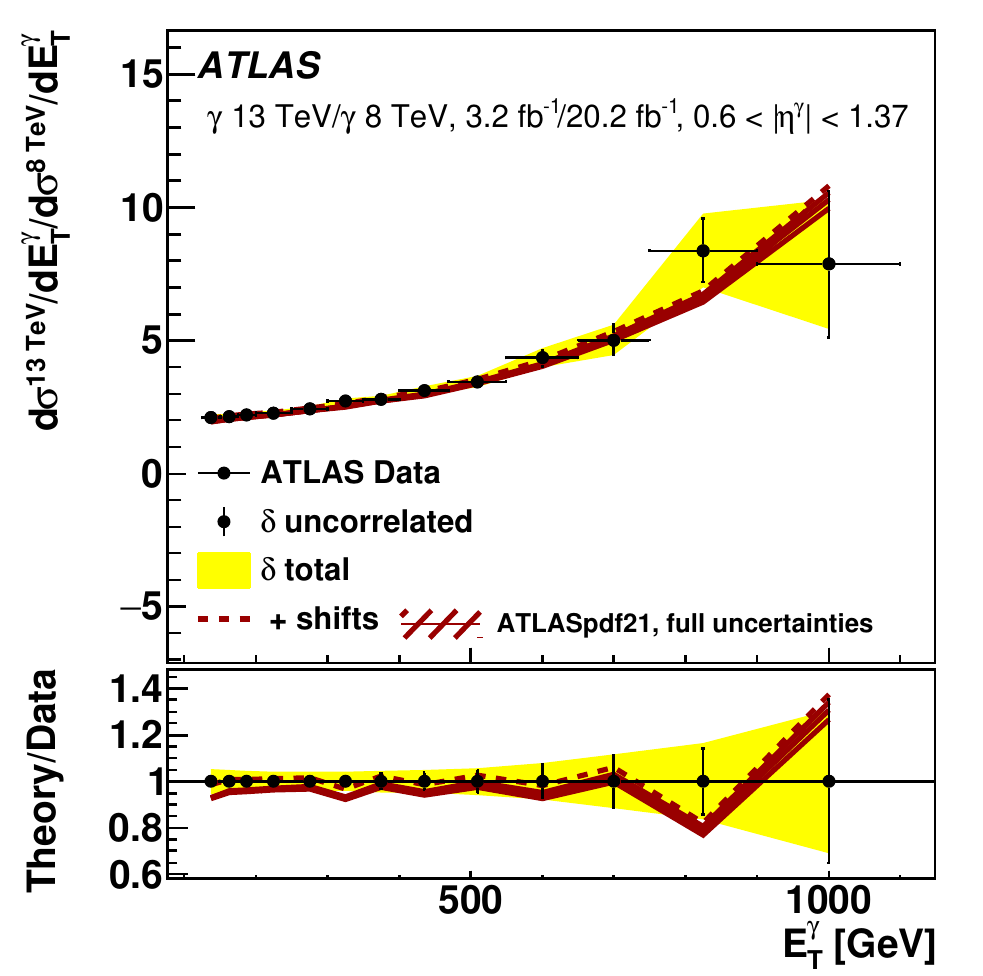}
\includegraphics[width=0.48\textwidth]{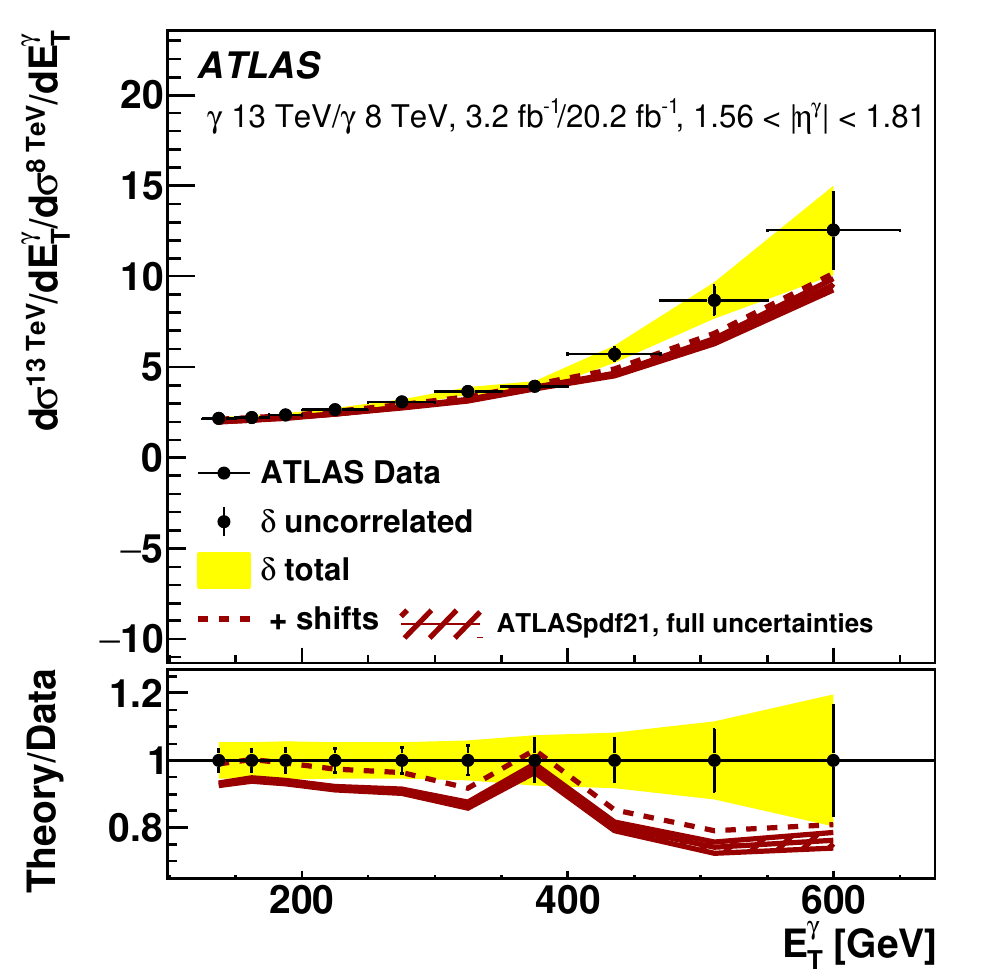}
\includegraphics[width=0.48\textwidth]{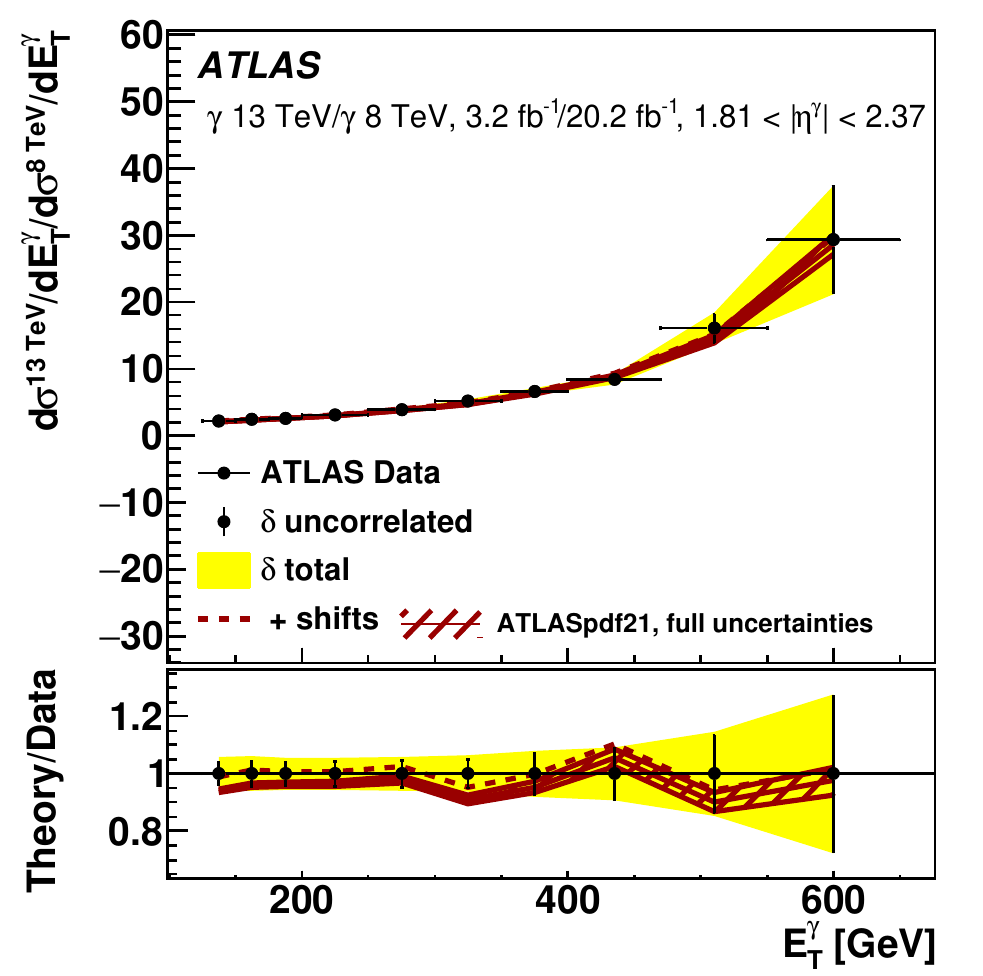}
\caption{The ratios of the cross sections for inclusive isolated-photon production at 8 and 13~\TeV\ in Ref.~\cite{terron} (black points) as functions of the photon transverse energy, $E_{\mathrm{T}}^{\gamma}$, for $E_{\mathrm{T}}^{\gamma}> 125$~\GeV\ in bins of photon absolute pseudorapidity, $|\eta^{\gamma}|$. Top left: $|\eta^{\gamma}| < 0.6$. Top right: $0.6 < |\eta^{\gamma}| < 1.37$. Bottom left: $1.56 < |\eta^{\gamma}| < 1.81$. Bottom right: $1.81 < |\eta^{\gamma}| < 2.37$. The bin-to-bin uncorrelated part of the data uncertainties is shown as black error bars, while the total uncertainties are shown as a yellow band. The cross sections are compared with the predictions computed with the PDFs resulting from the ATLASpdf21 fit. The solid line shows the predictions without shifts of the systematic uncertainties, while for the dashed line the $b_j$ parameters associated with the experimental systematic uncertainties as shown in Eq.~(\ref{eqn:chi2}) are allowed to vary to minimise the $\chi^{2}$. The red band represents the full uncertainty (experimental (evaluated with $T=3$) +~model +~parameterisation) of the fit prediction.
}
\end{centering}
\end{figure*}
\begin{figure*}
\begin{centering}
\includegraphics[width=0.38\textwidth]{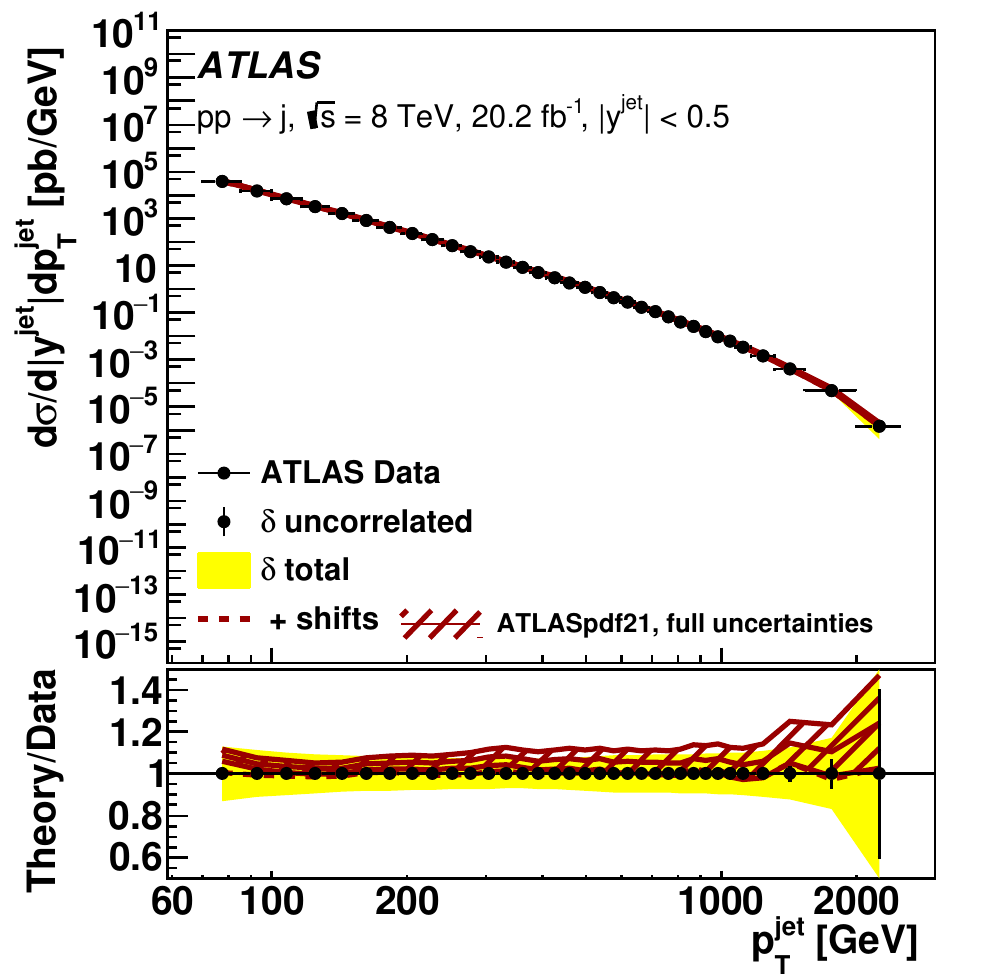}
\includegraphics[width=0.38\textwidth]{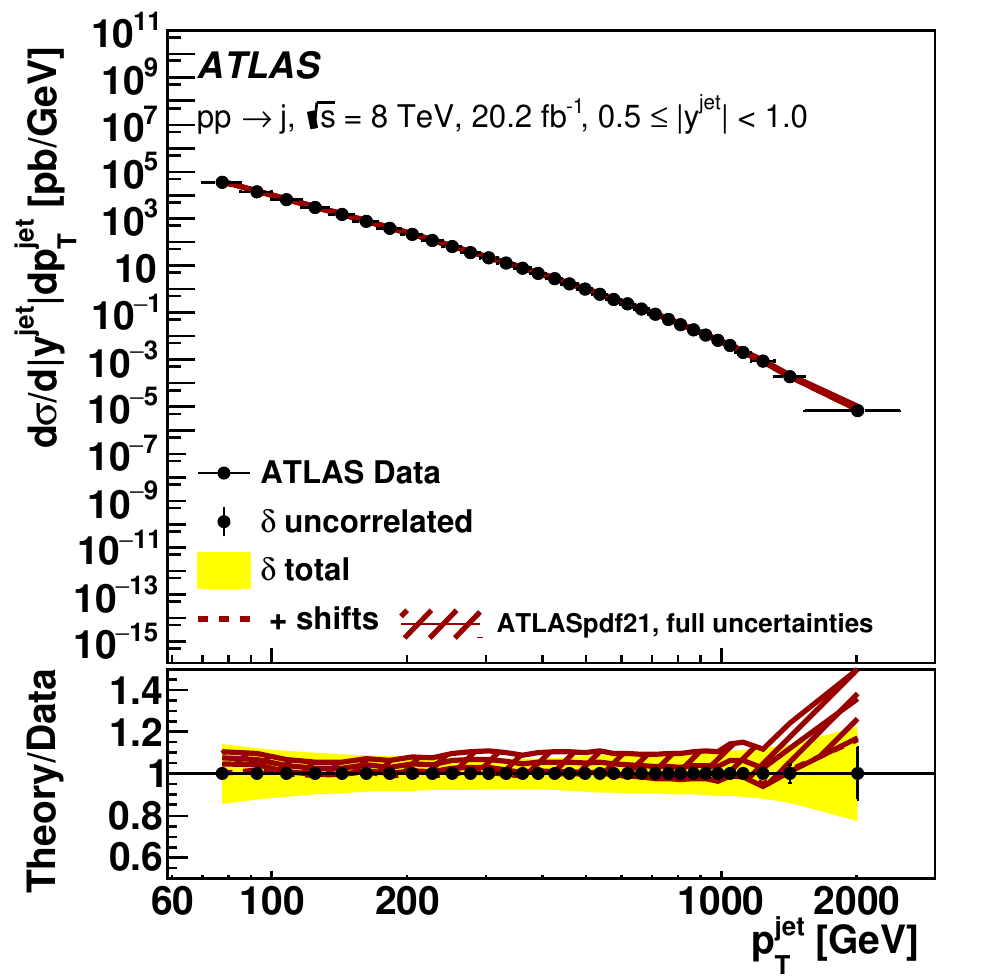}
\includegraphics[width=0.38\textwidth]{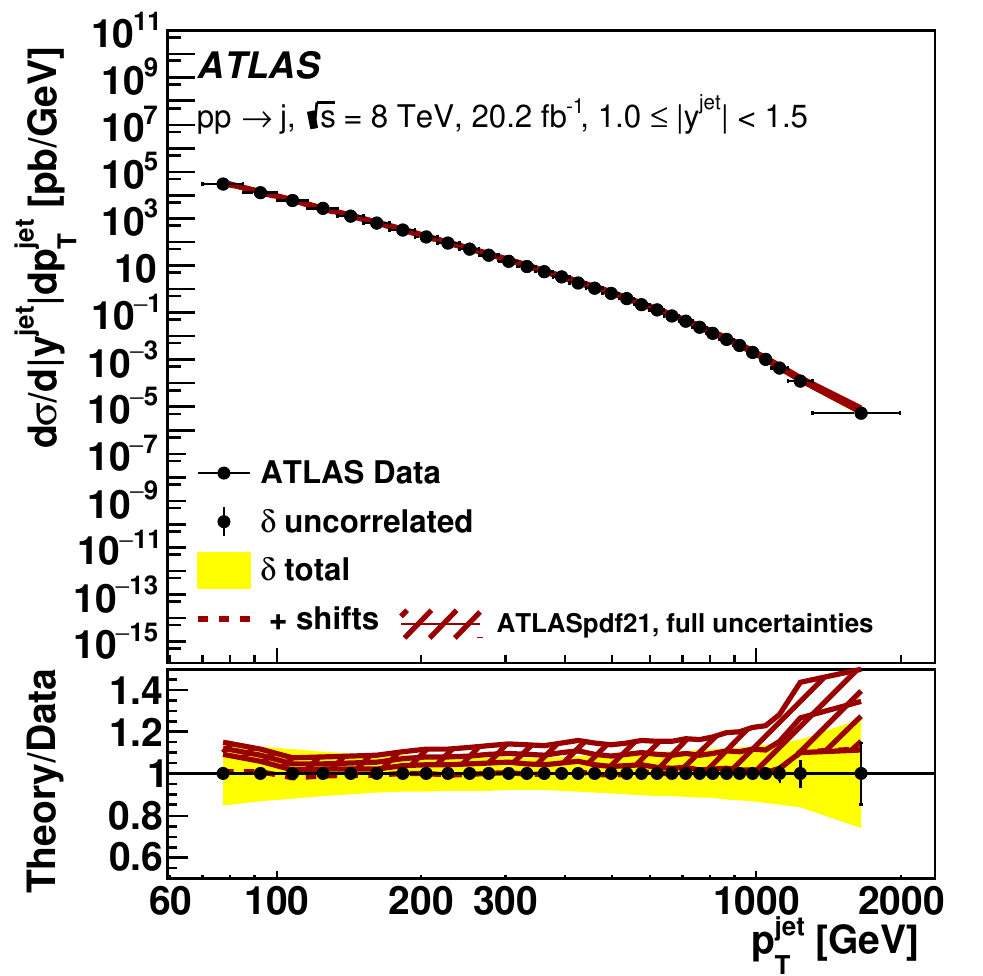}
\includegraphics[width=0.38\textwidth]{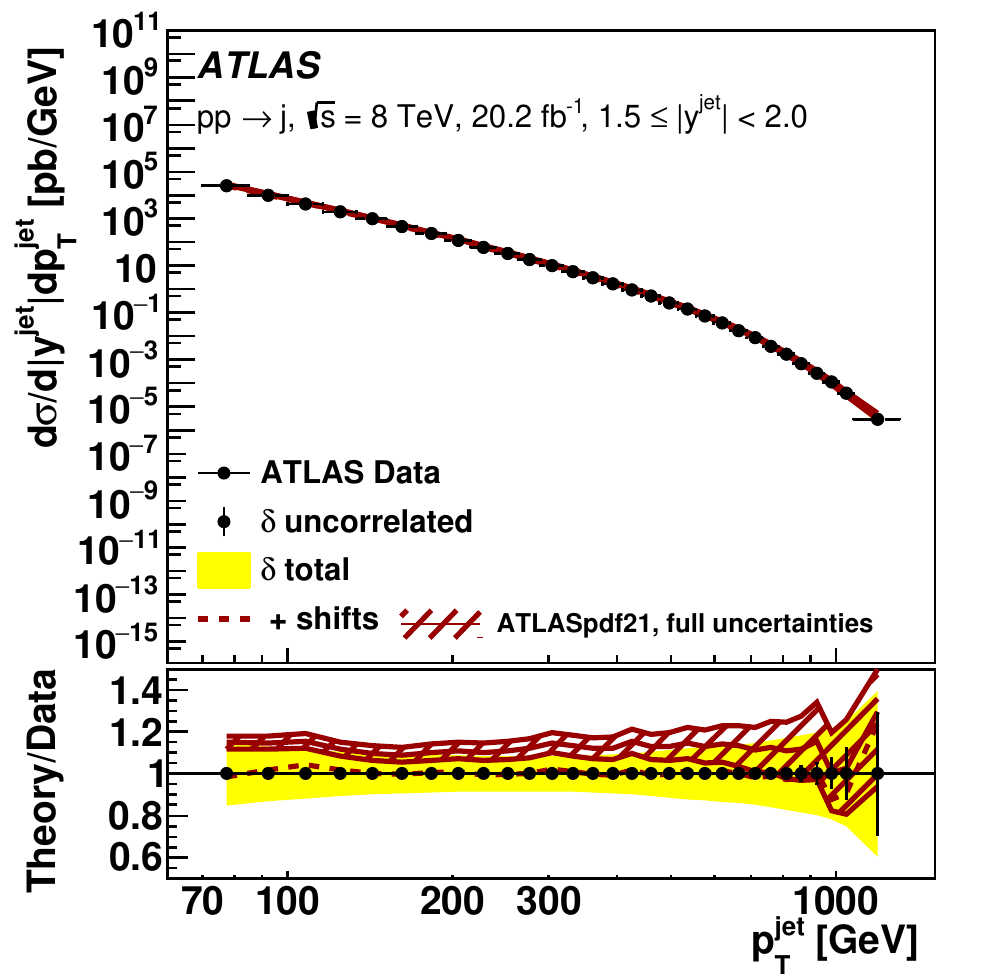}
\includegraphics[width=0.38\textwidth]{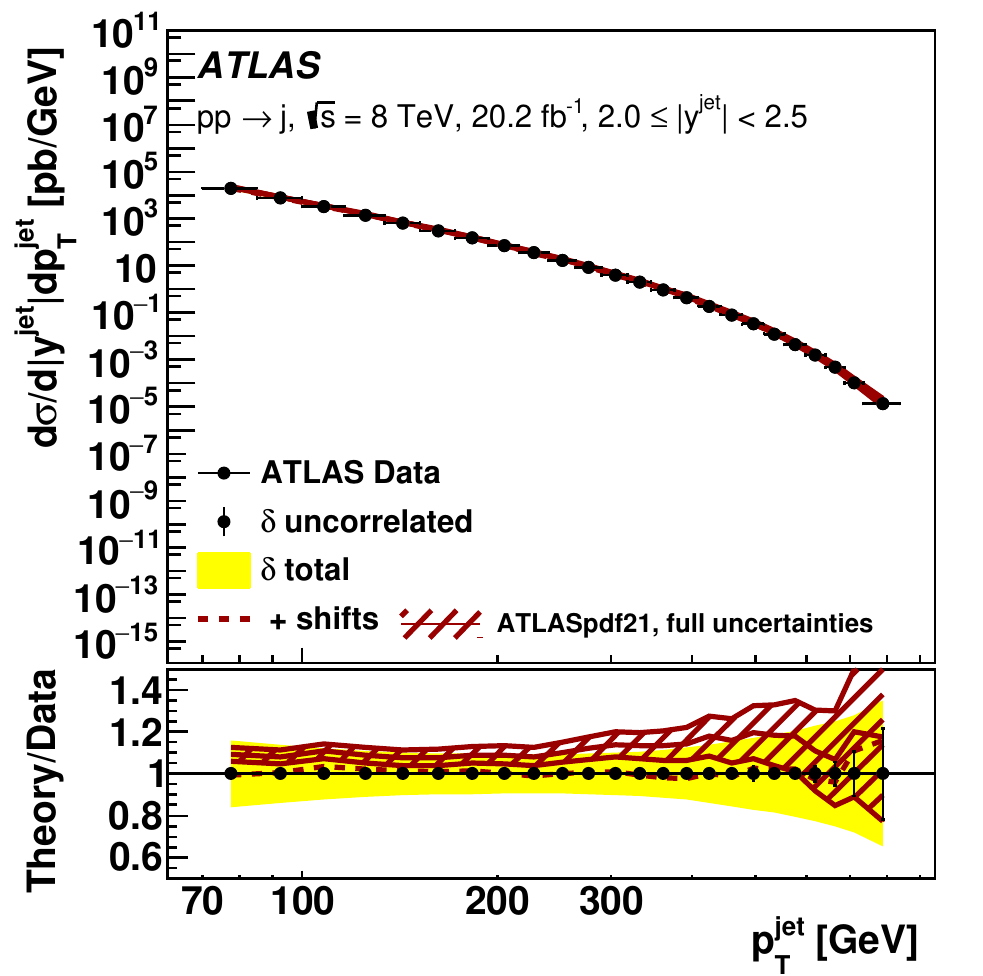}
\includegraphics[width=0.38\textwidth]{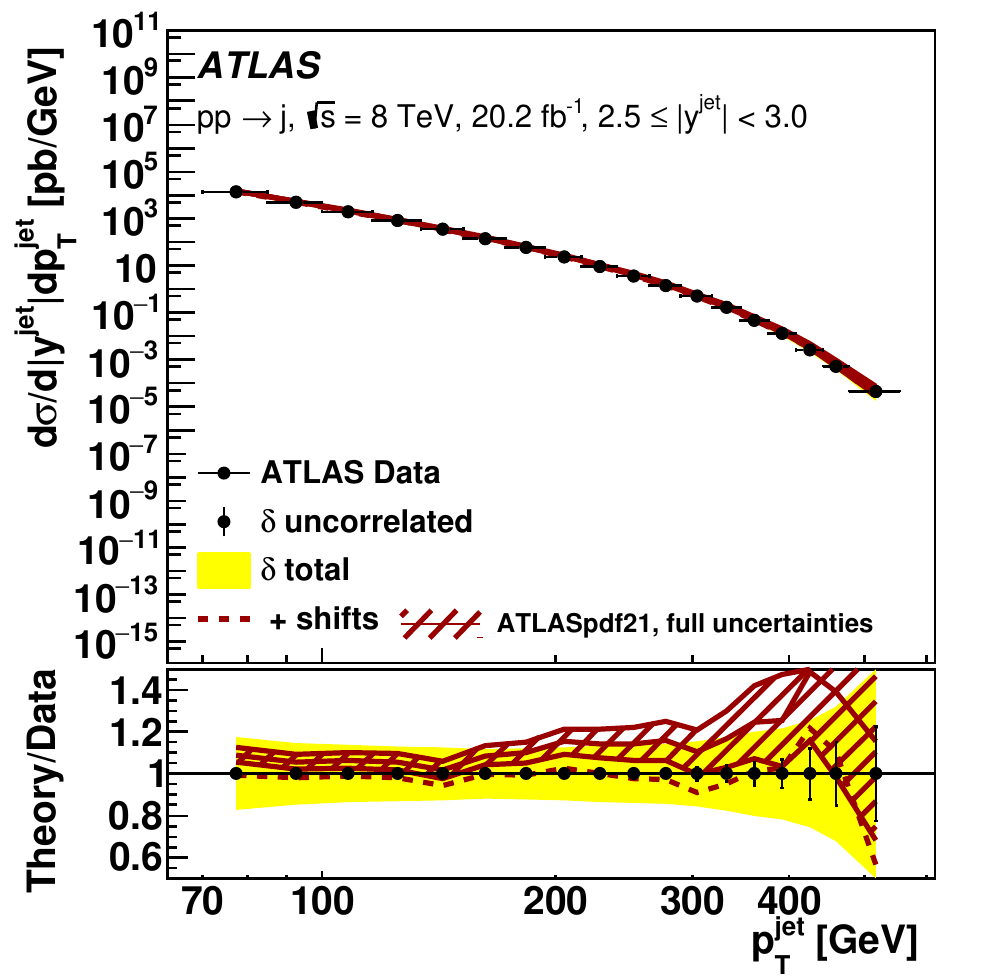}
\caption{ The differential cross-section measurements of inclusive jet production at 8~\TeV\ in Ref.~\cite{1706.03192} (black points), for $R=0.6$, as a function of the jet $p_{\mathrm{T}}^{\mathrm{jet}}$, in six bins of absolute rapidity, $|y^{\mathrm{jet}}|$. Top left: $|y^{\mathrm{jet}} | < 0.5$. Top right: $0.5 < |y^{\mathrm{jet}} | < 1.0$. Middle left: $1.0 < |y^{\mathrm{jet}} | < 1.5$. Middle right: $1.5 < |y^{\mathrm{jet}} | < 2.0$. Bottom left: $2.0 < |y^{\mathrm{jet}} | < 2.5$. Bottom right: $2.5 < |y^{\mathrm{jet}} | < 3.0$. The bin-to-bin uncorrelated part of the data uncertainties is shown as black error bars, while the total uncertainties are shown as a yellow band. The cross sections are compared wth the predictions computed for the scale choice $p_{\mathrm T}^{\mathrm{jet}}$, with the PDFs resulting from the ATLASpdf21 fit. The solid line shows the predictions without shifts of the systematic uncertainties, while for the dashed line the $b_j$ parameters associated with the experimental systematic uncertainties as shown in Eq.~(\ref{eqn:chi2}) are allowed to vary to minimise the $\chi^{2}$. The red band represents the full uncertainty (experimental (evaluated with $T=3$) +~model +~parameterisation) of the fit prediction.
\label{fig:IncJets_8TeVptjet}
}
\end{centering}
\end{figure*}
\clearpage
\section{Comparison with extra data sets not included in the fit}
\label{sec:extradatafitcomp}
 
The ATLASpdf21 fit does not include data from the Tevatron or from fixed-target DY data.
In this appendix the predictions of the ATLASpdf21 fit for some of these data sets, which were found to be most impactful in the global fits, are explored and found to be satisfactory.
 
\begin{figure*}[h!]
\begin{centering}
\includegraphics[width=0.48\textwidth]{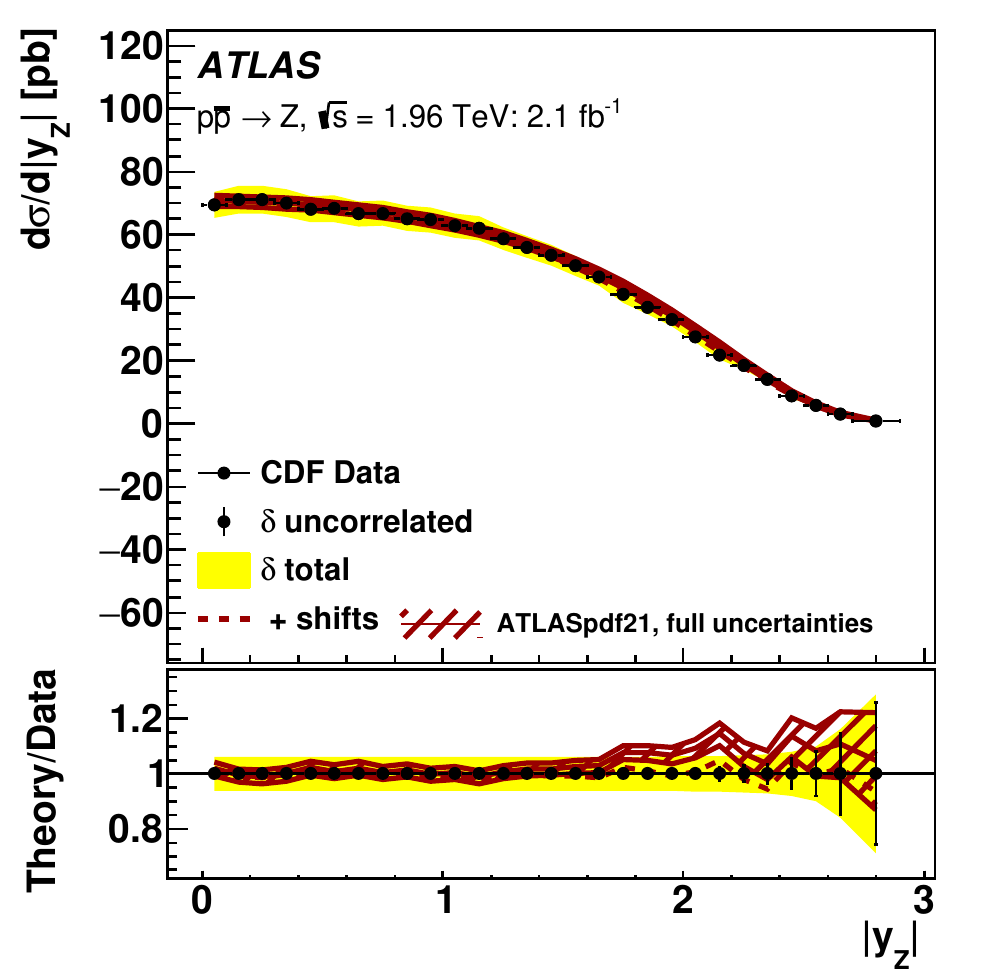}
\includegraphics[width=0.48\textwidth]{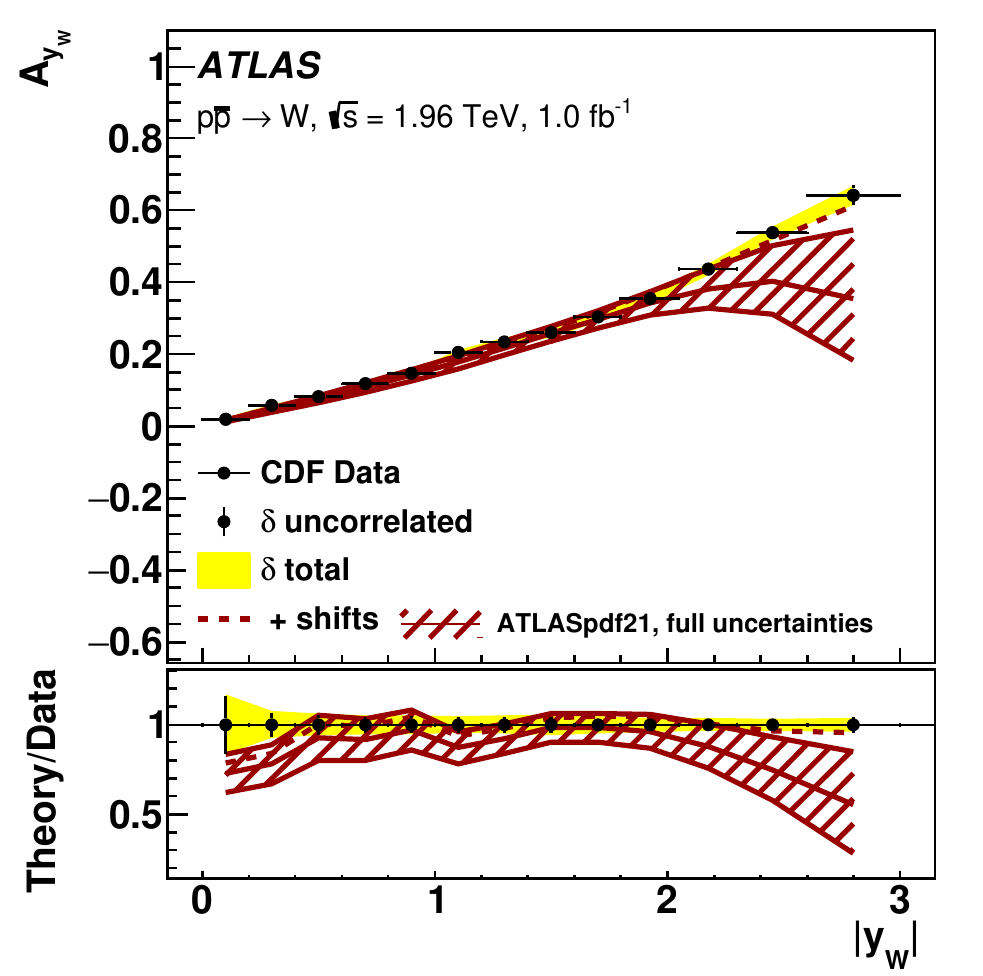}
\caption{ The differential cross-section measurements of (left) $Z$ and (right) $W$ bosons in Refs.~\cite{Aaltonen:2009ta,Aaltonen:2010zza} (black points) as a function of the absolute rapidity of the boson, $|y_{Z}|$ or $|y_{W}|$. The bin-to-bin uncorrelated part of the data uncertainties is shown as black error bars, while the total uncertainties are shown as a yellow band. The cross sections are compared with the predictions computed with the PDFs resulting from the ATLASpdf21 fit. The solid line shows the predictions without shifts of the systematic uncertainties, while for the dashed line the $b_j$ parameters associated with the experimental systematic uncertainties as shown in Eq.~(\ref{eqn:chi2}) are allowed to vary to minimise the $\chi^{2}$. The red band represents the full uncertainty (experimental (evaluated with $T=3$) +~model +~parameterisation) of the fit prediction.
\label{fig:CDF}
}
\end{centering}
\end{figure*}

\begin{figure*}[h!]
\begin{centering}
\includegraphics[width=0.48\textwidth]{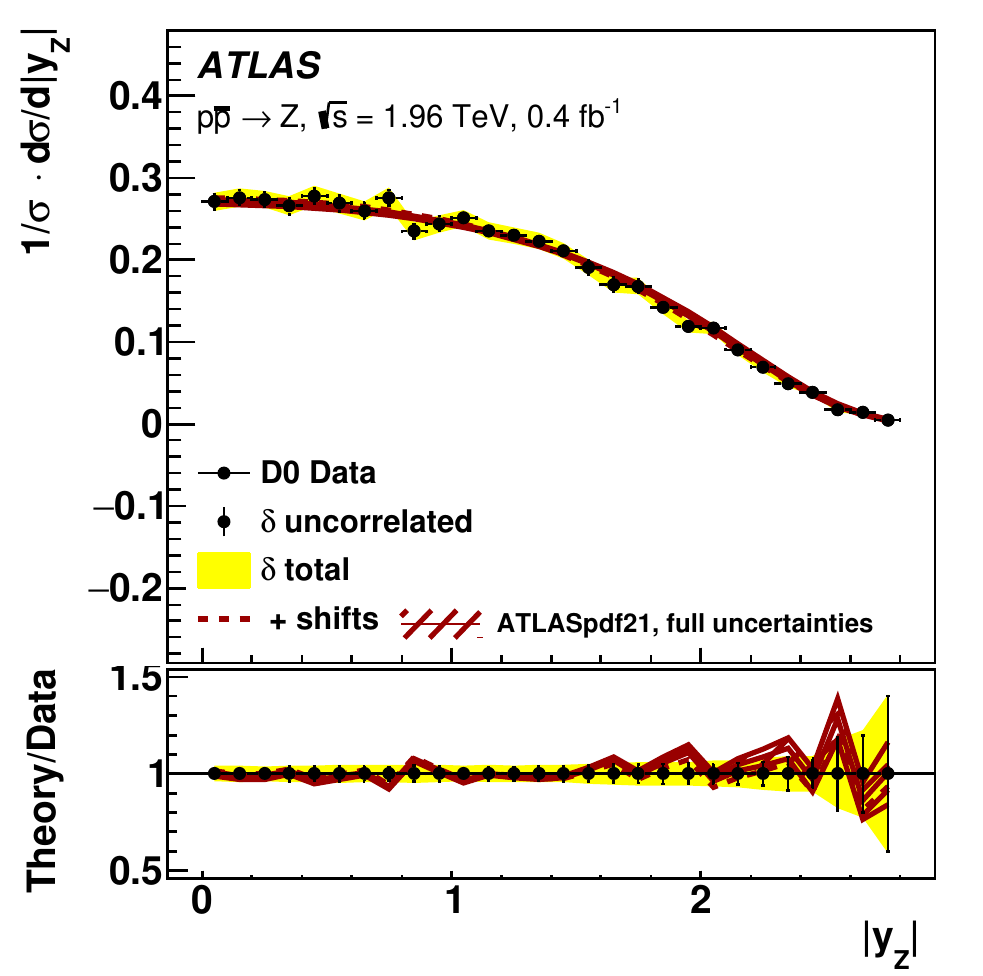}
\includegraphics[width=0.48\textwidth]{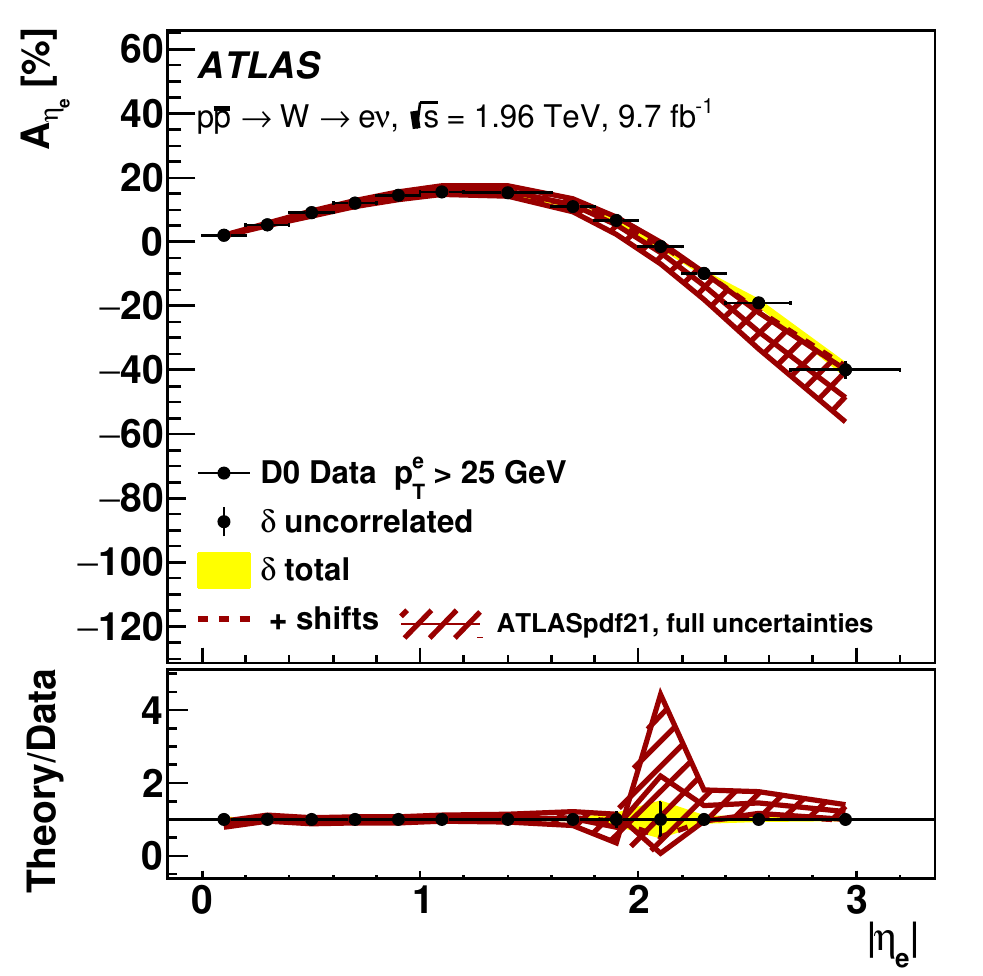}
\includegraphics[width=0.48\textwidth]{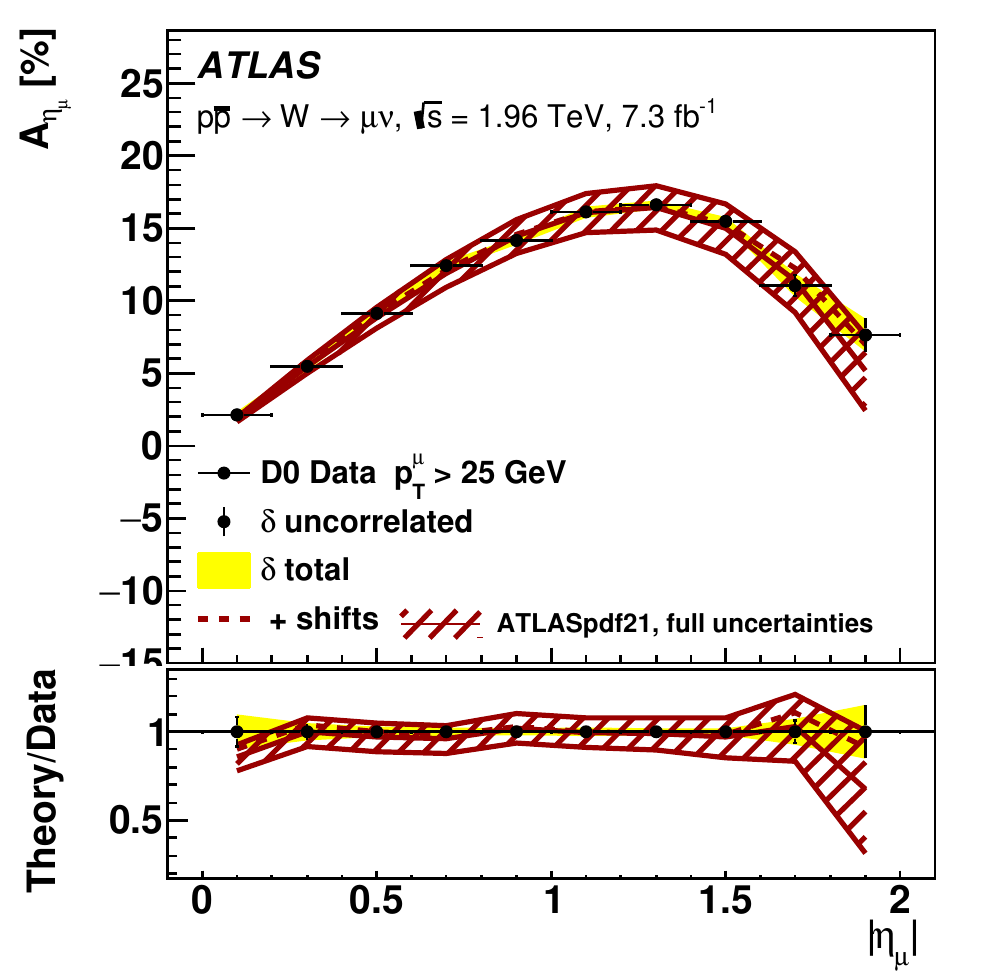}
\caption{ The differential cross-section measurements of (left) $Z$, (right) $W$ (electron decay channel) and (bottom) $W$ (muon decay channel) in Refs.~\cite{Abazov:2007jy,Abazov:2013rja,D0:2014kma} (black points) as a function of the absolute rapidity of the $Z$ boson or absolute pseudorapidity of the decay lepton. The bin-to-bin uncorrelated part of the data uncertainties is shown as black error bars, while the total uncertainties are shown as a yellow band. The cross sections are compared with the predictions computed with the PDFs resulting from the ATLASpdf21 fit. The solid line shows the predictions without shifts of the systematic uncertainties, while for the dashed line the $b_j$ parameters associated with the experimental systematic uncertainties as shown in Eq.~(\ref{eqn:chi2}) are allowed to vary to minimise the $\chi^{2}$. The red band represents the full uncertainty (experimental (evaluated with $T=3$) +~model +~parameterisation) of the fit prediction.
\label{fig:D0}
}
\end{centering}
\end{figure*}

In Figures~\ref{fig:CDF} and~\ref{fig:D0} the ATLASpdf21 fit is shown in comparison with Tevatron $W$ and $Z$ data.
The $\chi^{2}/\mathrm{NDF}$ values for the CDF data are 31/28 for the $Z$ data and 35/13 for the $W$-asymmetry data.
The $\chi^{2}/\mathrm{NDF}$ values for the D0 data are 23/28 for the $Z$ data, 25/13 for the $W$-electron asymmetry data and 13/10
for the $W$-muon asymmetry data. The ATLASpdf21 fit therefore provides a fair description, $\chi^{2}/\mathrm{NDF} = 126/92$, of these Tevatron data, which mostly influence the high-$x$ valence quarks.

\begin{figure*}[h!]
\begin{centering}
\includegraphics[width=0.48\textwidth]{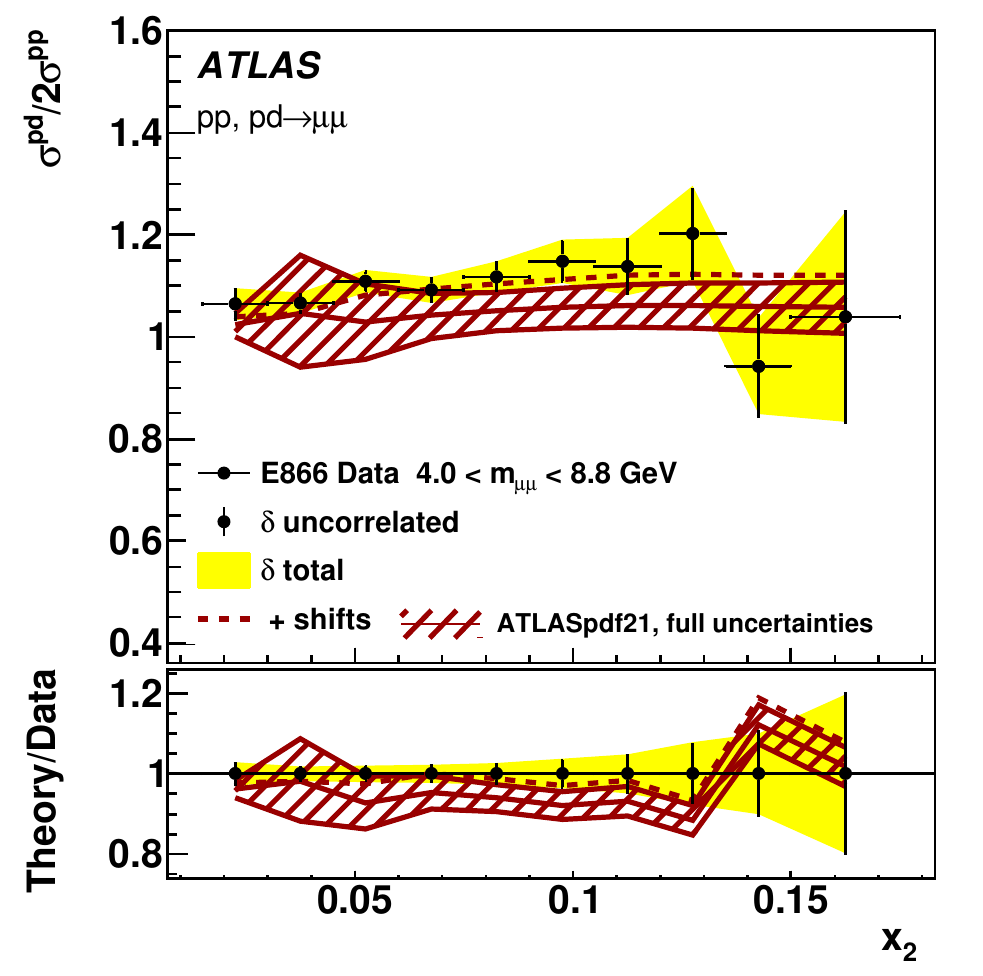}
\includegraphics[width=0.48\textwidth]{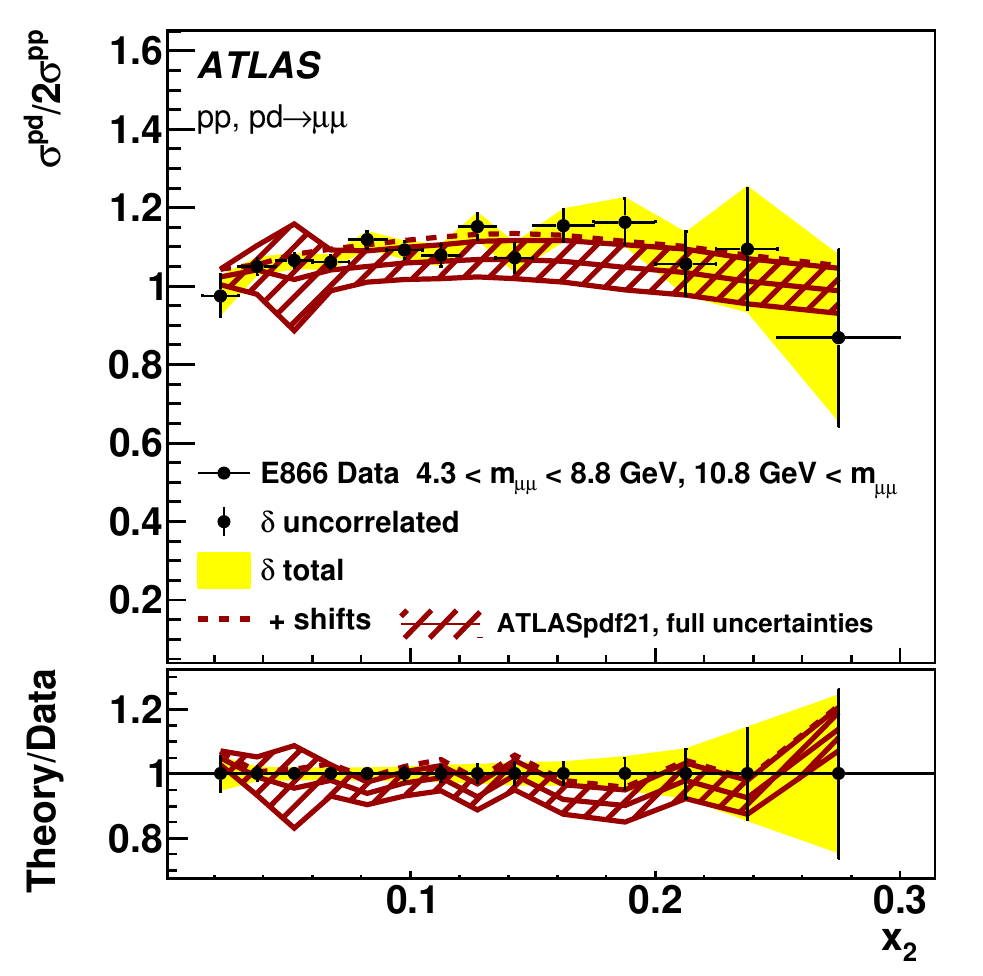}
\includegraphics[width=0.48\textwidth]{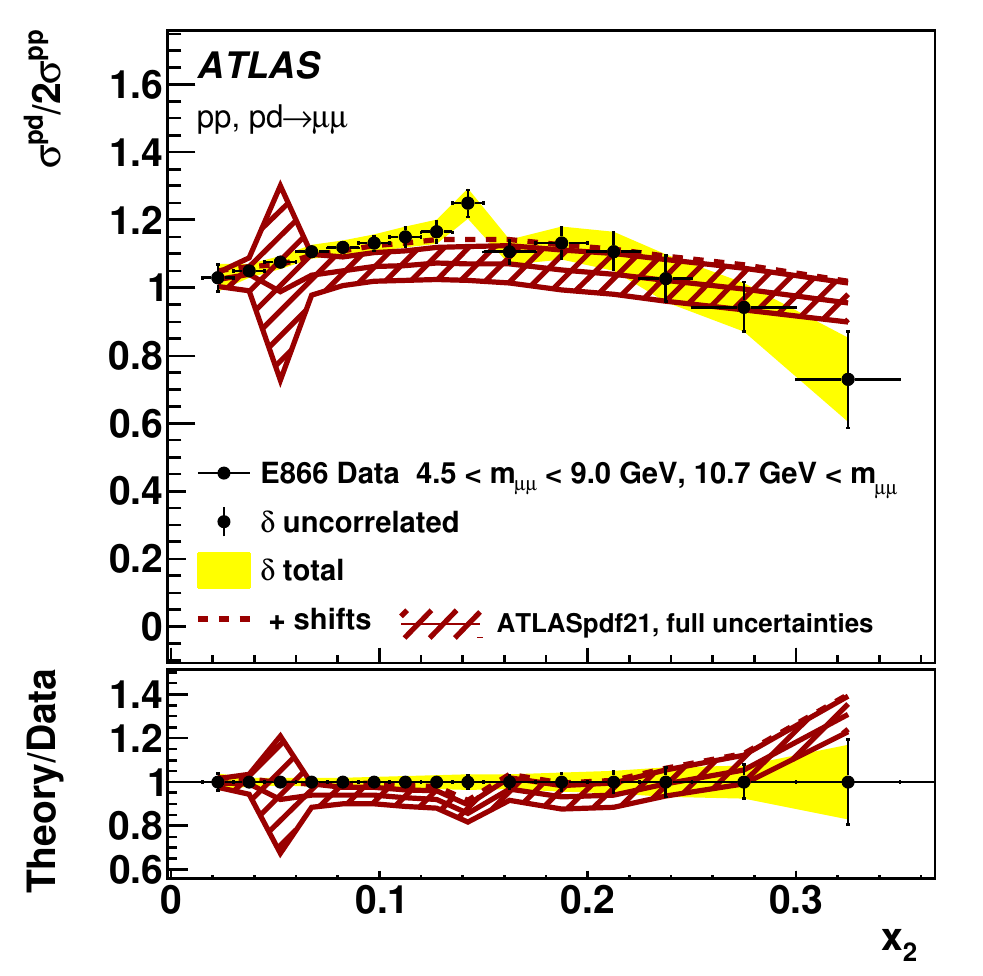}
\caption{ $\sigma^{pD}/2\sigma^{pp}$ from Ref.~\cite{Towell:2001nh} (black points) in the (top left) low-mass, (top right) intermediate-mass and (bottom) high-mass regions as a function of $x_{2}$. The bin-to-bin uncorrelated part of the data uncertainties is shown as black error bars, while the total uncertainties are shown as a yellow band. The cross sections are compared with the predictions computed with the PDFs resulting from the ATLASpdf21 fit. The solid line shows the predictions without shifts of the systematic uncertainties, while for the dashed line the $b_j$ parameters associated with the experimental systematic uncertainties as shown in Eq.~(\ref{eqn:chi2}) are allowed to vary to minimise the $\chi^{2}$. The red band represents the full uncertainty (experimental (evaluated with $T=3$) +~model +~parameterisation) of the fit prediction.
\label{fig:E866}
}
\end{centering}
\end{figure*}

\begin{figure*}[h]
\begin{centering}
\includegraphics[width=0.9\textwidth]{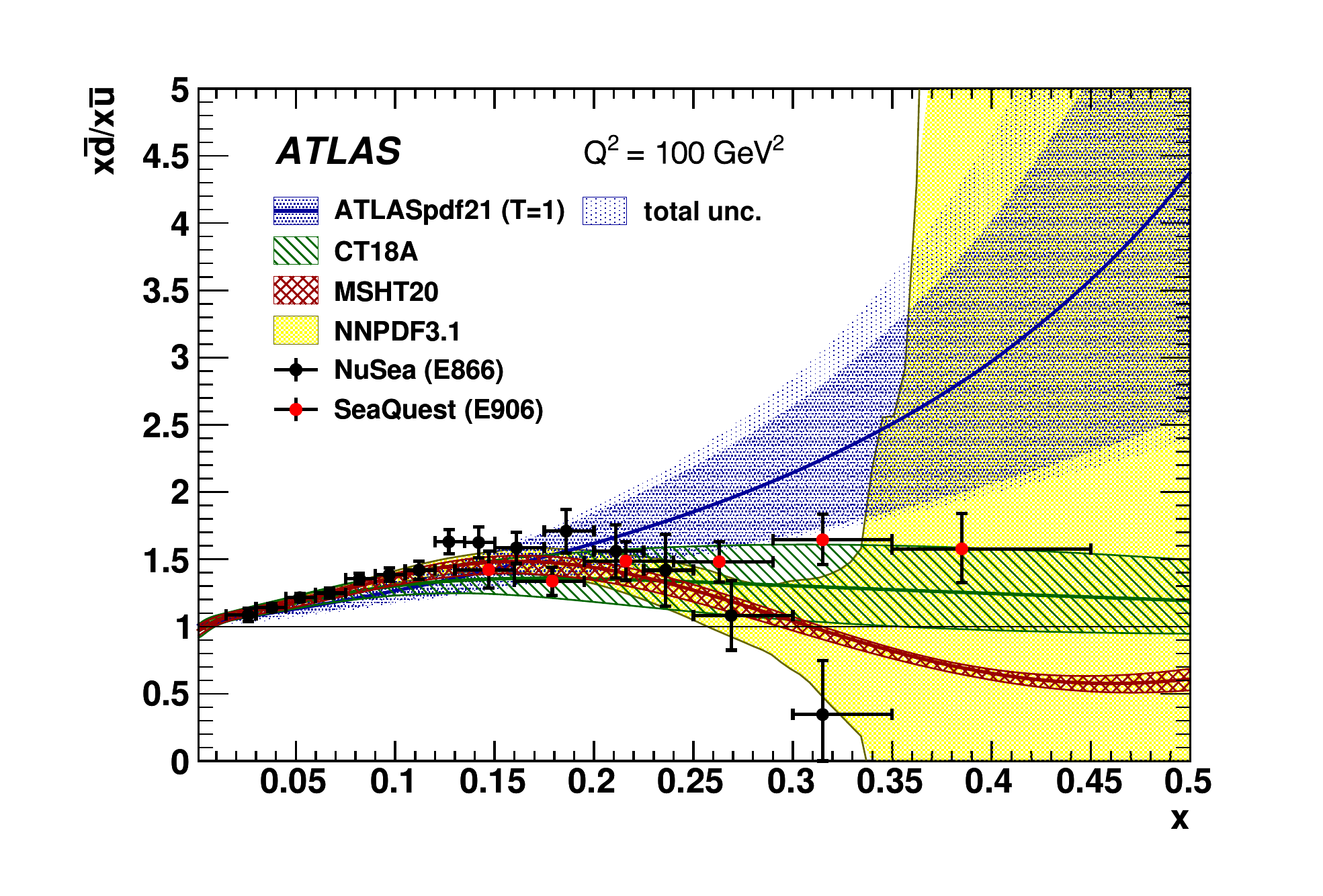}
\caption{A comparison of the $x\bar{d}/x\bar{u}$ ratios extracted from E866~\cite{Towell:2001nh} and E906~\cite{Dove:2021ejl}, together with the predictions of the ATLASpdf21, CT18A, MSHT20 and NNPDF3.1 PDFs.
\label{fig:E906}
}
\end{centering}
\end{figure*}

It is also interesting to consider the description of the fixed-target Drell--Yan data from E866 and E906, since
these are uniquely able to constrain the difference $x(\bar{d}-\bar{u})$ at high~$x$.
Figure~\ref{fig:E866} compares
the predictions of the ATLASpdf21 fit with the E866 $pD/pp$ data~\cite{Towell:2001nh} in three mass regions.
The $\chi^{2}/\mathrm{NDF}$ values are 9.6/10, 14/14 and 21/15 for the low-, intermediate- and
high-mass regions, respectively.  Thus, the description of the E866 data, which mostly give
information about high-$x$ sea quarks, is good. However, the high-mass region, which covers
larger Bjorken-$x$, is not as well fitted as the lower-mass regions. The ATLASpdf21 fit
is in better agreement with the new data from E906~\cite{Dove:2021ejl} in the high-mass region, as
seen in Figure~\ref{fig:E906}, which compares ATLASpdf21 and other PDFs with the
$x\bar{d}/x\bar{u}$ ratios extracted from E866 and E906.

 
\clearpage
\printbibliography

\clearpage
 
\begin{flushleft}
\hypersetup{urlcolor=black}
{\Large The ATLAS Collaboration}

\bigskip

\AtlasOrcid[0000-0002-6665-4934]{G.~Aad}$^\textrm{\scriptsize 99}$,    
\AtlasOrcid[0000-0002-5888-2734]{B.~Abbott}$^\textrm{\scriptsize 125}$,    
\AtlasOrcid[0000-0002-7248-3203]{D.C.~Abbott}$^\textrm{\scriptsize 100}$,    
\AtlasOrcid[0000-0002-2788-3822]{A.~Abed~Abud}$^\textrm{\scriptsize 35}$,    
\AtlasOrcid[0000-0002-1002-1652]{K.~Abeling}$^\textrm{\scriptsize 52}$,    
\AtlasOrcid[0000-0002-2987-4006]{D.K.~Abhayasinghe}$^\textrm{\scriptsize 92}$,    
\AtlasOrcid[0000-0002-8496-9294]{S.H.~Abidi}$^\textrm{\scriptsize 28}$,    
\AtlasOrcid[0000-0002-9987-2292]{A.~Aboulhorma}$^\textrm{\scriptsize 34e}$,    
\AtlasOrcid[0000-0001-5329-6640]{H.~Abramowicz}$^\textrm{\scriptsize 158}$,    
\AtlasOrcid[0000-0002-1599-2896]{H.~Abreu}$^\textrm{\scriptsize 157}$
% ...
\bigskip
\\

% ...
$^{28}$Physics Department, Brookhaven National Laboratory, Upton NY; United States of America.\\
% ...
$^{34}$$^{(a)}$Facult\'e des Sciences Ain Chock, R\'eseau Universitaire de Physique des Hautes Energies - Universit\'e Hassan II, Casablanca;$^{(b)}$Facult\'{e} des Sciences, Universit\'{e} Ibn-Tofail, K\'{e}nitra;$^{(c)}$Facult\'e des Sciences Semlalia, Universit\'e Cadi Ayyad, LPHEA-Marrakech;$^{(d)}$LPMR, Facult\'e des Sciences, Universit\'e Mohamed Premier, Oujda;$^{(e)}$Facult\'e des sciences, Universit\'e Mohammed V, Rabat;$^{(f)}$Mohammed VI Polytechnic University, Ben Guerir; Morocco.\\
$^{35}$CERN, Geneva; Switzerland.\\
% ...
$^{52}$II. Physikalisches Institut, Georg-August-Universit\"{a}t G\"ottingen, G\"ottingen; Germany.\\
% ...
$^{92}$Department of Physics, Royal Holloway University of London, Egham; United Kingdom.\\
% ...
$^{99}$CPPM, Aix-Marseille Universit\'e, CNRS/IN2P3, Marseille; France.\\
$^{100}$Department of Physics, University of Massachusetts, Amherst MA; United States of America.\\
% ...
$^{125}$Homer L. Dodge Department of Physics and Astronomy, University of Oklahoma, Norman OK; United States of America.\\
% ...
$^{157}$Department of Physics, Technion, Israel Institute of Technology, Haifa; Israel.\\
$^{158}$Raymond and Beverly Sackler School of Physics and Astronomy, Tel Aviv University, Tel Aviv; Israel.\\
% ...
$^{a}$ Also at Borough of Manhattan Community College, City University of New York, New York NY; United States of America.\\
$^{b}$ Also at Bruno Kessler Foundation, Trento; Italy.\\
$^{c}$ Also at Center for High Energy Physics, Peking University; China.\\
$^{d}$ Also at Centro Studi e Ricerche Enrico Fermi; Italy.\\
$^{e}$ Also at CERN, Geneva; Switzerland.\\
$^{f}$ Also at D\'epartement de Physique Nucl\'eaire et Corpusculaire, Universit\'e de Gen\`eve, Gen\`eve; Switzerland.\\
% ...
$^{*}$ Deceased

\end{flushleft}


\end{document}